\documentclass[12pt]{article}

\usepackage[a4paper, left=20mm, top=20mm, bottom=20mm, right=20mm]{geometry}
\usepackage{hyperref}
\usepackage{physics} 
\usepackage[noadjust]{cite}
\usepackage{siunitx}
\usepackage{changepage}
\usepackage{enumerate} 
\usepackage{pgfplots}
\usepackage{pgfplotstable}
\usepackage{tikz}
\usepackage{amssymb}
\usepackage{lettrine}
\usepackage{makecell}
\usepackage{twoopt}
\usepackage{footnote}
\usepackage{refcount}
\usepackage{tablefootnote}
\usepackage{longtable}
\usepackage{caption}
\usepackage{titlesec}
\usepackage{numprint}
\titleformat{\paragraph}
{\normalfont\normalsize\bfseries}{\theparagraph}{1em}{}
\titlespacing*{\paragraph}
{0pt}{3.25ex plus 1ex minus .2ex}{1.5ex plus .2ex}
\makesavenoteenv{tabular}

\newcommand{\pa}{\partial}

\newcommandtwoopt{\twographs}[4][][]{
    \begin{figure}[!h]
  \centering
  \begin{tabular}{ m{8cm} m{8cm} }
    \begin{minipage}[!ht]{8cm}
      \begin{center}
      \includegraphics[width=1.0\linewidth]{#3.png}
      \end{center}
    \end{minipage}
    &
    \begin{minipage}[!ht]{8cm}
      \begin{center}
      \includegraphics[width=1.0\linewidth]{#4.png}
      \end{center}
    \end{minipage}
    \\
    \begin{minipage}[!ht]{8cm}
    \begin{center}
      $(a)$
    \end{center}
    \end{minipage}
    &
    \begin{minipage}[!ht]{8cm}
      \begin{center}
      $(b)$
      \end{center}
    \end{minipage}
    \\
  \end{tabular}
  \caption{#1}
  \label{fig:#2}
\end{figure}
}

\newcommand{\twographsintable}[2]{
    \begin{minipage}[!ht]{8cm}
      \begin{center}
      \includegraphics[width=1.0\linewidth]{#1.png}
      \end{center}
    \end{minipage}
    &
    \begin{minipage}[!ht]{8cm}
      \begin{center}
      \includegraphics[width=1.0\linewidth]{#2.png}
      \end{center}
    \end{minipage}
    \\
}

\def\tablefootnotemark#1{\textsuperscript{\getrefnumber{#1}}}

\pgfplotsset{compat=1.14}

\usepackage{tikz} 
\usepackage{amsmath}
\usepackage{graphicx}

\newcommand{\hishtetraquark}[5][0.5]{
\begin{center}
	\begin{tikzpicture}[scale=#1]
    \coordinate [label=left:$$] (O) at (0,0);
    \coordinate [label=left:$$] (A) at (-2,0);
    \coordinate [label=right:$$] (B) at (2,0);
    \coordinate [label=right:{#2}] (C) at (3,1);
    \coordinate [label=right:{#3}] (D) at (3,-1);
    \coordinate [label=left:{#4}] (E) at (-3,1);
    \coordinate [label=left:{#5}] (F) at (-3,-1);
    \draw[transform canvas={xshift=0pt}] (A) -- (B);
    \draw[transform canvas={xshift=0pt}] (A) -- (E);
    \draw[transform canvas={xshift=0pt}] (A) -- (F);
    \draw[transform canvas={xshift=0pt}] (B) -- (C);
    \draw[transform canvas={xshift=0pt}] (B) -- (D);
    \filldraw[red,thick](A) circle (4pt);
    \filldraw[black,thick](B) circle (4pt);
    \filldraw[blue](C) circle (2pt);
    \filldraw[blue](D) circle (2pt);
    \filldraw[blue](E) circle (2pt);
    \filldraw[blue](F) circle (2pt);
\end{tikzpicture}
\end{center}}

\newcommand{\hishbaryon}[4][0.5]{
\begin{center}
	\begin{tikzpicture}[scale=#1]
    \coordinate [label=left:$$] (O) at (0,0);
    \coordinate [label=right:$$] (B) at (2,0);
    \coordinate [label=right:{#2}] (C) at (2.3,0.4);
    \coordinate [label=right:{#3}] (D) at (2.3,-0.4);
    \coordinate [label=left:{#4}] (E) at (-2.3,0);
    \draw[transform canvas={xshift=0pt}] (B) -- (E);
    \draw[transform canvas={xshift=0pt}] (B) -- (C);
    \draw[transform canvas={xshift=0pt}] (B) -- (D);
    \filldraw[red](B) circle (4pt);
    \filldraw[blue](C) circle (2pt);
    \filldraw[blue](D) circle (2pt);
    \filldraw[blue](E) circle (2pt);
\end{tikzpicture}
\end{center}}

\newcommand{\hishpentaquark}[5]{
\begin{center}
	\begin{tikzpicture}[scale=0.5]
    \coordinate [label=left:$$] (O) at (0,0);
    \coordinate [label=right:$$] (B) at (2,0);
    \coordinate [label=right:{#1}] (C) at (3,0.4);
    \coordinate [label=right:{#2}] (D) at (3,-0.4);
    \coordinate [label=left:{#3}] (E) at (-3,0);
    \coordinate [label=right:{#4}] (F) at (2.2,2);
    \coordinate [label=left:{#5}] (G) at (1.8,2);
    \draw[->,transform canvas={xshift=-2pt},thick] (E) -- (B);
    \draw[->,transform canvas={xshift=2pt},thick] (C) -> (B);
    \draw[->,transform canvas={xshift=2pt},thick] (D) -> (B);
    \draw [black, <-,transform canvas={yshift=2pt},thick] (B) to[out=45,in=-45] (F);
    \draw [black, ->,transform canvas={yshift=-1pt},thick] (B) to[out=135,in=-135] (G);
    \filldraw[red](B) circle (4pt);
    \filldraw[blue](C) circle (2pt);
    \filldraw[blue](D) circle (2pt);
    \filldraw[blue](E) circle (2pt);
    \filldraw[blue](F) circle (2pt);
    \filldraw[blue](G) circle (2pt);
\end{tikzpicture}
\end{center}}

\newcommand{\hishpentaquarkk}[5]{
\begin{center}
	\begin{tikzpicture}[scale=0.5]
    \coordinate [label=right:$$] (A) at (-2,0);
    \coordinate [label=left:$$] (O) at (0,0);
    \coordinate [label=right:$$] (B) at (2,0);    
    \coordinate [label=left:{#1}] (C) at (-3,0.4);
    \coordinate [label=left:{#2}] (D) at (-3,-0.4);
    \coordinate [label=right:{#3}] (F) at (3,0.4);
    \coordinate [label=right:{#4}] (G) at (3,-0.4);
    \coordinate [label=right:{#5}] (H) at (0.2,1.5);
    \draw[->,transform canvas={xshift=-2pt,yshift=1pt},thick] (C) -> (A);
    \draw[->,transform canvas={xshift=-2pt,yshift=-1pt},thick] (D) -> (A);
    \draw[->,transform canvas={xshift=2pt,yshift=1pt},thick] (F) -> (B);
    \draw[->,transform canvas={xshift=2pt,yshift=-1pt},thick] (G) -> (B);
    \draw[->,transform canvas={xshift=-2pt},thick] (O) -> (B);
    \draw[->,transform canvas={xshift=2pt},thick] (O) -> (A);
    \draw[->,transform canvas={yshift=-1pt},thick] (O) -> (H);
    \filldraw[red](A) circle (4pt);
    \filldraw[black,thick](O) circle (4pt);
    \filldraw[red](B) circle (4pt);
    \filldraw[blue](C) circle (2pt);
    \filldraw[blue](D) circle (2pt);
    \filldraw[blue](F) circle (2pt);
    \filldraw[blue](G) circle (2pt);
    \filldraw[blue](H) circle (2pt);
\end{tikzpicture}
\end{center}}

\newcommand{\baryonicdecay}[6][0.5]{
\begin{center}
	\begin{tikzpicture}[scale=#1]
    \coordinate [label=left:$$] (O) at (0,0);
    \coordinate [label=left:$$] (A) at (-2,0);
    \coordinate [label=right:$$] (B) at (2,0);
    \coordinate [label=right:$\bar{#2}$] (C) at (3,1);
    \coordinate [label=right:$\bar{#3}$] (D) at (3,-1);
    \coordinate [label=left:${#4}$] (E) at (-3,1);
    \coordinate [label=left:${#5}$] (F) at (-3,-1);
    \coordinate [label=above:$\bar{#6}$] (Q) at (0.6,0);
    \coordinate [label=above:${#6}$] (P) at (-0.6,0);
    \draw[transform canvas={xshift=0pt}] (A) -- (E);
    \draw[transform canvas={xshift=0pt}] (A) -- (F);
    \draw[transform canvas={xshift=0pt}] (A) -- (P);
    \draw[transform canvas={xshift=0pt}] (B) -- (C);
    \draw[transform canvas={xshift=0pt}] (B) -- (D);
    \draw[transform canvas={xshift=0pt}] (B) -- (Q);
    \filldraw[red,thick](A) circle (4pt);
    \filldraw[black,thick](B) circle (4pt);
    \filldraw[blue](C) circle (2pt);
    \filldraw[blue](D) circle (2pt);
    \filldraw[blue](E) circle (2pt);
    \filldraw[blue](F) circle (2pt);
    \filldraw[blue](P) circle (2pt);
    \filldraw[blue](Q) circle (2pt);
	\end{tikzpicture}
\end{center}}
\newcommand{\stringbreaking}[6][0.5]{\baryonicdecay[#1]{#2}{#3}{#4}{#5}{#6}}
\newcommand{\udecay}[5][0.5]{
    \stringbreaking[#1]{#2}{#3}{#4}{#5}{u}
}

\newcommand{\boundellipse}[3]{
\filldraw[yellow](#1) ellipse (#2 and #3)
}

\newcommand{\stringybranes}{
	\boundellipse{0,0}{2}{0.2};
	\boundellipse{0,-1}{2}{0.2};
	\boundellipse{0,-2}{2}{0.2};
}

\newcommand{\stringycoordinates}{
	\coordinate [label=right:$$] (OR) at (1.2,-1);
	\coordinate [label=left:$$] (OL) at (-1.2,-1);
	\coordinate [label=left:$c$] (C) at (-2,0);
	\coordinate [label=left:$s$] (S) at (-2,-1);
	\coordinate [label=left:$u/d$] (UD) at (-2,-2);
	\coordinate [label={[text=white]right:$c$}] (C) at (2,0);
	\coordinate [label={[text=white]right:$s$}] (S) at (2,-1);
	\coordinate [label={[text=white]right:$u/d$}] (UD) at (2,-2);
	\coordinate [label=right:$$] (CBVR) at (1.2,0);
	\coordinate [label=left:$$] (CBVL) at (-1.2,0);
	\coordinate [label=right:$$] (CR) at (0.8,0.12);
	\coordinate [label=right:$$] (CRR) at (1.9,0);
	\coordinate [label=left:$$] (CLL) at (-1.9,0);
	\coordinate [label=left:$$] (CL) at (-0.8,0.12);
	\coordinate [label=right:$$] (CBR) at (0.2,0);
	\coordinate [label=left:$$] (CBL) at (-0.2,0);
	\coordinate [label=right:$$] (SBVR) at (1.2,-1);
	\coordinate [label=left:$$] (SBVL) at (-1.2,-1);
	\coordinate [label=right:$$] (SR) at (0.8,-0.88);
	\coordinate [label=right:$$] (SRR) at (1.9,-1);
	\coordinate [label=left:$$] (SLL) at (-1.9,-1);
	\coordinate [label=left:$$] (SL) at (-0.8,-0.88);
	\coordinate [label=right:$$] (SBR) at (0.2,-1);
	\coordinate [label=left:$$] (SBL) at (-0.2,-1);
	\coordinate [label=right:$$] (SBR3D) at (0.2,-0.9);
	\coordinate [label=left:$$] (SBL3D) at (-0.2,-1.1);
	\coordinate [label=right:$$] (UDBVR) at (1.2,-2);
	\coordinate [label=left:$$] (UDBVL) at (-1.2,-2);
	\coordinate [label=right:$$] (UDR) at (0.8,-1.88);
	\coordinate [label=right:$$] (UDRR) at (1.9,-2);
	\coordinate [label=left:$$] (UDLL) at (-1.9,-2);
	\coordinate [label=left:$$] (UDL) at (-0.8,-1.88);
	\coordinate [label=right:$$] (UDBR) at (0.2,-2);
	\coordinate [label=left:$$] (UDBL) at (-0.2,-2);
	\coordinate [label=right:$$] (UDBR3D) at (0.2,-1.8);
	\coordinate [label=left:$$] (UDBL3D) at (-0.2,-2.3);
	\coordinate [label=right:$$] (UDBVR3D) at (0.2,-2.1);
	\coordinate [label=left:$$] (UDBVL3D) at (-0.2,-2.7);
	\coordinate [label=right:$$] (UDBR3DL) at (0.3,-1.8);
	\coordinate [label=left:$$] (UDBL3DL) at (-0.1,-2.0);
	\coordinate [label=right:$$] (UDBVR3DR) at (0.1,-2.1);
	\coordinate [label=left:$$] (UDBVL3DR) at (-1.1,-2.3);
	\coordinate [label=left:$$] (UDBVL3DL) at (0.0,-2.7);
	\coordinate [label=left:$$] (UDBVR3DL) at (1.0,-2.5);
	\coordinate [label=left:$$] (ERR) at (1.2,-2.4);
	\coordinate [label=left:$$] (ERL) at (0.2,-2.4);
	\coordinate [label=left:$$] (ELR) at (-0.2,-2.4);
	\coordinate [label=left:$$] (ELL) at (-1.2,-2.4);
	\coordinate [label=left:$$] (REL) at (-0.6,-2.4);
}

\newcommand{\stringytetraquark}[7][0.8]{
\begin{center}
	\begin{tikzpicture}[scale=#1]
	\stringycoordinates
	\stringybranes
	\draw[->,thick] (#3BVR) -- (#7R);
	\draw[->,thick] (#3BVR) -- (#6R);
	\draw[->,thick] (#5L) -- (#2BVL);
	\draw[->,thick] (#4L) -- (#2BVL);
	\draw[-,thick] (#2BVL) -- (ELL);
	\draw[->,thick] (#3BVR) -- (ERR);
	\draw [black, -,thick] (ERR) to[out=-160,in=-20] (ELL);
	\filldraw[black,thick](#3BVR) circle (1.5pt);
	\filldraw[red,thick](#2BVL) circle (1.5pt);
	\filldraw[blue](#7R) circle (1.5pt);
	\filldraw[blue](#6R) circle (1.5pt);
	\filldraw[blue](#5L) circle (1.5pt);
	\filldraw[blue](#4L) circle (1.5pt);
	\end{tikzpicture}
\end{center}}

\newcommand{\stringytetraquarkcomplex}[5][0.8]{
\begin{center}
	\begin{tikzpicture}[scale=#1]
	\stringycoordinates
	\stringybranes
	\filldraw[blue](#2BVR) circle (1.5pt);
	\filldraw[blue](#3BVL) circle (1.5pt);
	\filldraw[blue](#4BR3D) circle (1.5pt);
	\filldraw[blue](#5BL3D) circle (1.5pt);
	\draw[-,thick] (#2BVR) -- (UDBVR);
        \draw[-,thick] (UDBVR) -- (ERR);
        \draw[-,thick] (UDBVR) -- (UDBVR3DL);
	\draw[-,thick] (#3BVL) -- (UDBVL);
        \draw[-,thick] (UDBVL) -- (ELL);
        \draw[-,thick,opacity=0.3] (UDBVL) -- (UDBVL3DR);
	\draw[-,thick,opacity=0.3] (#4BR3D) -- (UDBR3D);
        \draw[-,thick,opacity=0.3] (UDBR3D) -- (UDBVR3D);
	\draw[-,thick] (#5BL3D) -- (UDBL3D);
        \draw[-,thick] (UDBL3D) -- (UDBVL3D);
        \draw[-,thick,opacity=0.3] (UDBR3D) -- (UDBVR3DR);
        \draw[-,thick] (UDBL3D) -- (UDBVL3DL);
        \draw[-,thick] (UDBVR3DL) -- (UDBVL3DL);
        \draw[-,thick,opacity=0.3] (UDBVR3DR) -- (UDBVL3DR);
        \draw[-,thick] (ELL) -- (UDBVL3D);
        \draw[-,thick,opacity=0.3] (ERR) -- (UDBVR3D);
	\filldraw[red](UDBVR) circle (1.5pt);
	\filldraw[red](UDBVL) circle (1.5pt);
	\filldraw[black,opacity=0.5](UDBR3D) circle (1.5pt);
	\filldraw[black](UDBL3D) circle (1.5pt);
	\end{tikzpicture}
\end{center}}

\newcommand{\stringytetraquarkbreak}[8]{
\begin{center}
	\begin{tikzpicture}[scale=0.8]
	\stringycoordinates
	\stringybranes
	\filldraw[red](#2BVR) circle (1.5pt);
	\filldraw[red](#1BVL) circle (1.5pt);
	\filldraw[blue](#6R) circle (1.5pt);
	\filldraw[blue](#5R) circle (1.5pt);
	\filldraw[blue](#4L) circle (1.5pt);
	\filldraw[blue](#3L) circle (1.5pt);
	\filldraw[blue](#7BR) circle (1.5pt);
	\filldraw[blue](#8BL) circle (1.5pt);
	\draw[->,thick] (#2BVR) -- (#6R);
	\draw[->,thick] (#2BVR) -- (#5R);
	\draw[->,thick] (#4L) -- (#1BVL);
	\draw[->,thick] (#3L) -- (#1BVL);
	\draw[-,thick] (#1BVL) -- (ELL);
	\draw[-,thick] (#2BVR) -- (ERR);
	\draw[->,thick] (ERL) -- (#7BR);
	\draw[->,thick] (#8BL) -- (ELR);
	\draw [black, -,thick] (ERR) to[out=-160,in=-20] (ERL);
	\draw [black, -,thick] (ELR) to[out=-160,in=-20] (ELL);
	\end{tikzpicture}
\end{center}}

\newcommand{\stringybaryon}[4]{
\begin{center}
	\begin{tikzpicture}[scale=0.8]
	\stringycoordinates
	\stringybranes
	\filldraw[blue](#2BVL) circle (1pt);
	\filldraw[red](#1BVR) circle (2pt);
	\filldraw[blue](#4R) circle (1pt);
	\filldraw[blue](#3) circle (1pt);
	\draw[->,thick,transform canvas={xshift=1pt}] (#4R) -- (#1BVR);
	\draw[->,thick,transform canvas={yshift=1pt}] (#3) -- (#1BVR);
	\draw[-,thick,transform canvas={yshift=-1pt}] (#1BVR) -- (ERR);
	\draw[-,thick] (#2BVL) -- (ELL);
    \draw [black, ->,thick] (ELL) to[out=-20,in=-160] (ERR);
	\end{tikzpicture}
\end{center}}

\newcommand{\stringyellipse}[3]
	{\draw (#1) ellipse [x radius=#2,y radius=#3]}

\newcommand{\stringymesonspair}[4]{
\begin{center}
	\begin{tikzpicture}[scale=0.8]
	\stringycoordinates
	\stringybranes	
	\filldraw[blue](#1L) circle (1.5pt);
	\filldraw[blue](#2L) circle (1.5pt);	
	\draw[->,thick]  (ELL) -- (#1L);
	\draw[->,thick] (#2L) -- (ELR);
	\draw [black, -,thick] (ELR) to[out=-160,in=-20] (ELL);
	\filldraw[blue](#3R) circle (1.5pt);
	\filldraw[blue](#4R) circle (1.5pt);	
	\draw[->,thick]  (ERL) -- (#4R);
	\draw[->,thick] (#3R) -- (ERR);
	\draw [black, -,thick] (ERR) to[out=-160,in=-20] (ERL);	
	\end{tikzpicture}
\end{center}}

\newcommand{\stringymeson}[2]{
\begin{center}
	\begin{tikzpicture}[scale=0.8]
	\stringycoordinates
	\stringybranes
	\filldraw[blue](#2BVL) circle (1.5pt);
	\filldraw[blue](#1BVR) circle (1.5pt);
	\draw[->,thick] (#2BVL) -- (ELL);
	\draw[->,thick] (ERR) -- (#1BVR);
	\draw [black, -,thick] (ERR) to[out=-160,in=-20] (ELL);
	\end{tikzpicture}
\end{center}}

\newcommand{\hishtetraquarkcomplex}[5][0.8]{
\begin{center}
    \begin{tikzpicture}[scale=#1]
    \coordinate [label=left:$$] (O) at (0,0);
    \coordinate [label=right:$$] (A) at (1,1);
    \coordinate [label=right:$$] (B) at (1,-1);
    \coordinate [label=right:$$] (D) at (-1,1);
    \coordinate [label=right:$$] (C) at (-1,-1);
    \coordinate [label=right:#5] (E) at (1.2,1.2);
    \coordinate [label=right:#3] (F) at (1.2,-1.2);
    \coordinate [label=left:#4] (G) at (-1.2,-1.2);
    \coordinate [label=left:#2] (H) at (-1.2,1.2);
    \draw[-,thick] (A) -- (B);
    \draw[-,thick] (B) -- (C);
    \draw[-,thick] (C) -- (D);
    \draw[-,thick] (D) -- (A);
    \draw[-,thick] (A) -- (E);
    \draw[-,thick] (B) -- (F);
    \draw[-,thick] (C) -- (G);
    \draw[-,thick] (D) -- (H);
    \filldraw[red](A) circle (2pt);
    \filldraw[black,thick](B) circle (2pt);
    \filldraw[red](C) circle (2pt);
    \filldraw[black,thick](D) circle (2pt);
    \filldraw[blue](E) circle (1pt);
    \filldraw[blue](F) circle (1pt);
    \filldraw[blue](G) circle (1pt);
    \filldraw[blue](H) circle (1pt);
    \end{tikzpicture}
\end{center}
}

\usepackage{tikz} 
\usepackage{twoopt}
\newcommandtwoopt{\hishmeson}[4][0.5][]{
\begin{center}
     \begin{tikzpicture}[scale=#1]
     \coordinate [label=right:{#2}] (O) at (-0.3,1);
     \coordinate [label=left:{#3}] (L) at (-2,0);
     \coordinate [label=right:{#4}] (R) at (2,0);
     \draw[transform canvas={xshift=0pt},thick] (L) -- (R);
     \filldraw[blue,thick](L) circle (2pt);
     \filldraw[blue,thick](R) circle (2pt);
	\end{tikzpicture}
\end{center}
}

\newcommandtwoopt{\hishmesondecay}[4][0.6][]{
\begin{center}
     \begin{tikzpicture}[scale=#1]
     \coordinate [label={[text=white]right:{#2}}] (O) at (-0.3,1);
     \coordinate [label=left:{#3}] (L) at (-2,0);
     \coordinate [label=right:{#4}] (R) at (2,0);
     \coordinate [label=left:{}] (RL) at (1,0);
     \coordinate [label=right:{}] (LR) at (-1,0);
     \draw[transform canvas={xshift=0pt},thick] (L) -- (R);
     \filldraw[blue](L) circle (2pt);
     \filldraw[blue](R) circle (2pt);
     \filldraw[red,thick](LR) circle (3pt);
     \filldraw[black,thick](RL) circle (3pt);
     \draw [black,-,transform canvas={yshift=1pt},thick,dashed] (LR) to[out=70,in=110] (RL);
     \end{tikzpicture}
\end{center}
}

\newcommand{\BV}[1]{{\it{V-baryonium }}\,}

\begin{document}

\title{\line(4,0){480}\\Taming the Zoo of Tetraquarks and Pentaquarks using the HISH Model\\\line(4,0){480}} 
\author{
  Michal Michael Green\\
  \text{mickeygreen@mail.tau.ac.il} \and 
  Jacob Sonnenschein\\
  \text{cobi@tauex.tau.ac.il}
}


\date{\centerline{{School of Physics and Astronomy,}}
\centerline{{\it The Raymond and Beverly Sackler Faculty of Exact Sciences, }} \centerline{{\it Tel Aviv University, Ramat Aviv 69978, Israel}} 
\today}  

\maketitle  



\begin{abstract} 
 In this paper we scan over all possible charmed tetraquarks and pentaquarks.  Using the holography inspired stringy hadron (HISH) model we determine the trajectories associated with each of the exotic hadron candidates. The trajectories include further exotic states with higher angular momentum or higher stringy excited states. A trajectory is a property of a genuine exotic hadron and can be used to distinguish between the latter and a molecule.  We examine 71 tetraquarks and 210 pentaquarks. Few of these states have already been found but most of the predicted   zoo have   yet not been discovered.    We analyze  the strong decay processes of these exotic hadrons and compute the corresponding decay widths of part of them.
 The predictions are summarized in tables \ref{table:charm_tetraquarks_predictions} and \ref{table:charm_pentaquarks_predictions}. The results for each of the exotic candidates are at sections \ref{section:tetra_cands} and \ref{section:pentaquarks_results}.
\end{abstract}
\newpage
\tableofcontents
\section{Introduction}

\lettrine[findent=2pt]{\textbf{T}}{he} 
Question of  whether  nature exhibits compact multiquark hadron states aside from mesons, baryons, and glueballs is still not fully resolved.  Observations  in recent years provide direct and indirect evidence for the existence of such exotic hadrons, yet there is no consensual picture about  exotic hadron states in the hadronic spectrum. For reviews of the experimental and theoretical  status see \cite{Chen:2016qju}, \cite{Esposito:2016noz}, \cite{Karliner:2017qhf}, \cite{Olsen:2017bmm},\cite{Brambilla:2019esw} and \cite{Brambilla:2022ura}.

The main challenges of any theoretical description of  exotic hadrons is the determinations of their spectra and decay widths. 
In previous works, the spectra of mesons \cite{Sonnenschein_2014_mesons}, baryons \cite{Sonnenschein:2014bia},  glueballs \cite{Sonnenschein_2015_glueballs} and tetraquarks \cite{Sonnenschein_2017_4630}  were analysed in the context of the HISH (holography inspired stringy hadron) model \cite{Sonnenschein_2017_HISH} \cite{Sonnenschein_quantization_2018}.  It was found that most of the known hadron resonances match nicely the HISH modified Regge trajectories (HMRT) generated by the model. The model also provides mechanisms for strong  hadronic decays \cite{Sonnenschein_decay_width_2018} from which one can predict not only the masses  of the resonances, but also their corresponding  decay width and branching ratios.

The description of hadrons in terms of strings was one of the  origins of string theory. In the ``old days" various aspects of hadron physics have been analyzed in terms of  a stringy behavior. For a review  and references therein see \cite{Collins:1977jy}. The ``modern" arena of stringy hadron physics has been sourced by the holographic string/gauge duality.  In particular 
the HISH framework\cite{Sonnenschein_2017_HISH} provides a phenomenological stringy  description of hadrons based on a map between  string configurations  that reside in  holographic confining  backgrounds  and strings in  four dimensional flat space-time. One should notice the difference between the holographic stringy  picture  of hadrons and the  field theory gravity/gauge duality approach where hadrons associate with  fluctuations of the bulk fields and flavor D branes.  For a review and references  about this approach see\cite{Erdmenger:2007cm}.  Several other  descriptions  of hadrons, and in particular exotic ones, in terms of  strings  have been proposed in the past, in fact  from the early days of strings and up to recent years. Such models    can be found for instance in \cite{Rossi:2020ezg},  \cite{Kikkawa:1978zt}, \cite{Bigazzi:2004ze},\cite{Ebert:2011kk} ,  \cite{Andreev:2022cax}, and  and \cite{Hayashi:2020ipd}.  

In the HISH model mesons are described as   open strings with massive   particles ( ``quarks") that carry  electric charge and spin  on their endpoints, baryons as strings stretched between a quark and a baryonic vertex (BV) to which a di-quark is attached and glueballs as closed strings. The stringy hadron can rotate thus  generating  an orbital angular momentum to a meson, baryon and glueball respectively.      The stringy hadrons furnish  the HMRT. These  trajectories, described in the plane of $M^2$  and $J$ the total angular momentum which is sum of the orbital angular momentum of the string and the spin of the endpoint particles. Another class of trajectories is  in the plane of  $M^2$ and $n$ the string excitation number. In both cases the trajectories  which are not linear are characterized by the masses of the endpoint particles, by the slope which relates to the string tension and by the intercept that relates to the quantum Casimir energy of the string. The latter depends on the properties of the endpoint particles, the masses  \cite{Sonnenschein_quantization_2018}, the electric charges \cite{Sonnenschein:2019bca} and the spins. 

Naturally the string description of hadrons can accommodate  on top of mesons, baryons and glueballs, also  ``exotic hadrons". These include states  without quarks, namely, exotic glueballs and with quarks  but  with vanishing baryon number:  tetraquarks,   hexaquarks  and hybrids of glueballs and mesons. States with unit baryon number: pentaquarks, septaquarks etc. including   hybrids with glueballs.   One can also easily construct configurations with baryon number equal to two or higher like sexaquarks\cite{Farrar:2017eqq} etc.
In fact,  
 trajectories of  tetraquarks candidates where analysed in the past in \cite{Sonnenschein_2017_4630} and \cite{Sonnenschein_2021_6900}. In the former  exotic states associated with  Y(4630) and  in the latter with X(6900).

Tetraquarks and pentaquarks have been thoroughly investigated  from  both the experimental and theoretical points of view.  For  reviews and references therein see for the  the former \cite{Esposito:2016noz}, \cite{Guo:2017jvc}and \cite{ParticleDataGroup:2022pth}.  Reviews and references  about pentaquark can be found in\cite{ParticleDataGroup:2022pthK} and \cite{Brambilla:2019esw}

Exotic   resonances can have light or heavy string endpoint quarks.  In recent years the focus of the study of exotic hadrons has been on states that include a heavy quark.
For that reason, in  the present  work we  systematically analyze all the  stringy  tetraquarks and pentaquarks  states that include at least one charmed quark. As a warm up we also describe several candidates of exotic hadrons that do not include a charm quark. 
 We analyse the  exotic hadron candidates in the context of the HISH model. 
 We sketch the possible arrangements constructed from the basic building blocks - open strings, massive endpoints ("quarks" or "di-quarks"), and  baryonic vertices. Using data extracted from the HMRT of relevant  mesons and baryons we determine the trajectories of the tetraquarks and pentaquark. This includes predicting the mass, angular momentum $J$  and string excitation number $n$   of the states that reside on the trajectories. 

 In this paper we study ``genuine" exotic hadrons. We do not study molecules of hadrons. The latter can be described as disconnected string configurations. The stringy description of molecules and mixtures of them and genuine connected exotic hadrons was analyzed in several papers. See for instance  \cite{Andreev:2022qdu} and \cite{Andreev:2023hmh}.
 
 Associated with each tetraquark state there are two thresholds: (i) The sum of the masses of the two stringy mesons that have the quark content of the original tetraquark (ii) The sum of the masses of the baryon and anti-baryon that can result from the breaking of the string of the tetraquark.
In fact the most natural decay mode of any  stringy hadron,if permitted by energy considerations,  is through breaking apart. In the case of the tetraquark simplest configuration (Fig-\ref{fig:tetraquark_simple}) it is the decay to a baryon-antibaryon pair (see Fig-\ref{fig:tetraquark_string_tear}). Obviously, this decay can occur only  provided the mass of the tetraquark is higher than the sum of the masses of the daughter baryon and anti-baryon. Due to the structure of the tetraquark, the computation of the decay width of such a process is similar to that of a meson decaying into two mesons \cite{Peeters_2006}. When the mass is  below the threshold for decay through tearing, a tetraquark decay can take place  through BV-anti-BV annihilation, which results in two mesons (see Fig-\ref{fig:tetraquark_annihilation} and \ref{Decay mechanisms of tetraquarks}). Most of the recently observed exotics were below the baryonia threshold, and instead reconstructed through a mesons pair channel.

For pentaquarks, the possible configurations are more complex. There are two possible structures, the first (see Fig-\ref{fig:pentaquark_simple}) is baryon-like, with the addition of a quark anti-quark pair attached to a BV. The second is constructed similarly to a tetraquark (see Fig-\ref{fig:pentaquark_simple2}), where there are two BVs with diquarks and an additional anti-BV attached to an antiquark between them. Both structures will be discussed in this paper along with the criteria with which one determines the structure that takes place.

An important  question regarding exotic candidates observations is whether they are genuine multiquarks or bound states of mesons and baryons. The HISH model, where the  stringy exotic hadrons are genuine,
provides two ways to distinguish between the genuine exotic hadrons  and the molecules:(i) If the exotic states furnish a HMRT it means that they all are string configuration and are genuine. (ii) If a state decays into a baryon anti-baryon it is probably a genuine tetraquark which decays via a breaking of its string. Similarly, an outcome of two baryon and anti- baryon  and in fact also of a baryon and two mesons are smoking guns of pentaquarks.

In this paper  We analyze 71 candidates of charmed tetraquarks. For each of them we determine the thresholds to decay into two mesons and into a baryon and anti-baryon. The masses of the excited states are determined. We group all the states into 31 separate cases for which we write down the HMRT, the and widths associated with breaking of the string and with an annihilation process.
210 different candidates of Pentaquarks are analyzed. Again we write down the threshold for decay, this time to a baryon and a meson, draw the HMRTs and compute the two different decay widths.

Our procedure of determining the HMRTs of all the tetraquark candidates  was based relating the spectra and the various decay channels. 
One  assumption that  we made was that the first  tetraquark state that can decay into a baryon and an anti-baryon is of a mass which  is 40-125 MEV higher than  the baryonic threshold  . This is based on the few cases with such a decay  that have been already discovered\cite{Sonnenschein_2017_4630} and \cite{Sonnenschein_2021_6900}. Using this assumption we determined the approximated value of the intercept. We also assumed  that in between the threshold of decaying into two mesons and that of a baryon anti-baryon decay two low lying states reside on the trajectory, so that the first state with the later decay is characterized by $J+n=3$. We determined the full HMRT. For that  we used the endpoint masses, the intercept and the slope determined from meson and baryon trajectories with  similar endpoint particles.  Altogether we determined on each HMRT 5 states. Out of them can decay only into two mesons and in the other three a decay to a baryon and anti-baryon is also possible.

The paper is organized as follows. In section \ref{section:HISH_review} we briefly review the HISH model and the important results that will aid confronting the data. It includes the basic ingredients, the simplest configurations of the different multiquarks, the classical and quantum  HISH modified Regge trajectories and decay channels and widths. The following section \ref{Exotic hadrons} is dedicated to the general structure of stringy exotics hadrons,  In section \ref{HISH_tetraquarks} we  discuss the tetraquark structure and decay mechanisms, as well as the mechanisms through which tetraquarks are created. In section \ref{HISH_pentaquarks} we  analyse the pentaquark structure and decays. At last, in section \ref{section:HISH_data} we confront experimental data. We will first present the fitting procedures and phenomenological methodology to be used to analyse the observations and predict further states on the HMRTs. Then we will discuss all possible charm tetraquarks configurations, we will examine the data for those with an observed candidates, and provide predictions for those that do not. We will also confront, for the first time, pentaquark candidates for the few that were observed. We end up with a summary section, a section of open questions and an appendix that includes fitting of the HISH to the meson spectra.
 
\section{Brief review of HISH (holography inspired string hadron) model}\label{section:HISH_review}

The AdS/CFT duality conjecture is an equivalence between bulk string theories and field theories that live on the boundary. Realistic field theories should be non-supersymmetric, non-conformal and confining. The dual string background on the other hand, should admit confining Wilson lines, dual to a boundary with a matter sector that is invariant under a chiral flavor symmetry that is spontaneously broken (an example review paper on how to achieve these requirements \cite{aharony2003nonadsnoncft}).

Unlike most of the applications of the holographic duality, in which hadrons loose their stringy properties due to taking the $\alpha^{'}\rightarrow0$ limit, the HISH framework preserves these characteristics. In particular, it admits the semi-linear Regge behavior that arise naturally for strings. Moreover, the model is mostly consistent with the data for mesons \cite{Sonnenschein_2014_mesons} and baryons \cite{Sonnenschein_2014_baryons}.

The idea  behind the  HISH model  is to construct a phenomenological unitary  string model that is in accordance with as much as possible experimental data of hadron physics   and then use it   to  predict properties and  phenomena that  have not been measured so far.
The model is in flat four dimensions but is based on mapping hadronic    string configurations of  holographic backgrounds.
 
The prescription for building and using the model  includes the following steps: 
\begin{itemize}
\item
  Determining a prototype  confining  holographic background with flavor branes. A confining holographic background means a background for which the Maldacena-Wilosn line admits an area law behavior\cite{Kinar:1998vq}.
\item
 Analyzing in the holographic confining  background  classical  strings,  in particular rotating ones, that correspond to hadrons: (i) Mesons- open strings between flavor D branes, (ii) Glueballs- closed strings, (iii) Baryons- systems that include a baryonic vertex (B.V) with two short strings (di-quark) connected to flavor branes attached to it   that is connected  with a  long string to a flavor D brane and  (iv) Exotic hadrons- stings between a B.V  and an anti- B.V to which a di-quark and an anti-diquark are attached respectively. 
\item 
 ``Mapping'' the   classical  holographic strings, in particular the  rotating ones,  to  strings with massive particles on their ends  in four flat dimensions. The endpoint particles ``quarks" carry electric and flavor charges but not a color charge and a spin.   The B.V is mapped into a string junction that carries a unit baryon number  with , a net number $N_c=3$ of  three strings coming out of it and a net number of  three strings  coming into an anti-string junction of baryon number -1.
\item
 Examining for the stringy baryons the classical stability against classical fluctuations  of the Y shape  versus the  single string configuration between a quark and a B.V with attached diquark. 
\item
 Solving the classical equations of motion for rotating string and computing the classical energy and angular momentum  of the strings with  massive particles on their ends. This constitutes the classical  HMRT  ( HISH modified Regge trajectory)
\item
Quantizing the  fluctuations  around the classical stringy hadrons subjected to the boundary conditions associated with the   endpoint particles with  masses, electric charges,  and spins. and  to the baryonic vertices. Determining the eigenfrequencies  of the fluctuations, namely, the spectrum. 

\item
 Renormalizing the world sheet Hamiltonian using a    ``Cauchy Casimir-like method''. Determining  the intercept and the quantum HMRT. This includes  the contribution of the Liouville mode associated  the fact taht the string resides in  d=4 non-critical dimension.
\item
Confronting the outcome of the model, in particular the HMRTs with  experimental spectra  examining the fitness of the theoretical model and   extracting the best fit values  for the string tension( slope), endpoint masses and intercepts from all  the hadron ( meson and baryon)  trajectories. Predicting the masses and quantum numbers of resonances that have yet not been detected.
\item
Determining the strong decay width of hadrons including (i) Computing the total decay width associated with possible breaking of the parent string into two daughter strings.
 (ii) Evaluating   the probability  of a particular pair creation at the breaking point and hence the branching ratios, (iii) Approximating the  width associated with annihilation of a b.v and an anti-b.v. Comparing the measured decay width to the corresponding results of the HISH and predicting yet unknown decay widths.  
\end{itemize}


\subsection{The HISH mapping}

\begin{figure}[t!] \centering
\includegraphics[width= 0.9\textwidth]{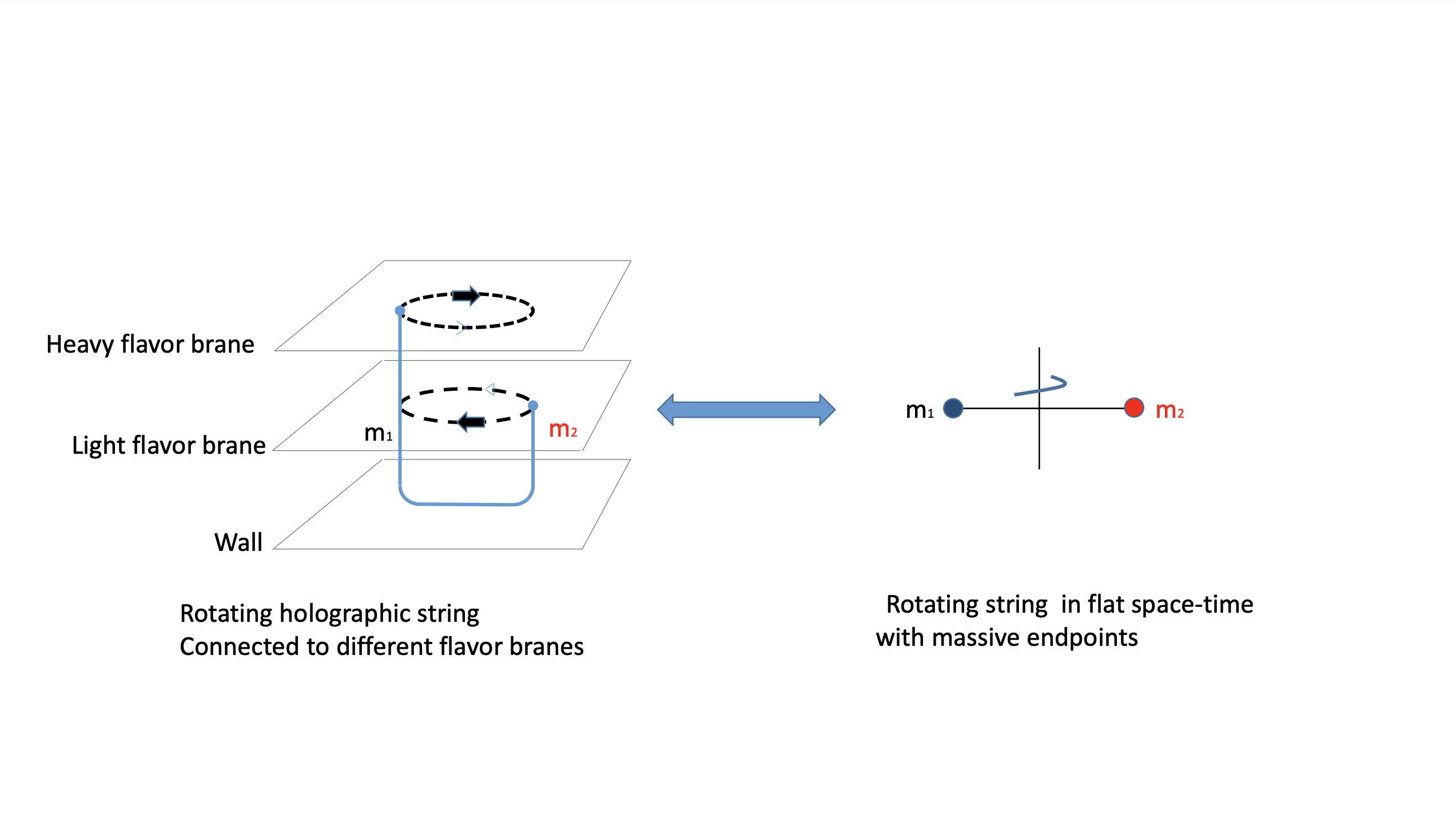}
					\caption{\label{fig:mapholflat} \textbf{Left:} Rotating holographic open string. \textbf{Right:} The corresponding open string with massive endpoints in flat spacetime. In this case we show a heavy-light meson.\label{fig:hishmap}}
		\end{figure}

Confining holographic backgrounds are characterized  by a ``wall'' that truncates in one way or another the range of the holographic radial direction. A common feature to all the holographic stringy hadrons is that there is a segment of the string that stretches along a constant radial coordinate in the vicinity of the ``wall'', as in figure \ref{fig:hishmap} \cite{Kruczenski:2004me,Sonnenschein_2017_HISH}. For a stringy glueball it is the whole folded closed string that rotates there. For an open string it is part of the string, the horizontal segment, that connects with vertical segments either to the boundary for a Wilson line or to flavor branes for a meson or baryon. This fact that the classical solutions of the flatly  rotating strings reside at fixed radial direction is a main rationale behind the map between rotating strings in curved spacetime and rotating strings in flat spacetime described in figure \ref{fig:hishmap}. In the HISH picture, the vertical segments are represented as $m_{sep}$- the endpoint masses of the string in flat spacetime. As mentioned, the HISH approximation improves as the horizontal string becomes longer with respect to its endpoint masses - hence, the higher the excitation, the better the HISH model describes the state.\\

A key player of the map is the ``string endpoint mass'', $m_{sep}$\cite{Kruczenski:2004me}, that provides in the flat space-time description the dual of the string action associated with the  vertical string segments. It is important to note that upon extracting this mass from the fits to experimental data,  it turns out that this mass is neither the QCD mass parameter (the current quark mass) nor the constituent quark mass. Notice also  that the massive endpoint as a map of an exactly vertical segment is an approximation that is more accurate the longer the horizontal string is.

\subsection{Massive Strings in Flat Spacetime}

The HISH model basic building blocks are:

\begin{itemize}
\item Open strings, characterized by a slope $\alpha^{'}={\frac{1}{2\pi T}}$ (T is the string tension). The string, which rotates in general, possesses mass and angular momentum. In the quantum level, the angular momentum has a contribution from an intercept \textit{$a$}, that can be interpreted as a "Casimir force" that prevents the string ground state from collapsing into itself.

\item Quarks - massive particles ($m_{u/d}$, $m_{s}$, $m_{c}$, $m_{b}$) that are attached to the ends of open strings. These masses are the parameters of the HISH model from a pure flat spacetime picture. They have been determined in previous works (\cite{Sonnenschein_2014_mesons}, \cite{Sonnenschein_2017_HISH}) by fitting.

\item "Baryonic vertex" (BV) - a massive vertex that is connected to an $N_c$ number of strings. Hadronic structures that include BVs connected to one long string and two short strings attached to quarks are called "diquarks". Unlike other models, these are not part of the basic building block of the theory, rather a configuration that is always attached to a BV.

\item Closed strings - the closed strings slope is half that of the open string ($\alpha^{'}_{closed}={{\frac{1}{2}}\alpha_{open}^{'}}$). Their angular momentum (hence, the excitation number) of the closed string is always even. Their intercept is also expected to be twice that of an open string.
\end{itemize}

The total mass of the hadron is determined by its presenting configuration of all of these building blocks.

\subsection{Hadron Constructions}\label{HadCon}

Each of the hadron types can be constructed from the above ingredients. There are multiple possible configurations for each type. The simplest constructions of the basic multiquark structures are as follows:

\begin{itemize}
    \item Meson - a single open string attached to a quark on one end and an antiquark on the other end.    
    \begin{figure}[!ht]
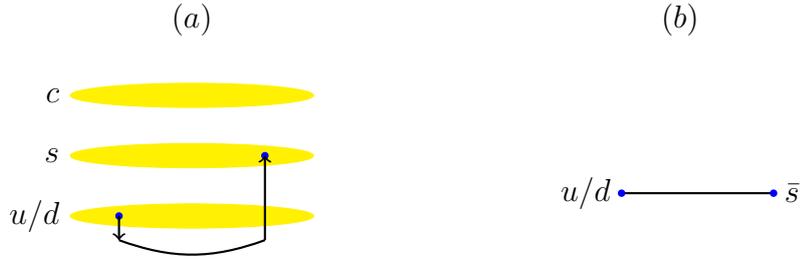

      \centering
      \begin{tabular}{ c m{6cm} m{6cm} }
        \begin{minipage}[!ht]{6cm}
        \begin{center}
          $(a)$
        \end{center}
        \end{minipage}
        &
        \begin{minipage}[!ht]{6cm}
          \begin{center}
          $(b)$
          \end{center}
        \end{minipage}
        \\\\
        \begin{minipage}{6cm}
          \stringymeson{S}{UD}
        \end{minipage}
        &
        \begin{minipage}{6cm}
          \hishmeson{$u/d$}{$\bar{s}$}
        \end{minipage}
        \\
      \end{tabular}
      \caption{Meson - (a) is a meson with content $u/d\bar{s}$ in the stringy picture. (b) is the mapping of the meson to the HISH picture}\label{fig:Meson}
    \end{figure}
    
    \item Baryon - Apriori there are several possible configurations for the HISH baryon. However, as will be discussed in subsection (\ref{baryonic_structures}), in nature it takes the form of a single open string attached to a quark on one endpoint and a diquark on the other as depiected in figure (\ref{fig:Baryon}).
    \begin{figure}[!ht]
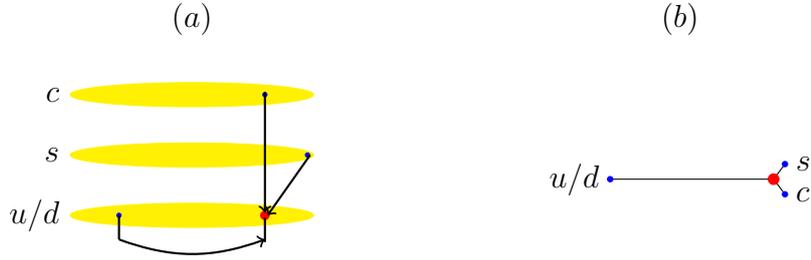

      \centering
      \begin{tabular}{ c m{6cm} m{6cm} }
        \begin{minipage}[!ht]{6cm}
        \begin{center}
          $(a)$
        \end{center}
        \end{minipage}
        &
        \begin{minipage}[!ht]{6cm}
          \begin{center}
          $(b)$
          \end{center}
        \end{minipage}
        \\\\
        \begin{minipage}{6cm}
          \stringybaryon{UD}{UD}{CBVR}{SR}
        \end{minipage}
        &
        \begin{minipage}{6cm}
          \hishbaryon{$s$}{$c$}{$u/d$}
        \end{minipage}
        \\
      \end{tabular}
      \caption{Baryon - (a) is a baryon with content $u/dsc$ in the stringy picture. (b) is the mapping of the baryon to the HISH picture}\label{fig:Baryon}
    \end{figure}
    
    \item Glueball - a single closed string is the simplest configuration for glueballs. A more complex configuration can be achieved using BVs and anti-BVs configurations that creates closed structures (see Fig-\ref{fig:Glueball}).
    
\end{itemize}

Since in this paper we concentrate on exotic hadrons, the configurations for tetraquarks and pentaquarks are analysed in dedicated sections \ref{HISH_tetraquarks}, \ref{HISH_pentaquarks}.


\subsection{Action and equations of motion} \label{sec:action}
The action of the HISH model includes the worldsheet action of an open bosoic string  and the action of the endpoint particles.For closed strings obviously the latter is not included. The action reads
\begin{equation}\label{action}
 S = S_{st} + \left(S_{pm}+ S_{pq} + S_{ps}\right )\vert_{\sigma=0} + \left( S_{pm}+ S_{pq}+ S_{ps}\right )\vert_{\sigma=\ell} .
\end{equation}
where $S_{st}$ is the string worldsheet action.
and $S_{pm}+ S_{pq} + S_{ps}$  are the particle action terms associated with the mass, electric charge and spin of the particle\footnote{In fact the charge generalized to the electroweak one. It is important to note that the endpoint charges do not carry QCD color charges}.

 For the action of the string we take the Nambu-Goto action. Viewing the action  as an effective action of a long string one can add additional higher order terms\cite{Aharony:2013ipa} and in particular an extrinsic curvature term \cite{Polyakov:1986cs}\footnote{ This is probably relevant for folded string solutions.} However here we use only the Nambu-Goto action given by  
\begin{equation}
 S_{st} = -T\int d\tau d\sigma \sqrt{-h} = -T\int d\tau d\sigma \sqrt{\dot X^2 X^\prime{}^2 - (\dot X\cdot X^\prime)^2}.
\end{equation}
where
 \(h_{\alpha\beta} = \eta_{\mu\nu}\pa_\alpha X^\mu \pa_\beta X^\nu\) is the induced metric on the worldsheet and \(h\) is its determinant, the indices \(\alpha\), \(\beta\) being either \(\tau\) or \(\sigma\). The target space is $d$ dimensional $\mu,\nu= 0,...d-1$. 
In this paper we take $d=4$. The  action is defined on the strip: \(-\infty<\tau<\infty\) and \(-l\leq\sigma\leq\ell\).

The particles located at both endpoints of the string $\sigma=-l$ and $\sigma=\ell$ have an action that is built from three parts. First is a mass term $S_{pm}$ given by 
\begin{equation}\label{massaction}
S_{pm} = m_i\int d\tau \sqrt{-\dot X^2}  
 \end{equation}
The second term on the boundary action $S_{pq}$ respectively 
\begin{equation}\label{chargeaction}
S_{pq}= T q_i \int d\tau A_\mu(X) \dot X^\mu 
\end{equation}
The parameters $m_i$ and $q_i$  are the mass and charge of the endpoint particles. The electromagnetic interaction of the charges follows the usual Maxwell action that has to be added:
\begin{equation} S \to S -\frac1{4g^2}\int d^4 x F_{\mu\nu}F^{\mu\nu} \end{equation}
 
The third term on the boundary, the spin term can be written in a form which is analogous to the string Green Schwarz  term as follows
 \begin{equation}
 S_{ps}=  \int d\tau \left [ \frac{-i \dot X^\mu(\bar \psi\gamma_\mu\pa_\tau\psi)}{2 \sqrt{-\dot X^2}} -m_i\sqrt{-\dot X^2}\bar\psi\psi\right]
 \end{equation}

The system of an open bosoinc  string with massive endpoints was considered first in \cite{Chodos:1973gt} and was addressed afterwards by many authors.

Next come the equations of motion. Since the charges and spins will not play a major role in the analysis of this paper we set them to zero from here on. 
 
  Using the  two dimensional reparameterization  symmetry we fix the orthogonal gauge 
  \begin{equation}
  \dot X^2+ {X'}^2= \dot X\cdot X'=0.
  \end{equation}
   In this gauge the bulk equations of motion read
\begin{equation} \pa^2_\sigma X^{\mu} - \ddot X^\mu = 0 \label{eq:bulkeom}
\end{equation}
and the boundary conditions
\begin{equation} T X^{\prime\mu} +  m_1\pa_\tau\frac{\dot X^\mu}{\sqrt{-\dot X^2}}   = 0 \qquad \sigma = -\ell \label{eq:bd0}\end{equation}
\begin{equation} T X^{\prime\mu}- m_2\pa_\tau\frac{\dot X^\mu}{\sqrt{-\dot X^2}} =0 \qquad \sigma = \ell \label{eq:bdl}\end{equation}

\begin{equation}\label{Sng}.\end{equation}

The equations of motion admit several types of solutions: (i) static solutions, (ii) Yo-Yo vibrating solutions and (iii) rotating solutions. 
Classically, in the static case the string collapses to zero size. This will not be the case in the full quantum picture. The vibrating solution was studied in \cite{Bars:1975dd}. Since it  it  does not play an important role for the spectra, we will not consider it here. 
\subsection{Classical Regge trajectories}
The  rotating  solution  takes the following form configuration 
\begin{equation}\label{Shish_bulk_sol}
\begin{aligned}
    X^{0}=\tau,X^{1}=R(\sigma)\cos{(\omega\tau)},X^{2}=R(\sigma)\sin{(\omega\tau)}
\end{aligned}
\end{equation}
for any choice of $R(\sigma)$.




The energy and angular momentum associated with this solution for an open string with masses $m_1$ and $m_2$ on its endpoints constitute the classical trajectories referred to as HMRT ( his modified Regge trajectory).
\begin{equation}\label{energy_cl}
    E = \sum_{i=1,2}{m_{i}{\left({\frac{\beta_{i}\arcsin{\beta_{i}}+\sqrt{1-\beta_{i}}}{1-\beta^{2}_{i}}}\right)}}
\end{equation}

\begin{equation}\label{J_cl}
    J = \sum_{i=1,2}{\pi\alpha^{'}m_{i}^{2}{\frac{\beta_{i}^{2}}{(1-\beta_{i}^{2})^2}}{\left({{\arcsin{\beta_{i}}+\beta_{i}\sqrt{1-\beta_{i}}}}\right)}}
\end{equation}

where $\beta_{i}=\omega l_{i}$ is the velocity of the i-th endpoint and $l_{i}$ is the length of the i-th string segment from the center of mass.

The velocities are related by the boundary condition:

\begin{equation}\label{velocities_relation}
    {\frac{T}{\omega}}={m_{1}{\frac{\beta_{1}}{1-\beta_{1}^{2}}}={m_{2}{\frac{\beta_{2}}{1-\beta_{2}^{2}}}}}
\end{equation}

By taking the relativistic limit ($\beta_{i}\rightarrow{1}$) where $m<<E$:

\begin{equation}
\begin{aligned}
    {J} = {\alpha^{'}E^{2}\times\left(1-\sum^{2}_{i=1}\left({\frac{4\sqrt{\pi}}{3}}\left({\frac{m_{i}}{E}}\right)^{{3}/{2}}+{\frac{2\sqrt{\pi^{3}}}{10\sqrt{2}}}{\left({\frac{m_{i}}{E}}\right)}^{5/2}+...\right)\right)}
\end{aligned}
\end{equation}

where in the limit $m_{i}\rightarrow{0}$ the linear Regge trajectory is recovered:

\begin{equation}\label{linear-regge}
    {J}={\alpha^{'}E^{2}}
\end{equation}

The high mass limit ($\beta_{i}\rightarrow{0}$) where 

$\Delta E\equiv E-m_{1}-m_{2}/(m_{1}+m_{2})<<1$:

\begin{align}
    {J} = {{\frac{4\pi}{3\sqrt{3}}}\alpha^{'}{\sqrt{\frac{m_{1}m_{2}}{m_{1}+m_{2}}}(\Delta E)^{3/2}}}+{{\frac{7\sqrt{2}\pi}{27\sqrt{3}}}\alpha^{'}{\frac{m_{1}^{2}-m_{1}m_{2}+m_{2}^{2}}{m_{1}m_{2}\sqrt{(m_{1}+m_{2})^{3}}}}(\Delta E)^{5/2}+...}
\end{align}

\subsection{The quantum static string}\label{the_quantum_static_string}
The mass of a static string with massive endpoints is the sum of the masses of the endpoints and the energy of the string, namely
\begin{equation}
M=m_{1} + m_{2} + TL
\end{equation}\label{Mesonmass}
Classically, unless the   endpoints are nailed,   the static  string collapses to zero length  since the force on the massive endpoints   due to the tension is not  balanced. In classical rotating strings the as was discussed above the  centrifugal force is balancing the tension. 
Quantum mechanically  a static string may have a non-vanishing length if the Casimir force acting on the endpoint particles is repulsive  opposing the tension. 
The Casimir force  is related to the intercept in the following form 
\begin{equation}
F_C= -2 \pi \frac{a}{L^2}
\end{equation}
If $a>0$, as it is in the ordinary bosonic string,  the force is attractive and it will add up to the tension and would not balance it. However, if $a<0$  we have a repulsive  force. The equilibrium  length of the string in that case is 
\begin{equation}
L= \sqrt{\frac{2\pi|a|}{T}},
\end{equation}
and hence the total mass associated with the ground state of the static string is given by
\begin{equation}\label{Massmeson}
M=m_{1} + m_{2} + \sqrt{\frac{|a|}{\alpha'}}
\end{equation}
For the $n$ excited state the relation between the mass and $n$  is given by
\begin{equation}
n=\alpha'(M-(m_1+m_2))^2 + a
\end{equation}









\subsection{Quantum Regge trajectories}
In the quantum string the classical rotating configuration is dressed by quantum fluctuations. In d=4 there are fluctuations along one transverse and one planar directions.
The eigenfrequencies $w_n$ of fluctuations when the endpoints carry the same mass $m$ are solutions of the transcendental equation
\begin{equation}
\tan(w_n L)=\frac{2mT w_n}{m^2 w_n^2-T^2}
\end{equation}
 The worldsheet hamiltonian associated with these fluctuations is characterized by the intercept.
\begin{equation}
a=-\frac{1}{w}<H_{ws}>
\end{equation}
The computation of the $<H_{ws}>$ requires a renormalization which we perform using a ``Cauchy-Casimir like''procedure\cite{Sonnenschein_quantization_2018}
The intercept gets a contribution also from the Polchinski-Stronminger term \cite{Polchinski:1991ax}. For the simplified case of two identical masses $m$ the intercept, which are  small, namely, $\frac{2m}{TL}<<1$ the intercept can be expanded in the form
\begin{equation}
a= a_t+a_p+ a_{PS}\sim 1-\frac{11}{6\pi}\left[\frac{2m}{TL}\right]^{1/2}+\frac{143}{240\pi}+ \left[\frac{2m}{TL}\right]^{3/2}
\end{equation}
The classical HMRT trajectory turns into a quantum one by

\begin{equation}\label{basic_regge}
    {J_{cl}}={\alpha^{'}E_{cl}^{2}}\rightarrow{J+w_n}={\alpha^{'}E^{2}}+a
\sim {J+n}={\alpha^{'}E^{2}}+a
\end{equation}
Since the eigenfruquencies  $w_n$ are close in value to $n$ we will use in what follows the latter expression.  


As was discussed above for the static string, the existence of the intercept implies a non-vanishing Casimir force, repulsive for $a<0$ and  attractive for $a>0$. The Casimir force modified the classical balancing between the tension and the centrifugal force. Quantum mechanically it reads


\begin{equation}\label{l_i}
    {\frac{T}{\gamma_{i}}}={{\frac{\gamma_{i}m_{i}\beta_{i}^{2}}{l_{i}}}-{\frac{2\pi a}{L^{2}}}}
\end{equation}

From relating the  angular velocities:

\begin{equation}\label{l_2}
    {\frac{\beta_{1}}{l_{1}}}={\frac{\beta_{2}}{l_{2}}}\rightarrow{l_{2}}={{\frac{\beta_{2}}{\beta_{1}}}l_{1}}
\end{equation}

The solution for $l_{1}$

\begin{equation}\label{string_length}
    {l_{1}}={{{\frac{\gamma_{1}}{2T}}    {\left({\gamma_{1}m_{1}\beta_{1}^{2}}+\sqrt{{\left({\gamma_{1}m_{1}\beta_{1}^{2}}\right)}^{2}-{\frac{8\pi aT}{\gamma_{1}\left({1+\beta_{2}/\beta_{1}}\right)}}}\right)}}}
\end{equation}

To obtain $\beta_{2}$ numerically:

\begin{equation}
    {\frac{T}{\gamma_{2}}}={\frac{\gamma_{2}m_{2}\beta_{1}\beta{2}}{l_{1}}}
    -{\frac{2a}{l_{1}^{2}(1+\beta_{2}/\beta_{1})^{2}}}
\end{equation}

From here we can estimate $L$ using the expression for $l_{1}$ and \eqref{l_2}.
\subsubsection{Repulsive Casimir force}
Ground state hadrons, mesons and baryons have zero orbital angular momentum. As was mentioned in (\ref{the_quantum_static_string}) these states do not collapse and  have non-vanishing length only provided that there is a repulsive Casimir force  that balances the string tension. A repulsive Casimir force means a negative intercept. Two questions immediately raise: (i) Is indeed the observed   experimental intercept always negative? (ii) Can we write down a theoretical string theory that admits a negative intercept. As for the experimental picture, the Casimir force is in fact not related to the intercept $a$ but rather to the following modified intercept 
\begin{equation}
\tilde a = a -S
\end{equation}
where $S$ is the spin ( not the total angular momentum $J$) of the hadron. This follows from the fact that $a$ is extracted from the relation between the hadron mass  $M$ and $J$, but what is relevant for the Casimir force  is the relation between $M$ and $L_{oam}$ the orbital angular momentum. For a string with no massive endpoints 
\begin{equation}
J = L_{oam}+S = \alpha^{'} M^{2} + a \rightarrow \qquad  L_{oam} = \alpha^{'} M^2 +a-S = \alpha^{'} M^{2} + \tilde a
\end{equation}
A similar situation occurs for a string with massive endpoints.
In \cite{Sonnenschein_2019_predictions}  the values of $\tilde a$ were extracted for all the hadronic trajectories and indeed it was found that 
\begin{equation}
\tilde a_{obs}<0
\end{equation}
As for the theoretical picture, the situation has not yet been fully clarified.  It is well known that for the ordinary critical open bosonic string $\tilde a =1$. In fact, as was firstly derived in \cite{Hellerman:2013kba},  and proved in a different way in \cite{Sonnenschein_quantization_2018}, $\tilde a =1$ also for any non-critical string at $d\neq 26$ dimendions upon incorporating the Liuville mode. For strings with massive endpoints the intercept $\tilde a $ depends on the endpoints masses. When the latter are equal  it is always positive but in an asymmetric setup, in particular when one side is massive and the other is massless, $\tilde a $ can be negative \cite{Sonnenschein_quantization_2018}. However, this of course cannot explain the situation for strings with the same endpoints as is the case for flavorless hadrons. 
Adding electric charges to the endpoints also modifies the value of the intercept $\tilde a $ but cannot make it negative \cite{Sonnenschein:2019bca}. Spins at the endpoints probably affect the intercept but that has not yet been determined. An option to get negative intercepts is to involve also fermionic strings. It is well known that for a Ramond  boundary condition a transverse mode contributes  $-1/24$ to the intercept. The problem with this possibility is that it affects the statistics of the hadron and so it can be relevant only for baryonic hadronic strings.

\subsection{Baryonic structures}\label{baryonic_structures}
 What is a baryon in holography? Since an end of a string connected to a flavor brane is a ``quark'', a baryon has to be built from $N_c$ strings. It turns out \cite{Witten:1998xy} that to  a $D_p$ brane that wraps 
a non-trivial p cycle with
a flux of an RR field of value $N_c$, there must be $N_c$ strings attached to it. Thus, such a structure in holography 
 provides the  baryonic vertex (BV). Correspondingly a baryon in a  confining holographic background  is a system that includes a BV and $N_c$ strings connecting it to various flavor branes. This idea was first implemented in a confining background in \cite{Brandhuber:1998xy}.
 A priori there could be different metastable configurations
of the BV and the  $N_c$  strings. The location of the BV in the holographic direction, and whole structure is determined minimizing the energy of the system \cite{Seki:2008mu}.

\begin{figure}[!ht]
  \centering
  \includegraphics[width=.6\columnwidth]{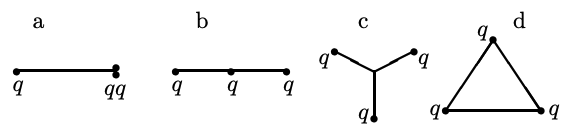}
  \caption{Baryonic configurations - (a) the quark-diquark model, (b) the linear structure, (c) the Y configuration and (d) the triangle.}\label{fig:baronic_structs}
\end{figure}

In HISH map described in figure (\ref{fig:Baryon}), for $N_c=3$ in four flat space-time  the baryon can have Y shape structure or it can be  a single string connected on one end to a quark and on the other end to a BV and a diquark. These configurations are two out of the four possible ones that connect three quarks. They are depicted in figure (\ref{fig:baronic_structs}).  

The different possible structures have been analysed in \cite{sharov1998_four_baryon_structs}, \cite{hooft2004minimal}, \cite{Sharov_2000} and \cite{petrov1998rotational_baryon_linear}. The conclusions were that the quark-diquark and delta structure are stable, whereas the Y model is unstable and with a small perturbation would collapse to the linear configuration (q-q-q) \cite{sharov1998_four_baryon_structs}. The linear configuration was analysed in \cite{sharov2001quasirotational} and was found to be unstable as well, but its periodic behavior prevent it from collapsing to the quark-diaqurk structure as was initially expected.

In the HISH picture, the delta and linear structure is not possible for baryons, since the model restricts BVs and anti-BVs connections to $N_{c}$, whereas quarks are not connected with more than one string.

The Y structure instability was deduced by assuming that all the quarks are identical. For baryons with quarks of different masses the analysis has to be repeated and a priori it is not obvious they are also unstable.
However, there is another argument in favor of the single string configuration.
Had  a symmetric  Y-shape string been stable, the baryon trajectory slope $\alpha_B$ should have been $\alpha_B= (2/3) \alpha_m$, where $\alpha_m$ is that of a meson. However, as was shown in \cite{Sonnenschein_2014_baryons}, the slope of the baryonic trajectory is within $5\%$ the same as that of a meson trajectory. 
This fact, that the slopes of baryons and mesons with similar endpoint masses, exclude the non-single string configurations also for non-symmetric baryons, namely with different quark masses. 


\subsection{The decay width of stringy hadrons}
The decays of  stringy hadrons  can involve 
 (i) breaking up  of a string, (ii) annihilation of endpoint particles, 
(iii) annihilation of a BV and  an anti-BV and (iv) electroweak decay of endpoint particles. For this paper only the first and third mechanisms will be relevant. 
 We briefly  summarize   in this section both mechanisms according to the analysis done in \cite{Sonnenschein_decay_width_2018} and \cite{Peeters_2006}.

\subsubsection{Hadronic decay through string breaking}\label{Hdtsb}
\begin{figure}[ht!]
			\centering
				\includegraphics[width=0.48\textwidth]{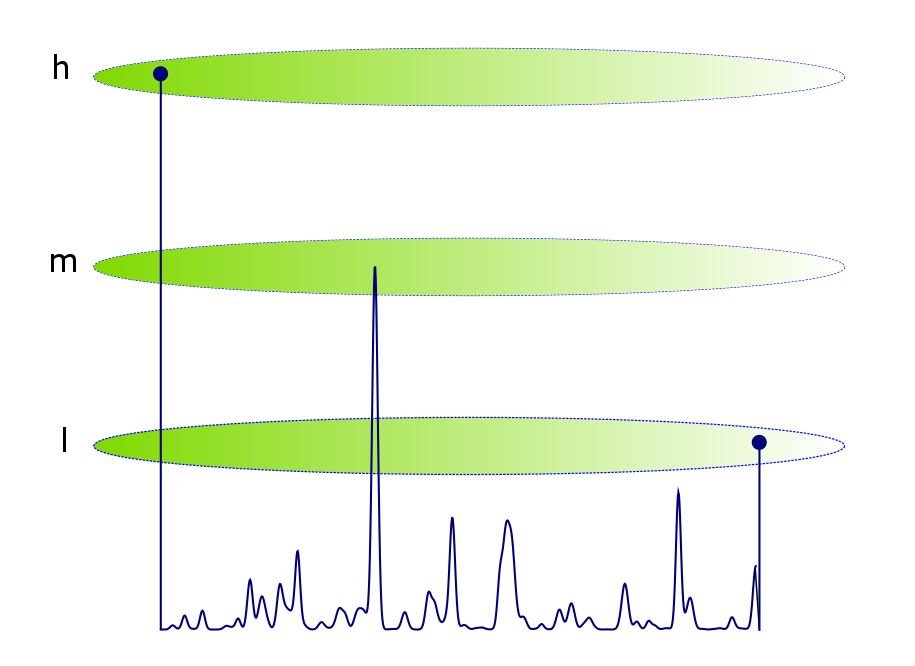}
				\caption{Quantum fluctuations of the horizontal segment of a heavy-light meson reach the medium flavor brane.}
				\label{quantumfluc}
	\end{figure}

An open string of Type I string theory can break apart at any point along the string. However, in holographic  backgrounds which are based on Type II string theory  strings cannot break in thin air,namely,  where the created endpoints are in the bulk but rather only when the later are on flavor branes. This can occur due to quantum fluctuations. 
This process is demonstrated in figure (\ref{quantumfluc}), where
fluctuations of a meson built from one heavy and one light endpoints reach a medium flavor brane.

\begin{figure}[ht!]
			\centering
				\includegraphics[width=0.90\textwidth]{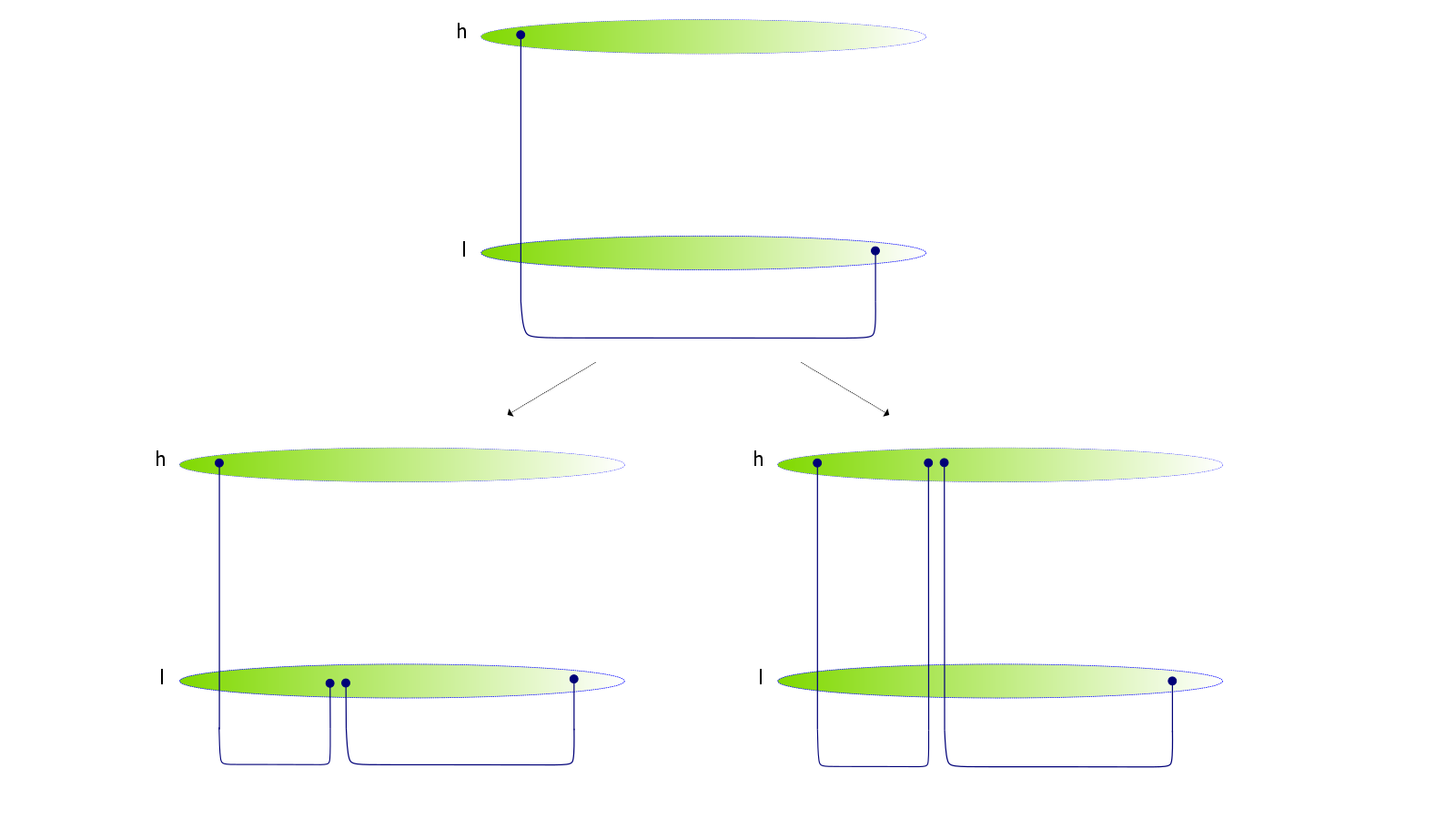}
				\caption{Possible decays of a heavy-light meson via light or heavy quark pair creation. The ratio between the two channels will include an exponential suppression term of the form \(e^{-C(m_h^2-m_l^2)/T}\).}
				\label{posholdecasy}
	\end{figure}

 In fact as shown in figure (\ref{posholdecasy}) there are more than one possibility for the decay of such a holographic 
meson. 

Quantum fluctuations can reach the light or heavier flavor branes. Of course, there
are also kinnematical constraints on whether or not a hadron can decay in a channel where
heavier quarks are created.
A calculation of the probability of the fluctuations of the horizontal segment of a hadronic
string to reach a flavor brane was first performed in [14].The result was found to be
\begin{equation}\label{CNN}
{\cal P}= Exp[-2\pi C \frac{m_q^2}{T}]
\end{equation}
where $m_q$ is the mass of the quark created at the breaking point and $C$ is a dimensionless coefficient of order 1 discussed in \cite{Sonnenschein_decay_width_2018}.

\begin{figure}[!ht]
  \centering
  \includegraphics[width=.6\columnwidth]{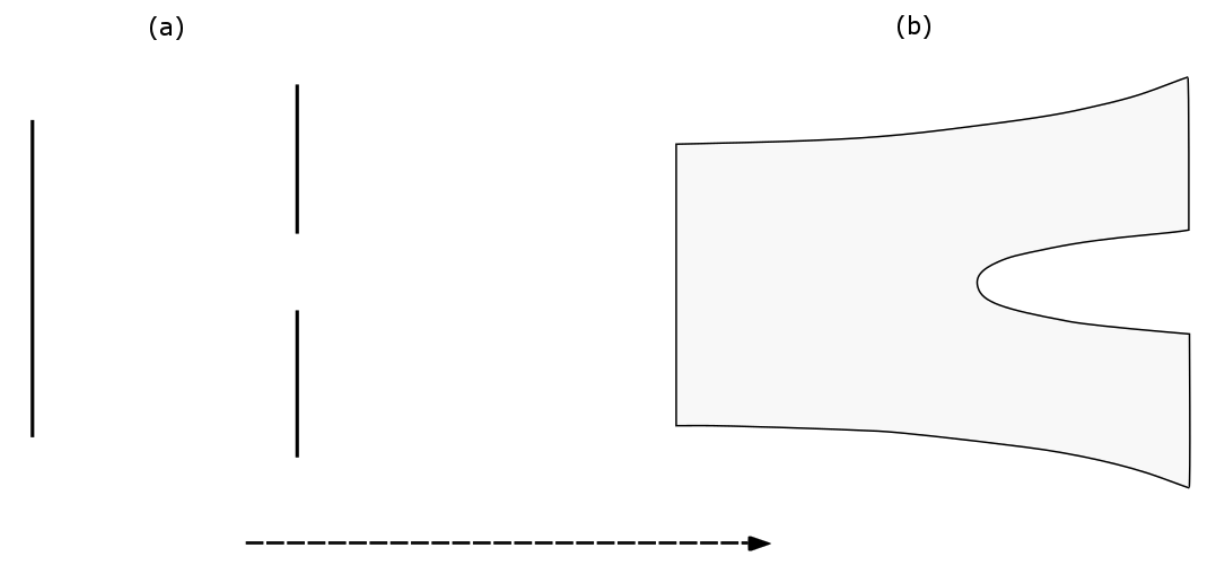}
  \caption{An open string breaks into two open strings. (a) Snapshots of before and after the
split. (b) The worldsheet diagram of the split. The arrow represents the time direction.}
\label{fig:string_tear}
\end{figure}

Once this probability is taken into account, one can compute  the width of the breaking mechanism as if it is an Type I open string. Moreover, since the profile of the holographic string is  of a flat horizontal string that stretches along a "wall" with additional vertical segments   we can consider it as a breaking of a string with massive endpoints in flat space-time. The decay width of a Type I open string  in flat space-time with no massive endpoints, described in figure (\ref{fig:string_tear}),  was computed in \cite{Dai:1989cp}and\cite{Mitchell:1989uc}.
This is done using the optical theorem  by computing the imaginary part of the string self energy diagram. In \cite{Dai:1989cp}  the corresponding  loop diagram  was mapped into a disk tree level diagram. The final result for the total decay width for long strings  takes the form 
\begin{equation}\label{width_tot}
\Gamma=\frac{\pi}{2} A TL (M,m_1,m_2,T) 
\end{equation}
namely, it admits  a linear dependence on the string length $L$, as expected since the string can break at any point. The constant $A$ is dimensionless, and proportional to the square of the string coupling .It is equal to the asymptotic ratio of $\frac{\Gamma}{L}$ for large $L$. It supposed to be universal for the decay of all stringy hadrons. 

The partial width to decay via a particular breaking  channel $i$  with a pair with a mass $m_{sep}$ and with the   probability given in (\ref{CNN})  takes the form
\begin{equation}\label{decay_tear_mq}
\frac{\Gamma_i}{\Gamma}= \Phi_i Exp[-2\pi C \frac{m_{q}^2}{T}]
\end{equation}
where $\Phi_i$ is the  phase space 
\begin{equation} \label{eq:phase_space}
\Phi(M,M_1,M_2) \equiv 2\frac{|p_f|}{M} = \sqrt{\left(1-(\frac{M_1+M_2}{M})^2\right)\left(1-(\frac{M_1-M_2}{M})^2\right)}\,.\end{equation}
and where $M$ is the mass of the parent hadron and $M_1$ and $M_2$ are the masses of the two daughter hadrons. The branching ratio formula is the HISH analog of the CNN formula of breaking of QCD flux tube \cite{Casher:1978wy}.
\subsubsection{Hadronic decay via annihilation}\label{Hdva}
As was mentioned above, in addition to decaying by string break up, a stringy hadron can decay also by annihilation either of the string endpoints or of a BV and an anti-BV. The former is relevant for flavorless mesons like quarkonia,  which in the holographic picture means that the two endpoints are on the same flavor brane. The latter is relevant, as will be discussed in section (\ref{Decay mechanisms of tetraquarks}), for the decay of tetraquarks when breaking is energenically not allowed. 

In \cite{Sonnenschein_decay_width_2018} the probability of the annihilation process was estimated to be 
\begin{equation}\label{decay_ann}
{\cal P}_{anh}\sim\sqrt{\frac{\pi}{2T_{av}}}e^{-T_{av}L^2/2} = \sqrt{\frac{\pi}{2T_{av}}}e^{\frac{-4(M-(m_1+m_2))^{2}}{9 T_{av}}}
\end{equation}
In the last expression we assumed that the string is not rotating 
For the decay of quarkonia $T_{av}$ is the tension  averaged  over the holographic direction from the wall to the location of the flavor brane.
For the annihilation of a BV anti-BV pair it is the ordinary tension, namely that at the vicinity of the wall. 
As discussed in \cite{Sonnenschein_decay_width_2018} the result of the annihilation of the endpoint quarks in the quarkonia case is a glueball that decay via breaking into two mesons.

\section{Exotic hadrons}\label{Exotic hadrons}
The toolkit for constructing exotic stringy hadronic configurations includes the following   building blocks: (i) Strings which can be  closed or open. In this note we discuss only the latter ones.  (ii)  ``Quarks", which are  massive  particles   carrying electric and  flavor charges ( but not color ones) and spins, located at the endpoints of open strings.  (iii) BV's and anti-BV's. These  string junctions can in general be connected to three strings  in a Y shape form, or to a diquark, namely  two short strings with a quark on their ends,  and a  regular string with a quark on its end or a tri-quark. Based on the baryonic  configurations found in nature, one may anticipate  that also for exotic hadrons only the second option exists. Never the less we check the two options for each exotic hadron separately.  

 With these building blocks  a variety of configurations can be constructed. 
 \begin{itemize}
 \item
 In general if  the number of BV's is the same as the number of anti-BV's namely, $n_{BV}=n_{\bar {BV}}$ the configuration has  the same number of external quarks and anti-quarks and hence it is non-baryonic.
 There are two types of such configurations:
 \item 
 Configurations with no quarks involved, namely generalizations of glueballs. Note that the basic glue ball is a simple closed string but, as is described in figure (\ref{fig:Glueball}), there are also ``glueballs" built from open strings. 
 \begin{figure}[!ht]
      \centering
      \includegraphics[width=.6\columnwidth]{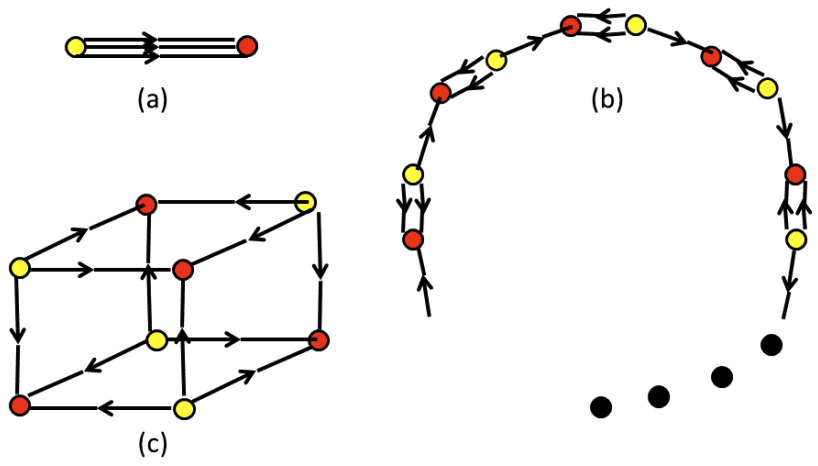}
      \caption{Gluballs - different configurations of BVs and anti-BVs. (a) Three strings connecting one BV and one anti-BV. (b) A planar polygon. (c) A cube.}
      \label{fig:Glueball}
    \end{figure}.
It is interesting to note that whereas the ordinary glueball that corresponds to a closed string has a tension that is twice the meson tension since for rotating strings it is a folded string \cite{Sonnenschein_2015_glueballs}, the glue ball depicted in (a) has a tension, that is three times, and in general $N_c$ times that of an ordinary meson tension. Thus, the mass of these objects will be linear in $N_c$ and hence claims of a similarity between $N_c=3$ and large $N_c$ are wrong for these states. 
\item 
Configurations that do include quarks at the ends of strings and hence are generalizations of mesons.   
 Examples  for such exotic hadrons are {\it tetraquarks} and  {\it hexaquarks} etc.
 \item
 If on one the other hand $n_{BV}=n_{\bar {BV}}+1 $ then $n_q-n_{\bar q} =3 $ and the configuration carries baryon number $B=1$. The simplest objects in this class are of course the baryons  which have more complicated cousines in the form of  {\it pentaquarks} and  {\it septaquarks} etc.

\item 
In a similar manner one can also construct objects with  $n_{BV}>n_{\bar BV}+1 $ which are multi-baryon states. There is a whole zoo of stringy multibaryons. It is easy to draw the corresponding figures.  For  example,  figure(\ref{Yss}) describes a di-baryon build from 3 BV's and one anti-BV. When the endpoint di-quarks contain two $u$, $d$ and $s$ the stringy configuration corresponds to the sexaquark\cite{Farrar:2017eqq}. In analogy to the collapse of the Y shape in the case of a baryon, one may expect also here a configuration with parallel double strings with 3 quarks on their ends.  
In this note we will not discuss these  multi-baryon states.
\item 
In the introduction we made a distinction between genuine exotic hadrons that can be described by a connected string diagram and molecules that take the form of disconnected diagram. 
 One can further  classify the genuine  exotic hadrons into two classes. The first will be referred as {\it not obviously  genuine exotic hadrons}. In this case  their decay products can also be the products of the decays of mesons and baryons. More specifically, two daughter mesons from a parent tetraquark that can be produced also by a decay of a meson, and a baryon and a meson that can be the result of a decay of a baryon.
 When the decay products cannot be a result of a decay of a meson, for tetraquarks, and a decay of a baryon for pentaquarks, these will be referred   as {\it  obvious genuine exotic hadrons} or for short {\it genuine exotic hadrons}. Genuine tetraquarks  can carry electric charges of +2 or -2 which clearly cannot be carried by a meson. Similarly genuine pentaquarks can carry charge of +3.
 This will be further elaborated in section (\ref{TetraquarksHMRTs}).
   In the following sections will discuss both types of  tetraquarks and  pentaquarks.

\begin{figure}[h!] \centering
					\includegraphics[width=.90\textwidth ]{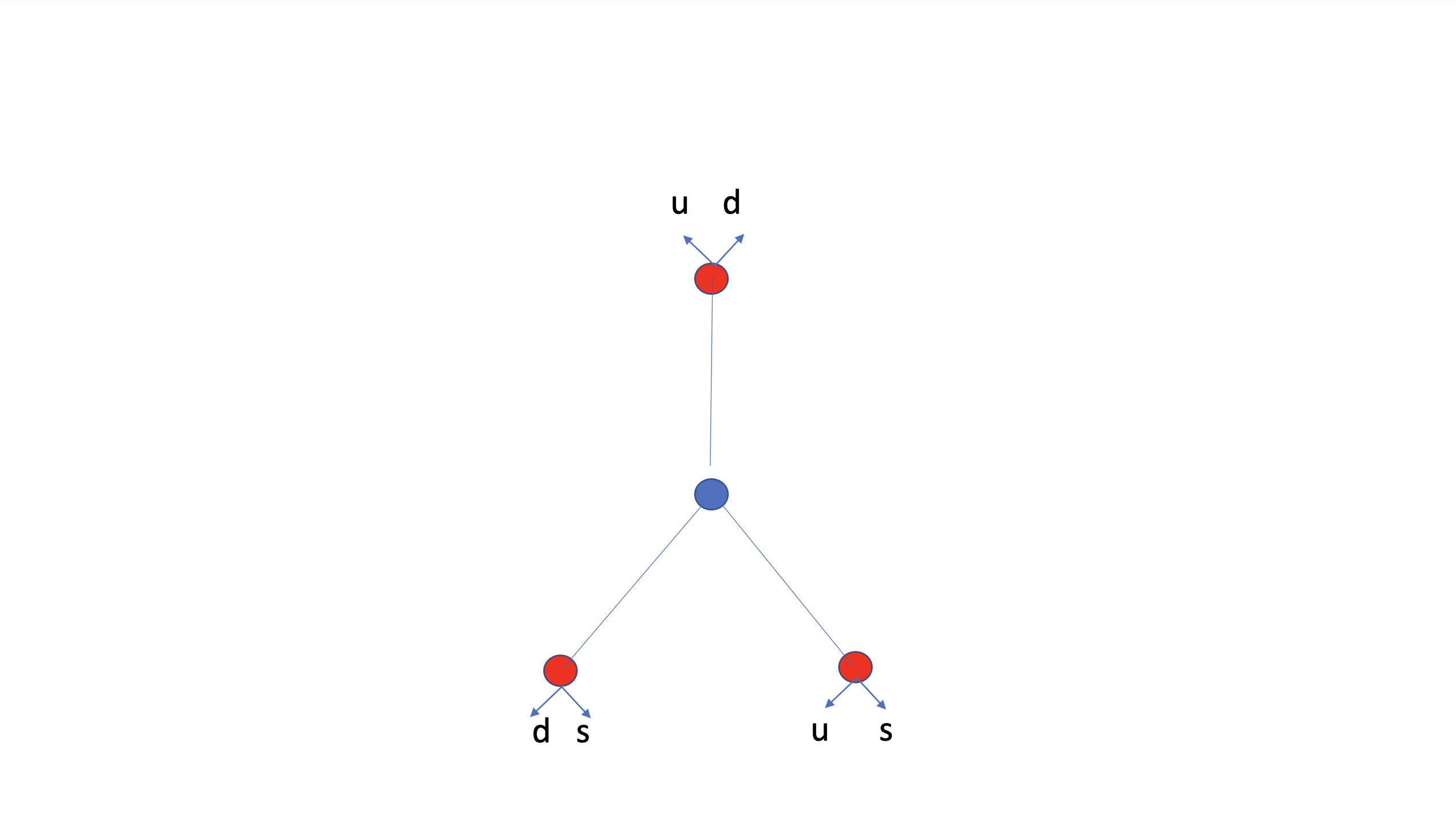}
					\caption{Sexaquark configuraiton. BV is denoted by red blobb and anti-BV by a blue one. \label{Yss}  }
\end{figure}

\item Another type of configurations to consider are exotic hadrons that decay through channels containing mesons that are a superposition of quark-antiquark compositions. These "mixed" products should be translated to mixed exotic hadrons that are also a superposition of a few configurations, such as $u\bar{u}u\bar{d}$ and $u\bar{d}d\bar{d}$, which is expected to decay to a product that includes for example $\pi^{0}$. The HISH picture does not deal directly with these mixture, but when confronting experimental data, we expect the only affect of it to be on the value of the endpoint masses for light quarks ($u$, $d$ and $s$). This means the results are not affected significantly when only $u$ and $d$ are involved, since $u$ and $d$ are not separated when we fit $m_{u/d}$. We will elaborate more on how we treated these cases in the \hyperref[section:HISH_data]{results section}.\label{section:mixed_configuration}

\end{itemize}
\section{Tetraquarks}\label{HISH_tetraquarks}

\begin{figure}[h!] \centering
					\includegraphics[width=.90\textwidth ]{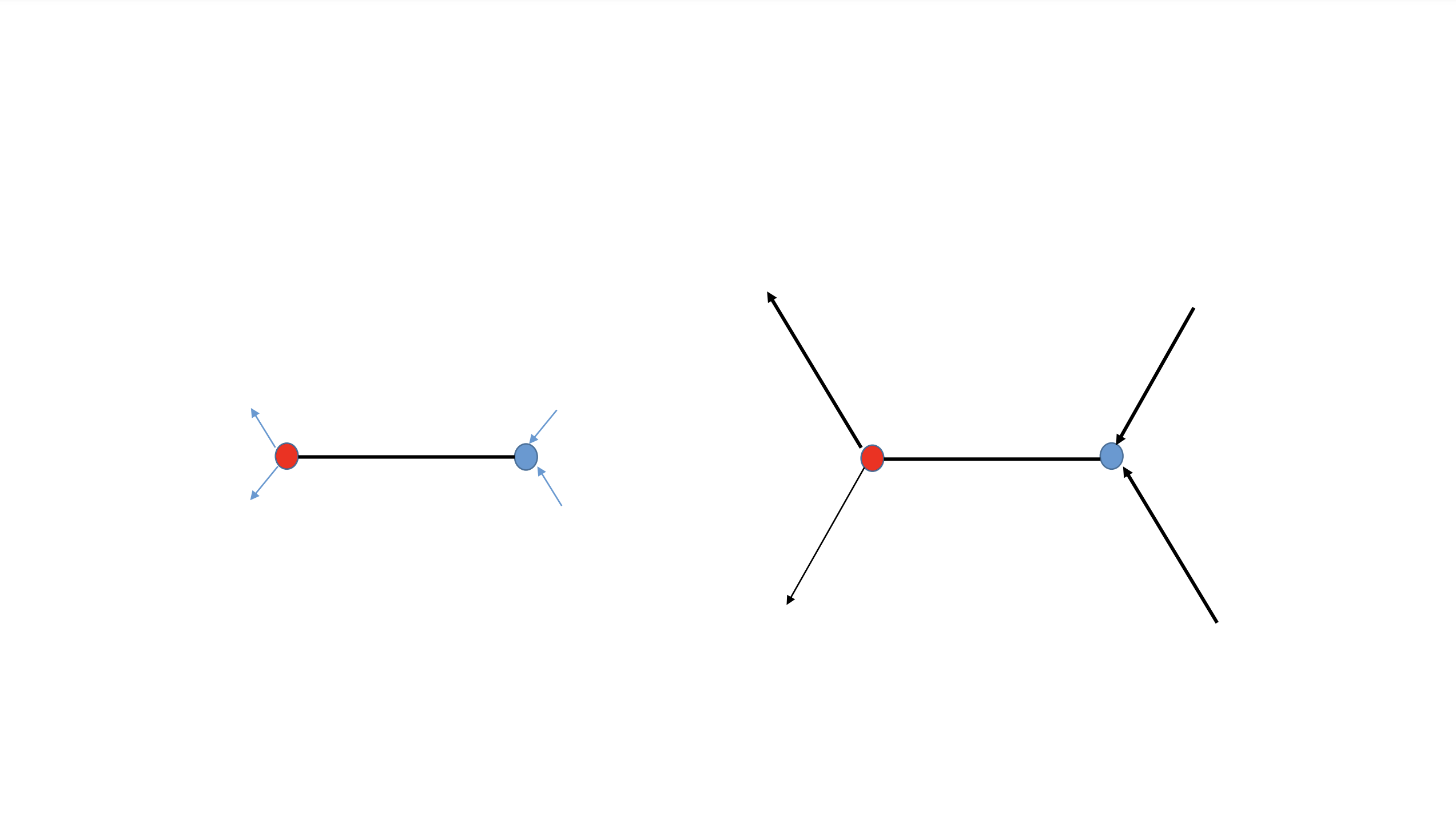}
					\caption{On the left a tetraquark built from a single string connecting a BV and an anti-BV. Adi-quark is attached to the BV and an anti-diquark to the anti-BV. On the right a 5 strings configuration with one string connecting the BV and the anti-BV\label{tetraquarkstructure}  }
\end{figure}

The simplest construction of a tetraquark is a single open string attached to a BV connected to a  diquark on one endpoint and an anti BV connected to an antidiquark at the other (\ref{tetraquarkstructure}) . As discussed above in (\ref{baryonic_structures}) the  stringy baryon   can be in principle in a Y shape or in a shape of a single string with a di-quark and a quark on both ends.  . In a similar manner the stringy tetraquark  can be a single string or  in the shape built from 5 strings as  described in figure (\ref{tetraquarkstructure}).

For the baryon the main reasons to adopt the single string picture were the theoretical stability of the configuration\} and \cite{Sharov_2000} and \cite{hooft2004minimal} and  more importantly  the fact that the slope  of the baryonic trajectories  is very close to that of mesons with similar endpoints. We will adopt the same criterion also for the tetraquarks for determining the preferred structure. It should be emphasized though  that  unlike for the baryons, for the tetraquarks  there have not been yet determined  trajectories  from which we would be able to examine the relations between their corresponding $\alpha'$  and that of the mesons.

\begin{figure}[!ht]
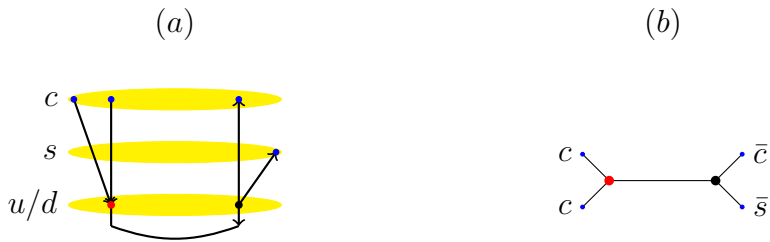

  \centering
  \begin{tabular}{ c m{6cm} m{6cm} }
    \begin{minipage}[!ht]{6cm}
    \begin{center}
      $(a)$
    \end{center}
    \end{minipage}
    &
    \begin{minipage}[!ht]{6cm}
      \begin{center}
      $(b)$
      \end{center}
    \end{minipage}
    \\\\
    \begin{minipage}{6cm}
      \stringytetraquark[0.7]{UD}{UD}{CBV}{CL}{CBV}{SR}
    \end{minipage}
    &
    \begin{minipage}{6cm}
      \hishtetraquark[0.35]{$\bar{c}$}{$\bar{s}$}{$c$}{$c$}
    \end{minipage}
    \\
  \end{tabular}
  \caption{Tetraquark - $(a)$ is a tetraquark with content $c\bar{c}c\bar{s}$ in the stringy picture. $(b)$ is the mapping of the tetraquark to the HISH picture. The red and black  vertices represents BV and  anti-BV respectively.}
  \label{fig:tetraquark_simple}
\end{figure}

\begin{figure}[!ht]
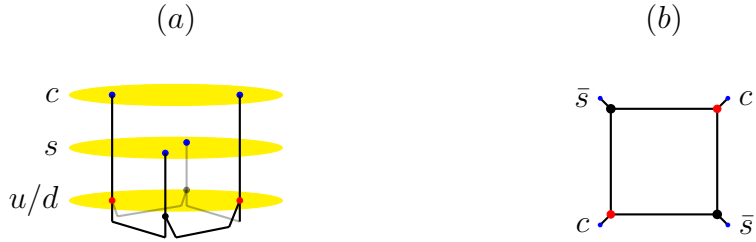

  \centering
  \begin{tabular}{ m{6cm} m{6cm} }
    \begin{minipage}{6cm}
    \begin{center}
      $(a)$
    \end{center}
    \end{minipage}
    &
    \begin{minipage}{6cm}
      \begin{center}
      $(b)$
      \end{center}
    \end{minipage}
    \\\\
    \begin{minipage}{6cm}
      \stringytetraquarkcomplex[0.7]{C}{C}{S}{S}
    \end{minipage}
    &
    \begin{minipage}{6cm}
        \hishtetraquarkcomplex[0.7]{$\bar{s}$}{$\bar{s}$}{$c$}{$c$}
    \end{minipage}
    \\
  \end{tabular}
  \caption{$(a)$ is an example of a holographic tetraquark built with two baryonic and two anti-baryonic vertices. $(b)$ is the HISH mapping of this tetraquark structure.}
  \label{fig:tetraquark_complex}
\end{figure}

In fig-\ref{fig:tetraquark_simple} an example of a $c\bar c c\bar s$ is depicted both in the holographic setup (a) and in the HISH one (b).
In fact the tetraquark may have an even
 more complex configuration as  the one in  Fig-\ref{fig:tetraquark_complex}.
 We will assume here that  it does not correspond to the low lying tetraquarks.
 
Following \cite {Sonnenschein_2017_4630} we refer to this configuration as the {\it V-baryonium}.
 We consider here  mainly  charmed tetraquarks which include  at least one charmed ( or anti-charmed ) quark and all the way to a $c\bar c c\bar c$ states. All together there are 125 different charmed tetraquarks including 40  with a single charmed quark and 40 with one anti-charmed quark, 10 with two charmed quarks and  10 with two anti-charmed quarks, 16 with $c\bar c$, 4  with $cc\bar c$, 4 with $\bar c\bar c c$ and one $c c \bar c \bar c$ state. The list of charmed tetraquarks  includes 5 symmetric states which have the form $q_1 \bar q_1 q_2,\bar q_2$. These state are obviously charge-less and with no flavor charge. There are 44 semi-symmetric states of the form $q_1\bar q_1 q_2\bar q_3$ which are  {\it not obviously  genuine tetraquarks} and  60 asymmetric   states of the form $q_1\bar q_2 q_3 \bar q_4$  which are {\it genuine tetarquarks} carrying flavor charges that cannot be carried by mesons and have electric charge $-2,-1,0,1,2$.

The {\it V-baryonium} tetraquarks are characterized by the following properties:

\begin{itemize}

\item
Since the {\it V-baryonium} is basically a  string, then like all other stringy hadrons it must have trajectories of higher excited states, one trajectory  of states with higher angular momentum, and one of higher radial excitation number. Here again we refer to such a  trajectory as a HISH modified Regge trajectory (HMRT) \cite{Sonnenschein:2018fph}. Since the mesons, baryons, and {\it V-baryonium} tetraquarks can all be seen as a single string with massive particles on its end,  the function describing the trajectories is the same for the different types of hadrons. The differences  between all of them are the  masses on the endpoints, the intercepts and whether there  is a BV or a pair of BV and anti-BV at the enedpoints.

\item The ground state which has  vanishing angular momentum has a mass given by
\begin{equation}\label{Tetramass}
M_T= M_{diquark_L} + M_{BV_L} + M_{diquark_R} + M_{BV_R} +\sqrt{\frac{|a_T|}{\alpha'}}
\end{equation}
where $M_{diquark_L}$, $M_{diquark_R}$, $M_{BV_L}$ and $M_{BV_R}$  are the masses of the left and right di-quarks and BVs respectively.



\end{itemize}

In  holography for the  case that the single string is preferred over the Y shapes, two out of the three strings that connect to a BV  are spanned  only along the holographic direction and not along the ordinary space direction. To be more precise their span along the latter direction is much shorter than the string that connects the BV and anti-BV.  
Thus the  endpoints in this picture are now composed from the BV (or anti-BV) contribution with the addition of the diquark contribution neglecting the contribution of the short strings.
Since the structure of the tetraquarks resembles that of the mesons, we expect the tetraquark $\alpha^{'}$ to resemble that of mesons with similar quark content (e.g $T_{c\bar{c}u\bar{u}}$ should have similar $\alpha^{'}$ to $\psi$).

The location of the BV along the holographic direction is determined  by the requirement of minimal energy. In figure (\ref{fig:tetraquark_simple}) it is placed on the light-quark flavor brane but it could also reside on the ``wall" or on the other flavor branes and in fact even between them.  The location of the BV was discussed in details in \cite{Seki:2008mu}. Here we briefly summarize  the issue. The total energy of the system is the sum of the energy of the baryonic vertex,which is a function of its location, and the energy associated with the strings that connect it to the flavor branes.
\begin{equation}
E_{sys}= E_{BV}(u_{BV}) + \sum_{i=1}^2 \int_{u_{BV}}^{u_{f_i}} T(u) du 
\end{equation}
where  $u_{BV}$ and $u_{f_i}$ are the locations of the BV and the $i^{th}$ flavor brane respectively.  $E_{BV}(u)$ is the energy of the BV wrapped brane. It is a function of the location $u$ but this dependence is not universal and varies in the different holographic models. $T(u)$ is the tension. Its dependence on $u$ is also model dependent. 
From previous fits to mesons and baryons\cite{Sonnenschein:2018fph} we will assume in this paper that the BV is located on the wall so that 
\begin{equation}
E_{sys}= M_{BV} + M_{diquark}= M_{BV}+\sum_{i=1}^2 m_{q_i} 
\end{equation}
where $m_{q_i}$ are the  holographic masses of the two quarks that built the di-quark. This means that we simply assume that the mass of the di-quark is the sum of the masses of its two quarks.
\subsection{The spin structure of tetraquarks }
It is well known that mesons with a give quark anti-quark content appear  with two possible spins. There are (pseudo) scalar mesons of spin $s=0$ and vector mesons of spin $s=1$. Correspondingly there are separate HMRT associated with the former and later spins. For instance there are different trajectories for $K$ and $K^*$ of strange mesons with spin zero and one respectively, for $D$ and $D^*$, $B$ and $B^*$
and so on. It was found out in \cite{Sonnenschein:2018fph} that the trajectories associated with the pseudo scalar and vector trajectories are different.

In a similar manner baryons can have for their ground states ( with zero orbital angular momentum ) spin $1/2$ and spin $3/2$ for instance the nucleons and the deltas.  Again here as well, there are different trajectories associated with the different spins with different intercepts.

The HISH  tetraquarks are  described by a string that has a di-quark and an anti di-quark on its ends. Thus, the spin of each end can be either $0$ or $1$ and the total spin of the system can take the values of $0,1$ and $2$. Given a tetraquark $T$ of spin zero there are also a spin one tetraquark that we denote by $T^*$ and a tetraquark of spin 2 $T^{**}$. We want to emphasize that we discuss here the spin and not the total angular momentum.  Naturally, we anticipate that there are  separate HMRTs  for the three different spins  with different corresponding intercepts. It is obvious from (\ref{Tetramass}) that the mass difference between  $M_{T^*}$ and $ M_T$  and similarly with $M_{T^{**}}$
\begin{equation}
M_{T^*}- M_T = \sqrt{\frac{|a_{T^*}|}{\alpha'}}- \sqrt{\frac{|a_{T}|}{\alpha'}}
\end{equation}

\subsection{Tetraquark Creation}\label{tetraquark_creation}
In this section we will heuristically describe the mechanism through which mesons may decay into tetraquarks in the HISH picture. In particular we will state the process for the formation of the X(3350) through B meson decay (in which this state was discovered), as well as the possible mechanisms in which baryon-antibaryon pairs are created from a meson. However, quantifying the decay width in these types of processes will require further research.

As mentioned above, the search for tetraquarks that would comply with the HISH model mainly includes searches for resonances that decay predominantly to baryon-antibaryon pairs and are above the mass threshold for this type of decay. This is since in the HISH model the string breaking is the main mechanism through which excited hadrons can decay, assuming they are above the mass threshold. While these are rare to find in the PDG data, there are quite a few baryonic decays that might be explained by an intermediate tetraquark state.

\subsubsection{Excited Meson Decay}
\begin{figure}[!ht]
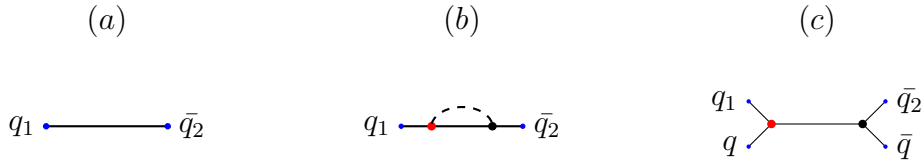

  \centering
  \begin{tabular}{ m{.250\textwidth} m{.250\textwidth} m{.250\textwidth} }
    \begin{minipage}{.250\textwidth}
    \begin{center}
      $(a)$
    \end{center}
    \end{minipage}
    &
    \begin{minipage}{.250\textwidth}
      \begin{center}
      $(b)$
      \end{center}
    \end{minipage}
    &
    \begin{minipage}{.250\textwidth}
      \begin{center}
      $(c)$
      \end{center}
    \end{minipage}
    \\\\
    \begin{minipage}{.250\textwidth}
      \hishmeson[0.4]{$q_{1}$}{$\bar{q_{2}}$}
    \end{minipage}
    &
    \begin{minipage}{.250\textwidth}
       \hishmesondecay[0.4]{$q_{1}$}{$\bar{q_{2}}$}
    \end{minipage}
    &
    \begin{minipage}{.250\textwidth}
        \hishtetraquark[0.3]{$\bar{q_{2}}$}{$\bar{q}$}{$q_{1}$}{$q$}
    \end{minipage}
    \\
  \end{tabular}
  \caption{the decay process of an excited meson to a tetraquark in the HISH picture (from $(a)$ to $(c)$). A BV-anti BV pair loop is created, and one of the strings breaks with a pair of light quark-antiquark (u, d and s with lower probability).}
  \label{fig:meson2tetraquark}
\end{figure}

Since a tetraquark contains a BV and an anti-BV, in the simplest decay, the meson has to go through a 1 loop BV-anti-BV diagram where one of the strings connecting the loop would break (\ref{fig:meson2tetraquark}).

Since the meson transformation to tetraquark includes energy conversion to the tetraquark mass, the meson has to be excited. For the specific decay $B\rightarrow(\Lambda_{c}^{+}\bar{p})_{s}\pi$ we expect the B meson to first go through a weak decay where two mesons would form, then one of the remaining mesons (the excited one) would go through a decay to a tetraquark, see fig-(\ref{fig:meson2tetraquark}).

\begin{figure}[!ht]
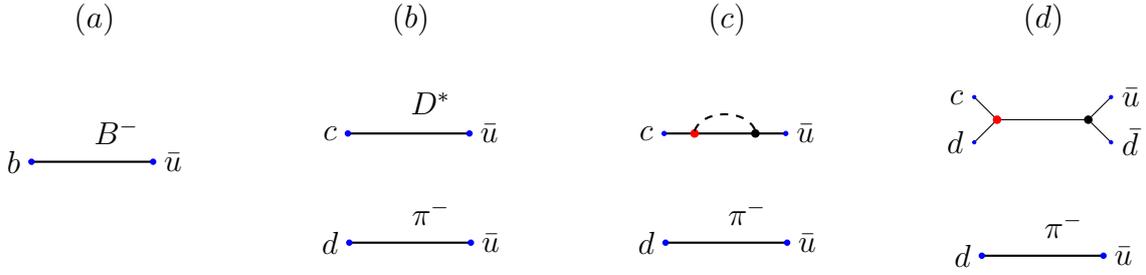

  \centering
  \begin{tabular}{ m{.220\textwidth} m{.220\textwidth} m{.220\textwidth} m{.220\textwidth} }
    \begin{minipage}{.220\textwidth}
    \begin{center}
      $(a)$
    \end{center}
    \end{minipage}
    &
    \begin{minipage}{.220\textwidth}
      \begin{center}
      $(b)$
      \end{center}
    \end{minipage}
    &
    \begin{minipage}{.220\textwidth}
      \begin{center}
      $(c)$
      \end{center}
    \end{minipage}
    &
    \begin{minipage}{.220\textwidth}
      \begin{center}
      $(d)$
      \end{center}
    \end{minipage}
    \\\\
    \begin{minipage}{.220\textwidth}
      \hishmeson[0.4][$B^{-}$]{$b$}{$\bar{u}$}
    \end{minipage}
    &
    \begin{minipage}{.220\textwidth}
       \hishmeson[0.4][$D^{*}$]{$c$}{$\bar{u}$}
       \hishmeson[0.4][$\pi^{-}$]{$d$}{$\bar{u}$}
    \end{minipage}
    &
    \begin{minipage}{.220\textwidth}
        \hishmesondecay[0.4][$D^{*}$]{$c$}{$\bar{u}$}
        \hishmeson[0.4][$\pi^{-}$]{$d$}{$\bar{u}$}
    \end{minipage}
    &
    \begin{minipage}{.220\textwidth}
    \hishtetraquark[0.3]{$\bar{u}$}{$\bar{d}$}{$c$}{$d$}
    \hishmeson[0.4][$\pi^{-}$]{$d$}{$\bar{u}$}
    \end{minipage}
  \end{tabular}
  \caption{The decay process of a B meson to a tetraquark in the HISH picture. First a $D^{*}$ meson is created and $\pi^{-}$ through a weak decay of the $b$ quark to $c\bar{u}d$. Then the $D^{*}$ decays to a tetraquark through the process described in \ref{fig:meson2tetraquark}.}
  \label{fig:Bmeson2tetraquark}
\end{figure}

The total probability to that type of decay is expected to be:
\begin{equation}\label{meson_tetraquark_decay}
    P_{tetra}=V_{bc}*V^{*}_{ud}*P_{BV-anti-BV}*P_{tear}
\end{equation}

Where $V_{bc}$ and $V^{*}_{ud}$ are the CKM matrix elements and $P_{tear}$ is described at \ref{Hdtsb}. In future work, we will try to estimate $P_{BV-anti-BV}$ (the probability for the creation of a BV-antiBV vertices) by analysing the baryonic decays of the $B$ mesons.

\subsubsection{Mesons Decay Through Scattering}
\begin{figure}[!ht]
  \centering
  \includegraphics[width=.6\columnwidth]{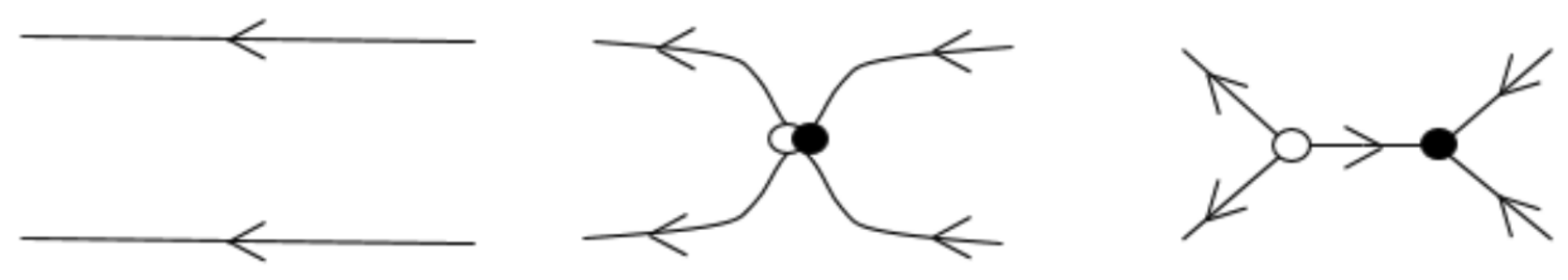}
  \caption{A scattering of two mesons to a tetraquark.}
  \label{fig:mesonsScattering2tetraquark}
\end{figure}

Another option for the formation of a BV-anti-BN vertex is through meson-meson scattering. In this process we need to calculate the S-Matrix element of a two mesons as the $|in>$  state, and a tetraquark as the $<out|$ state. The diagrams of the process are in Fig-(\ref{fig:mesonsScattering2tetraquark}). The details of this process were worked out in \cite{Nadav}.

\begin{figure}[!ht]
  \centering
  \includegraphics[width=.6\columnwidth]{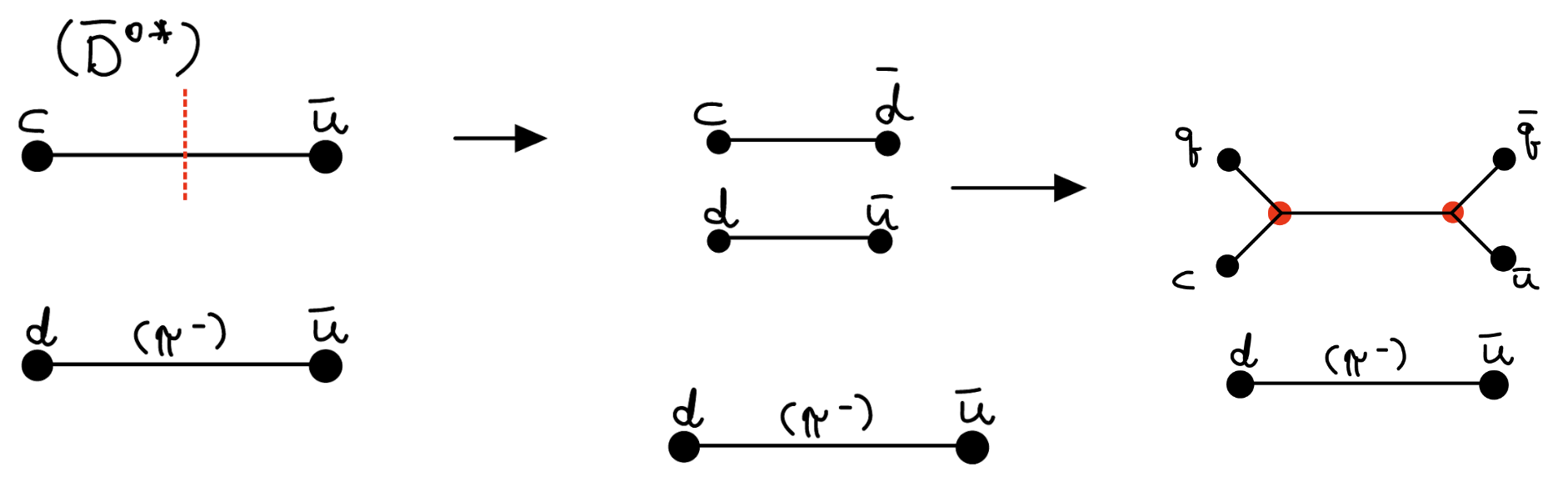}
  \caption{The decay of a B meson to a meson and a tetraquark through meson pair scattering (left to right). The B meson first decays to D*0 and - through weak decay of the b quark. Then the excited D*0 decay again to $\pi^-$ and D+ . The latter then scatter to a tetraquark.}
  \label{fig:B2X(3350)}
\end{figure}

For the specific case that we are reviewing, the process diagrams would look as in (\ref{fig:B2X(3350)}).

\subsection{Decay mechanisms of tetraquarks}\label{Decay mechanisms of tetraquarks}
Is it possible that a tetraquark state is stable against strong decays? For a tetraquark to be stable all its possible channels of decay should be forbidden. Above in sections \ref{Hdtsb} and \ref{Hdva} the mechanisms of string breaking and annihilation of a BV and an anti-BV were analyzed. Assuming that these are the only possible decays, stability of a tetraquark can occur only provided that its mass is smaller than the masses of the decay products. For the annihilation process the condition of stability takes the form
\begin{equation}
M_T-( M_{m_a}+ M_{m_b}) <0
\end{equation}
where the two mesons $m_a$ and $m_b$ are the  mesons strings that result from the annihilation which can be either with endpoints $(q_L^1 \bar q_R^1) and ((q_L^2 \bar q_R^2)$ or $(q_L^1 \bar q_R^2) and ((q_L^2 \bar q_R^1)$. 
Using (\ref{Tetramass}) for the
mass of the tetraquark and  (\ref{Mesonmass}) for the mass of a meson we get
\begin{equation}
M_T-( M_{m_a}+ M_{m_b})=  M_{{BV}_L} +M_{{BV}_R}\sqrt{\frac{|a_T|}{\alpha'_T}}- \sqrt{\frac{|a_{m_a}|}{\alpha'_{m_a}}}-\sqrt{\frac{|a_{m_b}|}{\alpha'_{m_b}}}
\end{equation}
where $M_{{BV}_L}$ and $M_{{BV}_R}$  are the masses of the BV and anti-BV respectively and  $(a_T,a_{m_a}, a_{m_b})$ and $(\alpha'_T,\alpha'_{m_a}, \alpha'_{m_b}) $ are the intercepts and slopes  associated with a tetraquark,  and the two  mesons respectively.
Note that the masses of the quarks cancel out in this expression. 
In fact, we can partially relate this difference to measured quantities. Using a baryon that includes a di-quark identical to the one in the tetraquark and a third quark denoted by $\tilde q_L$ and similarly for an anti-baryon on the right   we find

\begin{equation}
M_T-( M_{m_a}+ M_{m_b})=  +\sqrt{\frac{|a_T|}{\alpha'_T}}- (\sqrt{\frac{|a_{B_L}|}{\alpha'_{B_L}}} +\sqrt{\frac{|a_{B_R}|}{\alpha'_{B_R}}}) + (M_{B_L}+M_{B_R}) -(m_{\tilde q_L}+m_{\tilde q_L})-( M_{m_a}+ M_{m_b})
\end{equation}

\begin{figure}[!ht]
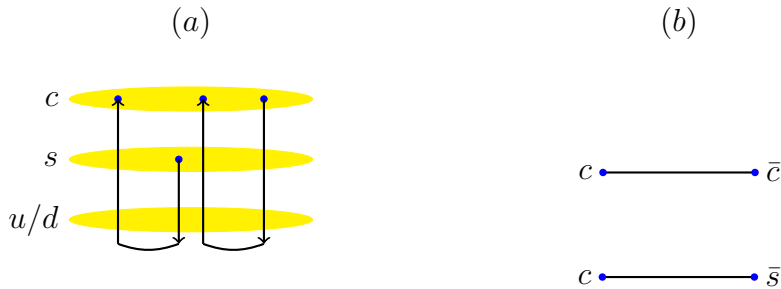

  \centering
  \begin{tabular}{ c m{6cm} m{6cm} }
    \begin{minipage}[!ht]{6cm}
    \begin{center}
      $(a)$
    \end{center}
    \end{minipage}
    &
    \begin{minipage}[!ht]{6cm}
      \begin{center}
      $(b)$
      \end{center}
    \end{minipage}
    \\\\
    \begin{minipage}{6cm}
      \stringymesonspair{CBV}{SB}{CBV}{CB}
    \end{minipage}
    &
    \begin{minipage}{6cm}
      \hishmeson[0.5]{$c$}{$\bar{c}$}
      \hishmeson[0.5]{$c$}{$\bar{s}$}
    \end{minipage}
    \\
  \end{tabular}
  \caption{Tetraquark annihilation - $(a)$ is a demonstration of the outcome of a tetraquark BV-antiBV annihilation with content $c\bar{c}c\bar{s}$ in the stringy picture, which results in two mesons. $(b)$ is the mapping of the mesons to the HISH picture.}
  \label{fig:tetraquark_annihilation}
\end{figure}

We can use in principle  this relation to check the stability and also in case that the tetraquark is unstable to strong disintegration and  to predict its ground state mass. However, for that we will need reliable information about the various intercepts that will be acquired when more tetraquarks are identified.
We demonstrate the decay  via annihilation  for the case of    a tetraquark candidate with content of $(cc\bar c \bar s)$
in figure
(\ref{fig:tetraquark_annihilation}). On the left is the holographic outcome and on the right the outcome in the HISH model.

\begin{figure}[!ht]
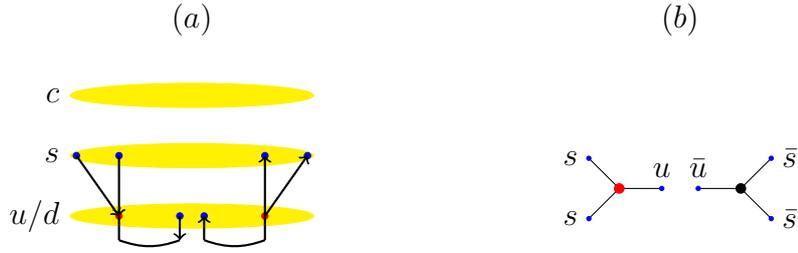

  \centering
  \begin{tabular}{ c m{6cm} m{6cm} }
    \begin{minipage}[!ht]{6cm}
    \begin{center}
      $(a)$
    \end{center}
    \end{minipage}
    &
    \begin{minipage}[!ht]{6cm}
      \begin{center}
      $(b)$
      \end{center}
    \end{minipage}
    \\\\
    \begin{minipage}{6cm}
      \stringytetraquarkbreak{UD}{UD}{SBV}{SL}{SBV}{SR}{UD}{UD}
    \end{minipage}
    &
    \begin{minipage}{6cm}
      \udecay[0.4]{s}{s}{s}{s}
    \end{minipage}
    \\
  \end{tabular}
  \caption{Tetraquark tear decay - $(a)$ is a demonstration of the tetraquark decay through string tear to two baryons with content $s\bar{s}s\bar{s}$ in the stringy picture. $(b)$ is the demonstration of the decay in the HISH picture.}
  \label{fig:tetraquark_string_tear}
\end{figure}

As was explained in (\ref{Hdtsb}) the most natural decay of a stringy hadron is a breakup of the string. This applies also to the string of the {\it V-baryonium}   tetraquark. However, often the mass of the parent tetraquark is below the threshold for a string breakup. Obviously the condition for a breakup is
\begin{equation}
M_T> M_{B_L} + M_{B_R}
\end{equation}
where $M_{B_L}$  is the mass of the  baryon produced at the  left  of the string and $M_{B_R}$ is the mass of the anti-baryon at the right of the string.  The total decay width of a tetra quark to any possible pair of a baryon and an anti-baryon is given in (\ref{width_tot})  and the partial decay width into a particular channel is given by (\ref{decay_tear_mq}). The process of a decay via  a breakup is drawn in figure (\ref{fig:tetraquark_string_tear}) for the case of a tetraquark candidate $(s,s,\bar s, \bar s)$ decaying into $\Xi^0$ and $\bar{\Xi^0}$

In the breaking mechanism, the content of the di-quark and anti-di-quark of the original tetraquark is the same as the that of the outcome baryon and anti-baryon. There are however, decays of tetraquarks to a baryon anti-baryon pair with quarks which are lighter than the ones of the exotic hadron. There are several decay processes that can yield 
such a situation:

(i) A  decay mechanism  through  an annihilation of a quark which is part of the di-quark and an anti-quark  which is part of the anti-di-quark. This will yield an exotic meson which includes a BV and and anti-BV connected with two strings.  This state may further decay into  a tetraquark with content which is different than the original one, and then depending on its mass decaying either via a breaking or via  an annihilation of the BV and anti-BV. 
   This mechanism is relevant only for not obviously  genuine states, which includes a  pair of quark anti-quark of the same flavor.  

(ii) One or more weak decays of a quark of heavy flavor ( like c or b) to a lighter quark ( like s or u/d), or similarly for an anti-quark.  After the weak decay the tetraquark will include a di-quark ( or an anti- diquark) with flavor lighter than the original one that can afterwards decay via breaking or annihilation. 
\section{Pentaquarks}\label{HISH_pentaquarks}
Tetraquarks are the simplest exotic hadrons that relate to mesons since they are of vanishing baryon number. In a similar manner there are also exotic hadrons related to baryons  which carry baryon number $B=1$. These are referred to as pentaquarks since they involve four quarks and one anti-quark. 

The total number of  possible configurations of pentaquarks (excluding  the top quark) is $350$ in total - there are five possibilities for the antiquark. Then for four different quarks there are 5 possible configurations to a total of $25$. For $\bar{q}_{1}q_{2}q_{2}q_{3}q_{4}$ there are $150$ configurations. For $\bar{q}_{1}q_{2}q_{2}q_{3}q_{3}$ we get $50$ configurations, for $\bar{q}_{1}q_{2}q_{2}q_{2}q_{3}$ there are $100$ configurations and there are another $25$ configurations for $\bar{q}_{1}q_{2}q_{2}q_{2}q_{2}$.

In this paper we put emphasis on charmed configurations. There are $210$ possible pentaquarks with $c/\bar{c}$ - $70$ with $\bar{c}$, $80$, $40$, $16$ and $4$ with one $c$, two, three and four respectively. Out of $210$ possible charm configurations, there are $115$ that are genuine (don't contain a quark and an antiquark with the same flavor) and another $50$ that are semi-genuine (contain a heavy pair of quark and antiquark).

\subsection{Pentaquark configurations}
A tetraquark is built from a mesonic string by adding a BV and anti-BV at its two endpoints. One can similarly construct a pentaquark by adding such a pair to the baryonic string. However, it turns out that for pentaquarks there is another possibility which does not include the adding of such a pair.
A BV by construction must have a net number of $N_c=3$ strings coming out of it. This does not imply that there are only 3 strings attached to it since the total number can be preserved also if there are additional pairs of string anti-string (opposite orientation) connected to it. The requirement is that 
$N_{out}-N_{in}=N_c$ where $N_{out}$ and $N_{in}$ are the outcome and incoming strings respectively.

Thus, the two simplest possible structures of a pentaquark are the following:

\subsubsection{ A single BV}

 A pentaquark structure based on a single baryonic vertex. One adds to the BV an incoming and an outgoing strings  with an antiquark and a quark attached to their  endpoints respectively.  A priori, there are several possible  structures depicted in figure (\ref{pentasetup}):
    \begin{figure}[!ht]
      \centering
      \includegraphics[width=1.0\columnwidth]{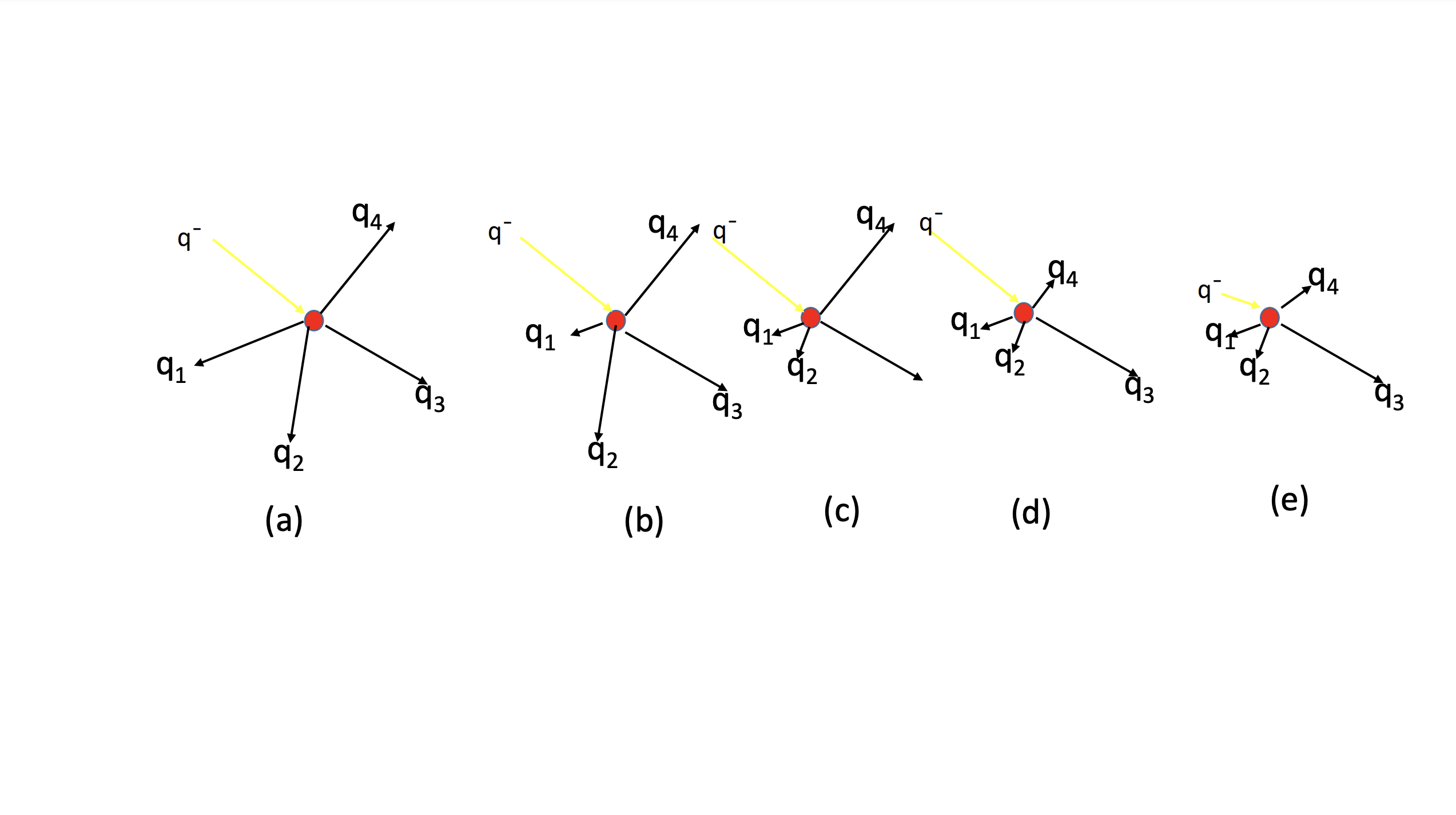}
      \caption{ Possible setups of pentaquarks: (a) 5 strings attached to the BV, 4 with outward orientation (quark) and one inward (anti-quark) (b) After shrinking of one string, the quark $q_1$ is attached to the BV. (c) After two shrinkings a diquark is attached to the BV. (d) a triquark attached to the BV. (e) A diaquark and an pair of a quark  antiquark.  The BV, a string attached to a quark and a string attached to an anti-quark are drawn in red, black and yellow respectively.\label{pentasetup} }
    \end{figure}.

    (a) The analog of   the Y shape   of a baryon, a configuration with a BV with 4 outcoming  strings  attached to a quark and one inward string attached to an anti-quark.  All the strings are of non-trivial lengths   (b) A structure with one string shrank to zero size so  the quark $q_1$ is attached to the BV.  (c) A diquark ($q_1,q_2)$ is attached to the BV  in addition to an outgoing and incoming strings. (d) The BV is attached to a triquark and an incoming string (e) In the last option after a shrinkage of the incoming string associated with the anti-quark there is only one string $q_3$ that stretches in ordinary space  and a diquark and a pair of quark anti-quark are attached to the BV. In the holographic picture The quarks $q_1,q_2,q_4$ and antiquark $\bar q$ are connected with strings that stretch only along the holographic direction. 

      Even though we have not   performed a stability analysis of these configuration, we assume  following \cite{sharov1998_four_baryon_structs}, \cite{sharov2001quasirotational} and \cite{hooft2004minimal} that the configuration (e) with  only one string along a space direction  will remain with finite length. On one end of this string there is an ordinary quark and on the other end a BV and  a composite object that can be viewed as a diquark plus a quark anti-quark pair or a triquark plus an antiquark.  From the point of view of the HISH the composite object is  relevant in determining  the total mass, electric charge, flavor charges and spin of this end of the string. 
    
    Another issue relevant to this construction is whether the  anti-quark $\bar q$  carries the same flavor  as one of the quarks $q_1,q_2,q_4$ .
    From the condition on the number of  strings that connect  to a given BV this  we cannot determine the answer to this question.  This issue is important for the possible decay channels of the petaquark as will be discussed in 
    (\ref{Decayspenta}).
    
    In a similar manner to the fact that for a baryon there are several possibilities of what is the single quark and what quarks form the di-quark, also for  
     a pentaquark of a general quark content of $(q_1,q_2, q_3,q_4,\bar q)$, there are four possibilities for the single quark which  is on the other side of the string. 
     Currently we do not have a way to determine which option out of these four is preferable and we will assume that all of them are possible.

\begin{figure}[!ht]
  \centering
  \includegraphics[width=.6\columnwidth]{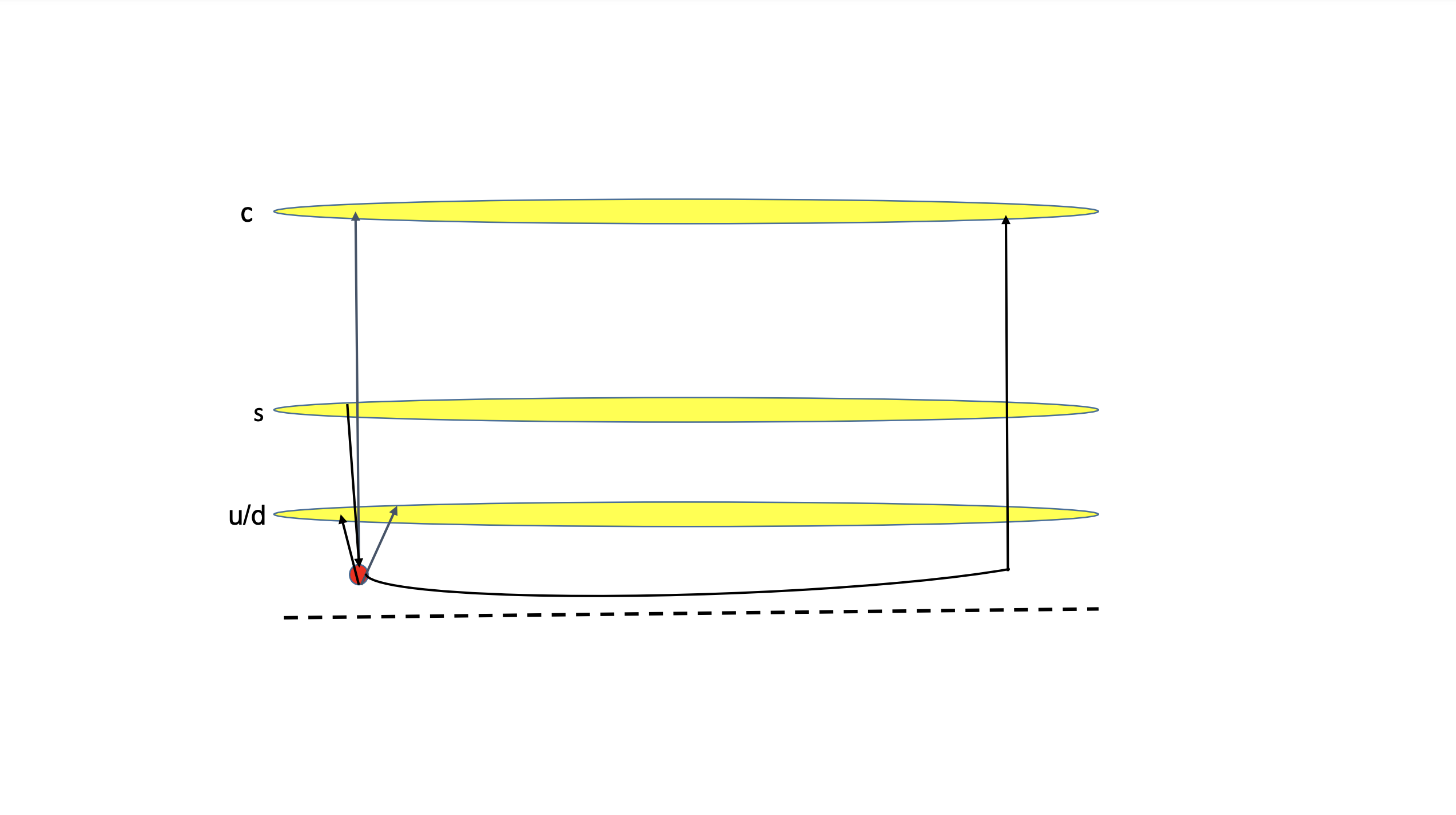}
  \caption{ Pentaquark with the structure of one string and one BV. This an example of  a content of $( c, c, \bar s u/d,u/d)$ and with $c$ the quark at the end of the string that does not attach directly to the BV.\label{penta1}}
\end{figure}

In (\ref{penta1}) the holographic structure of a pentaquark  of option (e) with a content of $(c, c, \bar s u/d,u/d)$ is drawn. In this example the single quark is a $c$ quark and the diquark and pair of quark anti-quark are either $(c ,u/d),(\bar s u/d)$  or $(u/d ,u/d),(\bar s c)$.

.

\begin{figure}[!ht]
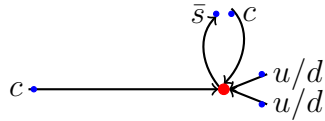
 
     \centering
     \hishpentaquark{$u/d$}{$u/d$}{$c$}{$c$}{$\bar{s}$}
     \caption{Pentaquark construction with one BV where an additional quark-antiquark pair is attached to it via two strings. \label{fig:pentaquark_simple}}
    \end{figure}

In (\ref{fig:pentaquark_simple}) the HISH structure of the same  pentaquark composed of $( c, c, \bar s u/d,u/d)$ is drawn. The diquark  is taken to be of $(u/d,u/d)$ and the additional pair connected to the BV is that of $(c,\bar c)$.

  \subsubsection{Configuration with BV anti-BV and a BV}
    \begin{itemize}
        \item The other option includes an addition of a BV and anti-BV to a baryonic configuration.  In this case  the pentaquark is build from  two  finite size strings 
     with diquarks on one of their endpoints and   an anti-BV on the other end which is also  attached to an antiquark through a string.  
     For a content of $(q_1,q_2,q_3,q_4,\bar q$ there are 6 possibilities of forming the two diquarks. 
    \end{itemize}
     
\begin{figure}[!ht]
      \centering
      \includegraphics[width=.6\columnwidth]{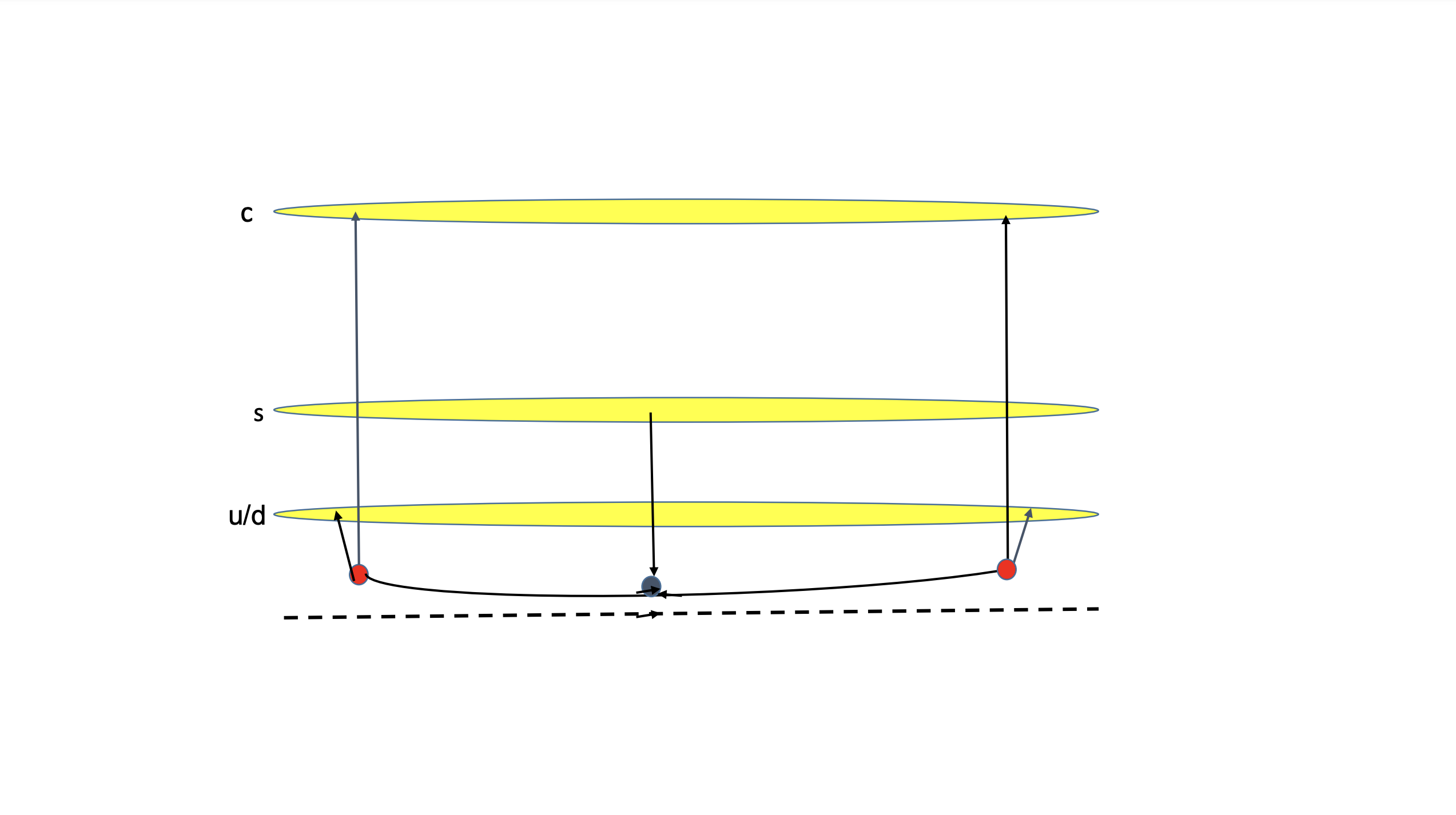}
      \caption{ Pentaquark with a structure based on two strings stretch between a BV and an anti-BV and between the latter and another BV.   This an example of  a content of $( c, c, \bar s u/d,u/d)$  there are diquarks of  $(u/d,c)$  attached to the two BVs. \label{fig:penta2}  }
    \end{figure}

\begin{figure}[!ht]
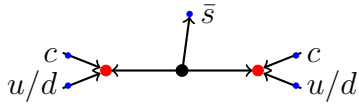

     \centering
     \hishpentaquarkk{$c$}{$u/d$}{$c$}{$u/d$}{$\bar{s}$}
     \caption{The HISH pentaquark construction with two BVs and one anti-BV.} \label{fig:pentaquark_simple2}
    \end{figure}

     In (Fig-\ref{fig:penta2}) and in (Fig-\ref{fig:pentaquark_simple2}) we show the holographic and HISH   structures of the same content of quarks as the one above. 
     .
     
     Three comments are in order:(i) It is not necessary that the two strings that connect a BV and anti-BV are on a line. More generally they can be non-collinear. Minimizing the total action of system will determine the particular structure. (ii) In this setup since the anti-quark in separated in space from any of the quarks, one can have a pentaquark even when there is a quark and anti-quark of the same flavor, unlike in option (c) of the first type of structure.  (iii) From fits to data of ordinary baryons, the contribution to the mass of the hadron a BV  was found out to be small. Thus the second type of pentaquarks is not necessarily heavier than the first one. A detailed calculation of the mass is needed in each case. 
     

\subsection{Decays of the pentaquarks}\label{Decayspenta}
Tetraquarks as we have seen in the previous section can decay either via an annihilation of a BV and an anti-BV or via a breakup of the string. For the pentaquarks these mechanism take place and there is another one, the detachment of a quark anti-quark from a BV. We discuss separately the decay of a structure based on a single BV and the a one with a BV anti-BV and a BV.
\subsubsection{Decay of a configuration with a single BV}
The petaquark that include a single BV can decay via two mechanisms: (i) A decay via breaking of the single long string producing a lower excited pentaquark and a meson. (ii) A decay via a detachment of the endpoints of an outgoing and incoming string on the BV. In holography as was described in (\ref{baryonic_structures}), the BV is a wrapped D brane. On this D brane the endpoint of an attached string is a point particle with a ``baryon number" charge, and that of a string with opposite orientation is a point particle with the opposite ``baryon number charge". Due to fluctuations the the two endpoints can get to the same point on the D brane and then annihilate.  It is plausible that an annihilation of such a pair can take place if the other side of the strings are connected to the same flavor brane. Once the pair is detached from the BV one gets a mesonic string. If  the annihilation, or detachment, is of a pair of the same flavor, the outcome meson will be flavorless like $\pi^0,\phi,J/\psi, botomonium$.

\begin{figure}[!ht]
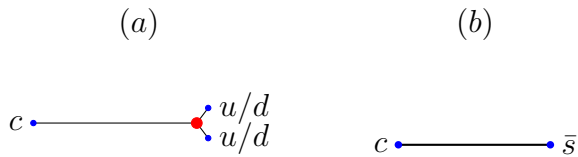

  \centering
  \begin{tabular}{ c m{4cm} c m{4cm} }
    \begin{minipage}[!ht]{4cm}
    \begin{center}
      $(a)$
    \end{center}
    \end{minipage}
    &
    \begin{minipage}[!ht]{4cm}
      \begin{center}
      $(b)$
      \end{center}
    \end{minipage}
    \\\\
    \begin{minipage}{4cm}
      \hishbaryon{$u/d$}{$u/d$}{$c$}
    \end{minipage}
    &
    \begin{minipage}{4cm}
      \hishmeson{$c$}{$\bar{s}$}
    \end{minipage}
    \\
  \end{tabular}
  \caption{Pentaquark decay through the detachment of a quark-antiquark pair.}
  \label{fig:pentaquark_detach}
\end{figure}

If indeed the pair is of the same flavor, there is also a probability of the annihilation and detachment from the flavor brane.  If the pentaquark configuration is of the setup (e) described in figure (\ref{pentasetup}), then the process should be immediate. This implies that this configuration is not of a pentaquark but of a single baryon with a closed string attached to it. To have a pentaquark the two endpoints on the flavor brane should not be at the same points, namely, the corresponding strings have some non-vanishing length along space coordinates  which occurs in configuration (d) for instance. 
An example of such a case is depicted  in figure (Fig.~\ref{fig:pentaquark_simple}).   The latter emerge from the detachment of the quark anti-quark pair from the BV. This process for a pentaquark with the content given in (Fig.~\ref{fig:pentaquark_simple}) is drawn in figure (\ref{fig:pentaquark_detach}). 

\subsubsection{Decay of a configuration  with BV anti-BV and a BV}
\begin{figure}[!ht]
      \centering
      \includegraphics[width=1.0\columnwidth]{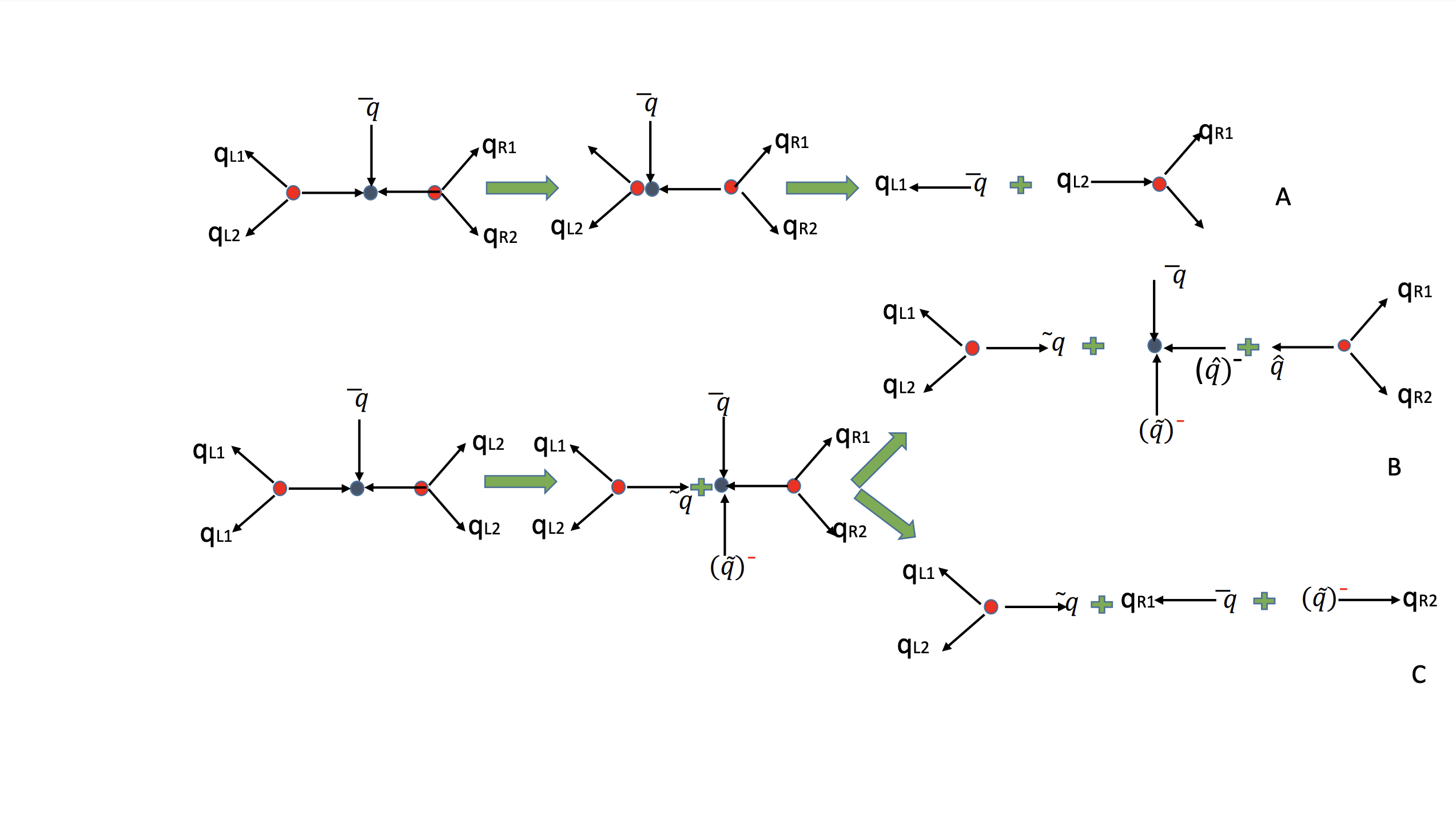}
      \caption{ The decay mechanisms of a pentaquark constructed from a BV anti-BV and another BV. Option A  follows from an annihilation of a BV and anti-BV. Option B and C follow from a breaking  into a baryon and a tetraquark that can yield two  baryons,  and an anti-baryon B and two mesons and a baryon C  \label{Pentadecay2f}  }
    \end{figure}

The second type of pentaquark that is built on two BV and an anti-BV  can decay via the following  mechanisms: 

(1)  Decays via the annihilation of one of the two BVs with the anti-BV leaving over a baryon and a meson. 
Notice that in this case the annihilation does not yield two mesons but only one and the other quark of the annihilated BV connects to the other BV.  This process is depicted in   A of fig (\ref{Pentadecay2f}).
We denote the  left diquark  by the quarks $(q_L^1,q_L^2)$ and similarly  for the right one by $(q_R^1,q_R^2)$ and the quark attached to the anti--BV by $\bar{ q}$.
.
    In figure A
the string with its ends $q_{L1}$ and $\bar q$ forms a meson whereas  the baryon is composed from $q_L^2,q_R^1,q_R^2)$. 
Obviously there are 4 possibilities of decay via such a process 
\begin{equation}   (q_L^1,\bar q), ( q_L^2,q_R^1,q_R^2)\qquad (q_L^2,\bar q),
( q_L^1,q_R^1,q_R^2)\qquad 
  (q_R^1,\bar q), ( q_L^1,q_L^2,q_R^2) \qquad 
(q_R^2,\bar q) ,( q_L^1,q_L^2,q_R^1) 
\end{equation}

(2) The other option is a decay via a breaking of a string yielding a  tetraquark  and a baryon.  In figure 
(\ref{Pentadecay2f}) the baryon associates with $(q_{L1},q_{L2}, \tilde q)$  and the tetra quark with ($(\bar q,\bar{\tilde q}, q_{R1},q_{R2})$. After this step there are two options denoted by B and C.
In option B  the tetraquark breaks apart into a baryon $(q_{L1},q_{L2}, \tilde q)$, anti baryon  $(\bar q, \bar {\tilde q},\bar{\hat q})$ and a baryon $(q_{R1},q_{R2}, \hat q)$. It should be emphasized that whereas the products of  the decay A can occur also in the decay of a molecule, the  decays drawn in B and C  can be an outcome only of a genuine pentaquark.   The third option depicted in C of (\ref{Pentadecay2f}) is a process where the tetraquark decays into two mesons. In the figure $(q_{R1} \bar q)$ and $(q_{R2} \bar{\tilde q} )$. Clearly there is also the option of $(q_{R2} \bar q)$ and $(q_{R1} \bar{\tilde q} )$  and also there is the possibility that the annihilation will be between the anti Bv and the left BV. A sign on this channel is the fact that one of the anti-quarks of two mesons must have the same flavor as one of the quarks of the baryon since in the process they emerge from a breakup of a string.  Note that this decay mechanism is also different from the one associated with the pentaquark built of a single BV. 



\section{Confronting Experimental Data}\label{HISH_confront_data}

In all our fits we assume the validity of \eqref{energy_cl} and \eqref{basic_regge}, where there is a long horizontal string, attached to massive endpoints. These massive endpoints may be diquarks (or antidiquarks) with very short strings that connect to the quarks (antiquarks).

From this assumption we derive that the slope $\alpha^{'}$ for a tetraquark should be similar to that of a meson  but with  the mass of an  endpoint being that of a di-quark ( anti- di-quark) plus  the mass of a BV (anti-BV) .
As for pentaquarks, if the structure is as described in figure  (\ref{fig:pentaquark_simple}), we expect the slope to be similar to that of baryons with similar quark content. If the structure is as described in  (\ref{fig:pentaquark_simple2}), we expect it to be similar to that of two strings attached at the anti-BV or maybe even an asymmetric Y shape where on one end there is a quark and on the other two there are BVs plus di-quarks.

Should the results admit an unexpected $\alpha^{'}$ value, it is an indication that this postulation is not correct for certain exotic hadrons. Moreover, as discussed in  subsections (\ref{HISH_pentaquarks}) and (\ref{baryonic_structures}), hadrons may have more complex structures and while some are proven to be unstable, it is not obvious that this instability arguments  can be applied to the exotic counterparts of these constructions.

\subsection{Fitting Procedure}

The fitting procedure will be done similarly to what was done in the previous works (\cite{Sonnenschein_2014_mesons}, \cite{Sonnenschein_2014_baryons}) and was called "the massive fit". We will treat $M^{2} = E^{2}(J + n)$ as the depended variable, where $E$ relates to $J + n$ through the velocities $\beta_{1}$ and $\beta_{2}$ - according to \eqref{energy_cl}:

\begin{equation}\label{energy_func_dep}
    {E}={E(m_{1}, m_{2}, \beta_{1}, \beta_{2}(m_{1}, m_{2}, \beta_{1}), \alpha^{'}, a)}
\end{equation}

Where the velocities are related through \eqref{velocities_relation}. The angular momentum according to \eqref{basic_regge}:

\begin{equation}\label{J_fit}
    {J+n}={(J+n)(\alpha^{'}, \beta_{1}, \beta_{2}(m_{1}, m_{2}, \beta_{1}), m_{1}, m_{2}, a)}
\end{equation}

Since $J + n$ is strictly monotonous in the domain we are interested in, assuming $m_1$, $m_2$, $\alpha^{'}$ and $a$ are known, we can calculate its inverse in terms of $\beta_1$:

\begin{equation}
    {\beta_{1}}={\beta_{1}\left(J + n, \alpha^{'}, a, m_{1}, m_{2}, \right)}
\end{equation}

Then we can rewrite (\ref{energy_func_dep}) as a direct dependency of the energy in the excitation level:

\begin{equation}\label{E_fit}
    {E}={E(J + n, m_{1}, m_{2}, \alpha^{'}, a)}\rightarrow\boxed{M^{2}={E^{2}(J + n, m_{1}, m_{2}, \alpha^{'}, a)}}
\end{equation}

The fit degree-of-conformity will be measured according to the regular $\chi^{2}_{r}$:

\begin{equation}\chi^{2}_{r}=\frac{1}{\nu}\sum_{i=1}^{n}\frac{(M_{i,exp}^{2}-M_{i,thry}^{2})^{2}}{\sigma_{i}^{2}}\end{equation}

Where $M_{i, exp}$ is the estimated measured mass of each state, $M_{i, thry}$ is the predicted mass of the fit, $\sigma$ is the total error of $M_{exp}^{2}$ and $\nu$ is the total number of DOFs which is the total number of measurements vs the number of fitting parameters that are not fixed. As usual, $\chi_{r}^{2}$ is expected to be as close to 1 as possible to reflect a good fit. Since we don't seek a high level of precision, we have taken the masses error to be the max between $60MeV$ and the total experimental error.

\subsection{Width estimation}\label{width_estimation}

The hadrons width is composed at least of the width of each of the decay mechanisms - annihilation and string tear. According to \eqref{decay_ann}, the annihilation decay width reduces exponentially with $L^{2}$. To estimate the annihilation width of resonances that can break to baryon anti baryon pair, we assume that we need to know the full width of one of the states that decays predominantly through the annihilation mechanism:

\begin{equation}\label{decay_ann_data}
{\Gamma_{ann}} = {\Gamma_{2}\frac{\exp{-{T_{1}L^{2}|_{tear}}}}{\exp{-{T_{2}L^{2}}|_{ann}}}}
\end{equation}

where $\Gamma_{2}$ is the width of one of the tetraquarks candidates.

The second mechanism is the tear of the string, which is calculated according to \eqref{width_tot} and \eqref{decay_tear_mq}, resulting in:
\begin{equation}\label{tear_width}
{\Gamma_{tear}} = \frac{\pi}{2}ATL(M, m_{1}, m_{2}, T)\Phi(M, M_{1}, M_{2})
\end{equation}

Where $M$ is the mass of the source hadron and $M_{1}$, $M_{2}$ are the masses of the two products - for tetraquarks it will be the masses of the baryon and antibaryon respectively.

The length of the string of each of these states can be estimated through:

\begin{equation}
L = l_{1}(E(\beta_{1}, m_{1}, \alpha^{'}, a)) + l_{2}(E(\beta_{2}(\beta_{1}, m_{2}, \alpha^{'}, a)))
\end{equation}

where the expression for $l_{i}$ is in accordance with \eqref{l_i} and \eqref{l_2}.

\subsection{Fitting Models}\label{fitting_models}

Tetraquarks and pentaquarks contain diquarks and anti-diquarks. These are connected to BVs and anti-BVs, and have a certain mass that is model dependant, as was shown in \cite{alon}. From \cite{Sonnenschein_2014_baryons} the results showed that the diquarks had mass that is usually the sum of the two quarks that are connected to the BV, implying that in the real world the BV is close to the wall where it's contribution to the endpoint mass is negligible with respect to the vertical strings that connect to the flavour branes and are mapped to the quarks masses at the flat-spacetime picture.

We will assume this conclusion also holds for the exotic hadrons, and expect the endpoint masses to be similar to the sum of the quarks (antiquarks) that compose the diquarks (anti-diquarks).
The other assumption, as mentioned above (\ref{HISH_confront_data}) is that the slopes should be similar to that of mesons/baryons with accordance to the compact-multiquarks structure and content.

In order to examine these assumptions, the fit of the exotic hadrons will be done according to \eqref{E_fit} in four ways:
\begin{itemize}
    \item Free fit - all parameters ($\alpha^{'}$, $a$, $m_{1}$ and $m_{2}$) will be free to vary. In this fit we expect to get values that are similar to these of our assumptions.
    \item Fixed masses - $m_{1}$ and $m_{2}$ will be fixed, each according to the sum of the quarks/antiquarks from which they are composed. $\alpha^{'}$ and $a$ on the other hand will be free to vary. The expectation is that the resulting $\alpha^{'}$ will be similar to that of the mesons for tetraquarks. 
    
    For pentaquarks, we will fix the masses in two separate fits:
    \begin{itemize}
        \item[o]{With fixed masses assuming the structure is similar to \ref{fig:pentaquark_simple}.}
        \item[o]{With fixed masses assuming the structure is similar to \ref{fig:pentaquark_simple2}. In this case we will ignore the contribution of the middle anti-quark, assuming its velocity is close to zero (located close to the CM of the pentaquark).}
    \end{itemize}
    \item Fixed slope - $\alpha^{'}$ will be fixed while the masses $m_{1}$ and $m_{2}$ will be free to vary. Here we expect again that the endpoint-masses will reflect our assumptions. Similarly to the fixed-masses fit, for pentaquarks, we will fix the slope once for each possible configuration (\ref{fig:pentaquark_simple}, \ref{fig:pentaquark_simple2}).
    \item Previous fit - $\alpha^{'}$, $m_{1}$ and $m_{2}$ will be fixed, while only the intercept $a$ is free to vary. Similarly to the fixed-masses fit, for pentaquarks, we will fix all the parameters once for each possible configuration (\ref{fig:pentaquark_simple}, \ref{fig:pentaquark_simple2}).
\end{itemize}

If our assumptions are correct, we expect to see no significant variations between the different fits in terms of $\chi^{2}$ and the predicted spectrum. We also expect the tetraquarks HMRTs slopes to be similar to these of the mesons.

Since there are two basic possible configurations - one that is similar to the tetraquarks structure, and another one that is similar to the baryonic configuration - the results for the pentaquarks may imply which configuration is a better description of these exotic hadrons.

The fit parameters will be bounded according to physical considerations - the endpoint masses will be bounded from below to zero and from above according to the mass of the lowest resonance in the trajectory.

From the fits we will calculate further resonances on the trajectory using \eqref{E_fit} and the decays widths as described in \ref{width_estimation}.

\subsection{Predictions of J and n}

Not all the observed states have J known. In addition, the data measured is not guaranteed to have a full trajectory with no gaps - there can be states in between that weren't observed. Therefore, predictions of the full MRT will be done in the following steps:

\begin{itemize}
    \item We will predict the trajectory for each quark content according to mesons/baryons thresholds - we will count the number of states that could rely on the same HMRT when building the trajectory around the thresholds.
    \item According to the fit, we will predict the value of J + n for the candidates that fit the trajectories and the decay channels.
\end{itemize}

Since some states were added to the data collected since the last works \cite{Sonnenschein_2014_mesons}, \cite{Sonnenschein_2014_baryons}, we are repeating the fits to verify the validity of the previous results, and also to align with the new fitting procedure.

\subsection{Mesons}\label{mesons_confronting_data}

In the previous HISH works \cite{Sonnenschein_2014_mesons} and \cite{Sonnenschein_2019_predictions}, the mesons data was fit in both $(J,M^{2})$ and $(n,M^{2})$ planes. Our assumption is that the tetraquarks structure is similar to that of mesons, where there is a long string with massive endpoints. In the case of the tetraquarks these massive endpoints are diquark (or anti diquark), connected to quarks (anti quarks) with very short strings. If this assumption is correct, we expect the slope of tetraquarks with similar quark content to resemble that of mesons.

This assumption is based on the results of \cite{Sonnenschein_2014_baryons}, where the baryons slopes were found to be similar to that of mesons. This observation reduces the chances of have a Y possible structure as detailed at \ref{baryonic_structures}, in which we expect the slope to be around $2/3$ of mesons slope with similar quark content. Hence, tetraquarks with slope similar to $2/3$ of their meson $\alpha^{'}$ counterparts may be explained by having such a structure.

In this paper, we systematically predict the tetraquarks trajectories for all possible charm configurations. One of the conclusions from \cite{Sonnenschein_2014_mesons} was that for heavy quarks the slopes of the $J$ and $n$ trajectories differ significantly. For the $(J,M^{2})$, a global fit was done for most of the mesons together, where the quarks masses are assumed to be global fit parameters of the model, that is common to all hadron configurations.

In order to provide trajectory predictions for tetraquarks in both planes - $J$ and $n$ excitations, we repeated the fit using the regular $\chi^{2}_{r}$ with global fit for each plane separately. At last we did a global fit for both planes together, assuming that the quark masses are constraint to be the same in both, where $\alpha^{'}$ wasn't constrained. The idea behind a separate global fit for each plane was to estimate the degree-of-conformity of the quark-masses-constraint assumption vs that of separate parameters of each plane. The results of the last global fit will be used for both planes tetraquarks spectrum predictions.

The full results of the global fits done for the mesons and the states used are in the appendix \ref{appendix:mesons_global_fit}. This includes the states used for the fits in tables \ref{table:mesons_states_J} and \ref{table:mesons_states_n}, as well as the graphs in \ref{fig:mesons_j_global_fit}-\ref{fig:mesons_all_global_fit} and fit values in table \ref{table:DOC_params_mesons_global}.

In this section at table \ref{table:mesons_global_fit} we summarize the global fits results which were used to predict the spectrum of all tetraquarks.

\begin{table}[!ht]
\begin{center}
\begin{tabular}{|c|c|c|c|c|c|c|c|}
\hline
\makecell{$\alpha^{'}_{n,light}$\\$[GeV^{-2}]$} & \makecell{$\alpha^{'}_{n,heavy}$\\$[GeV^{-2}]$} & \makecell{$\alpha^{'}_{j,light}$\\$[GeV^{-2}]$} & \makecell{$\alpha^{'}_{j,heavy}$\\$[GeV^{-2}]$} & \makecell{$m_{u/d}$\\$[GeV]$} & \makecell{$m_{s}$\\$[GeV]$} & \makecell{$m_{c}$\\$[GeV]$} & \makecell{$m_{b}$\\$[GeV]$}\\
\hline
\hline
$0.83$&$0.45$&$0.86$&$0.65$&$0.013$&$0.36$&$1.11$&$4.71$\\
\hline
\end{tabular}\caption{Mesons universal fit parameters. These will be used for calculating tetraquarks and tetraquarks-like configurations trajectories.}\label{table:mesons_global_fit}
\end{center} 
\end{table}

\subsection{Baryons}\label{section:baryons_n_global_fit_procedure}

The baryons fits were important for calculating the pentaquarks HMRTs, under the assumption of a baryonic structure (\ref{fig:baronic_structs}). For the $(J, M^{2})$ trajectories, we used the results from \cite{Sonnenschein_2014_baryons} and \cite{Sonnenschein_2019_predictions} as summarized in \ref{appendix:baryons_global_fit}. For the $(n, M^{2})$ trajectories, we did a universal fit (see \ref{appendix:baryons_n_global_fit}), similarly to the mesons fit. From this fit we extracted the light and heavy slopes of the baryons \ref{table:baryons_global_fit_n}. The procedure was done in the following way:

\begin{itemize}
    \item We assumed the universal values of the endpoint masses from the mesons fit holds for all configurations with quarks as endpoints.
    \item We went through all possible configurations for the baryons\footnote{By 'possible configurations' we mean that the baryons quark-diquark configuration is not always unambiguously established. For example $\Lambda_{c}$ which consists of $udc$ can either have endpoint masses $m_{BV_{u/d,u/d}}$ and $m_{c}$ or $m_{BV_{c,u/d}}$ and $m_{u/d}$. We fitted all the possible sets of configurations for the global $(n, M^{2})$ fit and selected the set that provides the best $\chi^{2}_{r}$.}, defining separately the diquarks masses.
    \item We fixed the quarks masses with the output of the masses obtained from the mesons global fit. But we let the diquarks masses vary with the restriction that their masses will be at least its heavy quark mass.
    \item The set of diquarks configurations that had the best universal fit was selected.
\end{itemize}

\subsection{Tetraquarks}

In the HISH framework, the natural decay of tetraquarks is to baryon-antibaryon due to the string breaking mechanism, given that they are above threshold. We searched for baryonia decays to find such candidates. In \ref{table:tetra_candidates} we enlist all possible candidates of charmed tetraquarks. We enlist the state, quark content (assumed), mass, width, $J^{PC}$ and the selected decay channels that match the HISH decay mechanisms (annihilation or string tear). 

In table \ref{table:charm_tetraquarks_predictions}, we enlist all the possible tetraquark charm configurations, and for each we detail the possible decays into two mesons, the  meson-meson threshold, decay into a baryon and anti-baryon and the corresponding threshold.
In fact, as discussed above  in (\ref{Decay mechanisms of tetraquarks}) the latter decay can, if energy permits, follow several channels. For instance the first item $(c,\bar{u}, u,\bar{u})$ can in principle decay via a $u,\bar{u}$ pair into $\Sigma_c^{++},\bar{\Delta}^{--}$, a pair of $d,\bar{d}$ into $\Lambda_{c}^{+}\bar{p}^-$, a pair of $s,\bar{s}$ into $\Xi_{c}^{+},\bar{\Sigma}^{-}$, and via  $c,\bar{c}$ and $b,\bar{b}$ but of course the latter two are very improbable. The type of multiplicities  may occur for all the states. In the table we write down only the most probable option when the string breaks by creating a light quark pair . The   decay  via annihilation of a given state is also not unique since there are two options of the quark anti-quark pairing. For instance the state $(c,d,\bar{u},\bar{d}) $ can decay into two mesons either via $(c,\bar{u}),(d\bar{d})$ or $(c,\bar{d}),(d\bar{u})$. Several of the candidates may have already been detected and they are mentioned in the prediction table of tetraquarks \ref{table:charm_tetraquarks_predictions}. 

\subsection{Pentaquarks}

The natural decay of an excited pentaquark with baryonic configuration is through string tear mechanism, into another pentraquark and a meson. If it isn't excited, we expect it to decay through the detachment mechanism to a meson and a baryon. In the available data no pentaquark candidates \ref{table:penta_candidates} decayed to a pentaquark and meson. The tetraquark-like configuration is expected to decay through string break into a tetraquark and a baryon. If below threshold it is expected to decay through annihilation into a baryon-meson pair.

While the detachment vs annihilation mechanism may not tell us which configuration is more likely according to the products (in both cases we expect baryon-meson decay), we can still separate the two configurations. In both cases, the lowest (in energy) final product may be of two mesons and a baryon, if the pentaquark is above the string tear configuration. The first configuration will be consistent with the results if the baryon and meson are of the same mother particle, while the second is expected if the mesons are of the same origin - created through annihilation of the tetraquark.

Another possible product of the second configuration is of two baryons and an antibaryon. This will also be a clear evidence that the latter configuration is more likely.

In table \ref{table:penta_candidates} we enlist all the possible candidates, and in table \ref{table:charm_pentaquarks_predictions} we calculate all possible predictions for pentaquarks based on both configurations and on the detachment/annihilation mechanism.

\section{Exotic Hadron Candidates}\label{section:HISH_data}

In this section we will analyse in detail specific resonances that are suspected as tetraquarks or pentaquarks according to the HISH model. In order to identify candidates, we scanned all PDG data, searching for states that correspond to the HISH criteria for exotic hadrons. Each of these are presented here.

The search for the data was done in two phases. At first we searched for states that fit our criteria. Than we built each trajectories and finally, searched the PDG again for states that may fit the same trajectory.

At last, we built all possible charmed configurations in tables [\ref{table:charm_tetraquarks_predictions}], than we went through the PDG data and searched for possible candidates with corresponding masses and decays that require more research.

This section is built as follows - I) we first describe the candidates selection criteria. II) Than we list all the exotic candidates, and elaborate on each of the candidates that have been observed. Each is confronted with the predictions and model assumptions. III) At last we list a table with the trajectories predictions for all the possible tetrauqarks configurations. We also elaborate on some of the pure tetraquarks predictions. IV) At last, we do the same for the pentaquarks.

\subsection{HISH Exotic Candidates selection Criteria}\label{exotic_selection_criteria}

In the HISH model, the natural decay of hadrons is through string tear. Hence, the best criteria for identifying tetraquarks was through scanning the PDG data for states that decay to baryon-antibaryon pairs. In addition, the candidate must admit the Modified Regge Trajectory behavior\footnote{The HMRT criteria can only be confirmed when there are at lease three states. If there are two, we require that the slope and masses will be similar to the mesonic case (see \ref{fitting_models}).}.

Another criteria are decays through annihilation, especially for states that are below the baryon-antibaryon threshold. These criteria can be divided to three types. The first is for candidates that are composed solely from heavy content. Since string-tear decay probability exponentially decreases by the mass of the created $q\bar{q}$ pair (\eqref{decay_tear_mq}), annihilation for a only heavy mesons is a strong indication that the mother particle is a tetraquark. The second is for states that do not  contain $q\bar{q}$ pair in their mesonic products, which implies that the products cannot come from a decay via breaking of a meson, hence these state are genuine tetraquark. The third are those states  that do not belong to each of the previous ones, but do not belong to any of the mesonic trajectories.

For pentaquarks, we relied on candidates that were identified as pentaquarks ( or molecules)  by the experiments that found them. The candidates that were observed so far are charmed and contain $c\bar{c}$ pair.
It is belived that  these states are molecules and not genuine pentaquarks\footnote{We thank M. Karliner for pointing this to us.} If on the other hand, it will be found in the future that there are excited states of them that furnish HMRT trajectories, then it will be a strong indication that they are genuine petaquarks. 

\subsection{Tetraquarks Candidates}\label{section:tetra_cands}

In this paper, we focused on states with charm flavor. But, there are a few non-charm candidates that decay to baryon-antibaryon pairs that we analysed. We begin from these light candidates, and proceed to our main goal, which are the charmed ones. Table \ref{table:tetra_candidates} summarizes the candidates and the measurements that were considered and used, followed by the detailed analysis of each state or trajectory.

The 'Selected Decay Channels' column includes the decay channels that matches the HISH decay possible mechanisms and imply the quark content of the states. As explained in section \ref{exotic_selection_criteria}, the decay channel is used as part of the selection criteria for the states that should be considered, and this was our main consideration for selecting certain channels over others, rather than the width value of the channels. 

The 'Quark Content' column is the assumed quark content according to the products. When there are more than one suggested compositions, it is either due to the products being a mixed ("superposition") configuration as explained in \ref{section:mixed_configuration}, or that the products minimal content is ambiguously established. When 'Molecule' is indicated, it means the state does not meet the HISH criteria for genuine compact-quark structure.

\begin{longtable}[c]{|c|c|c|c|c|c|c|}
\hline
\makecell{Candidate} & \makecell{Quarks \\ Content} & \makecell{Selected \\ Decay \\ Channels} & \makecell{$J^{PC}$} & \makecell{Mass \\ $[MeV]$} & \makecell{Width \\ $[MeV]$} & \makecell{Ref.} \\
\hline
\hline
$X(1855)$ & \makecell{Molecule} & \makecell{$\bar{p}n$} & $?^{??}$ & $1856.6\pm{5}$ & $20\pm{5}$ & \cite{X_1855} \\
\hline
$X(1835)$ & \makecell{Molecule} & \makecell{$p\bar{p}$} & $0^{-+}$ & $1826.5^{+13.0}_{-3.4}$ & $242^{+14}_{-15}$ & \makecell{\cite{X_1835_2005}, \cite{X_1835_2015}, \\ \cite{X_1835_2016}} \\
\hline
$f_{4}(2300)$ & \makecell{$u\bar{d}\bar{u}d$ \\ or \\ $d\bar{d}\bar{d}d$} & $N\bar{N}$ & $4^{++}$ & $2320\pm 60$ & $250\pm 80$ & \makecell{\cite{f4_2300_1970}, \cite{f4_2300_1973}, \\ \cite{f4_2300_1974}, \cite{f4_2300_2000}} \\
\hline
$f_{2}(1640)$ & \makecell{$u\bar{s}\bar{u}s$ \\ or \\ $d\bar{s}\bar{d}s$} & \makecell{$K\bar{K}$} & $2^{++}$ & $1639\pm{6}$ & $99^{+60}_{-40}$ & \makecell{\cite{f2_1640_1990}, \cite{f2_1640_1992}, \\ \cite{f2_1640_1995}} \\
\hline
$f_{2}(1750)$ & \makecell{$u\bar{s}\bar{u}s$ \\ or \\ $d\bar{s}\bar{d}s$} & \makecell{$K\bar{K}$} & $2^{++}$ & $1755\pm{10}$ & $67\pm{12}$ & \cite{f2_1750} \\
\hline
$f_{2}(2300)$ & \makecell{$u\bar{s}\bar{u}s$ \\ or \\ $d\bar{s}\bar{d}s$} & $\Lambda\bar{\Lambda}$ & $2^{++}$ & $2297\pm 28$ & $149\pm 40$ & \cite{f2_2300} \\
\hline
$\omega(2290)$ & \makecell{$u\bar{s}\bar{u}s$ \\ or \\ $d\bar{s}\bar{d}s$} & $\Lambda\bar{\Lambda}$ & $1^{--}$ & $2290\pm 20$ & $275\pm35$ & \makecell{\cite{omega_2290_2004}, \cite{omega_2290_2000}, \\ \cite{omega_2290_1990}} \\
\hline
$f_{3}(2300)$ & \makecell{$u\bar{s}\bar{u}s$ \\ or \\ $d\bar{s}\bar{d}s$} & $\Lambda\bar{\Lambda}$ & $3^{++}$ & $2334\pm25$ & $200\pm20$ & \cite{omega_2290_2004}, \cite{omega_2290_2000} \\
\hline
$X(3250)$ & \makecell{$\bar{u}su\bar{s}$ \\ or \\ $\bar{d}sd\bar{s}$} & $\Lambda\bar{p}K^{+}$ & $?^{??}$ & $3250\pm{8}\pm{20}$ & $45\pm{18}$ & \cite{K_tetra_1993} \\
\hline
$X(3250)$ & \makecell{$\bar{u}su\bar{s}$ \\ or \\ $\bar{d}sd\bar{s}$} & \makecell{$\Lambda\bar{p}K^{+}\pi^{\pm}$ \\ $K^{0}_{S}p\bar{p}K^{\pm}$} & $?^{??}$ & $3245\pm{8}\pm{20}$ & $25\pm{11}$ & \cite{K_tetra_1993} \\
\hline
$K(3100)^{0}$ & $u\bar{s}\bar{u}d$ & $\Sigma(1385)^{+}\bar{p}$ & $?^{??}$ & $\sim 3100$ & $\sim 10-70$ & \cite{K_tetra_1993} \\
\hline
$X(2632)$ & $c\bar{u}u\bar{s}$ & \makecell{$D^{0}K^{+}$ \\ $D^{+}_{s}\eta$} & $?^{??}$ & $2635.2\pm{3.3}$ & $<17$ & \cite{SELEX:2004drx} \\
\hline
$T_{cs0}(2900)^{0}$ & $\bar{c}du\bar{s}$ & $D^{-}K^{+}$ & $0^{+}$ & $2866\pm{7}$ & $57\pm{13}$ & \cite{LHCb:2020pxc} \\
\hline
$T_{cs1}(2900)^{0}$ & $\bar{c}d\bar{u}s$ & $D^{-}K^{+}$ & $1^{-}$ & $2904\pm{5}$ & $110\pm{12}$ & \cite{LHCb:2020pxc} \\
\hline
$T^{a}_{c\bar{s}0}(2900)^{0}$ & $c\bar{s}\bar{u}d$ & $D^{+}_{s}\pi^{-}$ & $0^{?}$ & $2892\pm{14}\pm{15}$ & $119\pm{26}\pm{13}$ & \cite{LHCb:2022sfr} \\
\hline
$T^{a}_{c\bar{s}0}(2900)^{++}$ & $c\bar{s}u\bar{d}$ & $D^{+}_{s}\pi^{+}$ & $0^{?}$ & $2921\pm{17}\pm{20}$ & $137\pm{32}\pm{17}$ & \cite{LHCb:2022sfr} \\
\hline
$X(3350)$ & $c\bar{d}d\bar{u}$ & $\Lambda_{c}^{+}\bar{p}$ & ${{0} or {1}}^{??}$ & $3350^{+10}_{-20}\pm{20}$ & $70^{+40}_{-30}\pm{40}$ & \makecell{\cite{X(3350)_Belle:2004dmq}, \cite{X(3350)_BaBar:2010eti}} \\
\hline
$X(3800)$\footnote{Needs confirmation, seen at \cite{X(3350)_Belle:2004dmq} but not at \cite{X(3350)_BaBar:2010eti}.} & $c\bar{d}d\bar{u}$ & $\Lambda_{c}^{+}\bar{p}$ & $?^{??}$ & $3840\pm{10}$ & $30\pm{30}$ & \cite{X(3350)_Belle:2004dmq} \\
\hline
$T_{cc}(3875)$ & $cc\bar{u}\bar{d}$ & \makecell{$D^{*+}D^{0}$ \\ $D^{*0}D^{+}$} & $?^{?}$ & $3874\pm{0.11}$ & $0.41\pm{0.17}$ &  \cite{LHCb:2021vvq} \\
\hline
$Z_{c}(3900)$ & \makecell{$c\bar{c}u\bar{d}$ \\ $c\bar{c}d\bar{u}$} & \makecell{$J/\psi \pi^{\pm}$ \\ ${D\bar{D^{*}}}^{\pm}$} & $1^{+-}$ & $3887.1\pm 2.6$ & $28.4\pm2.6$ & \makecell{Many, \\ including \\ \cite{BESIII:2020oph}} \\
\hline
\makecell{$Z_{c}(4200)^{\pm}$} & \makecell{$c\bar{c}u\bar{d}$ \\ or \\ $c\bar{c}d\bar{u}$} & \makecell{$J/\psi\pi^{+}$} & $1^{+-}$\footnote{\label{footnote:C_needs_confirmation}C needs confirmation.} & $4196^{+35}_{-32}$ & $370^{+100}_{-150}$ & \makecell{\cite{Belle:2014nuw}} \\ 
\hline
$Z_{c}(4430)^{\pm}$ & \makecell{$c\bar{c}u\bar{d}$ \\ or \\ $c\bar{c}d\bar{u}$} & \makecell{$\pi^{+}\psi(2S)$ \\ $\pi^{+}J/\psi$} & $1^{+-}$\tablefootnotemark{footnote:C_needs_confirmation} & $4478^{+15}_{-18}$ & $181\pm 31$ & \makecell{\cite{Belle:2007hrb}, \cite{Belle:2009lvn}, \\ \cite{Belle:2013shl}, \cite{LHCb:2014zfx}} \\
\hline
$X(4020)^{\pm}$ & \makecell{$c\bar{c}u\bar{d}$ \\ or \\ $c\bar{c}d\bar{u}$} & \makecell{$D^{*}\bar{D}^{*}$ \\ $h_{c}(1P)\pi^{\pm}$} & $?^{?-}$ & $4024.1\pm{1.9}$ & $13\pm 5$ & \makecell{\cite{BESIII:2013ouc}, \cite{BESIII:2014gnk} \\ \cite{BESIII:2013mhi}, \cite{BESIII:2013mhi}} \\
\hline
$X(4051)^{\pm}$ & \makecell{$c\bar{c}u\bar{d}$ \\ or \\ $c\bar{c}d\bar{u}$} & \makecell{$\pi^{+}\chi_{c1}(1P)$} & $?^{?+}$ & $4051^{+24}_{-40}$ & $82^{+50}_{-28}$ & \makecell{\cite{Belle:2008qeq}} \\
\hline
$X(4055)^{\pm}$ & \makecell{$c\bar{c}u\bar{d}$ \\ or \\ $c\bar{c}d\bar{u}$} & \makecell{$\pi^{+}\psi(2S)$} & $?^{?-}$ & $4054\pm 3.2$ & $45\pm 13$ & \makecell{\cite{Belle:2014wyt}, \cite{BESIII:2017vtc} \\ \cite{BESIII:2017tqk}} \\
\hline
$X(4100)^{\pm}$ & \makecell{$c\bar{c}u\bar{d}$ \\ or \\ $c\bar{c}d\bar{u}$} & \makecell{$\pi^{-}\eta(1S)$} & $?^{??}$ & $4096\pm 28$ & $152^{+80}_{-70}$ & \makecell{\cite{LHCb:2018oeg}} \\
\hline
$R_{c0}(4240)^{\pm}$ & \makecell{$c\bar{c}u\bar{d}$ \\ or \\ $c\bar{c}d\bar{u}$} & \makecell{$\pi^{-}\psi(2S)$} & $0^{--}$ & $4239^{+50}_{-21}$ & $220^{+120}_{-90}$ & \makecell{\cite{LHCb:2021uow}} \\
\hline
$X(4250)^{\pm}$ & \makecell{$c\bar{c}u\bar{d}$ \\ or \\ $c\bar{c}d\bar{u}$} & \makecell{$\pi^{+}\chi(1P)$} & $?^{?-}$\tablefootnotemark{footnote:C_needs_confirmation} & $4248^{+190}_{-50}$ & $177^{+320}_{-70}$ & \makecell{\cite{Belle:2008qeq}} \\
\hline
\makecell{$T_{\psi s1}^{\theta}(4000)^{+}$ \\ also \\ $Z_{cs}(4000)^{+}$} & $c\bar{c}u\bar{s}$ & \makecell{$J/\psi K^{+}$ \\ $D_{s}^{+}\bar{D}^{*0}$ \\ $D_{s}^{*+}\bar{D}^{0}$} & $1^{+}$ & $3980-4010$ & $5-150$ & \cite{LHCb:2021uow} \\ 
\hline
\makecell{$T_{\psi s1}(4220)^{+}$ \\ also \\ $Z_{cs}(4220)^{+}$} & $c\bar{c}u\bar{s}$ & \makecell{$J/\psi K^{+}$} & $1^{+}$ & $4216^{+50}_{-40}$ & $233^{+110}_{-90}$ & \cite{LHCb:2021uow} \\
\hline
\makecell{$T_{\psi s1}^{\theta}(4000)^{0}$} & $c\bar{c}d\bar{s}$ & \makecell{$J/\psi K^{0}_{S}$} & $1^{+}$ & $3991^{+14}_{-20}$ & $105^{+12}_{-10}{}^{+9}_{-17}$ & \makecell{\cite{LHCb:2023hxg}} \\
\hline
$X(3960)$ & $c\bar{c}s\bar{s}$ & $D_{s}^{+}$ $D_{s}^{-}$ & $0^{++}$ & $3956\pm 5\pm 10$ & $43\pm{13}\pm{8}$ & \makecell{\cite{LHCb:2022aki}} \\
$X(4140)$ & $c\bar{c}s\bar{s}$ & $J/\psi\phi$ & $1^{++}$ & $4146.5\pm 3.0$ & $19^{+7}_{-5}$ & \makecell{\cite{CMS:2013jru}} \\
$X(4274)$ & $c\bar{c}s\bar{s}$ & $J/\psi\phi$ & $1^{++}$ & $4286^{+8}_{-9}$ & $51\pm 7$ & \makecell{\cite{LHCb:2016axx}} \\
$\chi_{c0}(4500)$ & $c\bar{c}s\bar{s}$ & \makecell{$J/\psi\phi$} & $0^{++}$ & $4474\pm{4}$ & $77^{+12}_{-10}$ & \makecell{\cite{LHCb:2016axx}} \\
$\chi_{c1}(4685)$ & $c\bar{c}s\bar{s}$ & \makecell{$J/\psi\phi$} & $1^{++}$ & $4684^{+15}_{-17}$ & $126\pm{40}$ & \makecell{\cite{LHCb:2021uow}} \\
$\chi_{c0}(4700)$ & $c\bar{c}s\bar{s}$ & \makecell{$J/\psi\phi$} & $0^{++}$ & $4694^{+16}_{-5}$ & $87^{+18}_{-10}$ & \makecell{\cite{LHCb:2016axx}} \\
\hline
$X(4350)$\footnote{\label{footnote:state_needs_confirm}Needs confirmation.} & $c\bar{c}s\bar{s}$ & $J/\psi\phi$ & $?^{?+}$ & $4351\pm{5}$ & $13^{+18}_{-10}$ & \makecell{\cite{Belle:2009rkh}} \\
$X(4630)$ & $c\bar{c}s\bar{s}$ & \makecell{$J/\psi\phi$} & $1^{-+}$\footnote{$J^{P}$ needs confirmation.} & $4626^{+24}_{-110}$ & $174^{+140}_{-80}$ & \makecell{\cite{LHCb:2021uow}} \\
\hline
$\psi(4360)$ & $c\bar{c}s\bar{s}$ & $J/\psi\eta$ & $1^{--}$ & $4374\pm{7}$ & $118\pm 12$ & \makecell{\cite{BESIII:2020bgb}, \cite{Wang:2012bgc}} \\ \\
$\psi(4660)$ & $c\bar{s}\bar{c}s$ & \makecell{$\Lambda_{c}^{+}\bar{\Lambda_{c}}^{-}$ \\ $D^{+}_{s}D_{s1}^{-}(2536)$ \\ $D^{+}_{s}D_{s2}^{*-}(2573)$} & $1^{--}$ & $4630\pm{6}$ & $72^{+14}_{-12}$ & \makecell{\cite{Belle:2008xmh}} \\ \\
\hline
$T_{\psi\psi}(6600)$ & $c\bar{c}c\bar{c}$ & \makecell{$J/\psi J/\psi$} & $?^{??}$ & $6630\pm{90}$ & $350\pm{11}^{+110}_{-40}$ & \makecell{\cite{ATLAS:2023bft}} \\
$T_{\psi\psi}(6600)$ & $c\bar{c}c\bar{c}$ & \makecell{$J/\psi J/\psi$} & $?^{??}$ & $6552\pm{10}\pm{12}$ & $124^{+32}_{-26}\pm{33}$ & \makecell{\cite{CMS:2023owd}} \\
$T_{\psi\psi}(6900)$ & $c\bar{c}c\bar{c}$ & \makecell{$J/\psi J/\psi$} & $?^{??}$ & $6886\pm{16}$ & $168\pm{80}$ & \makecell{\cite{LHCb:2020bwg}} \\
$T_{\psi\psi}(6900)$ & $c\bar{c}c\bar{c}$ & \makecell{$J/\psi J/\psi$} & $?^{??}$ & $6927\pm{9}\pm{4}$ & $122^{+24}_{-21}\pm{18}$ & \makecell{\cite{CMS:2023owd}} \\
$T_{\psi\psi}(7300)$ & $c\bar{c}c\bar{c}$ & \makecell{$J/\psi J/\psi$} & $?^{??}$ & $7287^{+20}_{-18}\pm{5}$ & $95^{+59}_{-40}\pm{19}$ & \makecell{\cite{CMS:2023owd}} \\
\hline
\caption{Tetraquarks candidates summary. These states were analysed in detail in this paper and compared with the tetraquarks predicted HMRTs described at \ref{table:charm_tetraquarks_predictions}. The 'Selected Decay Channels' column includes the decay channels that matches the HISH decay mechanisms and imply the quark content. The 'Quark Content' column is the assumed quark composition, where 'Molecule' means it does not meet the HISH criteria for compact-quark structure.}\label{table:tetra_candidates}
\end{longtable}

For light tetraquarks candidates, we used the universal light slopes for $(J, M^{2})$ and $(n, M^{2})$ to calculate the spectrum for each HMRT. We set the sum of the diquark (antidiquark) masses as was outputed by the mesons fits to be the mass of each endpoint. We then extracted the intercepts $a_{J}$ and $a_{n}$ using all the other parameters along with the provided mass of one of the states on the trajectories.

Since we do not differentiate between $u$ and $d$ masses/D-brane, we treat contents containing them together. For example $T_{c\bar{d}d\bar{u}}$ is treated with $T_{c\bar{u}u\bar{u}}$ as the HISH model does not distinguish between them in terms of mass or preferable decay channel.

For heavy candidates, we did the same, but used the mesons heavy slopes for both planes.

\subsubsection{\texorpdfstring{Minimal content $T_{u\bar{u}\bar{u}d}$ or $T_{\bar{u}d\bar{d}d}$}{}}\label{section:X_1855}

In the available data there is the state $X(1855)$ (table \ref{table:tetra_mole_x_1855}) which corresponds to minimal quark content of either ${u\bar{u}\bar{u}d}$ or ${\bar{u}d\bar{d}d}$. This resonance was observed only in the baryonic channel $\bar p {n}$. In the HISH framework, the criteria for identifying a molecule is that it shouldn't fit an HMRT. Another condition for a tetraquark decay through string tear is that its energy is above the baryon-antibaryon pair threshold, which is $\sim 1877$ in this case. Therefore, the state $X(1855)$ which has a mass of $\sim 1857\pm{5}$ is expected to be a molecule and not a genuine compact-multiquark structure. Aside from these requirements, the description of molecules in the HISH framework requires more research as explained in the \hyperref[section:open_questions]{open-questions section}.

\begin{longtable}[c]{|c|c|c|c|c|c|c|}
\hline
\makecell{Candidate} & \makecell{Quarks \\ Content} & \makecell{Selected \\ Decay \\ Channels} & \makecell{$J^{PC}$} & \makecell{Mass \\ $[MeV]$} & \makecell{Width \\ $[MeV]$} & \makecell{Ref.} \\
\hline
\hline
$X(1855)$ & \makecell{Molecule} & \makecell{$\bar{p}n$} & $?^{??}$ & $1856.6\pm{5}$ & $20\pm{5}$ & \cite{X_1855} \\
\hline
\caption{$T_{u\bar{u}\bar{u}d}$ or $T_{\bar{u}d\bar{d}d}$ exotic candidate.}\label{table:tetra_mole_x_1855}
\end{longtable}

Table \ref{table:Tuudu_thresholds} summarizes the thresholds for $T_{\bar{u}d\bar{d}d}$ and the predicted annihilation and tear widths (in the 'Width' column the first value is the annihilation width, the second one is the tear width). Tables (\ref{table:Tuudu_j_predictions} and \ref{table:Tuudu_n_predictions}) provide the model HMRTs in planes $(J, M^{2})$ and $(n, M^{2})$ respectively. Indeed $X(1855)$ does not fit any of the trajectories. The predicted widths are also above the experimental value. Figure \ref{fig:Tuudu} contains the graphs of the predicted trajectories for $T_{\bar{u}d\bar{d}d}$.

\twographs[(a) is $(J, M^{2})$ $T_{\bar{u}d\bar{d}d}$ predicted HMRT, (b) $(n, M^{2})$ $T_{\bar{u}d\bar{d}d}$ HMRT. In both cases it is assumed to be a tetraquark due to its decay channel.][Tuudu]{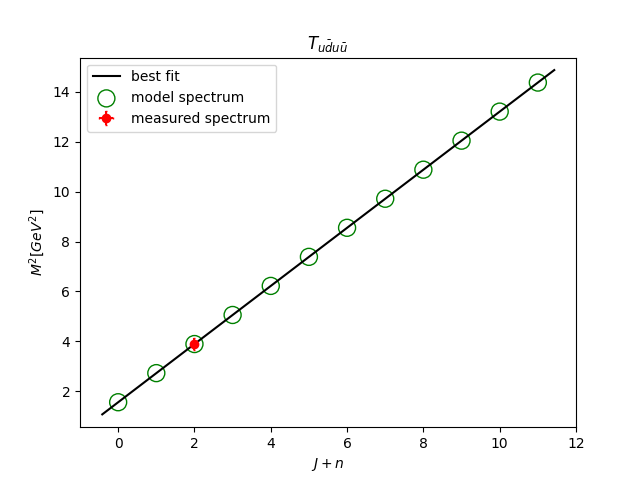}{tetraquarks/predictions/uu-du/by_n/T_uudu_one_by_n}

\begin{longtable}[c]{|c|c|c|c|c|c|}
  \hline
  \thead{Thry \\ Width \\ $[MeV]$} & \thead{Mesons \\ Pair} & \thead{Meson \\ Threshold \\ $[MeV]$} & \thead{Baryon \\ Anti-\\baryon} & \thead{Baryonic \\ Threshold \\ $[MeV]$} & \thead{Genuine} \\
  \hline
  \hline
\makecell{$24-115$\\$38-75$} & \makecell{$\pi^{+} \pi^{0}_{u}$ \\ $\pi^{0}_{u} \pi^{+}$} & \makecell{$275$ \\ $275$} & \makecell{$p \bar{n}$ \\ $\Delta^{++} \bar{p}$ \\ $\Sigma^{+} \bar{\Lambda}^{0}$} & \makecell{$1878$ \\ $2170$ \\ $2305$} &  \\
\hline
  \caption{$T_{u\bar{d}u\bar{u}}$ thresholds.} \label{table:Tuudu_thresholds}
 \end{longtable}

    \begin{center}
    \begin{minipage}{.49\linewidth}
        \begin{center}
            \begin{longtable}[c]{|c|c|}
  \hline
  \thead{$J$ \\ Spec} & \thead{$a_{T_{j}}$} \\
  \hline
  \hline
\makecell{$1161-1335$\\$1585-1717$\\$1918-2028$\\$2201-2297$\\$2451-2538$\\$2678-2758$\\$2887-2961$\\$3082-3152$\\$3266-3331$\\$3439-3502$\\$3604-3664$\\$3762-3819$} & \makecell{$-1.5$-$-1.1$} \\
\hline
  \caption{$T_{u\bar{d}u\bar{u}}$ $(J, M^2)$ predictions.} \label{table:Tuudu_j_predictions}
 \end{longtable}
        \end{center}
    \end{minipage}
    \begin{minipage}{.49\linewidth}
        \begin{center}
            \begin{longtable}[c]{|c|c|}
  \hline
  \thead{$n$ \\ Spec} & \thead{$a_{T_{n}}$} \\
  \hline
  \hline
\makecell{$1124-1303$\\$1572-1705$\\$1918-2028$\\$2210-2306$\\$2468-2555$\\$2702-2781$\\$2916-2990$\\$3116-3185$\\$3304-3369$\\$3482-3543$\\$3651-3710$\\$3812-3869$} & \makecell{$-1.4$-$-1.0$} \\
\hline
  \caption{$T_{u\bar{d}u\bar{u}}$ $(n, M^2)$ predictions.} \label{table:Tuudu_n_predictions}
 \end{longtable}
        \end{center}
    \end{minipage}
    \end{center}

\subsubsection{\texorpdfstring{$T_{u\bar{d}\bar{u}d}$ or $T_{d\bar{d}d\bar{d}}$}{}}

The $T_{u\bar{d}\bar{u}d}$ or $T_{d\bar{d}d\bar{d}}$ tetraquarks candidates that are considered are in the following table \ref{table:tetra_cand_udud}:

\begin{longtable}[c]{|c|c|c|c|c|c|c|}
\hline
\makecell{Candidate} & \makecell{Quarks \\ Content} & \makecell{Selected \\ Decay \\ Channels} & \makecell{$J^{PC}$} & \makecell{Mass \\ $[MeV]$} & \makecell{Width \\ $[MeV]$} & \makecell{Ref.} \\
\hline
\hline
$X(1835)$ & \makecell{Molecule} & \makecell{$p\bar{p}$ \\ $\eta^{'}\pi^{+}\pi^{-}$ \\ $\gamma\gamma$} & $0^{-+}$ & $1826.5^{+13.0}_{-3.4}$ & $242^{+14}_{-15}$ & \makecell{\cite{X_1835_2005}, \cite{X_1835_2015}, \\ \cite{X_1835_2016}} \\
\hline
$f_{4}(2300)$ & \makecell{$u\bar{d}\bar{u}d$ \\ or \\ $d\bar{d}\bar{d}d$} & \makecell{$N\bar{N}$ \\ $\eta\eta$ \\ $\omega\omega$ \\ $\pi\pi$ \\ $K\bar{K}$ \\ $\rho\rho$ \\ $\eta\pi\pi$} & $4^{++}$ & $2320\pm 60$ & $250\pm 80$ & \makecell{\cite{f4_2300_1970}, \cite{f4_2300_1973}, \\ \cite{f4_2300_1974}, \cite{f4_2300_2000}} \\
\hline
\caption{$T_{u\bar{d}\bar{u}d}$ or $T_{d\bar{d}d\bar{d}}$ tetraquark candidates.}\label{table:tetra_cand_udud}
\end{longtable}

Similar to $X(1855)$ from previous section \ref{section:X_1855}, the first state $X(1835)$ mass ($\sim 1826$) is below the threshold for $p\bar{p}$ decay ($\sim 1877$) by about $50 MeV$. Therefore, we expect it to not be on the predicted HMRT for $T_{u\bar{d}\bar{u}d}$ or $T_{d\bar{d}d\bar{d}}$.

The other state in this category, $f_{4}(2300)$ have been observed in $N\bar{N}$ decay channel, as well as in meson-meson pair channels and one 3-body meson channel as can be seen from the table \ref{table:tetra_cand_udud}. Since the strongest indication of a tetraquark candidate according to this model is a decay to baryon-antibaryon pair, we consider $f_{4}(2300)$ to be a candidate. Another indication is that we couldn't fit this state into a meson trajectory, implying it does not belong to this regime. However, the determination of the exact quark content is ambiguous since the products are mixes of all light quarks - in general the model allows annihilation of a quark-antiquark of the same flavour brane followed by the creation of a different pair on another flavor-brane. This could be interpreted either by implying that there are multiple decay channels, or by the suggestion that the model always offers a mix of states when this annihilation is possible. This is another open question of the model.

The $f_{4}(2300)$ mass is roughly $440$ MeV above the baryonic threshold (table \ref{table:Tdudu_thresholds}). Other tetraquarks candidates that are above the baryon-antibaryon threshold are closer to it with only about $20-130$ MeV above the threshold. Therefore, we assume it is not the first state on the trajectory that decay through $N\bar{N}$. In order to test it, we built the trajectories for $f_{4}(2300)$ in both $(J, M^{2})$ and $(n, M^{2})$ (tables \ref{table:Tdudu_j_predictions} and \ref{table:Tdudu_n_predictions}, figure \ref{fig:Tdduu_predictions}) to see if the $X(1835)$ state fits on any of the trajectories. We can see from table \ref{table:Tdudu_j_predictions} that the state fits on the J trajectory, and from \ref{table:Tdudu_thresholds} that the predicted width also fits the measured value. Moreover, the second row in table \ref{table:Tdudu_j_predictions} is the mass range of $J+n=3$, and row three is the mass range of $J+n=4$, which is consistent with the $J$ value of $f_{4}(2300)$ with $n=0$.

\twographs[(a) is $(J, M^{2})$ $T_{u\bar{d}\bar{u}d}$ or $T_{d\bar{d}d\bar{d}}$ predicted HMRT, (b) $(n, M^{2})$ $T_{u\bar{d}\bar{u}d}$ or $T_{d\bar{d}d\bar{d}}$ predicted HMRT. In both cases it is assumed to be a tetraquark due to its decay channel.][Tdduu_predictions]{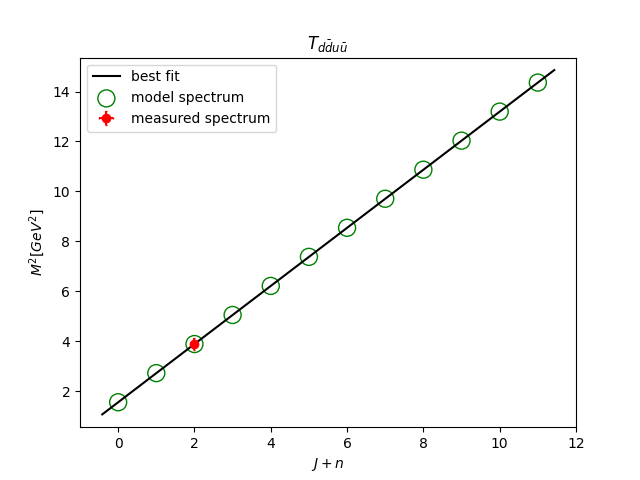}{tetraquarks/predictions/du-du/by_n/T_dudu_one_by_n}

\begin{longtable}[c]{|c|c|c|c|c|c|}
  \hline
  \thead{Thry \\ Width \\ $[MeV]$} & \thead{Mesons \\ Pair} & \thead{Meson \\ Threshold \\ $[MeV]$} & \thead{Baryon \\ Anti-\\baryon} & \thead{Baryonic \\ Threshold \\ $[MeV]$} & \thead{Genuine} \\
  \hline
  \hline
\makecell{$25-115$\\$38-75$} & \makecell{$\pi^{0}_{d} \pi^{0}_{u}$ \\ $\pi^{-} \pi^{+}$} & \makecell{$270$ \\ $280$} & \makecell{$p \bar{p}$ \\ $n \bar{n}$ \\ $\Lambda^{0} \bar{\Lambda}^{0}$} & \makecell{$1876$ \\ $1880$ \\ $2232$} &  \\
\hline
  \caption{$T_{d\bar{d}u\bar{u}}$ thresholds.} \label{table:Tdudu_thresholds}
 \end{longtable}

    \begin{center}
    \begin{minipage}{.49\linewidth}
        \begin{center}
            \begin{longtable}[c]{|c|c|}
  \hline
  \thead{$J$ \\ Spec} & \thead{$a_{T_{j}}$} \\
  \hline
  \hline
\makecell{$1157-1332$\\$1583-1715$\\$1916-2026$\\$2199-2295$\\$2450-2536$\\$2677-2756$\\$2886-2960$\\$3081-3150$\\$3264-3330$\\$3438-3500$\\$3603-3663$\\$3761-3818$} & \makecell{$-1.5$-$-1.1$} \\
\hline
  \caption{$T_{d\bar{d}u\bar{u}}$ $(J, M^2)$ predictions.} \label{table:Tdudu_j_predictions}
 \end{longtable}
        \end{center}
    \end{minipage}
    \begin{minipage}{.49\linewidth}
        \begin{center}
            \begin{longtable}[c]{|c|c|}
  \hline
  \thead{$n$ \\ Spec} & \thead{$a_{T_{n}}$} \\
  \hline
  \hline
\makecell{$1120-1300$\\$1570-1702$\\$1916-2026$\\$2209-2305$\\$2467-2553$\\$2700-2779$\\$2915-2988$\\$3115-3184$\\$3303-3368$\\$3481-3542$\\$3650-3709$\\$3811-3868$} & \makecell{$-1.4$-$-1.0$} \\
\hline
  \caption{$T_{d\bar{d}u\bar{u}}$ $(n, M^2)$ predictions.} \label{table:Tdudu_n_predictions}
 \end{longtable}
        \end{center}
    \end{minipage}
    \end{center}

\paragraph{$f_{4}(2300)$}{
    This state was observed in $f_{4}(2300)\rightarrow N\bar{N}$. 
    Tables \ref{table:f_4_J_HMRT} and
    \ref{table:f_4_n_HMRT} are the built HMRT that includes
    the measured $f_{4}(2300)$ and \ref{fig:f_4_2300} are the
    plot of these trajectories. These are consistent with table
    \ref{table:Tdudu_j_predictions}. The intercept values are
    $a_{J}=-0.52$ and $a_{n}=-0.37$.

    \begin{center}
    \begin{minipage}{.49\linewidth}
        \begin{center}
            \begin{longtable}[c]{|c|c|c|c|c|c|}
  \hline
  \thead{$State$} & \thead{$M$ \\ $[MeV]$} & \thead{Thry \\ $M$ \\ $[MeV]$} & \thead{$J$} & \thead{Thry \\ $n$} & \thead{Thry \\ $J + n$} \\
  \hline
  \hline
 &  & $1721$ &  &  & $2$ \\
\hline
 &  & $2031$ &  &  & $3$ \\
\hline
$f(2300)$ & $2300$ & $2300$ & $4$ & $0$ & $4$ \\
\hline
 &  & $2541$ &  &  & $5$ \\
\hline
 &  & $2760$ &  &  & $6$ \\
\hline
  \caption{$f_{4}(2300) (J, M^{2})$ HMRT.} \label{table:f_4_J_HMRT}
 \end{longtable}
        \end{center}
    \end{minipage}
    \begin{minipage}{.49\linewidth}
        \begin{center}
            \begin{longtable}[c]{|c|c|c|c|c|c|}
  \hline
  \thead{$State$} & \thead{$M$ \\ $[MeV]$} & \thead{Thry \\ $M$ \\ $[MeV]$} & \thead{$J$} & \thead{Thry \\ $n$} & \thead{Thry \\ $J + n$} \\
  \hline
  \hline
 &  & $1696$ &  &  & $2$ \\
\hline
 &  & $2021$ &  &  & $3$ \\
\hline
$f(2300)$ & $2300$ & $2300$ & $4$ & $0$ & $4$ \\
\hline
 &  & $2549$ &  &  & $5$ \\
\hline
 &  & $2775$ &  &  & $6$ \\
\hline
  \caption{$f_{4}(2300) (n, M^{2})$ HMRT.} \label{table:f_4_n_HMRT}
 \end{longtable}
        \end{center}
    \end{minipage}
    \end{center}

    \twographs[(a) is $(J, M^{2})$ $f_{4}(2300)$ MHRT, (b) $(n, M^{2})$ $f_{4}(2300)$ MHRT. In both cases it is assumed to be a tetraquark due to its decay channel.][f_4_2300]{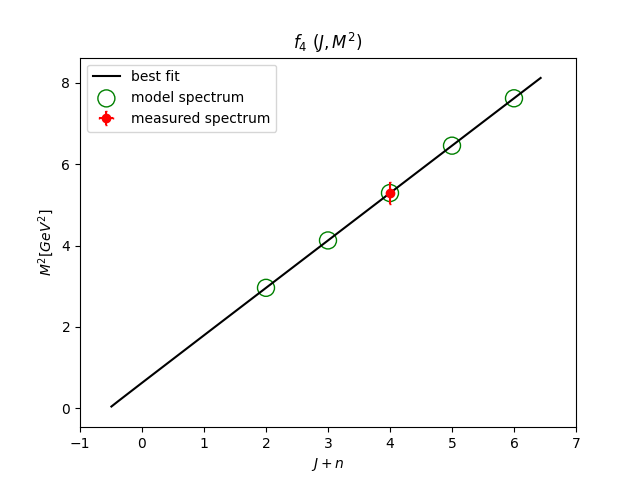}{tetraquarks/f4_2300_/by_n_one_point/fit-_f_f4_one_by_n}
}

\subsubsection{\texorpdfstring{$T_{u\bar{s}\bar{u}s}$ or $T_{d\bar{s}\bar{d}s}$}{}}

For a tetraquark with the quark content $u\bar{s}\bar{u}s$ or
$d\bar{s}\bar{d}s$ the mesonic and baryonic thresholds are
$\sim 990$ and $\sim 2230$ respectively.
The candidate resonances are presented in table
\ref{table:tetra_cand_usus}. Both states $X(3250)$ will be
discussed separately, since they are special cases of 3-body and
4-body decays. We generated the predicted HMRTs for this quark
content in tables \ref{table:Tsusu_j_predictions} and
\ref{table:Tsusu_n_predictions}, and the thresholds table
\ref{table:Tsusu_thresholds} includes the predicted width.
The HMRTs are also plotted in \ref{fig:Tdduu_predictions}.

\begin{longtable}[c]{|c|c|c|c|c|c|c|}
\hline
\makecell{Candidate} & \makecell{Quarks \\ Content} & \makecell{Selected \\ Decay \\ Channels} & \makecell{$J^{PC}$} & \makecell{Mass \\ $[MeV]$} & \makecell{Width \\ $[MeV]$} & \makecell{Ref.} \\
\hline
\hline
$f_{2}(1640)$ & \makecell{$u\bar{s}\bar{u}s$ \\ or \\ $d\bar{s}\bar{d}s$} & \makecell{$K\bar{K}$} & $2^{++}$ & $1639\pm{6}$ & $99^{+60}_{-40}$ & \makecell{\cite{f2_1640_1990}, \cite{f2_1640_1992}, \\ \cite{f2_1640_1995}} \\
\hline
$f_{2}(1750)$ & \makecell{$u\bar{s}\bar{u}s$ \\ or \\ $d\bar{s}\bar{d}s$} & \makecell{$K\bar{K}$} & $2^{++}$ & $1755\pm{10}$ & $67\pm{12}$ & \cite{f2_1750} \\
\hline
$f_{2}(2300)$ & \makecell{$u\bar{s}\bar{u}s$ \\ or \\ $d\bar{s}\bar{d}s$} & $\Lambda\bar{\Lambda}$ & $2^{++}$ & $2297\pm 28$ & $149\pm 40$ & \cite{f2_2300} \\
\hline
$\omega(2290)$ & \makecell{$u\bar{s}\bar{u}s$ \\ or \\ $d\bar{s}\bar{d}s$} & $\Lambda\bar{\Lambda}$ & $1^{--}$ & $2290\pm 20$ & $275\pm35$ & \makecell{\cite{omega_2290_2004}, \cite{omega_2290_2000}, \\ \cite{omega_2290_1990}} \\
\hline
$f_{3}(2300)$ & \makecell{$u\bar{s}\bar{u}s$ \\ or \\ $d\bar{s}\bar{d}s$} & $\Lambda\bar{\Lambda}$ & $3^{++}$ & $2334\pm25$ & $200\pm20$ & \cite{omega_2290_2004}, \cite{omega_2290_2000} \\
\hline
$X(3250)$ & \makecell{$\bar{u}su\bar{s}$ \\ or \\ $\bar{d}sd\bar{s}$} & $\Lambda\bar{p}K^{+}$ & $?^{??}$ & $3250\pm{8}\pm{20}$ & $45\pm{18}$ & \cite{K_tetra_1993} \\
\hline
$X(3250)$ & \makecell{$\bar{u}su\bar{s}$ \\ or \\ $\bar{d}sd\bar{s}$} & \makecell{$\Lambda\bar{p}K^{+}\pi^{\pm}$ \\ $K^{0}_{S}p\bar{p}K^{\pm}$} & $?^{??}$ & $3245\pm{8}\pm{20}$ & $25\pm{11}$ & \cite{K_tetra_1993} \\
\hline
\caption{$T_{u\bar{d}\bar{u}d}$ or $T_{d\bar{d}d\bar{d}}$ tetraquark candidates.}\label{table:tetra_cand_usus}
\end{longtable}

\begin{longtable}[c]{|c|c|c|c|c|c|}
  \hline
  \thead{Thry \\ Width \\ $[MeV]$} & \thead{Mesons \\ Pair} & \thead{Meson \\ Threshold \\ $[MeV]$} & \thead{Baryon \\ Anti-\\baryon} & \thead{Baryonic \\ Threshold \\ $[MeV]$} & \thead{Genuine} \\
  \hline
  \hline
\makecell{$79-364$\\$30-59$} & \makecell{$K^{-} K^{+}$ \\ $\phi(1020) \pi^{0}_{u}$} & \makecell{$988$ \\ $1154$} & \makecell{$\Lambda^{0} \bar{\Lambda}^{0}$ \\ $\Sigma^{+} \bar{\Sigma}^{-}$ \\ $\Xi^{0} \bar{\Xi}^{0}$} & \makecell{$2232$ \\ $2378$ \\ $2630$} &  \\
\hline
  \caption{$T_{s\bar{s}u\bar{u}}$ thresholds.} \label{table:Tsusu_thresholds}
 \end{longtable}

    \begin{center}
    \begin{minipage}{.49\linewidth}
        \begin{center}
            \begin{longtable}[c]{|c|c|}
  \hline
  \thead{$J$ \\ Spec} & \thead{$a_{T_{j}}$} \\
  \hline
  \hline
\makecell{$1591-1755$\\$1968-2097$\\$2272-2382$\\$2535-2633$\\$2770-2859$\\$2985-3067$\\$3184-3261$\\$3371-3443$\\$3547-3615$\\$3714-3779$\\$3873-3936$\\$4026-4086$} & \makecell{$-1.6$-$-1.2$} \\
\hline
  \caption{$T_{s\bar{s}u\bar{u}}$ $(J, M^2)$ predictions.} \label{table:Tsusu_j_predictions}
 \end{longtable}
        \end{center}
    \end{minipage}
    \begin{minipage}{.49\linewidth}
        \begin{center}
            \begin{longtable}[c]{|c|c|}
  \hline
  \thead{$n$ \\ Spec} & \thead{$a_{T_{n}}$} \\
  \hline
  \hline
\makecell{$1559-1727$\\$1956-2085$\\$2272-2382$\\$2544-2641$\\$2786-2875$\\$3008-3089$\\$3212-3288$\\$3403-3475$\\$3584-3651$\\$3755-3819$\\$3918-3980$\\$4075-4134$} & \makecell{$-1.5$-$-1.1$} \\
\hline
  \caption{$T_{s\bar{s}u\bar{u}}$ $(n, M^2)$ predictions.} \label{table:Tsusu_n_predictions}
 \end{longtable}
        \end{center}
    \end{minipage}
    \end{center}

\twographs[(a) is $(J, M^{2})$ $T_{u\bar{s}\bar{u}s}$ or $T_{d\bar{s}\bar{d}s}$ predicted HMRT, (b) $(n, M^{2})$ $T_{u\bar{s}\bar{u}s}$ or $T_{d\bar{s}\bar{d}s}$ predicted HMRT.][Tsusu_predictions]{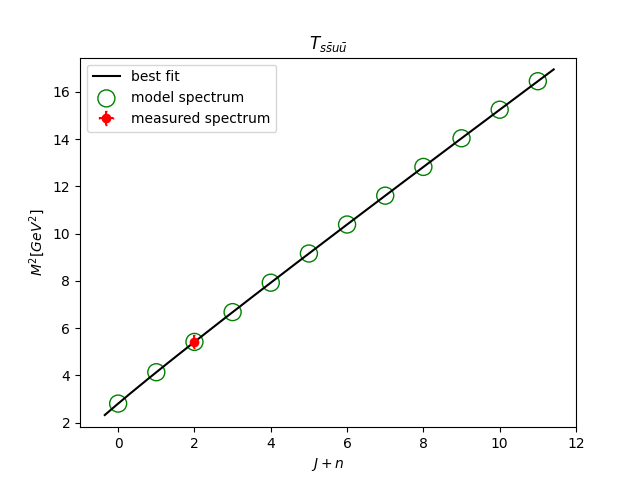}{tetraquarks/predictions/su-su/by_n/T_susu_one_by_n}

From table \ref{table:Tsusu_n_predictions} we can see that indeed
states $f_{2}(2300)$ and $f_{2}(1750)$ belong to the same predicted
$n$ trajectory. The second row is the first state above the baryon-antibaryon
threshold.

All states ($f_{2}(2300)$, $\omega(2290)$ and $f_{3}(2300)$) are on the predicted
trajectories, differ only by the composition of $J+n$. The predicted width is
a bit above the measured one for $f_{2}(2300)$ and $f_{3}(2300)$, but it
provides a good estimation.

We first create a trajectory in the $(n, M^{2})$ plane for $f_{2}(1750)$ and
$f_{2}(2300)$. The first state didn't match any $f$ or $\omega/f$ trajectory so far,
and it was selected since its meson-pair minimal quark content matches the
$f_{2}(2300)$. We expect the fit to match the mesons slope and the endpoint
masses to be similar to the sum of the $m_{s}$ and $m_{u/d}$.

\paragraph{$\omega(2290)$}{
    This state was observed in $\omega(2290)\rightarrow\Lambda\bar{\Lambda}$. This state was predicted to
    exist in \cite{Sonnenschein_2017_4630} as the analog of the decay of
    $\psi(4660)$ into a pair of $\Lambda_c\bar\Lambda_c$.

    Tables \ref{table:omega_J_HMRT} and \ref{table:omega_n_HMRT} are the built
    HMRT that includes the measured $\omega(2290)$ and \ref{fig:omega_2290} are the
    plot of these trajectories. These are consistent with tables
    \ref{table:Tsusu_j_predictions} and \ref{table:Tsusu_n_predictions}. The
    intercept values are $a_{J}=-0.24$ and $a_{n}=-1.13$.

    \begin{center}
    \begin{minipage}{.49\linewidth}
        \begin{center}
            \begin{longtable}[c]{|c|c|c|c|c|c|}
  \hline
  \thead{$State$} & \thead{$M$ \\ $[MeV]$} & \thead{Thry \\ $M$ \\ $[MeV]$} & \thead{$J$} & \thead{Thry \\ $n$} & \thead{Thry \\ $J + n$} \\
  \hline
  \hline
 &  & $1619$ &  &  & $1$ \\
\hline
 &  & $1989$ &  &  & $2$ \\
\hline
$\omega(2290)$ & $2290$ & $2290$ & $1$ & $2$ & $3$ \\
\hline
 &  & $2551$ &  &  & $4$ \\
\hline
 &  & $2785$ &  &  & $5$ \\
\hline
  \caption{$\omega(2290) (J, M^{2})$ HMRT.} \label{table:omega_J_HMRT}
 \end{longtable}
        \end{center}
    \end{minipage}
    \begin{minipage}{.49\linewidth}
        \begin{center}
            \begin{longtable}[c]{|c|c|c|c|c|c|}
  \hline
  \thead{$State$} & \thead{$M$ \\ $[MeV]$} & \thead{Thry \\ $M$ \\ $[MeV]$} & \thead{$J$} & \thead{Thry \\ $n$} & \thead{Thry \\ $J + n$} \\
  \hline
  \hline
 &  & $1587$ &  &  & $0$ \\
\hline
 &  & $1977$ &  &  & $1$ \\
\hline
$\omega(2290)$ & $2290$ & $2290$ & $1$ & $1$ & $2$ \\
\hline
 &  & $2560$ &  &  & $3$ \\
\hline
 &  & $2801$ &  &  & $4$ \\
\hline
  \caption{$\omega(2290) (n, M^{2})$ HMRT.} \label{table:omega_n_HMRT}
 \end{longtable}
        \end{center}
    \end{minipage}
    \end{center}

    \twographs[(a) is $(J, M^{2})$ $\omega(2290)$ MHRT, (b) $(n, M^{2})$ $\omega(2290)$ MHRT. In both cases it is assumed to be a tetraquark due to its decay channel.][omega_2290]{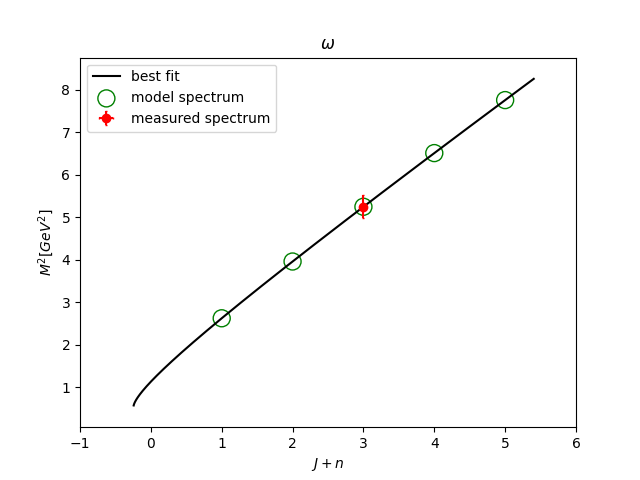}{tetraquarks/omega_2290_/by_n_one_point/fit-_omega_one_by_n}
}

\paragraph{$f_{2}(2300)$}{
    This state was observed in $f_{2}(2300)\rightarrow\Lambda\bar{\Lambda}$. Most of the states of
    $f_{2}$ do not match any meson trajectory that we built so far. The
    predicted $(n, M^{2})$ HMRT provides indication for the states that may be
    considered as tetraquarks, assuming $f_{2}(2300)$, which decays to
    $\Lambda\bar{\Lambda}$ just above threshold, is on the trajectory. We searched the
    $f_{2}$ known states for other states on the HMRT and found also
    $f_{2}(1640)$ that decays to $K\bar{K}$. Both $f_{2}(1750)$ and
    $f_{2}(1640)$ fit the $(n, M^{2})$ predicted tetraquark trajectory.

    Tables \ref{table:f_2_J_HMRT} and \ref{table:f_2_n_HMRT} are the built HMRT
    that includes the measured $f_{2}(2300)$ and \ref{fig:f_2_2300} are the plot
    of these trajectories. These are consistent with tables
    \ref{table:Tsusu_j_predictions} and \ref{table:Tsusu_n_predictions}.
    The intercept values are $a_{J}=-1.27$ and $a_{n}=-1.15$.

    \begin{center}
    \begin{minipage}{.49\linewidth}
        \begin{center}
            \begin{longtable}[c]{|c|c|c|c|c|c|}
  \hline
  \thead{$State$} & \thead{$M$ \\ $[MeV]$} & \thead{Thry \\ $M$ \\ $[MeV]$} & \thead{$J$} & \thead{Thry \\ $n$} & \thead{Thry \\ $J + n$} \\
  \hline
  \hline
 &  & $1629$ &  &  & $0$ \\
\hline
 &  & $1997$ &  &  & $1$ \\
\hline
$f(2300)$ & $2297$ & $2297$ & $2$ & $0$ & $2$ \\
\hline
 &  & $2557$ &  &  & $3$ \\
\hline
 &  & $2790$ &  &  & $4$ \\
\hline
  \caption{$f_{2}(2300) (J, M^{2})$ HMRT.} \label{table:f_2_J_HMRT}
 \end{longtable}
        \end{center}
    \end{minipage}
    \begin{minipage}{.49\linewidth}
        \begin{center}
            \begin{longtable}[c]{|c|c|c|c|c|c|}
  \hline
  \thead{$State$} & \thead{$M$ \\ $[MeV]$} & \thead{Thry \\ $M$ \\ $[MeV]$} & \thead{$J$} & \thead{Thry \\ $n$} & \thead{Thry \\ $J + n$} \\
  \hline
  \hline
 &  & $1598$ &  &  & $0$ \\
\hline
 &  & $1985$ &  &  & $1$ \\
\hline
$f(2300)$ & $2297$ & $2297$ & $2$ & $0$ & $2$ \\
\hline
 &  & $2566$ &  &  & $3$ \\
\hline
 &  & $2806$ &  &  & $4$ \\
\hline
  \caption{$f_{2}(2300) (n, M^{2})$ HMRT.} \label{table:f_2_n_HMRT}
 \end{longtable}
        \end{center}
    \end{minipage}
    \end{center}

    \twographs[(a) is $(J, M^{2})$ $f_{2}(2300)$ MHRT, (b) $(n, M^{2})$ $f_{2}(2300)$ MHRT. In both cases it is assumed to be a tetraquark due to its decay channel.][f_2_2300]{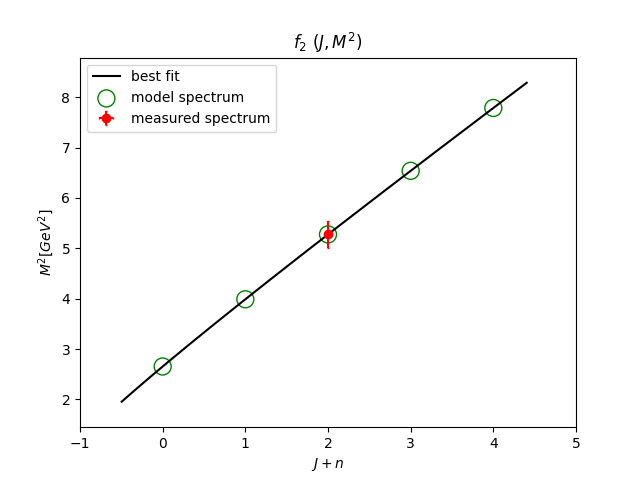}{tetraquarks/f2_2300_/by_n_one_point/fit-_f_f2_one_by_n}
}

\paragraph{$f_{3}(2300)$}{
    This state was observed in $f_{3}(2300)\rightarrow\Lambda\bar{\Lambda}$. Tables
    \ref{table:f_3_J_HMRT} and \ref{table:f_3_n_HMRT} are the built HMRT that
    includes the measured $f_{3}(2300)$ and \ref{fig:f_3_2300} are the plot of
    these trajectories. These are consistent with tables
    \ref{table:Tsusu_j_predictions} and \ref{table:Tsusu_n_predictions}.
    The intercept values are $a_{J}=-0.40$ and $a_{n}=-0.28$.

    \begin{center}
    \begin{minipage}{.49\linewidth}
        \begin{center}
            \begin{longtable}[c]{|c|c|c|c|c|c|}
  \hline
  \thead{$State$} & \thead{$M$ \\ $[MeV]$} & \thead{Thry \\ $M$ \\ $[MeV]$} & \thead{$J$} & \thead{Thry \\ $n$} & \thead{Thry \\ $J + n$} \\
  \hline
  \hline
 &  & $1685$ &  &  & $1$ \\
\hline
 &  & $2041$ &  &  & $2$ \\
\hline
$f(2300)$ & $2334$ & $2334$ & $3$ & $0$ & $3$ \\
\hline
 &  & $2590$ &  &  & $4$ \\
\hline
 &  & $2820$ &  &  & $5$ \\
\hline
  \caption{$f_{3}(2300) (J, M^{2})$ HMRT.} \label{table:f_3_J_HMRT}
 \end{longtable}
        \end{center}
    \end{minipage}
    \begin{minipage}{.49\linewidth}
        \begin{center}
            \begin{longtable}[c]{|c|c|c|c|c|c|}
  \hline
  \thead{$State$} & \thead{$M$ \\ $[MeV]$} & \thead{Thry \\ $M$ \\ $[MeV]$} & \thead{$J$} & \thead{Thry \\ $n$} & \thead{Thry \\ $J + n$} \\
  \hline
  \hline
 &  & $1655$ &  &  & $1$ \\
\hline
 &  & $2029$ &  &  & $2$ \\
\hline
$f(2300)$ & $2334$ & $2334$ & $3$ & $0$ & $3$ \\
\hline
 &  & $2599$ &  &  & $4$ \\
\hline
 &  & $2836$ &  &  & $5$ \\
\hline
  \caption{$f_{3}(2300) (n, M^{2})$ HMRT.} \label{table:f_3_n_HMRT}
 \end{longtable}
        \end{center}
    \end{minipage}
    \end{center}

    \twographs[(a) is $(J, M^{2})$ $f_{3}(2300)$ MHRT, (b) $(n, M^{2})$ $f_{3}(2300)$ MHRT. In both cases it is assumed to be a tetraquark due to its decay channel.][f_3_2300]{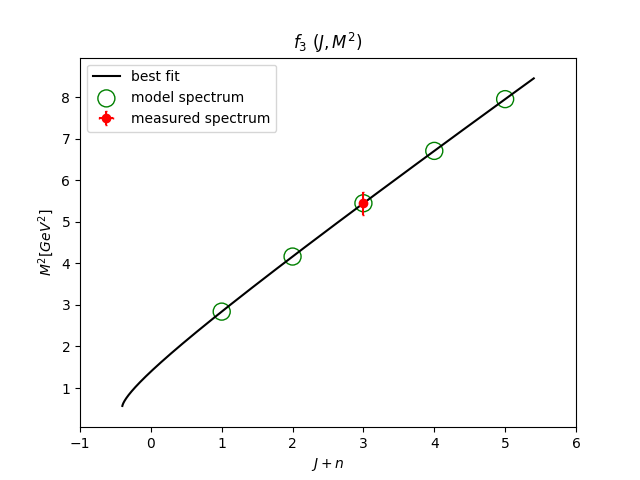}{tetraquarks/f3_2300_/by_n_one_point/fit-_f_f3_one_by_n}
}

\paragraph{$X(3250)$}{
    This state is categorized as a meson in PDG database. 
    Table \ref{table:Tsusu_j_predictions} indeed predicts the mass of both
    $X(3250)$ states, implying they may be on the J trajectory of
    $T_{u\bar{s}\bar{u}s}$.
    
    However, the decay product combination $\Lambda \bar{p}K^{+}$ could originate from a tetraquark with a configuration as in \ref{fig:tetraquark_complex}, that decay through string tear to a tetra-like pentaquark and a baryon.
    In this paper we focused on the simple tetraquark configuration, therefore we leave the categorization of this state as an open question.
}

\subsubsection{\texorpdfstring{$K(3100)$ ($T_{s\bar{u}u\bar{d}}$)}{}}

This state (table \ref{table:tetra_k_3100_cand}) was observed only in baryonic
decays, and decays also to $K(3100)^{0}\rightarrow\Sigma^{+}\bar{p}$ \cite{K_tetra_1993}.
Tables \ref{table:Tdusu_j_predictions} and \ref{table:Tdusu_n_predictions} are
the predicted HMRTs in both planes. The thresholds and predicted width are at
table \ref{table:Tdusu_thresholds}.

\begin{longtable}[c]{|c|c|c|c|c|c|c|}
\hline
\makecell{Candidate} & \makecell{Quarks \\ Content} & \makecell{Selected \\ Decay \\ Channels} & \makecell{$J^{PC}$} & \makecell{Mass \\ $[MeV]$} & \makecell{Width \\ $[MeV]$} & \makecell{Ref.} \\
\hline
\hline
$K(3100)^{0}$ & $u\bar{s}\bar{u}d$ & $\Sigma(1385)^{+}\bar{p}$ & $?^{??}$ & $\sim 3100$ & $\sim 10-70$ & \cite{K_tetra_1993} \\
\hline
\caption{$T_{u\bar{u}\bar{u}d}$ or $T_{\bar{u}d\bar{d}d}$ exotic candidate.}\label{table:tetra_k_3100_cand}
\end{longtable}

\begin{longtable}[c]{|c|c|c|c|c|c|}
  \hline
  \thead{Thry \\ Width \\ $[MeV]$} & \thead{Mesons \\ Pair} & \thead{Meson \\ Threshold \\ $[MeV]$} & \thead{Baryon \\ Anti-\\baryon} & \thead{Baryonic \\ Threshold \\ $[MeV]$} & \thead{Genuine} \\
  \hline
  \hline
\makecell{$45-209$\\$34-66$} & \makecell{$K^{0} \pi^{0}_{u}$ \\ $\pi^{-} K^{+}$} & \makecell{$633$ \\ $634$} & \makecell{$n \bar{\Lambda}^{0}$ \\ $p \bar{\Sigma}^{-}$ \\ $\Lambda^{0} \bar{\Xi}^{0}$} & \makecell{$2056$ \\ $2127$ \\ $2431$} &  \\
\hline
  \caption{$T_{d\bar{s}u\bar{u}}$ thresholds.} \label{table:Tdusu_thresholds}
 \end{longtable}

    \begin{center}
    \begin{minipage}{.49\linewidth}
        \begin{center}
            \begin{longtable}[c]{|c|c|}
  \hline
  \thead{$J$ \\ Spec} & \thead{$a_{T_{j}}$} \\
  \hline
  \hline
\makecell{$1378-1547$\\$1778-1908$\\$2096-2206$\\$2369-2466$\\$2611-2699$\\$2832-2913$\\$3036-3112$\\$3227-3298$\\$3407-3474$\\$3577-3641$\\$3739-3800$\\$3894-3953$} & \makecell{$-1.5$-$-1.2$} \\
\hline
  \caption{$T_{d\bar{s}u\bar{u}}$ $(J, M^2)$ predictions.} \label{table:Tdusu_j_predictions}
 \end{longtable}
        \end{center}
    \end{minipage}
    \begin{minipage}{.49\linewidth}
        \begin{center}
            \begin{longtable}[c]{|c|c|}
  \hline
  \thead{$n$ \\ Spec} & \thead{$a_{T_{n}}$} \\
  \hline
  \hline
\makecell{$1344-1517$\\$1765-1896$\\$2096-2206$\\$2378-2475$\\$2628-2715$\\$2855-2935$\\$3065-3139$\\$3260-3330$\\$3444-3511$\\$3619-3682$\\$3785-3845$\\$3944-4001$} & \makecell{$-1.4$-$-1.0$} \\
\hline
  \caption{$T_{d\bar{s}u\bar{u}}$ $(n, M^2)$ predictions.} \label{table:Tdusu_n_predictions}
 \end{longtable}
        \end{center}
    \end{minipage}
    \end{center}

We built its HMRTs in $(n, M^{2})$ and $(J, M^{2})$ (tables \ref{table:k_3100_J} and \ref{table:k_3100_n}) according to the mesonic case slopes, which are consistent with \ref{table:k_3100_J} and \ref{table:k_3100_n}. The intercepts are $a_{J}=-4.22$ and $a_{n}=-3.81$.

    \begin{center}
    \begin{minipage}{.49\linewidth}
        \begin{center}
            \begin{longtable}[c]{|c|c|c|c|c|c|}
  \hline
  \thead{$State$} & \thead{$M$ \\ $[MeV]$} & \thead{Thry \\ $M$ \\ $[MeV]$} & \thead{$J$} & \thead{Thry \\ $n$} & \thead{Thry \\ $J + n$} \\
  \hline
  \hline
 &  & $2686$ &  &  & $1$ \\
\hline
 &  & $2901$ &  &  & $2$ \\
\hline
$K(3100)^{+}$ & $3100$ & $3100$ &  &  & $3$ \\
\hline
 &  & $3287$ &  &  & $4$ \\
\hline
 &  & $3463$ &  &  & $5$ \\
\hline
  \caption{$K(3100)$ $(J, M^{2})$ HMRT.} \label{table:k_3100_J}
 \end{longtable}
        \end{center}
    \end{minipage}
    \begin{minipage}{.49\linewidth}
        \begin{center}
            \begin{longtable}[c]{|c|c|c|c|c|c|}
  \hline
  \thead{$State$} & \thead{$M$ \\ $[MeV]$} & \thead{Thry \\ $M$ \\ $[MeV]$} & \thead{$J$} & \thead{Thry \\ $n$} & \thead{Thry \\ $J + n$} \\
  \hline
  \hline
 &  & $2669$ &  &  & $1$ \\
\hline
 &  & $2893$ &  &  & $2$ \\
\hline
$K(3100)^{+}$ & $3100$ & $3100$ &  &  & $3$ \\
\hline
 &  & $3293$ &  &  & $4$ \\
\hline
 &  & $3476$ &  &  & $5$ \\
\hline
  \caption{$K(3100)$ $(n, M^{2})$ HMRT.} \label{table:k_3100_n}
 \end{longtable}
        \end{center}
    \end{minipage}
    \end{center}

\twographs[(a) is $(J, M^{2})$ $K(3100)$ MHRT, (b) $(n, M^{2})$ $K(3100)$ MHRT. In both cases it is assumed to be a tetraquark due to its decay channel.][K_3100]{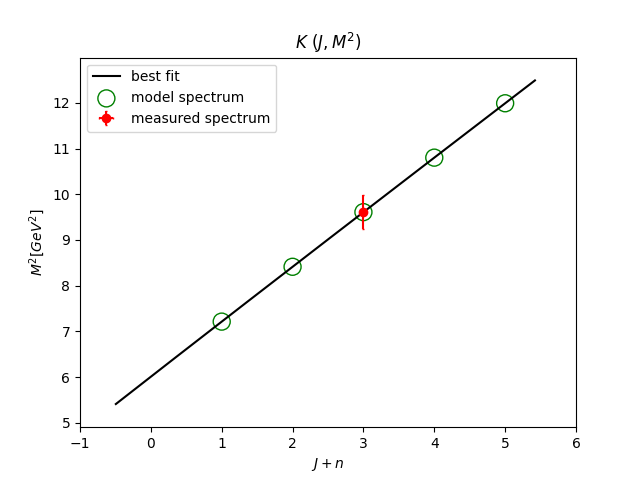}{tetraquarks/K_3100_/by_n_one_point/fit-_K__k_tetra_one_by_n}

\subsubsection{\texorpdfstring{$T_{c\bar{u}u\bar{s}}$ or $T_{c\bar{d}u\bar{s}}$}{}}

For a tetraquarks  with the quark content $u\bar{s}\bar{u}s$ or $d\bar{s}\bar{d}s$ (table \ref{table:tetra_cand_dscu}). The candidate resonances are presented in table. We generated the predicted HMRTs for this quark content in tables \ref{table:Tdscu_j_predictions} and \ref{table:Tdscu_n_predictions}, and the thresholds table \ref{table:Tdscu_thresholds} includes the predicted width. The HMRTs are also plotted in \ref{fig:Tdscu_predictions}.

\begin{longtable}[c]{|c|c|c|c|c|c|c|}
\hline
\makecell{Candidate} & \makecell{Quarks \\ Content} & \makecell{Selected \\ Decay \\ Channels} & \makecell{$J^{PC}$} & \makecell{Mass \\ $[MeV]$} & \makecell{Width \\ $[MeV]$} & \makecell{Ref.} \\
\hline
\hline
$X(2632)$ & $c\bar{u}u\bar{s}$ & \makecell{$D^{0}K^{+}$ \\ $D^{+}_{s}\eta$} & $?^{??}$ & $2635.2\pm{3.3}$ & $<17$ & \cite{SELEX:2004drx} \\
\hline
$T_{cs0}(2900)^{0}$ & $\bar{c}du\bar{s}$ & $D^{-}K^{+}$ & $0^{+}$ & $2866\pm{7}$ & $57\pm{13}$ & \cite{LHCb:2020pxc} \\
\hline
$T_{cs1}(2900)^{0}$ & $\bar{c}d\bar{u}s$ & $D^{-}K^{+}$ & $1^{-}$ & $2904\pm{5}$ & $110\pm{12}$ & \cite{LHCb:2020pxc} \\
\hline
$T^{a}_{c\bar{s}0}(2900)^{0}$ & $c\bar{s}\bar{u}d$ & $D^{+}_{s}\pi^{-}$ & $0^{?}$ & $2892\pm{14}\pm{15}$ & $119\pm{26}\pm{13}$ & \cite{LHCb:2022sfr} \\
\hline
$T^{a}_{c\bar{s}0}(2900)^{++}$ & $c\bar{s}u\bar{d}$ & $D^{+}_{s}\pi^{+}$ & $0^{?}$ & $2921\pm{17}\pm{20}$ & $137\pm{32}\pm{17}$ & \cite{LHCb:2022sfr} \\
\hline
\caption{$T_{c\bar{u}u\bar{s}}$ or $T_{c\bar{d}u\bar{s}}$ tetraquark candidates.}\label{table:tetra_cand_dscu}
\end{longtable}

\begin{longtable}[c]{|c|c|c|c|c|c|}
  \hline
  \thead{Thry \\ Width \\ $[MeV]$} & \thead{Mesons \\ Pair} & \thead{Meson \\ Threshold \\ $[MeV]$} & \thead{Baryon \\ Anti-\\baryon} & \thead{Baryonic \\ Threshold \\ $[MeV]$} & \thead{Genuine} \\
  \hline
  \hline
\makecell{$98-147$\\$29-59$} & \makecell{$\pi^{-} D^{-}_{s}$ \\ $D^{-} K^{-}$} & \makecell{$2108$ \\ $2364$} & \makecell{$\Sigma^{-} \bar{\Lambda}^{-}_{c}$ \\ $\Lambda^{0} \bar{\Sigma}^{--}_{c}$ \\ $\Xi^{-} \bar{\Xi}^{-}_{c}$} & \makecell{$3483$ \\ $3570$ \\ $3790$} & $**$ \\
\hline
  \caption{$T_{d\bar{c}s\bar{u}}$ thresholds.} \label{table:Tdscu_thresholds}
 \end{longtable}

    \begin{center}
    \begin{minipage}{.49\linewidth}
        \begin{center}
            \begin{longtable}[c]{|c|c|}
  \hline
  \thead{$J$ \\ Spec} & \thead{$a_{T_{j}}$} \\
  \hline
  \hline
\makecell{$2946-3085$\\$3252-3374$\\$3523-3633$\\$3769-3870$\\$3996-4091$\\$4209-4298$\\$4409-4493$\\$4599-4679$\\$4780-4856$\\$4953-5027$\\$5120-5191$\\$5281-5349$} & \makecell{$-2.1$-$-1.6$} \\
\hline
  \caption{$T_{d\bar{c}s\bar{u}}$ $(J, M^2)$ predictions.} \label{table:Tdscu_j_predictions}
 \end{longtable}
        \end{center}
    \end{minipage}
    \begin{minipage}{.49\linewidth}
        \begin{center}
            \begin{longtable}[c]{|c|c|}
  \hline
  \thead{$n$ \\ Spec} & \thead{$a_{T_{n}}$} \\
  \hline
  \hline
\makecell{$2628-2791$\\$3121-3250$\\$3523-3633$\\$3872-3970$\\$4186-4275$\\$4473-4556$\\$4740-4818$\\$4991-5064$\\$5228-5297$\\$5453-5519$\\$5668-5731$\\$5874-5935$} & \makecell{$-1.5$-$-1.2$} \\
\hline
  \caption{$T_{d\bar{c}s\bar{u}}$ $(n, M^2)$ predictions.} \label{table:Tdscu_n_predictions}
 \end{longtable}
        \end{center}
    \end{minipage}
    \end{center}

\twographs[(a) is $(J, M^{2})$ $T_{c\bar{u}u\bar{s}}$ or $T_{c\bar{d}u\bar{s}}$ predicted HMRT, (b) $(n, M^{2})$ $T_{c\bar{u}u\bar{s}}$ or $T_{c\bar{d}u\bar{s}}$ predicted HMRT.][Tdscu_predictions]{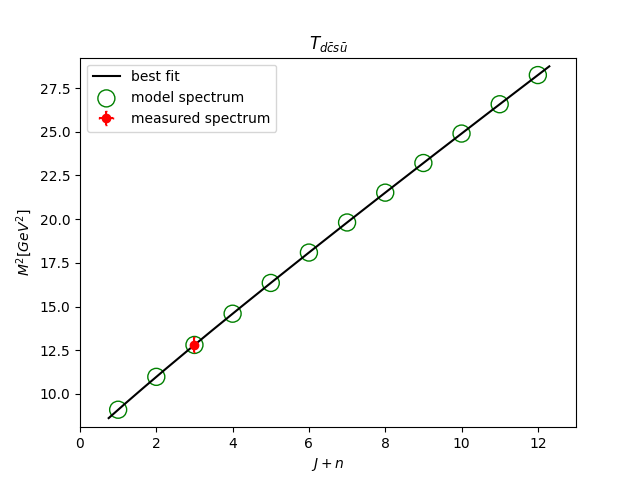}{tetraquarks/predictions/ds-cu/by_n/T_dscu_one_by_n}

We can see from tables \ref{table:Tdscu_j_predictions},
\ref{table:Tdscu_n_predictions} that $X(2632)$ fit the n HMRT,
while the other states are within error range to the J trajectory.
The difference between the actual mass of the states and the predicted
one is related to the difference in quantum numbers and the actual quark
content. The HISH model does not differentiate between $u/d$ in terms of
$m$.

In the next section, we analysed and fitted the $T_{cs}$ measured states.

\paragraph{$X(2632)$}{
    This state was observed in $X(2632)\rightarrow D^{0}K^{+}$ and $X(2632)\rightarrow D^{+}_{s}\eta$.

    Tables \ref{table:X_2632_J_HMRT} and \ref{table:X_2632_n_HMRT} are the built HMRT that includes the measured $X(2632)$ and \ref{fig:X_2632} are the plot of these trajectories. These are consistent with tables \ref{table:Tdscu_j_predictions} and \ref{table:Tdscu_n_predictions}. The intercept values are $a_{J}=-0.77$ and $a_{n}=-1.23$.

    \begin{center}
    \begin{minipage}{.49\linewidth}
        \begin{center}
            \begin{longtable}[c]{|c|c|c|c|c|c|}
  \hline
  \thead{$State$} & \thead{$M$ \\ $[MeV]$} & \thead{Thry \\ $M$ \\ $[MeV]$} & \thead{$J$} & \thead{Thry \\ $n$} & \thead{Thry \\ $J + n$} \\
  \hline
  \hline
$X(2632)$ & $2635$ & $2635$ &  &  & $1$ \\
\hline
 &  & $2988$ &  &  & $2$ \\
\hline
 &  & $3289$ &  &  & $3$ \\
\hline
 &  & $3556$ &  &  & $4$ \\
\hline
 &  & $3799$ &  &  & $5$ \\
\hline
  \caption{$X(2632)$ $(J, M^{2})$ HMRT.} \label{table:X_2632_J_HMRT}
 \end{longtable}
        \end{center}
    \end{minipage}
    \begin{minipage}{.49\linewidth}
        \begin{center}
            \begin{longtable}[c]{|c|c|c|c|c|c|}
  \hline
  \thead{$State$} & \thead{$M$ \\ $[MeV]$} & \thead{Thry \\ $M$ \\ $[MeV]$} & \thead{$J$} & \thead{Thry \\ $n$} & \thead{Thry \\ $J + n$} \\
  \hline
  \hline
$X(2632)$ & $2635$ & $2635$ &  &  & $0$ \\
\hline
 &  & $3127$ &  &  & $1$ \\
\hline
 &  & $3528$ &  &  & $2$ \\
\hline
 &  & $3876$ &  &  & $3$ \\
\hline
 &  & $4190$ &  &  & $4$ \\
\hline
  \caption{$X(2632)$ $(n, M^{2})$ HMRT.} \label{table:X_2632_n_HMRT}
 \end{longtable}
        \end{center}
    \end{minipage}
    \end{center}

    \twographs[(a) is $(J, M^{2})$ $X(2632)$ MHRT, (b) $(n, M^{2})$ $X(2632)$ MHRT. In both cases it is assumed to be a tetraquark due to its decay channel.][X_2632]{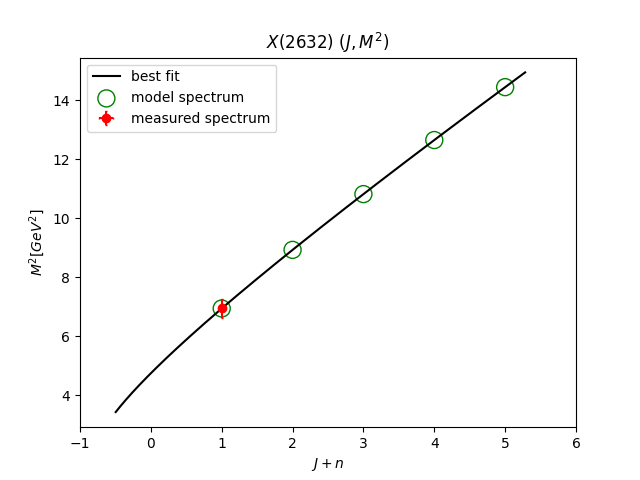}{tetraquarks/X_2632_/by_n_one_point/fit-_X_X_2632__tetra_one_by_n}
}

\pagebreak

\subsubsection{\texorpdfstring{$T_{cs}$ (with quark content ${c\bar{d}s\bar{u}}$)}{}}

In \cite{Aaij_2020_Tcdsu} the states $T_{cs0}$ and $T_{cs1}$ were observed. Both tetraquarks candidates have roughly the same mass. This implies that the total quantity $J + n$ is the same. The decay channel in which these states were observed is $D^{-}K^{+}$, which is according to the annihilation mechanism. This is expected, since the string tear threshold for baryonia ($\Xi_{c}p$ pair) creation is higher at around $3.4GeV$. Both trajectories are built using the $D_{s}$ slopes as calculated in \cite{Sonnenschein_2014_mesons}.

\paragraph{$T^{0}_{cs0}$ HMRTs}{

\twographs[(a) is $(J, M^{2})$ $T^{0}_{cs0}$ MHRT, (b) $(n, M^{2})$ $T^{0}_{cs0}$ MHRT by a single point. The trajectory was built with fixed massive endpoints according to the universal fit done for mesons.][T_cs0]{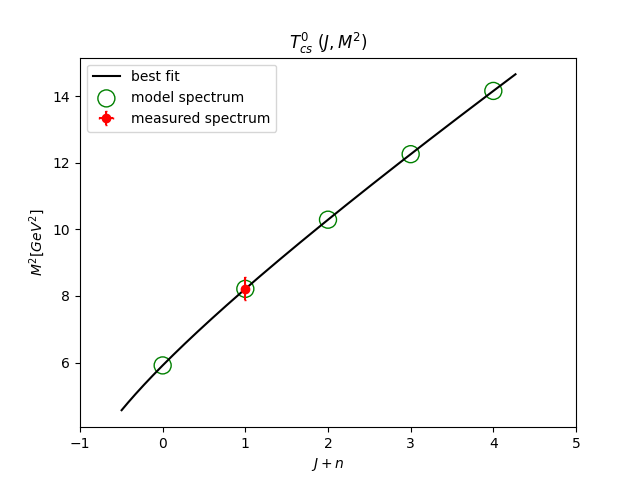}{tetraquarks/T_cs/Tcs0/by_n_one_point/fit-_T0_cs_Tcs0_by_n}

    \begin{center}
    \begin{minipage}{.49\linewidth}
        \begin{center}
            \begin{longtable}[c]{|c|c|c|c|c|c|}
  \hline
  \thead{$State$} & \thead{$M$ \\ $[MeV]$} & \thead{Thry \\ $M$ \\ $[MeV]$} & \thead{$J$} & \thead{Thry \\ $n$} & \thead{Thry \\ $J + n$} \\
  \hline
  \hline
    $T(2900)^{0}_{cs0}$ & $2866$ & $2866$ & $0$ & $1$ & $1$ \\
    \hline
     &  & $3208$ &  &  & $2$ \\
    \hline
     &  & $3501$ &  &  & $3$ \\
    \hline
     &  & $3762$ &  &  & $4$ \\
    \hline
     &  & $4001$ &  &  & $5$ \\
    \hline
    \caption{$T_{cs}^{0}$ $(J, M^{2})$ HMRT.} \label{table:Tcs0_by_j_one_point}
\end{longtable}
        \end{center}
    \end{minipage}
    \begin{minipage}{.49\linewidth}
        \begin{center}
            \begin{longtable}[c]{|c|c|c|c|c|c|}
  \hline
  \thead{$State$} & \thead{$M$ \\ $[MeV]$} & \thead{Thry \\ $M$ \\ $[MeV]$} & \thead{$J$} & \thead{Thry \\ $n$} & \thead{Thry \\ $J + n$} \\
  \hline
  \hline
$T(2900)^{0}_{cs0}$ & $2866$ & $2866$ & $0$ & $0$ & $0$ \\
\hline
 &  & $3343$ &  &  & $1$ \\
\hline
 &  & $3734$ &  &  & $2$ \\
\hline
 &  & $4076$ &  &  & $3$ \\
\hline
 &  & $4384$ &  &  & $4$ \\
\hline
  \caption{$T_{cs}^{0}$ $(n, M^{2})$ HMRT.} \label{table:Tcs0_by_n_one_point}
 \end{longtable}
        \end{center}
    \end{minipage}
    \end{center}

The intercept values are $a_{j}=-0.99$ and $a_{n}=-1.38$. The decays widths are $\Gamma_{tear}\lesssim 127MeV$ and $\Gamma_{annihilation}\approx 5MeV$. The sum of them is a good approximation for all the states widths except $T_{cs0}$}

\paragraph{$T^{0}_{cs1}$ HMRTs}{

\twographs[(a) is $(J, M^{2})$ $T^{0}_{cs1}$ MHRT, (b) $(n, M^{2})$ $T^{0}_{cs1}$ MHRT by a single point. The trajectory was built with fixed massive endpoints according to the universal fit done for mesons.][T_cs1]{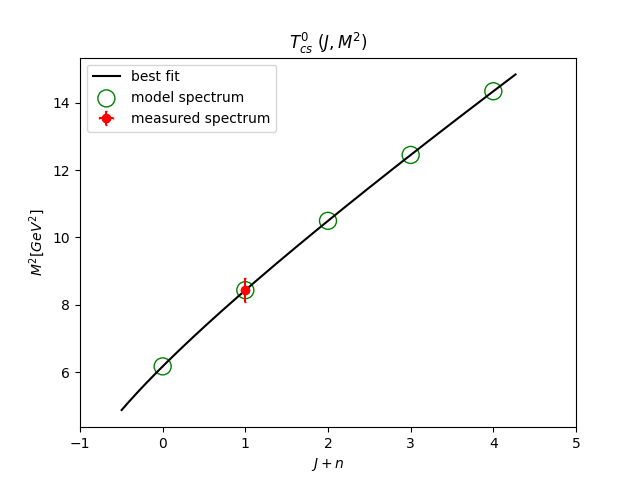}{tetraquarks/T_cs/Tcs1/by_n_one_point/fit-_T0_cs_Tcs1_by_n}

    \begin{center}
    \begin{minipage}{.49\linewidth}
        \begin{center}
            \begin{longtable}[c]{|c|c|c|c|c|c|}
  \hline
  \thead{$State$} & \thead{$M$ \\ $[MeV]$} & \thead{Thry \\ $M$ \\ $[MeV]$} & \thead{$J$} & \thead{Thry \\ $n$} & \thead{Thry \\ $J + n$} \\
  \hline
  \hline
$T(2900)^{0}_{cs1}$ & $2904$ & $2904$ & $1$ & $0$ & $1$ \\
\hline
 &  & $3240$ &  &  & $2$ \\
\hline
 &  & $3529$ &  &  & $3$ \\
\hline
 &  & $3788$ &  &  & $4$ \\
\hline
 &  & $4024$ &  &  & $5$ \\
\hline
  \caption{$T_{cs}^{0}$ $(J, M^{2})$ HMRT.} \label{table:Tcs1_by_j_one_point}
 \end{longtable}
        \end{center}
    \end{minipage}
    \begin{minipage}{.49\linewidth}
        \begin{center}
            \begin{longtable}[c]{|c|c|c|c|c|c|}
  \hline
  \thead{$State$} & \thead{$M$ \\ $[MeV]$} & \thead{Thry \\ $M$ \\ $[MeV]$} & \thead{$J$} & \thead{Thry \\ $n$} & \thead{Thry \\ $J + n$} \\
  \hline
  \hline
$T(2900)^{0}_{cs1}$ & $2904$ & $2904$ & $1$ & $-1$ & $0$ \\
\hline
 &  & $3373$ &  &  & $1$ \\
\hline
 &  & $3760$ &  &  & $2$ \\
\hline
 &  & $4099$ &  &  & $3$ \\
\hline
 &  & $4405$ &  &  & $4$ \\
\hline
  \caption{$T_{cs}^{0}$ $(n, M^{2})$ HMRT.} \label{table:Tcs1_by_n_one_point}
 \end{longtable}
        \end{center}
    \end{minipage}
    \end{center}

The intercept values are $a_{j}=-1.10$ and $a_{n}=-1.45$. The decays widths are $\Gamma_{tear}\lesssim 107MeV$ and $\Gamma_{annihilation}\approx 22MeV$.

\subsubsection{\texorpdfstring{$T_{c\bar{s}}^{a}$ (with quark content ${c\bar{s}d\bar{u}}$ or ${c\bar{s}u\bar{d}}$)}{}}

In \cite{lhcbcollaboration2022observation_Tcs_a} the states $T^{a^{0}}_{c\bar{s}0}$ and $T^{a^{++}}_{c\bar{s}0}$ were observed. Both tetraquarks candidates have roughly the same mass. This implies that the total quantity $J + n$ is the same. All four trajectories in this section are built using the mesonic universal heavy $n$ or $J$ slopes as calculated in the appendix \ref{mesons_confronting_data}.}

\paragraph{$T^{a^{0}}_{c\bar{s}0}$}{

The neutral state was reconstructed in the decay channel $D_{s}^{+}\pi^{-}$. This is in accordance with the annihilation mechanism, which is expected, due to the string tear threshold for baryonia ($\Lambda\Lambda_{c}$ pair) creation which is around $3.4GeV$. The intercept values are $a_{j}=-1.48$ and $a_{n}=-0.72$.

\twographs[(a) is $(J, M^{2})$ $T^{a^{0}}_{c\bar{s}0}$ MHRT, (b) $(n, M^{2})$ $T^{a^{0}}_{c\bar{s}0}$ MHRT by a single point. The trajectory was built with fixed massive endpoints according to the universal fit done for mesons.][fig:T_a0_c_bar{s}]{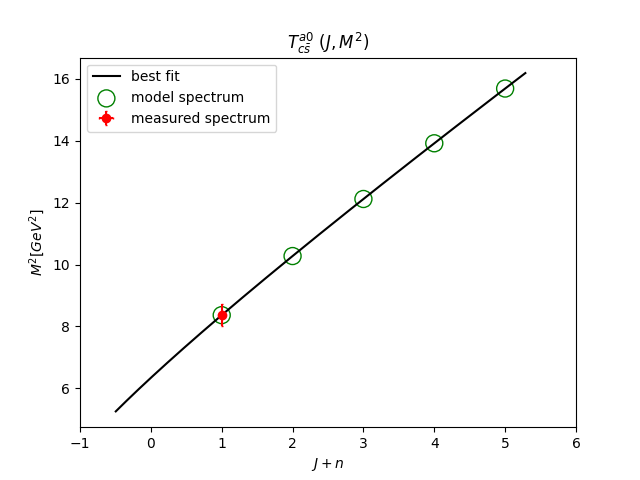}{tetraquarks/T_a_cs/Tacs0/by_n_one_point/fit-_Ta0_c_bs__bs_bar_s__Tcsdu_by_n}

    \begin{center}
    \begin{minipage}{.49\linewidth}
        \begin{center}
            \begin{longtable}[c]{|c|c|c|c|c|c|}
  \hline
  \thead{$State$} & \thead{$M$ \\ $[MeV]$} & \thead{Thry \\ $M$ \\ $[MeV]$} & \thead{$J$} & \thead{Thry \\ $n$} & \thead{Thry \\ $J + n$} \\
  \hline
  \hline
$T(2900)^{a0}_{c\bar{s}0}$ & $2892$ & $2892$ & $0$ & $1$ & $1$ \\
\hline
 &  & $3205$ &  &  & $2$ \\
\hline
 &  & $3481$ &  &  & $3$ \\
\hline
 &  & $3731$ &  &  & $4$ \\
\hline
 &  & $3961$ &  &  & $5$ \\
\hline
  \caption{$T_{c\bar{s}}^{a0}$ $(J, M^{2})$ HMRT.} \label{table:T^a0_cs0_by_j_one_point}
 \end{longtable}
        \end{center}
    \end{minipage}
    \begin{minipage}{.49\linewidth}
        \begin{center}
            \begin{longtable}[c]{|c|c|c|c|c|c|}
  \hline
  \thead{$State$} & \thead{$M$ \\ $[MeV]$} & \thead{Thry \\ $M$ \\ $[MeV]$} & \thead{$J$} & \thead{Thry \\ $n$} & \thead{Thry \\ $J + n$} \\
  \hline
  \hline
$T(2900)^{a0}_{c\bar{s}0}$ & $2892$ & $2892$ & $0$ & $1$ & $1$ \\
\hline
 &  & $3332$ &  &  & $2$ \\
\hline
 &  & $3704$ &  &  & $3$ \\
\hline
 &  & $4034$ &  &  & $4$ \\
\hline
 &  & $4333$ &  &  & $5$ \\
\hline
  \caption{$T_{c\bar{s}}^{a0}$ $(n, M^{2})$ HMRT.} \label{table:T^a0_cs0_by_n_one_point}
 \end{longtable}
        \end{center}
    \end{minipage}
    \end{center}
}

\paragraph{$T^{a^{++}}_{c\bar{s}0}$}{

The positive state was reconstructed in the decay channel $D_{s}^{+}\pi^{+}$. This is in accordance with the annihilation mechanism, which is expected, due to the string tear threshold for baryonia ($\Lambda\Lambda_{c}$ pair) creation which is around $3.4GeV$.  The intercept values are $a_{j}=-1.57$ and $a_{n}=-0.78$.

\twographs[(a) is $(J, M^{2})$ $T^{a^{++}}_{c\bar{s}0}$ MHRT, (b) $(n, M^{2})$ $T^{a^{++}}_{c\bar{s}0}$ MHRT by a single point. The trajectory was built with fixed massive endpoints according to the universal fit done for mesons.][fig:T_a++_c_bar{s}]{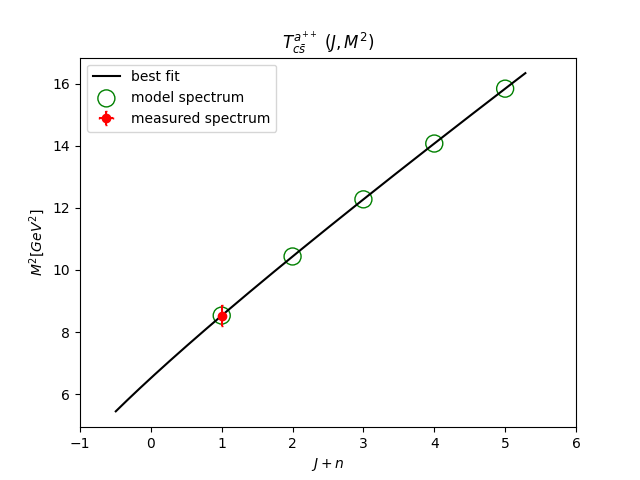}{tetraquarks/T_a_cs/Tacs__/by_n_one_point/fit-_Ta___c_bs__bs_bar_s__Tcsdu_plus_by_n}

    \begin{center}
    \begin{minipage}{.49\linewidth}
        \begin{center}
            \begin{longtable}[c]{|c|c|c|c|c|c|}
  \hline
  \thead{$State$} & \thead{$M$ \\ $[MeV]$} & \thead{Thry \\ $M$ \\ $[MeV]$} & \thead{$J$} & \thead{Thry \\ $n$} & \thead{Thry \\ $J + n$} \\
  \hline
  \hline
$T(2900)^{a++}_{c\bar{s}0}$ & $2921$ & $2921$ & $0$ & $1$ & $1$ \\
\hline
 &  & $3230$ &  &  & $2$ \\
\hline
 &  & $3504$ &  &  & $3$ \\
\hline
 &  & $3751$ &  &  & $4$ \\
\hline
 &  & $3980$ &  &  & $5$ \\
\hline
  \caption{$T_{c\bar{s}}^{a^{++}}$ $(J, M^{2})$ HMRT.} \label{table:T^a++_cs0_by_j_one_point}
 \end{longtable}
        \end{center}
    \end{minipage}
    \begin{minipage}{.49\linewidth}
        \begin{center}
            \begin{longtable}[c]{|c|c|c|c|c|c|}
  \hline
  \thead{$State$} & \thead{$M$ \\ $[MeV]$} & \thead{Thry \\ $M$ \\ $[MeV]$} & \thead{$J$} & \thead{Thry \\ $n$} & \thead{Thry \\ $J + n$} \\
  \hline
  \hline
$T(2900)^{a++}_{c\bar{s}0}$ & $2921$ & $2921$ & $0$ & $1$ & $1$ \\
\hline
 &  & $3355$ &  &  & $2$ \\
\hline
 &  & $3725$ &  &  & $3$ \\
\hline
 &  & $4053$ &  &  & $4$ \\
\hline
 &  & $4351$ &  &  & $5$ \\
\hline
  \caption{$T_{c\bar{s}}^{a^{++}}$ $(n, M^{2})$ HMRT.} \label{table:T^a0_cs++_by_n_one_point}
 \end{longtable}
        \end{center}
    \end{minipage}
    \end{center}
}

\subsubsection{\texorpdfstring{$X(3350)$ ($T_{c\Bar{d}d\Bar{u}}$ or $T_{c\Bar{u}u\Bar{u}}$)}{}}

In \cite{X(3350)_Belle:2004dmq} and \cite{X(3350)_BaBar:2010eti} the X(3350) have been observed, both in the decay channel of $\Lambda_{c}^{+}\bar{p}$. There was another pick observed in the same channel in \cite{X(3350)_Belle:2004dmq} at $3.8$GeV\footnote[2]{This possible resonance was found in \cite{X(3350)_Belle:2004dmq} with statistical significance of $2.8\sigma$, and wasn't analysed further. We assumed it exists in order to build the HMRT.}, but wasn't observed in \cite{X(3350)_BaBar:2010eti}. The states analysed in this section are summarized in table \ref{table:tetra_Tcddu_cand}.

Tables \ref{table:Tcddu_j_predictions} and \ref{table:Tcddu_n_predictions} are the predicted HMRTs in both planes. The thresholds and predicted width are at table \ref{table:Tcddu_thresholds}.

\begin{longtable}[c]{|c|c|c|c|c|c|c|}
\hline
\makecell{Candidate} & \makecell{Quarks \\ Content} & \makecell{Selected \\ Decay \\ Channels} & \makecell{$J^{PC}$} & \makecell{Mass \\ $[MeV]$} & \makecell{Width \\ $[MeV]$} & \makecell{Ref.} \\
\hline
\hline
$X(3350)$ & $c\bar{d}d\bar{u}$ & $\Lambda_{c}^{+}\bar{p}$ & ${{0} or {1}}^{??}$ & $3350^{+10}_{-20}\pm{20}$ & $70^{+40}_{-30}\pm{40}$ & \cite{X(3350)_Belle:2004dmq}, \cite{X(3350)_BaBar:2010eti} \\
\hline
$X(3800)$\tablefootnotemark{footnote:state_needs_confirm} & $c\bar{d}d\bar{u}$ & $\Lambda_{c}^{+}\bar{p}$ & $?^{??}$ & $3840\pm{10}$ & $30\pm{30}$ & \cite{X(3350)_Belle:2004dmq} \\
\hline
\caption{$T_{u\bar{u}\bar{u}d}$ or $T_{\bar{u}d\bar{d}d}$ exotic candidate.}\label{table:tetra_Tcddu_cand}
\end{longtable}

\begin{longtable}[c]{|c|c|c|c|c|c|}
  \hline
  \thead{Thry \\ Width \\ $[MeV]$} & \thead{Mesons \\ Pair} & \thead{Meson \\ Threshold \\ $[MeV]$} & \thead{Baryon \\ Anti-\\baryon} & \thead{Baryonic \\ Threshold \\ $[MeV]$} & \thead{Genuine} \\
  \hline
  \hline
\makecell{$79-118$\\$30-61$} & \makecell{$D^{0} \pi^{0}_{d}$ \\ $D^{+} \pi^{-}$} & \makecell{$2000$ \\ $2010$} & \makecell{$\Lambda^{+}_{c} \bar{p}$ \\ $\Sigma^{0}_{c} \bar{n}$ \\ $\Xi^{0}_{c} \bar{\Lambda}^{0}$} & \makecell{$3224$ \\ $3394$ \\ $3586$} &  \\
\hline
  \caption{$T_{c\bar{d}d\bar{u}}$ thresholds.} \label{table:Tcddu_thresholds}
 \end{longtable}

    \begin{center}
    \begin{minipage}{.49\linewidth}
        \begin{center}
            \begin{longtable}[c]{|c|c|}
  \hline
  \thead{$J$ \\ Spec} & \thead{$a_{T_{j}}$} \\
  \hline
  \hline
\makecell{$2634-2779$\\$2972-3095$\\$3264-3374$\\$3526-3626$\\$3766-3858$\\$3988-4075$\\$4197-4278$\\$4394-4471$\\$4581-4655$\\$4760-4831$\\$4931-5000$\\$5097-5162$} & \makecell{$-1.6$-$-1.2$} \\
\hline
  \caption{$T_{c\bar{d}d\bar{u}}$ $(J, M^2)$ predictions.} \label{table:Tcddu_j_predictions}
 \end{longtable}
        \end{center}
    \end{minipage}
    \begin{minipage}{.49\linewidth}
        \begin{center}
            \begin{longtable}[c]{|c|c|}
  \hline
  \thead{$n$ \\ Spec} & \thead{$a_{T_{n}}$} \\
  \hline
  \hline
\makecell{$2270-2448$\\$2829-2961$\\$3264-3374$\\$3635-3731$\\$3964-4051$\\$4263-4344$\\$4540-4615$\\$4799-4869$\\$5042-5109$\\$5273-5337$\\$5494-5554$\\$5705-5763$} & \makecell{$-1.2$-$-0.92$} \\
\hline
  \caption{$T_{c\bar{d}d\bar{u}}$ $(n, M^2)$ predictions.} \label{table:Tcddu_n_predictions}
 \end{longtable}
        \end{center}
    \end{minipage}
    \end{center}

Since $X(3800)$ needs confirmation, we built the HMRTs for the case in which it exists, and for the case in which we have only one point on the trajectory ($X(3350)$ ). We calculated both $(n,M^{2})$ and $(J, M^{2})$ HMRT.
According to the predicted spectrum (\ref{table:Tcddu_j_predictions} and \ref{table:Tcddu_n_predictions}), we do expect a state to exist around $3.8GeV$ on the $(J, M^{2})$ trajectory. Hence, it should have the same momentum value of $X(3350)$ plus 2.

\paragraph{$X(3350)$ $(J, M^{2})$ fit}{

\twographs[(a) $X(3350) (J, M^{2})$ HMRT. The fit was done with fixed massive endpoints according to the universal fit for mesons \ref{table:mesons_global_fit}. (b) $(J, M^{2})$ $X(3350)$ HMRT best fit is with the fixed masses.][X_3350_J]{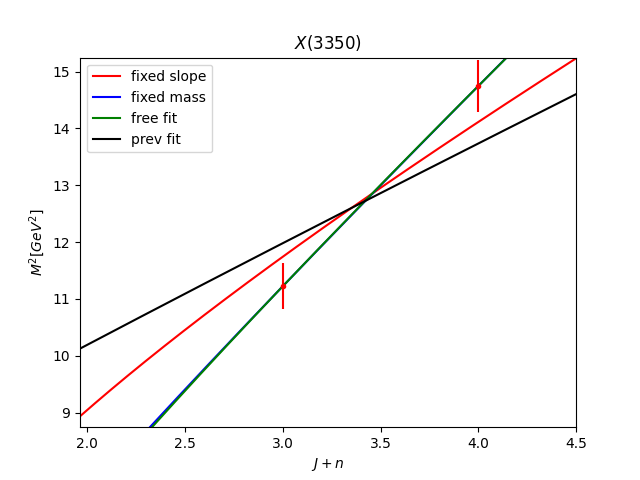}{tetraquarks/X_3350_/by_J/fit-_X_X_3350__by_j_fixed_m_j}

The masses were taken to be $1.33GeV$ and $0.03GeV$ in accordance with the mesons universal fit, assuming the diquarks/antidiquarks weighing as the sum of the two quarks in each end. The results of all the fits are summarized at table \ref{table:X_3350_J_results}. The spectrum of this trajectory can be found in table [\ref{table:charm_tetraquarks_predictions}] including lower and higher states.
\vskip 1cm
Since there are only two point on the trajectories, the best "fit" is the one that produced the lowest value of $chi^{2}_{r}$ (where the points are on the trajectory).

\begin{longtable}[c]{|c|c|c|c|c|c|c|}
  \hline
  \thead{$\chi^{2}_{r}$} & \thead{$\chi^{2}_{hish}$} & \thead{$\alpha^{'}$} & \thead{$a$} & \thead{$m_{1}$} & \thead{$m_{2}$} & \thead{Fit Type} \\
  \hline
  \hline
0.00 & 0.00 & 0.34 & 0.98 & 1.34 & 0.04 & free fit \\
\hline
8.37 & 0.06 & 0.65 & -1.96 & 1.12 & 0.03 & fixed $m_{1}$, $m_{2}$ and $\alpha^{'}$ \\
\hline
3.56 & 0.03 & 0.65 & 1.00 & 2.07 & 0.01 & fixed $\alpha^{'}$ \\
\hline
0.00 & 0.00 & 0.32 & 0.74 & 1.12 & 0.03 & fixed $m_{1}$ and $m_{2}$ \\
\hline
  \caption{$X(3350) (J, M^{2})$ output for each fit.} \label{table:X_3350_J_results}
 \end{longtable}}

\paragraph{$X(3350)$ $(n, M^{2})$ fit}{
    \twographs[(a) $X(3350) (n, M^{2})$ MHRT. The fit was done with fixed massive endpoints according to the universal fit done for mesons. (b) $(n, M^{2})$ $X(3350)$ MHRT best fit which was the free fit. ][X_3350_n]{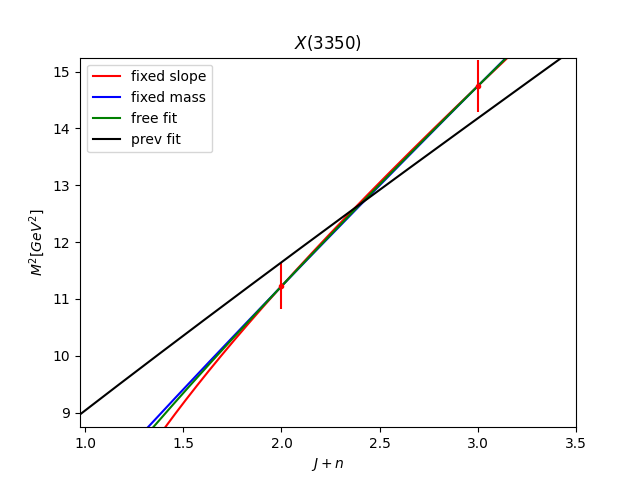}{tetraquarks/X_3350_/by_n/fit-_X_X_3350__by_n_new_n}
}

The masses were taken to be $1.33GeV$ and $0.03GeV$ with accordance to the universal fit, assuming the diquarks/antidiquarks weighing the same as the heavy quark. The spectrum of this trajectory can be found in table [\ref{table:X_3350_by_n}] including lower and higher states.

\begin{longtable}[c]{|c|c|c|c|c|c|c|}
  \hline
  \thead{$\chi^{2}_{r}$} & \thead{$\chi^{2}_{hish}$} & \thead{$\alpha^{'}$} & \thead{$a$} & \thead{$m_{1}$} & \thead{$m_{2}$} & \thead{Fit Type} \\
  \hline
  \hline
0.00 & 0.00 & 0.36 & 0.31 & 1.49 & 0.30 & free fit \\
\hline
2.62 & 0.02 & 0.45 & -1.30 & 1.12 & 0.03 & fixed $m_{1}$, $m_{2}$ and $\alpha^{'}$ \\
\hline
0.00 & 0.00 & 0.45 & 0.88 & 2.11 & 0.09 & fixed $\alpha^{'}$ \\
\hline
0.00 & 0.00 & 0.32 & -0.26 & 1.12 & 0.03 & fixed $m_{1}$ and $m_{2}$ \\
\hline
  \caption{$X(3350) (n, M^{2})$ output for each fit.} \label{table:X_3350_all_fits_output}
 \end{longtable}

    \begin{center}
    \begin{minipage}{.49\linewidth}
        \begin{center}
            \begin{longtable}[c]{|c|c|c|c|c|c|}
  \hline
  \thead{$State$} & \thead{$M$ \\ $[MeV]$} & \thead{Thry \\ $M$ \\ $[MeV]$} & \thead{$J$} & \thead{Thry \\ $n$} & \thead{Thry \\ $J + n$} \\
  \hline
  \hline
 &  & $2076$ &  &  & $1$ \\
\hline
 &  & $3008$ &  &  & $2$ \\
\hline
$X(3350)$ & $3350$ & $3427$ & $0$ & $3$ & $3$ \\
\hline
$X(3800)$ & $3840$ & $3757$ & $1$ & $3$ & $4$ \\
\hline
 &  & $4040$ &  &  & $5$ \\
\hline
 &  & $4292$ &  &  & $6$ \\
\hline
  \caption{$X(3350) (J, M^{2})$ HMRT.} \label{table:X_3350_by_one_j}
 \end{longtable}
        \end{center}
    \end{minipage}
    \begin{minipage}{.49\linewidth}
        \begin{center}
            \begin{longtable}[c]{|c|c|c|c|c|c|}
  \hline
  \thead{$State$} & \thead{$M$ \\ $[MeV]$} & \thead{Thry \\ $M$ \\ $[MeV]$} & \thead{$J$} & \thead{Thry \\ $n$} & \thead{Thry \\ $J + n$} \\
  \hline
  \hline
 &  & $2200$ &  &  & $0$ \\
\hline
 &  & $2531$ &  &  & $1$ \\
\hline
$X(3350)$ & $3350$ & $3350$ & $0$ & $2$ & $2$ \\
\hline
$X(3800)$ & $3840$ & $3840$ & $0$ & $3$ & $3$ \\
\hline
 &  & $4233$ &  &  & $4$ \\
\hline
 &  & $4573$ &  &  & $5$ \\
\hline
  \caption{$X(3350) (n, M^{2})$ HMRT.} \label{table:X_3350_by_n}
 \end{longtable}
        \end{center}
    \end{minipage}
    \end{center}

\paragraph{$X(3350)$ spectrum by one point}{

Since the second state ($X(3800)$) wasn't observed in another experiment, we also calculated the trajectories with a single point shown in the next figure.
    The intercept values are $a_{j}=-1.54$ and $a_{n}=-1.14$.

\twographs[(a) is $(J, M^{2})$ $X(3350)$ MHRT, (b) $(n, M^{2})$ $X(3350)$ MHRT by a single point. In both cases it is assumed to be a tetraquark due to its decay channel.][X_3350_one_point]{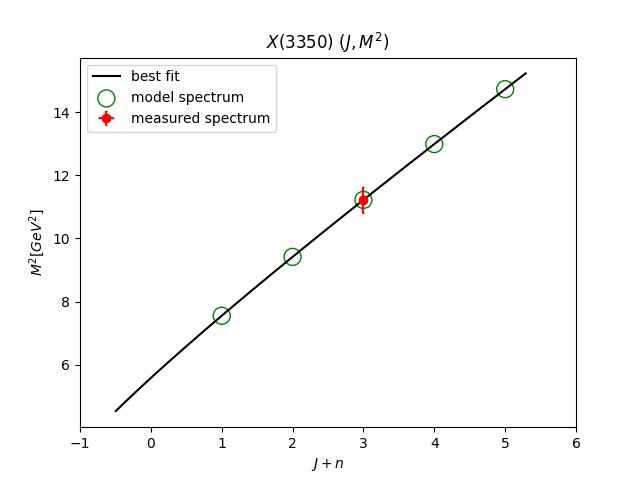}{tetraquarks/X_3350_/by_n_one_point/fit-_X_X_3350__one_by_n}

    \begin{center}
    \begin{minipage}{.49\linewidth}
        \begin{center}
            \begin{longtable}[c]{|c|c|c|c|c|c|}
  \hline
  \thead{$State$} & \thead{$M$ \\ $[MeV]$} & \thead{Thry \\ $M$ \\ $[MeV]$} & \thead{$J$} & \thead{Thry \\ $n$} & \thead{Thry \\ $J + n$} \\
  \hline
  \hline
 &  & $2748$ &  &  & $1$ \\
\hline
 &  & $3069$ &  &  & $2$ \\
\hline
$X(3350)$ & $3350$ & $3350$ &  &  & $3$ \\
\hline
 &  & $3604$ &  &  & $4$ \\
\hline
 &  & $3838$ &  &  & $5$ \\
\hline
  \caption{$X(3350)$ $(J, M^{2})$ HMRT by ignoring the $X(3800)$ state. The slope used is the $\alpha^{'}$ value for heavy mesons for $J$ excitation trajectory.} \label{table:X_3350_by_J_one_point}
 \end{longtable}
        \end{center}
    \end{minipage}
    \begin{minipage}{.49\linewidth}
        \begin{center}
            \begin{longtable}[c]{|c|c|c|c|c|c|}
  \hline
  \thead{$State$} & \thead{$M$ \\ $[MeV]$} & \thead{Thry \\ $M$ \\ $[MeV]$} & \thead{$J$} & \thead{Thry \\ $n$} & \thead{Thry \\ $J + n$} \\
  \hline
  \hline
 &  & $2410$ &  &  & $0$ \\
\hline
 &  & $2932$ &  &  & $1$ \\
\hline
$X(3350)$ & $3350$ & $3350$ &  &  & $2$ \\
\hline
 &  & $3710$ &  &  & $3$ \\
\hline
 &  & $4032$ &  &  & $4$ \\
\hline
  \caption{$X(3350)$ $(n, M^{2})$ HMRT by ignoring the $X(3800)$ state. The slope used is the $\alpha^{'}$ value for heavy mesons for $n$ excitations trajectory.} \label{table:X_3350_by_n_one_point}
 \end{longtable}
        \end{center}
    \end{minipage}
    \end{center}

The decay width is estimated to be $\Gamma_{tear}\lesssim 45MeV$. The ratio is:

\begin{equation}
\begin{aligned}
    \frac{\Gamma_{annihilation}}{\Gamma[\Psi(2710)]}&=\frac{exp(-\frac{TL^{2}}{2}|_{X(3350)})}{exp(-\frac{TL^{2}}{2}|_{X(2710)})}\approx{0.41}
\end{aligned}
\end{equation}

Based on the predicted state which is lower than the string-tear threshold, this might be the higher state of $D^{*}(2760)$, also according to its decay channel ($D^{+}\pi^{-}$). By taking the width of this state, we can estimate the width of the X(3350):

\begin{align}
   \Gamma_{annihilation}\lesssim 73MeV
\end{align}}

\subsubsection{\texorpdfstring{$T_{c\Bar{d}c\Bar{u}}$}{}}

For a tetraquarks  with the quark content $T_{c\Bar{d}c\Bar{u}}$ (table \ref{table:tetra_cand_ccdu}). The candidate resonance are presented in table \ref{table:tetra_cand_ccdu}. We generated the predicted HMRTs for this quark content in tables \ref{table:Tccdu_j_predictions} and \ref{table:Tccdu_n_predictions}, and the thresholds table \ref{table:Tccdu_thresholds} includes the predicted width. The HMRTs are also plotted in \ref{fig:Tccdu_predictions}.

\begin{longtable}[c]{|c|c|c|c|c|c|c|}
\hline
\makecell{Candidate} & \makecell{Quarks \\ Content} & \makecell{Selected \\ Decay \\ Channels} & \makecell{$J^{PC}$} & \makecell{Mass \\ $[MeV]$} & \makecell{Width \\ $[MeV]$} & \makecell{Ref.} \\
\hline
\hline
$T_{cc}(3875)$ & $cc\bar{u}\bar{d}$ & \makecell{$D^{*+}D^{0}$ \\ $D^{*0}D^{+}$} & $?^{?}$ & $3874\pm{0.11}$ & $0.41\pm{0.17}$ &  \cite{LHCb:2021vvq} \\
\hline
\caption{$T_{c\Bar{d}c\Bar{u}}$ tetraquark candidates.}\label{table:tetra_cand_ccdu}
\end{longtable}

\begin{longtable}[c]{|c|c|c|c|c|c|}
  \hline
  \thead{Thry \\ Width \\ $[MeV]$} & \thead{Mesons \\ Pair} & \thead{Meson \\ Threshold \\ $[MeV]$} & \thead{Baryon \\ Anti-\\baryon} & \thead{Baryonic \\ Threshold \\ $[MeV]$} & \thead{Genuine} \\
  \hline
  \hline
 & \makecell{$D^{+} D^{0}$ \\ $D^{0} D^{+}$} & \makecell{$3735$ \\ $3735$} & \makecell{$\Xi^{+}_{cc} \bar{n}$ \\ $\Xi^{++}_{cc} \bar{p}$ \\ $\Omega^{+}_{cc} \bar{\Lambda}^{0}$} & \makecell{$4459$ \\ $4560$ \\ } & $**$ \\
\hline
  \caption{$T_{c\bar{d}c\bar{u}}$ thresholds.} \label{table:Tccdu_thresholds}
 \end{longtable}

    \begin{center}
    \begin{minipage}{.49\linewidth}
        \begin{center}
            \begin{longtable}[c]{|c|c|}
  \hline
  \thead{$J$ \\ Spec} & \thead{$a_{T_{j}}$} \\
  \hline
  \hline
\makecell{$3983-4118$\\$4255-4375$\\$4499-4609$\\$4723-4825$\\$4932-5027$\\$5128-5218$\\$5313-5399$\\$5490-5572$\\$5659-5738$\\$5822-5898$\\$5978-6052$\\$6130-6201$} & \makecell{$-3.7$-$-3.3$} \\
\hline
  \caption{$T_{c\bar{d}c\bar{u}}$ $(J, M^2)$ predictions.} \label{table:Tccdu_j_predictions}
 \end{longtable}
        \end{center}
    \end{minipage}
    \begin{minipage}{.49\linewidth}
        \begin{center}
            \begin{longtable}[c]{|c|c|}
  \hline
  \thead{$n$ \\ Spec} & \thead{$a_{T_{n}}$} \\
  \hline
  \hline
\makecell{$4499-4609$\\$4818-4917$\\$5107-5198$\\$5373-5458$\\$5622-5702$\\$5857-5932$\\$6080-6151$\\$6292-6361$\\$6496-6561$\\$6691-6755$\\$6880-6941$\\$7062-7122$} & \makecell{$-3.0$-$-2.6$} \\
\hline
  \caption{$T_{c\bar{d}c\bar{u}}$ $(n, M^2)$ predictions.} \label{table:Tccdu_n_predictions}
 \end{longtable}
        \end{center}
    \end{minipage}
    \end{center}

\twographs[(a) is $(J, M^{2})$ $T_{c\Bar{d}c\Bar{u}}$ predicted HMRT, (b) $(n, M^{2})$ $T_{c\Bar{d}c\Bar{u}}$ predicted HMRT.][Tccdu_predictions]{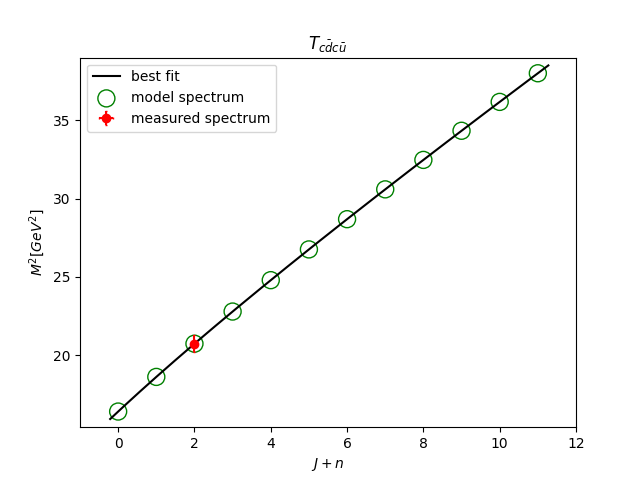}{tetraquarks/predictions/cc-du/by_n/T_ccdu_one_by_n}

In \cite{Tcc_2022} the state $T^{+}_{cc}$ was observed. It was also analysed in \cite{Gelman:2002wf}. It was reconstructed in the decay channel $D^{0}D^{0}\pi^{+}$ in accordance with the annihilation mechanism. This is expected since the string tear threshold for baryonia ($\Xi_{cc}^{+}p$ pair) creation is around $4.56GeV$. The trajectories are built using the mesons heavy slopes as calculated at \ref{mesons_confronting_data}. The intercepts values are $a_{j}=-2.89$ and $a_{n}=-2.00$

\twographs[(a) is $(J, M^{2})$ $T^{+}_{cc}$ MHRT, (b) $(n, M^{2})$ $T^{+}_{cc}$ MHRT by a single point. The trajectory was built with fixed massive endpoints according to the universal fit done for mesons.][T_cc]{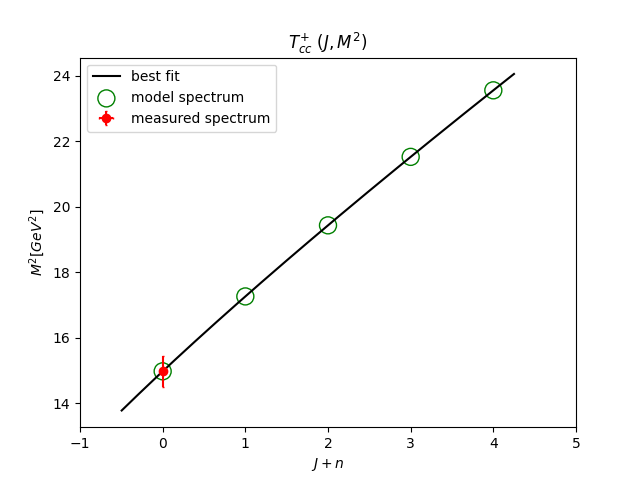}{tetraquarks/T_cc/by_n_one_point/fit-_T__cc_Tcc_by_n}

    \begin{center}
    \begin{minipage}{.49\linewidth}
        \begin{center}
            \begin{longtable}[c]{|c|c|c|c|c|c|}
  \hline
  \thead{$State$} & \thead{$M$ \\ $[MeV]$} & \thead{Thry \\ $M$ \\ $[MeV]$} & \thead{$J$} & \thead{Thry \\ $n$} & \thead{Thry \\ $J + n$} \\
  \hline
  \hline
$T(3875)^{+}_{cc0}$ & $3870$ & $3870$ & $0$ & $0$ & $0$ \\
\hline
 &  & $4155$ &  &  & $1$ \\
\hline
 &  & $4409$ &  &  & $2$ \\
\hline
 &  & $4640$ &  &  & $3$ \\
\hline
 &  & $4854$ &  &  & $4$ \\
\hline
  \caption{$T_{cc}^{+}$ $(J, M^{2})$ HMRT.} \label{table:Tcc_by_j_one_point}
 \end{longtable}
        \end{center}
    \end{minipage}
    \begin{minipage}{.49\linewidth}
        \begin{center}
            \begin{longtable}[c]{|c|c|c|c|c|c|}
  \hline
  \thead{$State$} & \thead{$M$ \\ $[MeV]$} & \thead{Thry \\ $M$ \\ $[MeV]$} & \thead{$J$} & \thead{Thry \\ $n$} & \thead{Thry \\ $J + n$} \\
  \hline
  \hline
$T(3875)^{+}_{cc0}$ & $3870$ & $3870$ & $0$ & $0$ & $0$ \\
\hline
 &  & $4271$ &  &  & $1$ \\
\hline
 &  & $4615$ &  &  & $2$ \\
\hline
 &  & $4922$ &  &  & $3$ \\
\hline
 &  & $5203$ &  &  & $4$ \\
\hline
  \caption{$T_{cc}^{+}$ $(n, M^{2})$ HMRT.} \label{table:Tcc_by_n_one_point}
 \end{longtable}
        \end{center}
    \end{minipage}
    \end{center}

The interesting thing about this state is its width, which is very narrow (\ref{table:Tccdu_thresholds}), meaning it is stable. The HISH annihilation decay mechanism suggests that the BV and anti-BV will annihilate in a certain probability, that is dependant on the length of the horizontal string (\cite{Sonnenschein_decay_width_2018}). In the case of $X(4630)/\psi(4660)$ and $X(6900)$ (\cite{Sonnenschein_2017_4630} and \cite{Sonnenschein_2021_6900}) it proved to be a good approximation to rely on the same $P_{annihilation}$ value for both states. In the case of $T_{cc}$, it seems the probability is much lower, despite the length of the string being similar. The conditions in which the annihilation can occur is still an open question. One possible explanation is that the asymmetry between the diquark and antidiquark affects the location of the vertices on the holographic space on the $u$ direction, which reduces the annihilation probability. Anyhow, this requires more research.

\subsubsection{\texorpdfstring{$T_{c\Bar{c}u\Bar{d}}$ or $T_{c\Bar{c}d\Bar{u}}$}{}}

For a tetraquark with the quark content $T_{c\Bar{c}u\Bar{d}}$ or $T_{c\Bar{c}d\Bar{u}}$ (table \ref{table:tetra_cand_cucd}). The candidate resonances are presented in table \ref{table:tetra_cand_cucd}. We generated the predicted HMRTs for this quark content in tables \ref{table:Tcucd_j_predictions} and \ref{table:Tcucd_n_predictions}, and the thresholds table \ref{table:Tcucd_thresholds} includes the predicted width. The HMRTs are also plotted in \ref{fig:Tcucd_predictions}.

\begin{longtable}[c]{|c|c|c|c|c|c|c|}
\hline
\makecell{Candidate} & \makecell{Quarks \\ Content} & \makecell{Selected \\ Decay \\ Channels} & \makecell{$J^{PC}$} & \makecell{Mass \\ $[MeV]$} & \makecell{Width \\ $[MeV]$} & \makecell{Ref.} \\
\hline
\hline
$Z_{c}(3900)$ & \makecell{$c\bar{c}u\bar{d}$ \\ $c\bar{c}d\bar{u}$} & \makecell{$J/\psi \pi^{\pm}$ \\ ${D\bar{D^{*}}}^{\pm}$} & $1^{+-}$ & $3887.1\pm 2.6$ & $28.4\pm2.6$ & \makecell{Many, \\ including \\ \cite{BESIII:2020oph}} \\
\hline
\makecell{$Z_{c}(4200)^{\pm}$} & \makecell{$c\bar{c}u\bar{d}$ \\ or \\ $c\bar{c}d\bar{u}$} & \makecell{$J/\psi\pi^{+}$} & $1^{+-}$\tablefootnotemark{footnote:C_needs_confirmation} & $4196^{+35}_{-32}$ & $370^{+100}_{-150}$ & \makecell{\cite{Belle:2014nuw}} \\ 
\hline
$Z_{c}(4430)^{\pm}$ & \makecell{$c\bar{c}u\bar{d}$ \\ or \\ $c\bar{c}d\bar{u}$} & \makecell{$\pi^{+}\psi(2S)$ \\ $\pi^{+}J/\psi$} & $1^{+-}$\tablefootnotemark{footnote:C_needs_confirmation} & $4478^{+15}_{-18}$ & $181\pm 31$ & \makecell{\cite{Belle:2007hrb}, \cite{Belle:2009lvn}, \\ \cite{Belle:2013shl}, \cite{LHCb:2014zfx}} \\
\hline
$X(4020)^{\pm}$ & \makecell{$c\bar{c}u\bar{d}$ \\ or \\ $c\bar{c}d\bar{u}$} & \makecell{$D^{*}\bar{D}^{*}$ \\ $h_{c}(1P)\pi^{\pm}$} & $?^{?-}$ & $4024.1\pm{1.9}$ & $13\pm 5$ & \makecell{\cite{BESIII:2013ouc}, \cite{BESIII:2014gnk} \\ \cite{BESIII:2013mhi}, \cite{BESIII:2013mhi}} \\
\hline
$X(4051)^{\pm}$ & \makecell{$c\bar{c}u\bar{d}$ \\ or \\ $c\bar{c}d\bar{u}$} & \makecell{$\pi^{+}\chi_{c1}(1P)$} & $?^{?+}$ & $4051^{+24}_{-40}$ & $82^{+50}_{-28}$ & \makecell{\cite{Belle:2008qeq}} \\
\hline
$X(4055)^{\pm}$ & \makecell{$c\bar{c}u\bar{d}$ \\ or \\ $c\bar{c}d\bar{u}$} & \makecell{$\pi^{+}\psi(2S)$} & $?^{?-}$ & $4054\pm 3.2$ & $45\pm 13$ & \makecell{\cite{Belle:2014wyt}, \cite{BESIII:2017vtc} \\ \cite{BESIII:2017tqk}} \\
\hline
$X(4100)^{\pm}$ & \makecell{$c\bar{c}u\bar{d}$ \\ or \\ $c\bar{c}d\bar{u}$} & \makecell{$\pi^{-}\eta(1S)$} & $?^{??}$ & $4096\pm 28$ & $152^{+80}_{-70}$ & \makecell{\cite{LHCb:2018oeg}} \\
\hline
$R_{c0}(4240)^{\pm}$ & \makecell{$c\bar{c}u\bar{d}$ \\ or \\ $c\bar{c}d\bar{u}$} & \makecell{$\pi^{-}\psi(2S)$} & $0^{--}$ & $4239^{+50}_{-21}$ & $220^{+120}_{-90}$ & \makecell{\cite{LHCb:2021uow}} \\
\hline
$X(4250)^{\pm}$ & \makecell{$c\bar{c}u\bar{d}$ \\ or \\ $c\bar{c}d\bar{u}$} & \makecell{$\pi^{+}\chi(1P)$} & $?^{?-}$\tablefootnotemark{footnote:C_needs_confirmation} & $4248^{+190}_{-50}$ & $177^{+320}_{-70}$ & \makecell{\cite{Belle:2008qeq}} \\
\hline
\caption{$T_{c\Bar{c}u\Bar{d}}$ or $T_{c\Bar{c}d\Bar{u}}$ tetraquark candidates.}\label{table:tetra_cand_cucd}
\end{longtable}

\begin{longtable}[c]{|c|c|c|c|c|c|}
  \hline
  \thead{Thry \\ Width \\ $[MeV]$} & \thead{Mesons \\ Pair} & \thead{Meson \\ Threshold \\ $[MeV]$} & \thead{Baryon \\ Anti-\\baryon} & \thead{Baryonic \\ Threshold \\ $[MeV]$} & \thead{Genuine} \\
  \hline
  \hline
\makecell{$11-31$\\$34-67$} & \makecell{$\eta_{c}(1S) \pi^{+}$ \\ $D^{+} \bar{D}^{0}$} & \makecell{$3124$ \\ $3735$} & \makecell{$\Sigma^{++}_{c} \bar{\Lambda}^{-}_{c}$ \\ $\Lambda^{+}_{c} \bar{\Sigma}^{0}_{c}$ \\ $\Xi^{+}_{c} \bar{\Xi}^{0}_{c}$} & \makecell{$4740$ \\ $4740$ \\ $4938$} & $*$ \\
\hline
  \caption{$T_{c\bar{c}u\bar{d}}$ thresholds.} \label{table:Tcucd_thresholds}
 \end{longtable}

    \begin{center}
    \begin{minipage}{.49\linewidth}
        \begin{center}
            \begin{longtable}[c]{|c|c|}
  \hline
  \thead{$J$ \\ Spec} & \thead{$a_{T_{j}}$} \\
  \hline
  \hline
\makecell{$4373-4497$\\$4583-4699$\\$4780-4890$\\$4967-5072$\\$5145-5245$\\$5316-5412$\\$5480-5572$\\$5638-5727$\\$5790-5877$\\$5938-6023$\\$6082-6164$\\$6222-6301$} & \makecell{$-2.6$-$-2.0$} \\
\hline
  \caption{$T_{c\bar{c}u\bar{d}}$ $(J, M^2)$ predictions.} \label{table:Tcucd_j_predictions}
 \end{longtable}
        \end{center}
    \end{minipage}
    \begin{minipage}{.49\linewidth}
        \begin{center}
            \begin{longtable}[c]{|c|c|}
  \hline
  \thead{$n$ \\ Spec} & \thead{$a_{T_{n}}$} \\
  \hline
  \hline
\makecell{$4174-4306$\\$4491-4611$\\$4780-4890$\\$5047-5150$\\$5297-5394$\\$5533-5625$\\$5757-5844$\\$5971-6054$\\$6176-6256$\\$6373-6450$\\$6563-6638$\\$6747-6819$} & \makecell{$-2.0$-$-1.6$} \\
\hline
  \caption{$T_{c\bar{c}u\bar{d}}$ $(n, M^2)$ predictions.} \label{table:Tcucd_n_predictions}
 \end{longtable}
        \end{center}
    \end{minipage}
    \end{center}

\twographs[(a) is $(J, M^{2})$ $T_{c\Bar{c}u\Bar{d}}$ or $T_{c\Bar{c}d\Bar{u}}$ predicted HMRT, (b) $(n, M^{2})$ $T_{c\Bar{c}u\Bar{d}}$ or $T_{c\Bar{c}d\Bar{u}}$ predicted HMRT.][Tcucd_predictions]{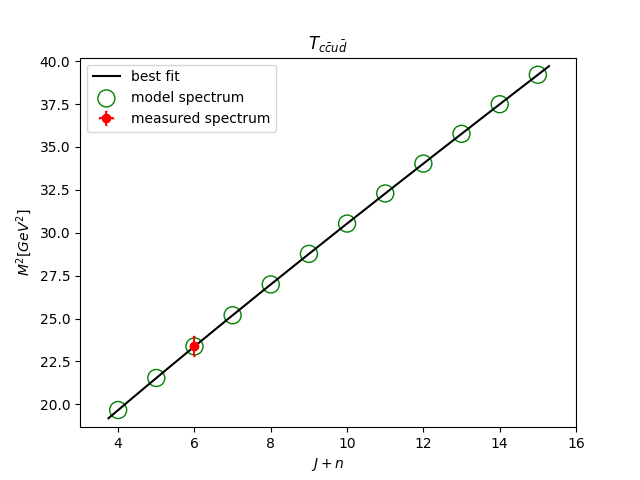}{tetraquarks/predictions/cu-cd/by_n/T_cucd_one_by_n}

\paragraph{$Z_{c}$ $(n, M^{2})$}{
    All $Z_{c}$ states were observed in $Z_{c}\rightarrow J/\psi\pi$ decay channels, but more channels are detailed in table \ref{table:tetra_cand_cucd}. Table \ref{table:Zc_n_HMRT_fit_output} is the built HMRTs output that includes all three states $Z_{c}(3900)$, $Z_{c}(4200)$ and $Z_{c}(4430)$. \ref{fig:Zc_n_best_fit} are the plot of these trajectories. These are consistent with tables \ref{table:Tsusu_j_predictions} and \ref{table:Tsusu_n_predictions}. Moreover, the predictions of the states in table \ref{table:Tcucd_n_predictions} is close to the $Z_{c}$ sample. The intercept value of the best fit is $a_{n}=-2.10$.

    \begin{longtable}[c]{|c|c|c|c|c|c|c|}
  \hline
  \thead{$\chi^{2}_{r}$} & \thead{$\chi^{2}_{hish}$} & \thead{$\alpha^{'}$} & \thead{$a$} & \thead{$m_{1}$} & \thead{$m_{2}$} & \thead{Fit Type} \\
  \hline
  \hline
0.00 & 0.00 & 0.51 & -2.14 & 1.11 & 1.11 & free fit \\
\hline
0.41 & 0.00 & 0.45 & -1.64 & 1.12 & 1.12 & fixed $m_{1}$, $m_{2}$ and $\alpha^{'}$ \\
\hline
0.75 & 0.00 & 0.45 & -1.70 & 1.11 & 1.11 & fixed $\alpha^{'}$ \\
\hline
0.00 & 0.00 & 0.51 & -2.10 & 1.12 & 1.12 & fixed $m_{1}$ and $m_{2}$ \\
\hline
  \caption{The fit for $Z_{c}$ states (\ref{table:tetra_cand_cucd}) in the $(n, M^{2})$ plane. The best fit is with fixed $\alpha^{'}$, but all the fits provide similar results and are consistent with the assumption that tetraquarks should admit near slope as the mesons with similar flavour content.} \label{table:Zc_n_HMRT_fit_output}
 \end{longtable}

    \begin{longtable}[c]{|c|c|c|c|c|c|}
  \hline
  \thead{$State$} & \thead{$M$ \\ $[MeV]$} & \thead{Thry \\ $M$ \\ $[MeV]$} & \thead{$J$} & \thead{Thry \\ $n$} & \thead{Thry \\ $J + n$} \\
  \hline
  \hline
$Z(3900)_{c1}$ & $3887$ & $3851$ & $1$ & $0$ & $1$ \\
\hline
$Z(4200)_{c1}$ & $4196$ & $4202$ & $1$ & $1$ & $2$ \\
\hline
$Z(4430)_{c1}$ & $4478$ & $4515$ & $1$ & $2$ & $3$ \\
\hline
 &  & $4801$ &  &  & $4$ \\
\hline
 &  & $5066$ &  &  & $5$ \\
\hline
 &  & $5314$ &  &  & $6$ \\
\hline
 &  & $5548$ &  &  & $7$ \\
\hline
  \caption{$Z_{c}$ $(n, M^{2})$ HMRT.}
 \end{longtable}

    \clearpage

    \twographs[(a) $Z_{c}$ $(n, M^{2})$ HMRTs output, (b) $Z_{c}$ $(n, M^{2})$ best fit (fixed $\alpha{'}=0.46$).][Zc_n_best_fit]{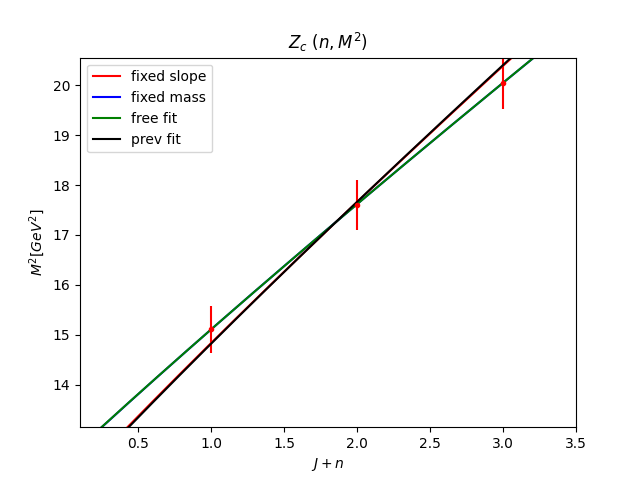}{tetraquarks/Zc/fit-_Z_c_Zc_by_n_fixed_slope_n}
}

\paragraph{$Z_{c}(3900)$ $(J, M^{2})$}{
    This state was observed in $Z_{c}(3900)\rightarrow J/\psi\pi$ and $Z_{c}(3900)\rightarrow D\bar{D}^{*}$. Table \ref{table:Zc_3900_J_HMRT} includes the measured $Z_{c}(3900)$ and \ref{fig:Zc_3900_J} is the plot of the HMRT. It is consistent with table \ref{table:Tcucd_j_predictions}. The intercept value is $a_{J}=-1.96$.

    \begin{longtable}[c]{|c|c|c|c|c|c|}
  \hline
  \thead{$State$} & \thead{$M$ \\ $[MeV]$} & \thead{Thry \\ $M$ \\ $[MeV]$} & \thead{$J$} & \thead{Thry \\ $n$} & \thead{Thry \\ $J + n$} \\
  \hline
  \hline
 &  & $3301$ &  &  & $0$ \\
\hline
 &  & $3614$ &  &  & $1$ \\
\hline
$Z(3900)_{c1}$ & $3887$ & $3887$ & $1$ & $1$ & $2$ \\
\hline
 &  & $4133$ &  &  & $3$ \\
\hline
 &  & $4359$ &  &  & $4$ \\
\hline
  \caption{$Z_{c}(3900)$ $(J, M^{2})$ HMRT.} \label{table:Zc_3900_J_HMRT}
 \end{longtable}

    \begin{figure}[!ht] 
     \centering
     \includegraphics[width=.6\columnwidth]{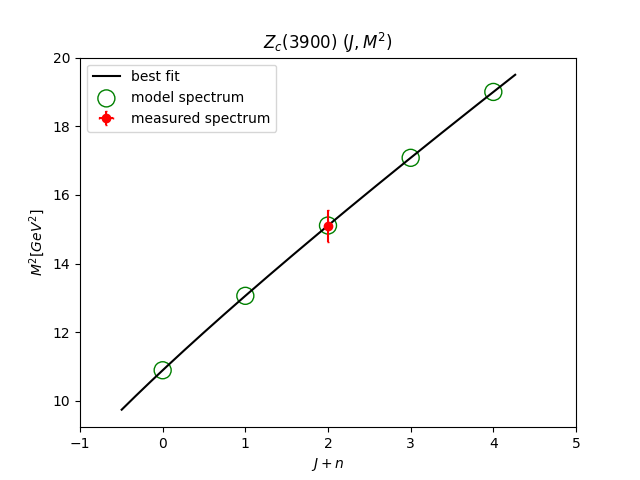}
     \caption{$Z_{c}(3900)$ $(J, M^{2})$ HMRT.} \label{fig:Zc_3900_J}
    \end{figure}

    The preferred decay channel of this state is to $D\bar{D^{*}}$ as opposed to $J/\psi\pi$. One explanation is that it is a molecule and not a compact-multiquark structure. We leave it as an open question for future work.
}

\paragraph{$Z_{c}(4200)$ $(J, M^{2})$}{
    This state was observed in $Z_{c}(4200)\rightarrow J/\psi\pi$. Table \ref{table:Zc_4200_J_HMRT} includes the measured $Z_{c}(4200)$ and \ref{fig:Zc_4200_J} is the plot of the HMRT. It is consistent with table \ref{table:Tcucd_j_predictions}. The intercept value is $a_{J}=-2.23$.

    \begin{longtable}[c]{|c|c|c|c|c|c|}
  \hline
  \thead{$State$} & \thead{$M$ \\ $[MeV]$} & \thead{Thry \\ $M$ \\ $[MeV]$} & \thead{$J$} & \thead{Thry \\ $n$} & \thead{Thry \\ $J + n$} \\
  \hline
  \hline
 &  & $3691$ &  &  & $1$ \\
\hline
 &  & $3956$ &  &  & $2$ \\
\hline
$Z(4200)_{c1}$ & $4196$ & $4196$ & $1$ & $2$ & $3$ \\
\hline
 &  & $4417$ &  &  & $4$ \\
\hline
 &  & $4624$ &  &  & $5$ \\
\hline
  \caption{$Z_{c}(4200)$ $(J, M^{2})$ HMRT.} \label{table:Zc_4200_J_HMRT}
 \end{longtable}

    \begin{figure}[!ht] 
     \centering
     \includegraphics[width=.6\columnwidth]{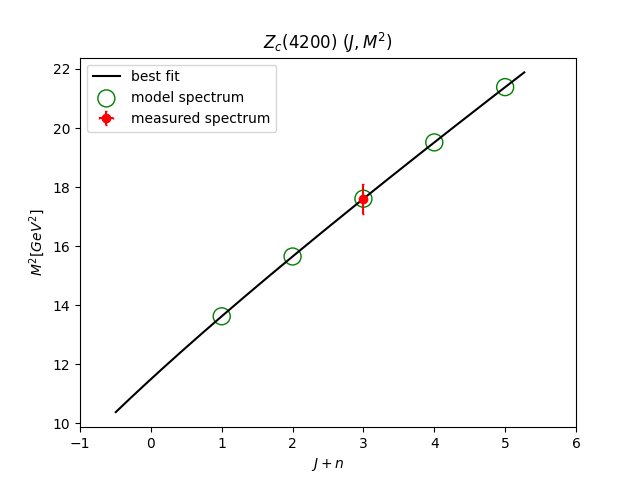}
     \caption{$Z_{c}(4200)$ $(J, M^{2})$ HMRT.} \label{fig:Zc_4200_J}
    \end{figure}
}

\paragraph{$Z_{c}(4430)$ $(J, M^{2})$}{
    This state was observed in $Z_{c}(4430)\rightarrow J/\psi\pi$ and $Z_{c}(4430)\rightarrow \psi(2S)\pi$. Table \ref{table:Zc_4430_J_HMRT} includes the measured $Z_{c}(4430)$ and \ref{fig:Zc_4430_J} is the plot of the HMRT. It is consistent with table \ref{table:Tcucd_j_predictions}. The intercept value is $a_{J}=-1.52$.

    \begin{longtable}[c]{|c|c|c|c|c|c|}
  \hline
  \thead{$State$} & \thead{$M$ \\ $[MeV]$} & \thead{Thry \\ $M$ \\ $[MeV]$} & \thead{$J$} & \thead{Thry \\ $n$} & \thead{Thry \\ $J + n$} \\
  \hline
  \hline
 &  & $4027$ &  &  & $3$ \\
\hline
 &  & $4261$ &  &  & $4$ \\
\hline
$Z(4430)_{c1}$ & $4478$ & $4478$ & $1$ & $4$ & $5$ \\
\hline
 &  & $4681$ &  &  & $6$ \\
\hline
 &  & $4873$ &  &  & $7$ \\
\hline
  \caption{$Z_{c}(4430)$ $(J, M^{2})$ HMRT.} \label{table:Zc_4430_J_HMRT}
 \end{longtable}

    \begin{figure}[!ht] 
     \centering
     \includegraphics[width=.6\columnwidth]{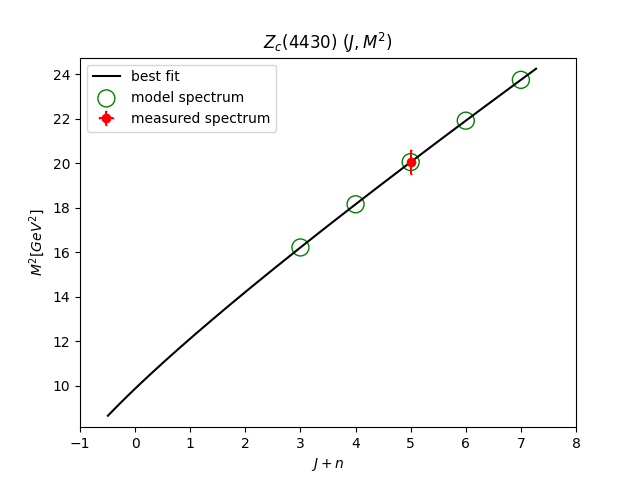}
     \caption{$Z_{c}(4430)$ $(J, M^{2})$ HMRT.} \label{fig:Zc_4430_J}
    \end{figure}
}

\paragraph{$X(4020)^{\pm}$}{
    $X(4020)$ was observed in $X(4020)\rightarrow D^{*}\bar{D^{*}}$ and $X(4020)\rightarrow h_{c}(1P)\pi^{\pm}$. Tables \ref{table:X_4020_J_HMRT} and \ref{table:X_4020_n_HMRT} are the built HMRT that includes the measured $X(4020)^{\pm}$ and \ref{fig:X_4020} are the plot of these trajectories. These are consistent with tables \ref{table:Tcucd_j_predictions} and \ref{table:Tcucd_n_predictions}. The intercept values are $a_{J}=-1.51$ and $a_{n}=-1.12$.

    \begin{center}
    \begin{minipage}{.49\linewidth}
        \begin{center}
            \begin{longtable}[c]{|c|c|c|c|c|c|}
  \hline
  \thead{$State$} & \thead{$M$ \\ $[MeV]$} & \thead{Thry \\ $M$ \\ $[MeV]$} & \thead{$J$} & \thead{Thry \\ $n$} & \thead{Thry \\ $J + n$} \\
  \hline
  \hline
 &  & $3478$ &  &  & $1$ \\
\hline
 &  & $3767$ &  &  & $2$ \\
\hline
$X(4020)^{+}$ & $4024$ & $4024$ &  &  & $3$ \\
\hline
 &  & $4258$ &  &  & $4$ \\
\hline
 &  & $4476$ &  &  & $5$ \\
\hline
  \caption{$X(4020)$ $(J, M^{2})$ HMRT.} \label{table:X_4020_J_HMRT}
 \end{longtable}
        \end{center}
    \end{minipage}
    \begin{minipage}{.49\linewidth}
        \begin{center}
            \begin{longtable}[c]{|c|c|c|c|c|c|}
  \hline
  \thead{$State$} & \thead{$M$ \\ $[MeV]$} & \thead{Thry \\ $M$ \\ $[MeV]$} & \thead{$J$} & \thead{Thry \\ $n$} & \thead{Thry \\ $J + n$} \\
  \hline
  \hline
 &  & $3179$ &  &  & $0$ \\
\hline
 &  & $3643$ &  &  & $1$ \\
\hline
$X(4020)^{+}$ & $4024$ & $4024$ &  &  & $2$ \\
\hline
 &  & $4357$ &  &  & $3$ \\
\hline
 &  & $4657$ &  &  & $4$ \\
\hline
  \caption{$X(4020)$ $(n, M^{2})$ HMRT.} \label{table:X_4020_n_HMRT}
 \end{longtable}
        \end{center}
    \end{minipage}
    \end{center}

    \twographs[(a) is $(J, M^{2})$ $X(4020)$ MHRT, (b) $(n, M^{2})$ $X(4020)$ MHRT.][X_4020]{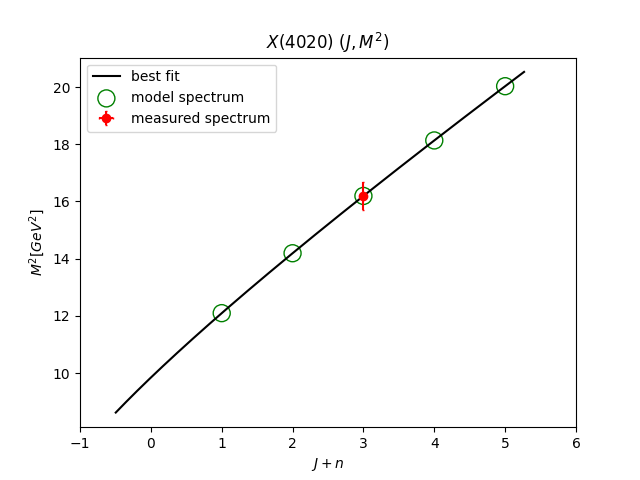}{tetraquarks/X_4020_/by_n_one_point/fit-_X__X_4020__tetra_one_by_n}
}

\paragraph{$X(4051)^{\pm}$}{
    $X(4051)$ was observed in $X(4051)\rightarrow\pi^{+}\chi_{c1}(1P)$. Tables \ref{table:X_4051_J_HMRT} and \ref{table:X_4051_n_HMRT} are the built HMRT that includes the measured $X(4051)^{\pm}$ and \ref{fig:X_4051} are the plot of these trajectories. These are consistent with tables \ref{table:Tcucd_j_predictions} and \ref{table:Tcucd_n_predictions}. The intercept values are $a_{J}=-1.62$ and $a_{n}=-1.20$.

    \begin{center}
    \begin{minipage}{.49\linewidth}
        \begin{center}
            \begin{longtable}[c]{|c|c|c|c|c|c|}
  \hline
  \thead{$State$} & \thead{$M$ \\ $[MeV]$} & \thead{Thry \\ $M$ \\ $[MeV]$} & \thead{$J$} & \thead{Thry \\ $n$} & \thead{Thry \\ $J + n$} \\
  \hline
  \hline
 &  & $3512$ &  &  & $1$ \\
\hline
 &  & $3797$ &  &  & $2$ \\
\hline
$X(4050)^{+}$ & $4051$ & $4051$ &  &  & $3$ \\
\hline
 &  & $4283$ &  &  & $4$ \\
\hline
 &  & $4499$ &  &  & $5$ \\
\hline
  \caption{$X(4051)$ $(J, M^{2})$ HMRT.} \label{table:X_4051_J_HMRT}
 \end{longtable}
        \end{center}
    \end{minipage}
    \begin{minipage}{.49\linewidth}
        \begin{center}
            \begin{longtable}[c]{|c|c|c|c|c|c|}
  \hline
  \thead{$State$} & \thead{$M$ \\ $[MeV]$} & \thead{Thry \\ $M$ \\ $[MeV]$} & \thead{$J$} & \thead{Thry \\ $n$} & \thead{Thry \\ $J + n$} \\
  \hline
  \hline
 &  & $3219$ &  &  & $0$ \\
\hline
 &  & $3675$ &  &  & $1$ \\
\hline
$X(4050)^{+}$ & $4051$ & $4051$ &  &  & $2$ \\
\hline
 &  & $4381$ &  &  & $3$ \\
\hline
 &  & $4679$ &  &  & $4$ \\
\hline
  \caption{$X(4051)$ $(n, M^{2})$ HMRT.} \label{table:X_4051_n_HMRT}
 \end{longtable}
        \end{center}
    \end{minipage}
    \end{center}

    \twographs[(a) is $(J, M^{2})$ $X(4051)$ MHRT, (b) $(n, M^{2})$ $X(4051)$ MHRT.][X_4051]{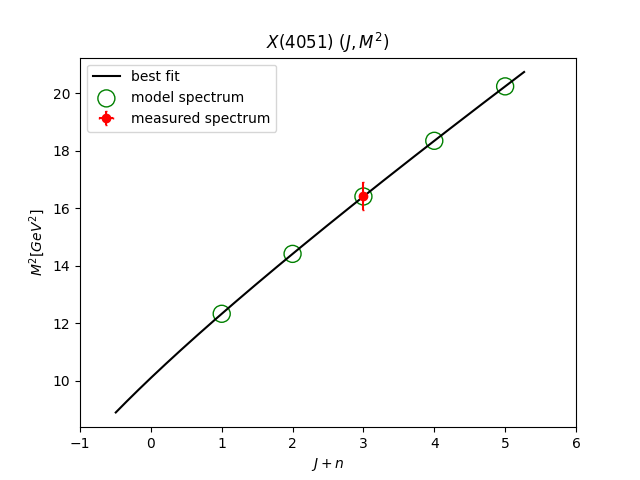}{tetraquarks/X_4051_/by_n_one_point/fit-_X__X_4051__tetra_one_by_n}
}

\paragraph{$X(4055)^{\pm}$}{
    $X(4055)$ was observed in $X(4055)\rightarrow\pi^{+}\psi(2S)$. Tables \ref{table:X_4055_J_HMRT} and \ref{table:X_4055_n_HMRT} are the built HMRT that includes the measured $X(4055)^{\pm}$ and \ref{fig:X_4055} are the plot of these trajectories. These are consistent with tables \ref{table:Tcucd_j_predictions} and \ref{table:Tcucd_n_predictions}. The intercept values are $a_{J}=-1.63$ and $a_{n}=-1.20$.

    \begin{center}
    \begin{minipage}{.49\linewidth}
        \begin{center}
            \begin{longtable}[c]{|c|c|c|c|c|c|}
  \hline
  \thead{$State$} & \thead{$M$ \\ $[MeV]$} & \thead{Thry \\ $M$ \\ $[MeV]$} & \thead{$J$} & \thead{Thry \\ $n$} & \thead{Thry \\ $J + n$} \\
  \hline
  \hline
 &  & $3516$ &  &  & $1$ \\
\hline
 &  & $3800$ &  &  & $2$ \\
\hline
$X(4055)^{+}$ & $4054$ & $4054$ &  &  & $3$ \\
\hline
 &  & $4286$ &  &  & $4$ \\
\hline
 &  & $4501$ &  &  & $5$ \\
\hline
  \caption{$X(4055)$ $(J, M^{2})$ HMRT.} \label{table:X_4055_J_HMRT}
 \end{longtable}
        \end{center}
    \end{minipage}
    \begin{minipage}{.49\linewidth}
        \begin{center}
            \begin{longtable}[c]{|c|c|c|c|c|c|}
  \hline
  \thead{$State$} & \thead{$M$ \\ $[MeV]$} & \thead{Thry \\ $M$ \\ $[MeV]$} & \thead{$J$} & \thead{Thry \\ $n$} & \thead{Thry \\ $J + n$} \\
  \hline
  \hline
 &  & $3224$ &  &  & $0$ \\
\hline
 &  & $3678$ &  &  & $1$ \\
\hline
$X(4055)^{+}$ & $4054$ & $4054$ &  &  & $2$ \\
\hline
 &  & $4383$ &  &  & $3$ \\
\hline
 &  & $4681$ &  &  & $4$ \\
\hline
  \caption{$X(4055)$ $(n, M^{2})$ HMRT.} \label{table:X_4055_n_HMRT}
 \end{longtable}
        \end{center}
    \end{minipage}
    \end{center}

    \twographs[(a) is $(J, M^{2})$ $X(4055)$ MHRT, (b) $(n, M^{2})$ $X(4055)$ MHRT.][X_4055]{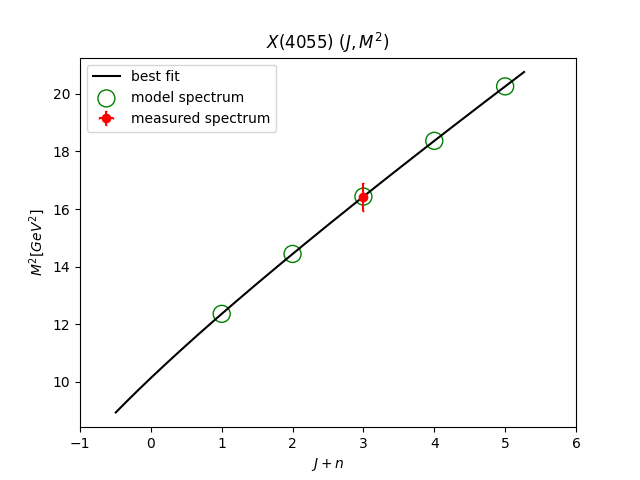}{tetraquarks/X_4055_/by_n_one_point/fit-_X__X_4055__tetra_one_by_n}
}

\paragraph{$X(4100)^{\pm}$}{
    $X(4100)$ was observed in $X(4100)\rightarrow\pi^{-}\eta(1S)$. Tables \ref{table:X_4100_J_HMRT} and \ref{table:X_4100_n_HMRT} are the built HMRT that includes the measured $X(4100)^{\pm}$ and \ref{fig:X_4100} are the plot of these trajectories. These are consistent with tables \ref{table:Tcucd_j_predictions} and \ref{table:Tcucd_n_predictions}. The intercept values are $a_{J}=-1.80$ and $a_{n}=-1.36$.

    \begin{center}
    \begin{minipage}{.49\linewidth}
        \begin{center}
            \begin{longtable}[c]{|c|c|c|c|c|c|}
  \hline
  \thead{$State$} & \thead{$M$ \\ $[MeV]$} & \thead{Thry \\ $M$ \\ $[MeV]$} & \thead{$J$} & \thead{Thry \\ $n$} & \thead{Thry \\ $J + n$} \\
  \hline
  \hline
 &  & $3568$ &  &  & $1$ \\
\hline
 &  & $3846$ &  &  & $2$ \\
\hline
$X(4100)^{+}$ & $4096$ & $4096$ &  &  & $3$ \\
\hline
 &  & $4325$ &  &  & $4$ \\
\hline
 &  & $4537$ &  &  & $5$ \\
\hline
  \caption{$X(4100)$ $(J, M^{2})$ HMRT.} \label{table:X_4100_J_HMRT}
 \end{longtable}
        \end{center}
    \end{minipage}
    \begin{minipage}{.49\linewidth}
        \begin{center}
            \begin{longtable}[c]{|c|c|c|c|c|c|}
  \hline
  \thead{$State$} & \thead{$M$ \\ $[MeV]$} & \thead{Thry \\ $M$ \\ $[MeV]$} & \thead{$J$} & \thead{Thry \\ $n$} & \thead{Thry \\ $J + n$} \\
  \hline
  \hline
 &  & $3285$ &  &  & $0$ \\
\hline
 &  & $3727$ &  &  & $1$ \\
\hline
$X(4100)^{+}$ & $4096$ & $4096$ &  &  & $2$ \\
\hline
 &  & $4421$ &  &  & $3$ \\
\hline
 &  & $4716$ &  &  & $4$ \\
\hline
  \caption{$X(4100)$ $(n, M^{2})$ HMRT.} \label{table:X_4100_n_HMRT}
 \end{longtable}
        \end{center}
    \end{minipage}
    \end{center}

    \twographs[(a) is $(J, M^{2})$ $X(4100)$ MHRT, (b) $(n, M^{2})$ $X(4100)$ MHRT.][X_4100]{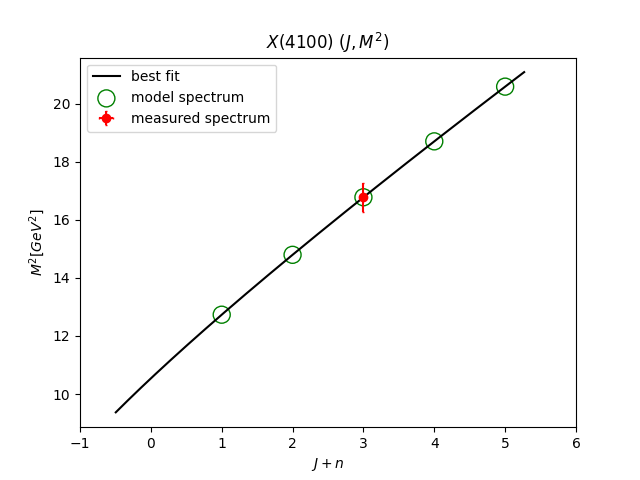}{tetraquarks/X_4100_/by_n_one_point/fit-_X__X_4100__tetra_one_by_n}
}

\paragraph{$R_{c0}(4240)^{\pm}$}{
    $R_{c0}(4240)^{\pm}$ was observed in $R_{c0}(4240)\rightarrow\pi^{-}\psi(2S)$. Tables \ref{table:R_4240_J_HMRT} and \ref{table:R_4240_n_HMRT} are the built HMRT that includes the measured $R_{c0}(4240)^{\pm}$ and \ref{fig:Rc_4240} are the plot of these trajectories. These are consistent with tables \ref{table:Tcucd_j_predictions} and \ref{table:Tcucd_n_predictions}. The intercept values are $a_{J}=-2.42$ and $a_{n}=-1.75$.

    \begin{center}
    \begin{minipage}{.49\linewidth}
        \begin{center}
            \begin{longtable}[c]{|c|c|c|c|c|c|}
  \hline
  \thead{$State$} & \thead{$M$ \\ $[MeV]$} & \thead{Thry \\ $M$ \\ $[MeV]$} & \thead{$J$} & \thead{Thry \\ $n$} & \thead{Thry \\ $J + n$} \\
  \hline
  \hline
 &  & $3744$ &  &  & $1$ \\
\hline
 &  & $4003$ &  &  & $2$ \\
\hline
$R(4240)_{c0}$ & $4239$ & $4239$ & $0$ & $3$ & $3$ \\
\hline
 &  & $4457$ &  &  & $4$ \\
\hline
 &  & $4662$ &  &  & $5$ \\
\hline
  \caption{$R_{c0}(4240)$ $(J, M^{2})$ HMRT.} \label{table:R_4240_J_HMRT}
 \end{longtable}
        \end{center}
    \end{minipage}
    \begin{minipage}{.49\linewidth}
        \begin{center}
            \begin{longtable}[c]{|c|c|c|c|c|c|}
  \hline
  \thead{$State$} & \thead{$M$ \\ $[MeV]$} & \thead{Thry \\ $M$ \\ $[MeV]$} & \thead{$J$} & \thead{Thry \\ $n$} & \thead{Thry \\ $J + n$} \\
  \hline
  \hline
 &  & $3486$ &  &  & $0$ \\
\hline
 &  & $3891$ &  &  & $1$ \\
\hline
$R(4240)_{c0}$ & $4239$ & $4239$ & $0$ & $2$ & $2$ \\
\hline
 &  & $4550$ &  &  & $3$ \\
\hline
 &  & $4834$ &  &  & $4$ \\
\hline
  \caption{$R_{c0}(4240)$ $(n, M^{2})$ HMRT.} \label{table:R_4240_n_HMRT}
 \end{longtable}
        \end{center}
    \end{minipage}
    \end{center}

    \twographs[(a) is $(J, M^{2})$ $Rc0(4240)$ MHRT, (b) $(n, M^{2})$ $Rc0(4240)$ MHRT.][Rc_4240]{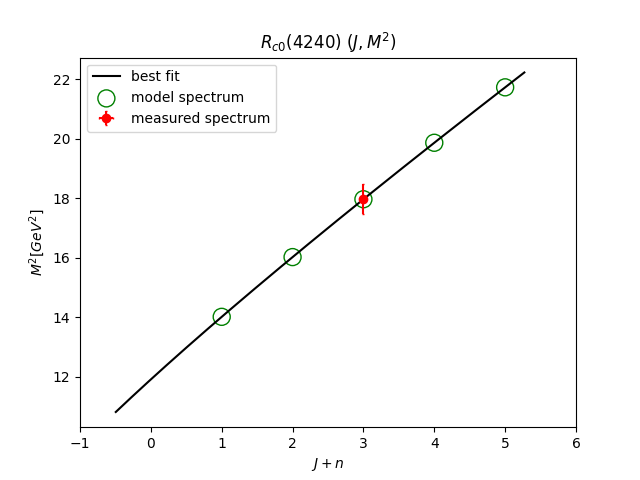}{tetraquarks/Rc0_4240_/by_n_one_point/fit-_R_c_Rc0_4240__tetra_one_by_n}
}

\clearpage

\paragraph{$X(4250)^{\pm}$}{
    $X(4250)$ was observed in $X(4250)\rightarrow\pi^{+}\chi(1P)$. Tables \ref{table:X_4250_J_HMRT} and \ref{table:X_4250_n_HMRT} are the built HMRT that includes the measured $X(4250)^{\pm}$ and \ref{fig:X_4250} are the plot of these trajectories. These are consistent with tables \ref{table:Tcucd_j_predictions} and \ref{table:Tcucd_n_predictions}. The intercept values are $a_{J}=-2.42$ and $a_{n}=-1.75$.

    \begin{center}
    \begin{minipage}{.49\linewidth}
        \begin{center}
            \begin{longtable}[c]{|c|c|c|c|c|c|}
  \hline
  \thead{$State$} & \thead{$M$ \\ $[MeV]$} & \thead{Thry \\ $M$ \\ $[MeV]$} & \thead{$J$} & \thead{Thry \\ $n$} & \thead{Thry \\ $J + n$} \\
  \hline
  \hline
 &  & $3744$ &  &  & $1$ \\
\hline
 &  & $4003$ &  &  & $2$ \\
\hline
$R(4240)_{c0}$ & $4239$ & $4239$ & $0$ & $3$ & $3$ \\
\hline
 &  & $4457$ &  &  & $4$ \\
\hline
 &  & $4662$ &  &  & $5$ \\
\hline
  \caption{$X(4250)$ $(J, M^{2})$ HMRT.} \label{table:X_4250_J_HMRT}
 \end{longtable}
        \end{center}
    \end{minipage}
    \begin{minipage}{.49\linewidth}
        \begin{center}
            \begin{longtable}[c]{|c|c|c|c|c|c|}
  \hline
  \thead{$State$} & \thead{$M$ \\ $[MeV]$} & \thead{Thry \\ $M$ \\ $[MeV]$} & \thead{$J$} & \thead{Thry \\ $n$} & \thead{Thry \\ $J + n$} \\
  \hline
  \hline
 &  & $3486$ &  &  & $0$ \\
\hline
 &  & $3891$ &  &  & $1$ \\
\hline
$R(4240)_{c0}$ & $4239$ & $4239$ & $0$ & $2$ & $2$ \\
\hline
 &  & $4550$ &  &  & $3$ \\
\hline
 &  & $4834$ &  &  & $4$ \\
\hline
  \caption{$X(4250)$ $(n, M^{2})$ HMRT.} \label{table:X_4250_n_HMRT}
 \end{longtable}
        \end{center}
    \end{minipage}
    \end{center}

    \twographs[(a) is $(J, M^{2})$ $X(4250)$ MHRT, (b) $(n, M^{2})$ $X(4250)$ MHRT.][X_4250]{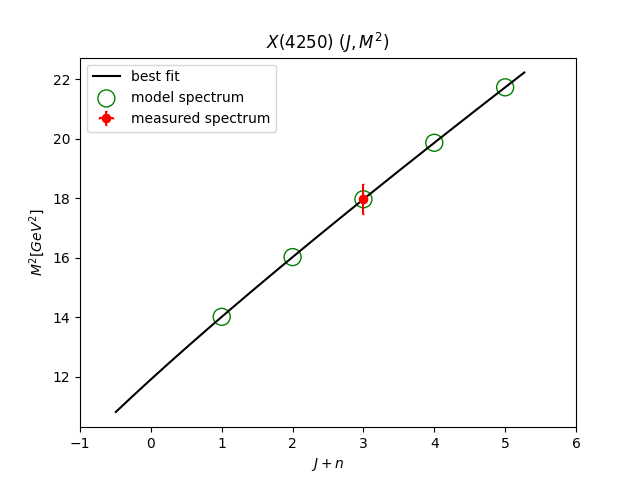}{tetraquarks/X_4250_/by_n_one_point/fit-_R_c_X_4250__tetra_one_by_n}
}

\subsubsection{\texorpdfstring{$T_{c\Bar{c}u\Bar{s}}$ or $T_{c\Bar{c}d\Bar{s}}$}{}}

For a tetraquarks  with the quark content $T_{c\Bar{c}u\Bar{s}}$ (table \ref{table:tetra_cand_cucs}). The candidate resonances are presented in table. We generated the predicted HMRTs for this quark content in tables \ref{table:Tcucs_j_predictions} and \ref{table:Tcucs_n_predictions}, and the thresholds table \ref{table:Tcucs_thresholds} includes the predicted width. The HMRTs are also plotted in \ref{fig:Tcucs_predictions}.

\begin{longtable}[c]{|c|c|c|c|c|c|c|}
\hline
\makecell{Candidate} & \makecell{Quarks \\ Content} & \makecell{Selected \\ Decay \\ Channels} & \makecell{$J^{PC}$} & \makecell{Mass \\ $[MeV]$} & \makecell{Width \\ $[MeV]$} & \makecell{Ref.} \\
\hline
\hline
\makecell{$T_{\psi s1}^{\theta}(4000)^{+}$ \\ also \\ $Z_{cs}(4000)^{+}$} & $c\bar{c}u\bar{s}$ & \makecell{$J/\psi K^{+}$ \\ $D_{s}^{+}\bar{D}^{*0}$ \\ $D_{s}^{*+}\bar{D}^{0}$} & $1^{+}$ & $3980-4010$ & $5-150$ & \makecell{} \\ 
\hline
\makecell{$T_{\psi s1}(4220)^{+}$ \\ also \\ $Z_{cs}(4220)^{+}$} & $c\bar{c}u\bar{s}$ & \makecell{$J/\psi K^{+}$} & $1^{+}$ & $4216^{+50}_{-40}$ & $233^{+110}_{-90}$ & \makecell{} \\
\hline
\makecell{$T_{\psi s1}^{\theta}(4000)^{0}$} & $c\bar{c}d\bar{s}$ & \makecell{$J/\psi K^{0}_{S}$} & $1^{+}$ & $3991^{+14}_{-20}$ & $105^{+12}_{-10}{}^{+9}_{-17}$ & \makecell{} \\
\hline
\caption{$T_{c\bar{u}u\bar{s}}$ or $T_{c\bar{d}u\bar{s}}$ tetraquark candidates.}\label{table:tetra_cand_cucs}
\end{longtable}

\begin{longtable}[c]{|c|c|c|c|c|c|}
  \hline
  \thead{Thry \\ Width \\ $[MeV]$} & \thead{Mesons \\ Pair} & \thead{Meson \\ Threshold \\ $[MeV]$} & \thead{Baryon \\ Anti-\\baryon} & \thead{Baryonic \\ Threshold \\ $[MeV]$} & \thead{Genuine} \\
  \hline
  \hline
\makecell{$47-79$\\$29-59$} & \makecell{$\eta_{c}(1S) K^{+}$ \\ $D^{+}_{s} \bar{D}^{0}$} & \makecell{$3478$ \\ $3833$} & \makecell{$\Lambda^{+}_{c} \bar{\Xi}^{0}_{c}$ \\ $\Sigma^{++}_{c} \bar{\Xi}^{-}_{c}$ \\ $\Xi^{+}_{c} \bar{\Omega}^{0}_{c}$} & \makecell{$4756$ \\ $4922$ \\ $5163$} & $*$ \\
\hline
  \caption{$T_{c\bar{c}u\bar{s}}$ thresholds.} \label{table:Tcucs_thresholds}
 \end{longtable}

    \begin{center}
    \begin{minipage}{.49\linewidth}
        \begin{center}
            \begin{longtable}[c]{|c|c|}
  \hline
  \thead{$J$ \\ Spec} & \thead{$a_{T_{j}}$} \\
  \hline
  \hline
\makecell{$4362-4488$\\$4587-4704$\\$4796-4906$\\$4993-5097$\\$5179-5278$\\$5356-5450$\\$5525-5616$\\$5688-5776$\\$5845-5930$\\$5997-6079$\\$6144-6223$\\$6287-6364$} & \makecell{$-3.3$-$-2.7$} \\
\hline
  \caption{$T_{c\bar{c}u\bar{s}}$ $(J, M^2)$ predictions.} \label{table:Tcucs_j_predictions}
 \end{longtable}
        \end{center}
    \end{minipage}
    \begin{minipage}{.49\linewidth}
        \begin{center}
            \begin{longtable}[c]{|c|c|}
  \hline
  \thead{$n$ \\ Spec} & \thead{$a_{T_{n}}$} \\
  \hline
  \hline
\makecell{$4145-4282$\\$4489-4610$\\$4796-4906$\\$5076-5178$\\$5337-5432$\\$5580-5670$\\$5811-5896$\\$6030-6111$\\$6240-6317$\\$6441-6515$\\$6634-6706$\\$6821-6891$} & \makecell{$-2.0$-$-1.6$} \\
\hline
  \caption{$T_{c\bar{c}u\bar{s}}$ $(n, M^2)$ predictions.} \label{table:Tcucs_n_predictions}
 \end{longtable}
        \end{center}
    \end{minipage}
    \end{center}

\twographs[(a) is $(J, M^{2})$ $T_{c\Bar{d}c\Bar{u}}$ predicted HMRT, (b) $(n, M^{2})$ $T_{c\Bar{d}c\Bar{u}}$ predicted HMRT.][Tcucs_predictions]{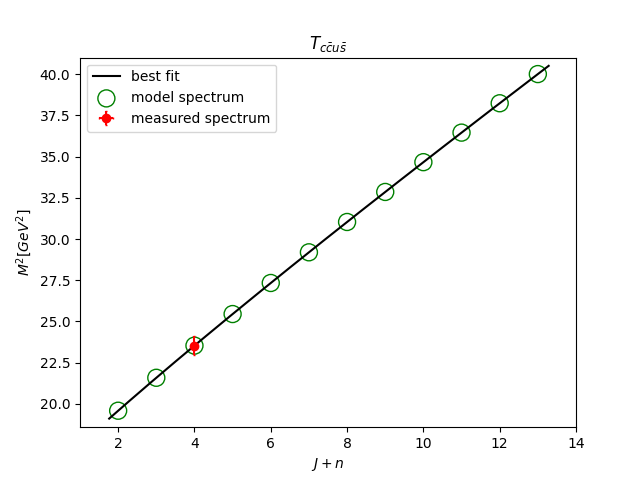}{tetraquarks/predictions/cu-cs/by_n/T_cucs_one_by_n}

\paragraph{$Z_{cs}$ $(n, M^{2})$ fit}{

    For $J^{PC}=1^{+}$, the best fit is with the fixed endpoint masses. The fits are shown in [Fig. \ref{fig:Zcs_n}] along with the predicted spectrum. The best fit parameters are: $\alpha^{'}_{Z_{cs}}=0.79$, $a_{Z_{cs}}=-2.98$.
    
    \begin{longtable}[c]{|c|c|c|c|c|c|c|}
  \hline
  \thead{$\chi^{2}_{r}$} & \thead{$\chi^{2}_{hish}$} & \thead{$\alpha^{'}$} & \thead{$a$} & \thead{$m_{1}$} & \thead{$m_{2}$} & \thead{Fit Type} \\
  \hline
  \hline
0.00 & 0.00 & 0.74 & -3.65 & 1.11 & 1.24 & free fit \\
\hline
3.58 & 0.03 & 0.45 & -1.08 & 1.12 & 1.47 & fixed $m_{1}$, $m_{2}$ and $\alpha^{'}$ \\
\hline
2.25 & 0.02 & 0.45 & -1.95 & 1.11 & 1.11 & fixed $\alpha^{'}$ \\
\hline
0.00 & 0.00 & 0.79 & -2.98 & 1.12 & 1.47 & fixed $m_{1}$ and $m_{2}$ \\
\hline
  \caption{}
 \end{longtable}
    
    \twographs[$Z_{cs}$ with $J^{P}=1^{+}$ HMRT. (a) Comparison between the fits, (b) $Z_{cs}$ with fixed mass HMRT, which is the best fit.][Zcs_n]{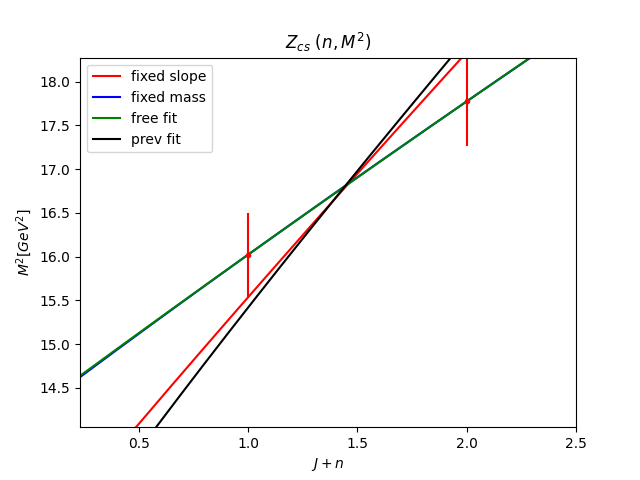}{tetraquarks/Zcs/fit-_Z__cs_Zcs_by_n_fixed_m_n}
    
    \begin{longtable}[c]{|c|c|c|c|c|c|}
  \hline
  \thead{$State$} & \thead{$M$ \\ $[MeV]$} & \thead{Thry \\ $M$ \\ $[MeV]$} & \thead{$J$} & \thead{Thry \\ $n$} & \thead{Thry \\ $J + n$} \\
  \hline
  \hline
$Z(4000)^{+}_{cs1}$ & $4003$ & $4003$ & $1$ & $0$ & $1$ \\
\hline
$Z(4220)^{+}_{cs1}$ & $4216$ & $4216$ & $1$ & $1$ & $2$ \\
\hline
 &  & $4412$ &  &  & $3$ \\
\hline
 &  & $4596$ &  &  & $4$ \\
\hline
 &  & $4768$ &  &  & $5$ \\
\hline
 &  & $4932$ &  &  & $6$ \\
\hline
  \caption{$Z_{cs}$ $(n, M^{2})$ HMRT.}
 \end{longtable}
}

\subsubsection{\texorpdfstring{$T_{\psi\phi}$ (quark content ${c\Bar{c}s\Bar{s}}$)}{}}

For tetraquarks with the quark content $T_{c\Bar{c}s\Bar{s}}$ The candidate resonances are presented in table \ref{table:tetra_cand_cscs}. We generated the predicted HMRTs for this quark content in tables \ref{table:Tcscs_j_predictions} and \ref{table:Tcscs_n_predictions}, and the thresholds table \ref{table:Tcscs_thresholds} includes the predicted width. The HMRTs are also plotted in \ref{fig:Tcscs_predictions}.

\begin{longtable}[c]{|c|c|c|c|c|c|c|}
\hline
\makecell{Candidate} & \makecell{Quarks \\ Content} & \makecell{Selected \\ Decay \\ Channels} & \makecell{$J^{PC}$} & \makecell{Mass \\ $[MeV]$} & \makecell{Width \\ $[MeV]$} & \makecell{Ref.} \\
\hline
\hline
$X(3960)$ & $c\bar{c}s\bar{s}$ & $D_{s}^{+}$ $D_{s}^{-}$ & $0^{++}$ & $3956\pm 5\pm 10$ & $43\pm{13}\pm{8}$ & \makecell{\cite{LHCb:2022aki} \\ another \\ approach in \\ \cite{PhysRevD.107.094018}} \\
$X(4140)$ & $c\bar{c}s\bar{s}$ & $J/\psi\phi$ & $1^{++}$ & $4146.5\pm 3.0$ & $19^{+7}_{-5}$ & \makecell{\cite{CMS:2013jru}} \\
$X(4274)$ & $c\bar{c}s\bar{s}$ & $J/\psi\phi$ & $1^{++}$ & $4286^{+8}_{-9}$ & $51\pm 7$ & \makecell{\cite{LHCb:2016axx}} \\
$\chi_{c0}(4500)$ & $c\bar{c}s\bar{s}$ & \makecell{$J/\psi\phi$} & $0^{++}$ & $4474\pm{4}$ & $77^{+12}_{-10}$ & \makecell{\cite{LHCb:2016axx}} \\
$\chi_{c1}(4685)$ & $c\bar{c}s\bar{s}$ & \makecell{$J/\psi\phi$} & $1^{++}$ & $4684^{+15}_{-17}$ & $126\pm{40}$ & \makecell{\cite{LHCb:2021uow}} \\
$\chi_{c0}(4700)$ & $c\bar{c}s\bar{s}$ & \makecell{$J/\psi\phi$} & $0^{++}$ & $4694^{+16}_{-5}$ & $87^{+18}_{-10}$ & \makecell{\cite{LHCb:2016axx}} \\
\hline
$X(4350)$\tablefootnotemark{footnote:state_needs_confirm} & $c\bar{c}s\bar{s}$ & $J/\psi\phi$ & $?^{?+}$ & $4351\pm{5}$ & $13^{+18}_{-10}$ & \makecell{} \\
$X(4630)$ & $c\bar{c}s\bar{s}$ & \makecell{$J/\psi\phi$} & $1^{-+}$\footnote{$J^{P}$ needs confirmation.} & $4626^{+24}_{-110}$ & $174^{+140}_{-80}$ & \makecell{} \\
\hline
$\psi(4360)$ & $c\bar{c}s\bar{s}$ & $J/\psi\eta$ & $1^{--}$ & $4374\pm{7}$ & $118\pm 12$ & \makecell{} \\ \\
$\psi(4660)$ & $c\bar{c}s\bar{s}$ & \makecell{$\Lambda_{c}^{+}\bar{\Lambda_{c}}^{-}$ \\ $D^{+}_{s}D_{s1}^{-}(2536)$ \\ $D^{+}_{s}D_{s2}^{*-}(2573)$} & $1^{--}$ & $4630\pm{6}$ & $72^{+14}_{-12}$ & \makecell{} \\ \\
\hline
\caption{$T_{c\bar{c}s\bar{s}}$ tetraquark candidates.}\label{table:tetra_cand_cscs}
\end{longtable}

\begin{longtable}[c]{|c|c|c|c|c|c|}
  \hline
  \thead{Thry \\ Width \\ $[MeV]$} & \thead{Mesons \\ Pair} & \thead{Meson \\ Threshold \\ $[MeV]$} & \thead{Baryon \\ Anti-\\baryon} & \thead{Baryonic \\ Threshold \\ $[MeV]$} & \thead{Genuine} \\
  \hline
  \hline
\makecell{$89-134$\\$27-54$} & \makecell{$D^{+}_{s} D^{-}_{s}$ \\ $\eta_{c}(1S) \phi(1020)$} & \makecell{$3936$ \\ $4003$} & \makecell{$\Xi^{+}_{c} \bar{\Xi}^{-}_{c}$ \\ $\Xi^{0}_{c} \bar{\Xi}^{0}_{c}$ \\ $\Omega^{0}_{c} \bar{\Omega}^{0}_{c}$} & \makecell{$4936$ \\ $4940$ \\ $5390$} &  \\
\hline
  \caption{$T_{c\bar{c}s\bar{s}}$ thresholds.} \label{table:Tcscs_thresholds}
 \end{longtable}

    \begin{center}
    \begin{minipage}{.49\linewidth}
        \begin{center}
            \begin{longtable}[c]{|c|c|}
  \hline
  \thead{$J$ \\ Spec} & \thead{$a_{T_{j}}$} \\
  \hline
  \hline
\makecell{$4536-4664$\\$4765-4882$\\$4976-5086$\\$5174-5278$\\$5361-5459$\\$5538-5632$\\$5708-5798$\\$5871-5958$\\$6028-6112$\\$6179-6260$\\$6326-6404$\\$6468-6545$} & \makecell{$-3.8$-$-3.2$} \\
\hline
  \caption{$T_{c\bar{c}s\bar{s}}$ $(J, M^2)$ predictions.} \label{table:Tcscs_j_predictions}
 \end{longtable}
        \end{center}
    \end{minipage}
    \begin{minipage}{.49\linewidth}
        \begin{center}
            \begin{longtable}[c]{|c|c|}
  \hline
  \thead{$n$ \\ Spec} & \thead{$a_{T_{n}}$} \\
  \hline
  \hline
\makecell{$4315-4454$\\$4666-4787$\\$4976-5086$\\$5258-5359$\\$5519-5613$\\$5763-5852$\\$5993-6078$\\$6212-6293$\\$6421-6498$\\$6622-6696$\\$6814-6886$\\$7001-7070$} & \makecell{$-2.7$-$-2.3$} \\
\hline
  \caption{$T_{c\bar{c}s\bar{s}}$ $(n, M^2)$ predictions.} \label{table:Tcscs_n_predictions}
 \end{longtable}
        \end{center}
    \end{minipage}
    \end{center}

\twographs[(a) is $(J, M^{2})$ $T_{c\Bar{c}s\Bar{s}}$ predicted HMRT, (b) $(n, M^{2})$ $T_{c\Bar{c}s\Bar{s}}$ predicted HMRT.][Tcscs_predictions]{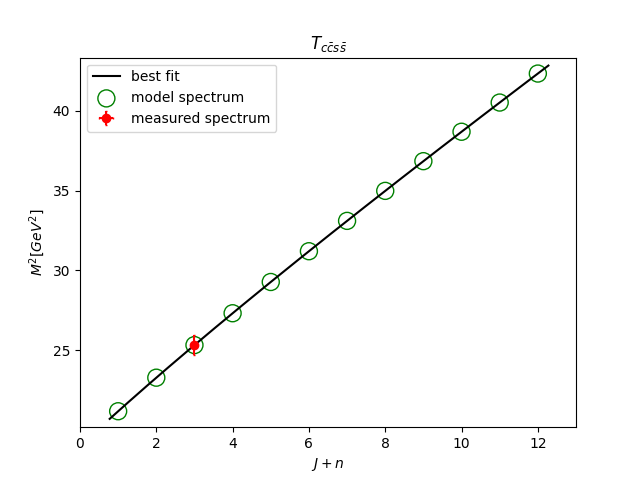}{tetraquarks/predictions/cs-cs/by_n/T_cscs_one_by_n}

In \cite{Tccss1}-\cite{Tccss4} there were seven resonances observed that are tetraquark candidates with content ${c\Bar{c}s\Bar{s}}$. The decays in which these states were observed are $J/\psi\phi$ and $D_{s}D_{s}$. 

We expect the natural decay of these states to be through a baryon-antibaryon pair for higher states that are above threshold. In this case, the annihilation threshold is $\approx3940$ which is slightly below X(3960) found in \cite{Tccss1}, while the threshold for a decay through string tear is $\approx4940$ to $\Xi_{c}\bar{\Xi_{c}}$.

The fit of the data was done separately for $J^{PC}=0^{++}$ and $J^{PC}=1^{++}$ in the $(n, M^{2})$ plane. For each trajectory we fit the data four times - once with free parameters, another with fixed endpoint masses, one with fixed $\alpha^{'}$ and at last we fixed both $\alpha^{'}$ and the endpoint masses, letting only the intercept to change. 

\paragraph{$T_{\psi\phi 0}$ $(n, M^{2})$ fit}{

    For $J^{PC}=0^{++}$, the best fit is with the fixed slope and endpoint masses, with $\chi^{2}_{r}=2.00$, but the spectrum prediction is similar for all fits. The fits are shown in [Fig. \ref{fig:T_cscs_0j_n}] along with the predicted spectrum.
    
    \begin{longtable}[c]{|c|c|c|c|c|c|c|}
  \hline
  \thead{$\chi^{2}_{r}$} & \thead{$\chi^{2}_{hish}$} & \thead{$\alpha^{'}$} & \thead{$a$} & \thead{$m_{1}$} & \thead{$m_{2}$} & \thead{Fit Type} \\
  \hline
  \hline
2.09 & 0.01 & 0.67 & -0.16 & 1.90 & 1.90 & free fit \\
\hline
2.00 & 0.02 & 0.45 & -1.55 & 1.47 & 1.47 & fixed $m_{1}$, $m_{2}$ and $\alpha^{'}$ \\
\hline
3.75 & 0.02 & 0.45 & -1.16 & 1.56 & 1.56 & fixed $\alpha^{'}$ \\
\hline
3.89 & 0.02 & 0.43 & -1.47 & 1.47 & 1.47 & fixed $m_{1}$ and $m_{2}$ \\
\hline
  \caption{$T_{\psi\phi 0}$ $(n, M^{2})$ fit results.} \label{table:T_ccss_n}
 \end{longtable}
    
    \twographs[$T_{\psi\phi}$ with $J^{PC}=0^{++}$ HMRT. (a) Comparison between the fits, (b) $T_{\psi\phi}$ with $J^{PC}=0^{++}$ free fit HMRT which is the best fit.][T_cscs_0j_n]{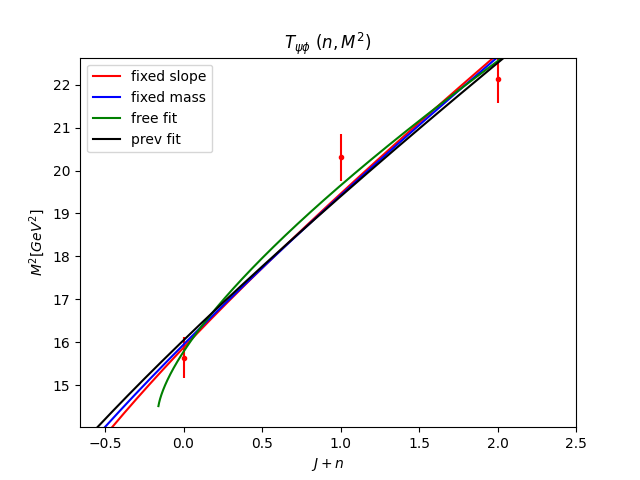}{tetraquarks/T_cscs/by_n/Tcscs0/fit-_X_0j_Tccss_prev_n}

    \clearpage
    
    \begin{longtable}[c]{|c|c|c|c|c|c|}
  \hline
  \thead{$State$} & \thead{$M$ \\ $[MeV]$} & \thead{Thry \\ $M$ \\ $[MeV]$} & \thead{$J$} & \thead{Thry \\ $n$} & \thead{Thry \\ $J + n$} \\
  \hline
  \hline
$X(3960)$ & $3955$ & $3973$ & $0$ & $0$ & $0$ \\
\hline
$X(4500)$ & $4506$ & $4433$ & $0$ & $1$ & $1$ \\
\hline
$X(4700)$ & $4704$ & $4748$ & $0$ & $2$ & $2$ \\
\hline
 &  & $5012$ &  &  & $3$ \\
\hline
 &  & $5245$ &  &  & $4$ \\
\hline
 &  & $5458$ &  &  & $5$ \\
\hline
 &  & $5656$ &  &  & $6$ \\
\hline
  \caption{$T_{\psi\phi}$ Regge trajectory.}
 \end{longtable}
}

\paragraph{$T_{\psi\phi 1}$ $(n, M^{2})$ fit}{
    For $J^{PC}=1^{++}$, the best fit is with the fixed masses and $\alpha^{'}$ parameters with $\chi^{2}_{r}=4.30$. The fits are shown in [Fig. \ref{fig:T_cscs_all_j1}]. The predicted spectrum is at \ref{table:T_cscs_spectrum}.
    
    \begin{longtable}[c]{|c|c|c|c|c|c|c|}
  \hline
  \thead{$\chi^{2}_{r}$} & \thead{$\chi^{2}_{hish}$} & \thead{$\alpha^{'}$} & \thead{$a$} & \thead{$m_{1}$} & \thead{$m_{2}$} & \thead{Fit Type} \\
  \hline
  \hline
4.38 & 0.02 & 0.48 & -4.20 & 0.81 & 0.81 & free fit \\
\hline
5.62 & 0.05 & 0.45 & -0.53 & 1.47 & 1.47 & fixed $m_{1}$, $m_{2}$ and $\alpha^{'}$ \\
\hline
4.30 & 0.02 & 0.45 & -5.71 & 0.35 & 0.35 & fixed $\alpha^{'}$ \\
\hline
4.69 & 0.02 & 0.64 & -1.50 & 1.47 & 1.47 & fixed $m_{1}$ and $m_{2}$ \\
\hline
  \caption{$T_{\psi\phi 1}$ $(n, M^{2})$ fit results.} \label{table:T_ccss1_n}
 \end{longtable}
    
    \twographs[$T_{\psi\phi}$ with $J^{PC}=1^{++}$ HMRT. (a) Comparison between the fits, (b) $T_{\psi\phi}$ with $J^{PC}=1^{++}$ fixed fit HMRT which is the best fit.][T_cscs_all_j1]{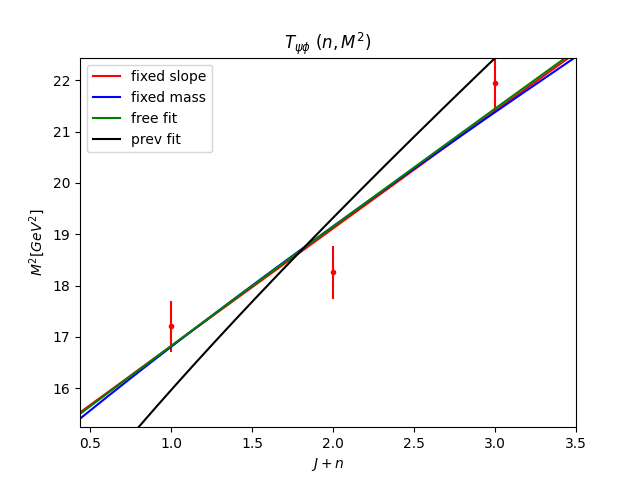}{tetraquarks/T_cscs/by_n/Tcscs1/fit-_X_1j_Tccss_new_n}
    
    \begin{longtable}[c]{|c|c|c|c|c|c|}
  \hline
  \thead{$State$} & \thead{$M$ \\ $[MeV]$} & \thead{Thry \\ $M$ \\ $[MeV]$} & \thead{$J$} & \thead{Thry \\ $n$} & \thead{Thry \\ $J + n$} \\
  \hline
  \hline
$X(4140)$ & $4148$ & $3996$ & $1$ & $0$ & $1$ \\
\hline
$X(4274)$ & $4273$ & $4396$ & $1$ & $1$ & $2$ \\
\hline
$X(4685)$ & $4684$ & $4736$ & $1$ & $2$ & $3$ \\
\hline
 &  & $5040$ &  &  & $4$ \\
\hline
 &  & $5317$ &  &  & $5$ \\
\hline
 &  & $5573$ &  &  & $6$ \\
\hline
 &  & $5814$ &  &  & $7$ \\
\hline
  \caption{$T_{\psi\phi}$ HMRT.} \label{table:T_cscs_spectrum}
 \end{longtable}
}

\paragraph{$X(4350)$ and $X(4630)$ $(n, M^{2})$ fit\label{section:X_4630_HMRT}}{
    For $J^{PC}=1^{-+}$ \footnote{We assume that the $X(4350)$ is with the same
    quantum numbers, although not all of them are known yet} the best fit is
    with the free parameters with $\chi^{2}_{r}=3.01$, but the spectrum
    prediction is similar for all fits. The fits are shown in
    [Fig. \ref{fig:X_4350_n}] along with the predicted spectrum.

    \begin{longtable}[c]{|c|c|c|c|c|c|c|}
  \hline
  \thead{$\chi^{2}_{r}$} & \thead{$\chi^{2}_{hish}$} & \thead{$\alpha^{'}$} & \thead{$a$} & \thead{$m_{1}$} & \thead{$m_{2}$} & \thead{Fit Type} \\
  \hline
  \hline
0.00 & 0.00 & 0.44 & -5.92 & 0.73 & 0.73 & free fit \\
\hline
0.79 & 0.01 & 0.45 & -2.31 & 1.47 & 1.47 & fixed $m_{1}$, $m_{2}$ and $\alpha^{'}$ \\
\hline
0.00 & 0.00 & 0.45 & -5.69 & 0.80 & 0.80 & fixed $\alpha^{'}$ \\
\hline
0.00 & 0.00 & 0.57 & -3.07 & 1.47 & 1.47 & fixed $m_{1}$ and $m_{2}$ \\
\hline
  \caption{}
 \end{longtable}

    \begin{longtable}[c]{|c|c|c|c|c|c|}
  \hline
  \thead{$State$} & \thead{$M$ \\ $[MeV]$} & \thead{Thry \\ $M$ \\ $[MeV]$} & \thead{$J$} & \thead{Thry \\ $n$} & \thead{Thry \\ $J + n$} \\
  \hline
  \hline
$X(4350)$ & $4351$ & $4351$ & $1$ & $-1$ & $0$ \\
\hline
$X(4630)$ & $4626$ & $4626$ & $1$ & $0$ & $1$ \\
\hline
 &  & $4883$ &  &  & $2$ \\
\hline
 &  & $5126$ &  &  & $3$ \\
\hline
 &  & $5356$ &  &  & $4$ \\
\hline
 &  & $5575$ &  &  & $5$ \\
\hline
  \caption{$X(4350)$, $X(4630)$ $(n, M^{2})$ HMRT.} \label{table:X_4350_HMRT_n}
 \end{longtable}

    \twographs[$X(4350)$ and $X(4630)$ with $J^{PC}=1^{-+}$ HMRT. (a) Comparison between the fits, (b) $X(4350)$ and $X(4630)$ fixed slope fit HMRT which is the best fit.][X_4350_n]{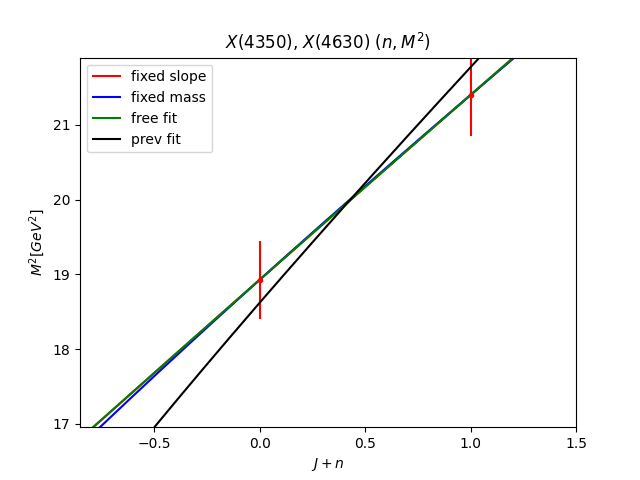}{tetraquarks/X_4350_/fit-_X_1j_X_4350_fixed_slope_n}
}

\paragraph{$\psi$ $(n, M^{2})$}{
    The candidates $\psi(4360)$ and $\psi(4660)$ were analysed in previous
    paper \cite{Sonnenschein_2017_4630} as candidates on the same trajectory.
    In this paper, the $\psi(4660)$ and $X(4630)$ were considered as the same
    state. But it is now known that they have different quantum numbers.
    Hence, we separate their HMRTs in this paper. Also, we assumed that $X
    (4350)$ is on the same HMRT as the $X(4630)$ (section
    \ref{section:X_4630_HMRT}).

    For $J^{PC}=1^{--}$ the best fit is with the free parameters with $\chi^{2}_{r}=3.01$, but the spectrum prediction is similar for all fits. The fits are shown in [Fig. \ref{fig:X_4350_n}] along with the predicted spectrum.
    
    \begin{longtable}[c]{|c|c|c|c|c|c|c|}
  \hline
  \thead{$\chi^{2}_{r}$} & \thead{$\chi^{2}_{hish}$} & \thead{$\alpha^{'}$} & \thead{$a$} & \thead{$m_{1}$} & \thead{$m_{2}$} & \thead{Fit Type} \\
  \hline
  \hline
0.00 & 0.00 & 0.53 & -3.68 & 1.16 & 1.16 & free fit \\
\hline
1.16 & 0.01 & 0.45 & -1.35 & 1.47 & 1.47 & fixed $m_{1}$, $m_{2}$ and $\alpha^{'}$ \\
\hline
0.00 & 0.00 & 0.45 & -6.24 & 0.47 & 0.47 & fixed $\alpha^{'}$ \\
\hline
0.03 & 0.00 & 0.57 & -2.12 & 1.47 & 1.47 & fixed $m_{1}$ and $m_{2}$ \\
\hline
  \caption{}
 \end{longtable}

    \begin{longtable}[c]{|c|c|c|c|c|c|}
  \hline
  \thead{$State$} & \thead{$M$ \\ $[MeV]$} & \thead{Thry \\ $M$ \\ $[MeV]$} & \thead{$J$} & \thead{Thry \\ $n$} & \thead{Thry \\ $J + n$} \\
  \hline
  \hline
$\psi(4360)$ & $4372$ & $4372$ & $1$ & $0$ & $1$ \\
\hline
$\psi(4660)$ & $4630$ & $4630$ & $1$ & $1$ & $2$ \\
\hline
 &  & $4873$ &  &  & $3$ \\
\hline
 &  & $5105$ &  &  & $4$ \\
\hline
 &  & $5325$ &  &  & $5$ \\
\hline
 &  & $5537$ &  &  & $6$ \\
\hline
  \caption{$\psi$ $(n, M^{2})$ HMRT.}
 \end{longtable}

    \twographs[$X(4350)$ and $X(4630)$ with $J^{PC}=1^{-+}$ HMRT. (a) Comparison between the fits, (b) $X(4350)$ and $X(4630)$ fixed slope fit HMRT which is the best fit.][psi_n]{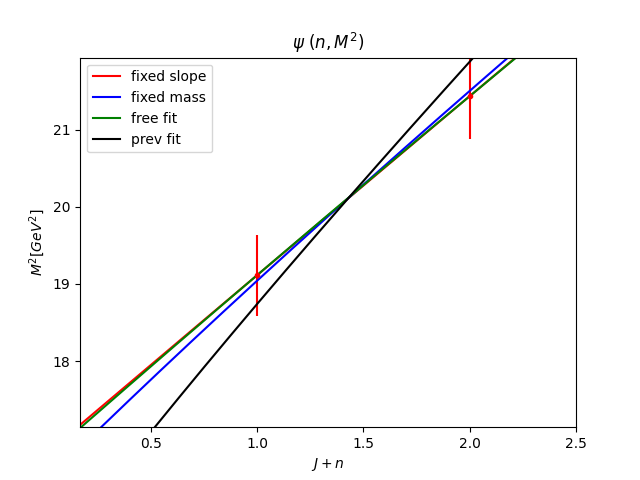}{tetraquarks/psi/by_n/fit-_psi_1j_psi_fixed_slope_n}
}

\subsubsection{\texorpdfstring{$T_{\psi\psi}$ (quark content $c\bar{c}c\bar{c}$)}{}}

In \cite{Sonnenschein_2021_6900} the state $X(6900)$ was analysed in the context of the HISH model. This note relied on \cite{LHCb_X(6900)_2020} where two additional resonance candidates at around $6.6GeV$ and $7.2GeV$ were observed. Now these candidates were observed in an additional note \cite{LHCb_X(6900)_2022} with higher statistical significance. The states that we analyse in this section are listed in \ref{table:tetra_cand_cccc}.

There are two measurements of $X(6600)$ and $X(6900)$, without quantum numbers. Not all of the states rely on the same trajectory. Hence, we fit the second, fourth and fifth  in an HMRT.

\begin{longtable}[c]{|c|c|c|c|c|c|c|}
\hline
\makecell{Candidate} & \makecell{Quarks \\ Content} & \makecell{Selected \\ Decay \\ Channels} & \makecell{$J^{PC}$} & \makecell{Mass \\ $[MeV]$} & \makecell{Width \\ $[MeV]$} & \makecell{Ref.} \\
\hline
\hline
$T_{\psi\psi}(6600)$ & $c\bar{c}c\bar{c}$ & \makecell{$J/\psi J/\psi$} & $?^{??}$ & $6630\pm{90}$ & $350\pm{11}^{+110}_{-40}$ & \makecell{} \\
$T_{\psi\psi}(6600)$ & $c\bar{c}c\bar{c}$ & \makecell{$J/\psi J/\psi$} & $?^{??}$ & $6552\pm{10}\pm{12}$ & $124^{+32}_{-26}\pm{33}$ & \makecell{} \\
$T_{\psi\psi}(6900)$ & $c\bar{c}c\bar{c}$ & \makecell{$J/\psi J/\psi$} & $?^{??}$ & $6886\pm{16}$ & $168\pm{80}$ & \makecell{} \\
$T_{\psi\psi}(6900)$ & $c\bar{c}c\bar{c}$ & \makecell{$J/\psi J/\psi$} & $?^{??}$ & $6927\pm{9}\pm{4}$ & $122^{+24}_{-21}\pm{18}$ & \makecell{} \\
$T_{\psi\psi}(7300)$ & $c\bar{c}c\bar{c}$ & \makecell{$J/\psi J/\psi$} & $?^{??}$ & $7287^{+20}_{-18}\pm{5}$ & $95^{+59}_{-40}\pm{19}$ & \makecell{} \\
\hline
\caption{$T_{c\bar{c}c\bar{c}}$ tetraquark candidates.}\label{table:tetra_cand_cccc}
\end{longtable}

\begin{longtable}[c]{|c|c|c|c|c|c|}
  \hline
  \thead{Thry \\ Width \\ $[MeV]$} & \thead{Mesons \\ Pair} & \thead{Meson \\ Threshold \\ $[MeV]$} & \thead{Baryon \\ Anti-\\baryon} & \thead{Baryonic \\ Threshold \\ $[MeV]$} & \thead{Genuine} \\
  \hline
  \hline
\makecell{$6-21$\\$29-58$} & \makecell{$\eta_{c}(1S) \eta_{c}(1S)$} & \makecell{$5968$} & \makecell{$\Xi^{+}_{cc} \bar{\Xi}^{-}_{cc}$ \\ $\Xi^{++}_{cc} \bar{\Xi}^{--}_{cc}$ \\ $\Omega^{+}_{cc} \bar{\Omega}^{-}_{cc}$} & \makecell{$7038$ \\ $7244$ \\ } & $*$ \\
\hline
  \caption{$T_{c\bar{c}c\bar{c}}$ thresholds.} \label{table:Tcccc_thresholds}
 \end{longtable}

    \begin{center}
    \begin{minipage}{.49\linewidth}
        \begin{center}
            \begin{longtable}[c]{|c|c|}
  \hline
  \thead{$J$ \\ Spec} & \thead{$a_{T_{j}}$} \\
  \hline
  \hline
\makecell{$6768-6887$\\$6926-7040$\\$7078-7188$\\$7225-7331$\\$7366-7469$\\$7504-7604$\\$7637-7735$\\$7767-7862$\\$7894-7987$\\$8018-8108$\\$8139-8227$\\$8257-8344$} & \makecell{$-5.9$-$-5.1$} \\
\hline
  \caption{$T_{c\bar{c}c\bar{c}}$ $(J, M^2)$ predictions.} \label{table:Tcccc_j_predictions}
 \end{longtable}
        \end{center}
    \end{minipage}
    \begin{minipage}{.49\linewidth}
        \begin{center}
            \begin{longtable}[c]{|c|c|}
  \hline
  \thead{$n$ \\ Spec} & \thead{$a_{T_{n}}$} \\
  \hline
  \hline
\makecell{$6622-6745$\\$6857-6973$\\$7078-7188$\\$7288-7393$\\$7489-7589$\\$7681-7778$\\$7866-7959$\\$8045-8135$\\$8218-8305$\\$8385-8470$\\$8549-8631$\\$8707-8788$} & \makecell{$-4.2$-$-3.7$} \\
\hline
  \caption{$T_{c\bar{c}c\bar{c}}$ $(n, M^2)$ predictions.} \label{table:Tcccc_n_predictions}
 \end{longtable}
        \end{center}
    \end{minipage}
    \end{center}

\twographs[(a) is $(J, M^{2})$ $T_{c\bar{c}c\bar{c}}$ predicted HMRT, (b) $(n, M^{2})$ $T_{c\bar{c}c\bar{c}}$ predicted HMRT.][Tcccc_predictions]{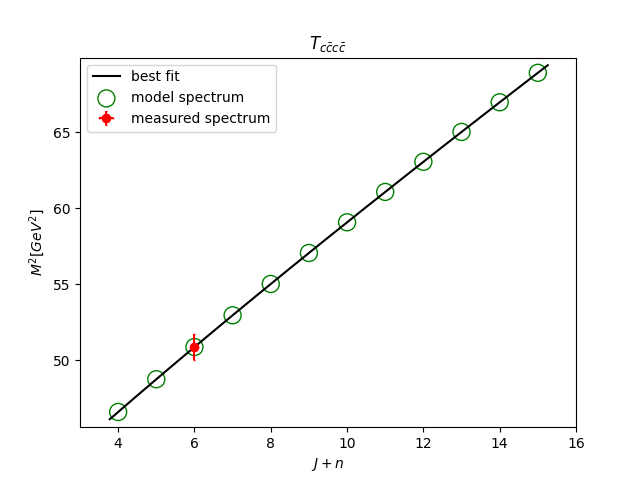}{tetraquarks/predictions/cc-cc/by_n/T_cccc_one_by_n}

The decay channel in which these states were observed
is $J/\psi J/\psi$. As before, we fit the data four times in
the $(n, M^{2})$ plane with $\alpha^{'}=0.45$. For this trajectory,
the mesons slope was too high for both planes.

The comparison between the different fits are shown in \ref{fig:T_cccc_fit}. The best fit, which is consistent with the universal description of the model is present in \ref{fig:T_cccc_fit} on (b). The spectrum prediction is shown in \ref{table:X(6900)_spectrum}.

\begin{longtable}[c]{|c|c|c|c|c|c|c|}
  \hline
  \thead{$\chi^{2}_{r}$} & \thead{$\chi^{2}_{hish}$} & \thead{$\alpha^{'}$} & \thead{$a$} & \thead{$m_{1}$} & \thead{$m_{2}$} & \thead{Fit Type} \\
  \hline
  \hline
0.12 & 0.00 & 0.21 & -4.43 & 0.70 & 0.70 & free fit \\
\hline
5.68 & 0.05 & 0.45 & -2.93 & 2.22 & 2.22 & fixed $m_{1}$, $m_{2}$ and $\alpha^{'}$ \\
\hline
3.05 & 0.01 & 0.45 & 1.00 & 2.78 & 2.78 & fixed $\alpha^{'}$ \\
\hline
0.84 & 0.00 & 0.31 & -0.86 & 2.22 & 2.22 & fixed $m_{1}$ and $m_{2}$ \\
\hline
  \caption{$T_{\psi\psi}$ HMRT.} \label{table:X(6900)_spectrum}
 \end{longtable}

\twographs[$X(6900)$ $(n, M^{2})$ HMRT. The graph in (a) shows the comparison between the fits, (b) $X(6900)$ fixed fit HMRT which is the best fit.][T_cccc_fit]{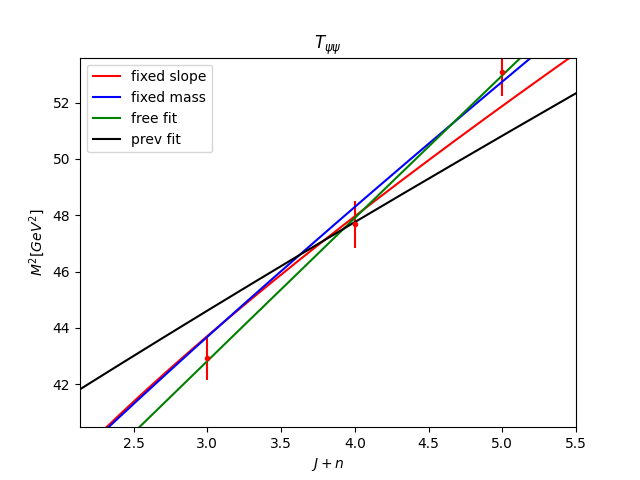}{tetraquarks/X_6900_/by_n/fit-_T_psi_bs__bs_psi_X_6900_by_n_fixed_m_n}

\clearpage

\begin{longtable}[c]{|c|c|c|c|c|c|}
  \hline
  \thead{$State$} & \thead{$M$ \\ $[MeV]$} & \thead{Thry \\ $M$ \\ $[MeV]$} & \thead{$J$} & \thead{Thry \\ $n$} & \thead{Thry \\ $J + n$} \\
  \hline
  \hline
 &  & $5800$ &  &  & $1$ \\
\hline
 &  & $6232$ &  &  & $2$ \\
\hline
$T(6600)_{\psi\psi0}$ & $6630$ & $6609$ &  &  & $3$ \\
\hline
$T(6900)_{\psi\psi0}$ & $6905$ & $6950$ &  &  & $4$ \\
\hline
$T(7300)_{\psi\psi0}$ & $7287$ & $7263$ &  &  & $5$ \\
\hline
 &  & $7555$ &  &  & $6$ \\
\hline
 &  & $7830$ &  &  & $7$ \\
\hline
  \caption{$T_{\psi\psi}$ HMRT.}
 \end{longtable}

The decays width ratio:

\begin{align}
    \Gamma_{tear}\lesssim 31MeV
\end{align}

\begin{align}
    \Gamma_{annihilation}=47MeV
\end{align}

Adding the two widths, we get a result width - $78MeV$ - which is a quite good approximation to the measured $95MeV$ in \cite{LHCb_X(6900)_2022}. Changing the $A$ value used to $0.15$ would give a closer result.

\subsection{Tetraquarks and Mesons Global Fit}\label{section:tetra_mesons_global}

To conclude this section, we checked the coherency of our assumptions by fitting
all tetraquarks and mesons data together, constraining the slopes of the heavy
tetraquarks to be the same as that of the charmed and bottoms mesons.

The results are of $\chi_{r}^{2} = 1.95$. The graphs are in the following figure
\ref{fig:mesons_tetra_global_fit}, the fit results are in table
\ref{table:mesons_tetra_global_fit}.

\begin{longtable}[c]{ m{8cm} m{8cm} }
\twographsintable{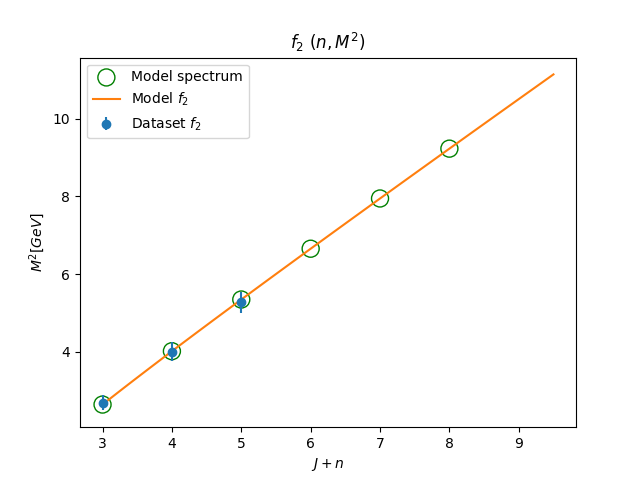}{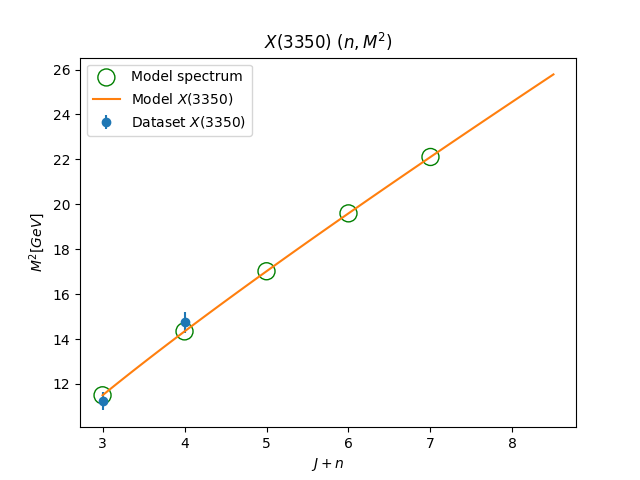}
\twographsintable{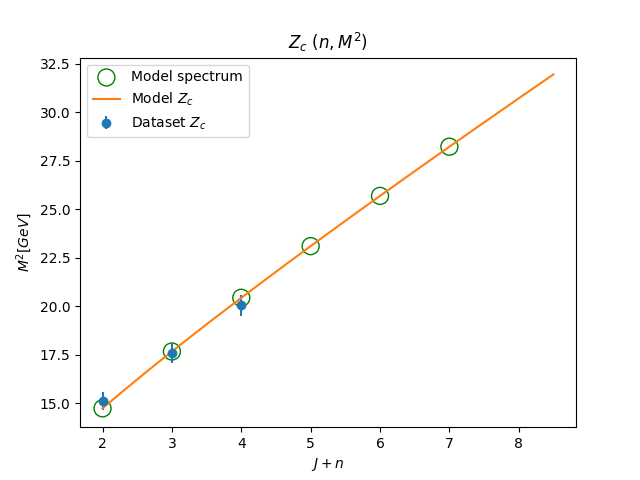}{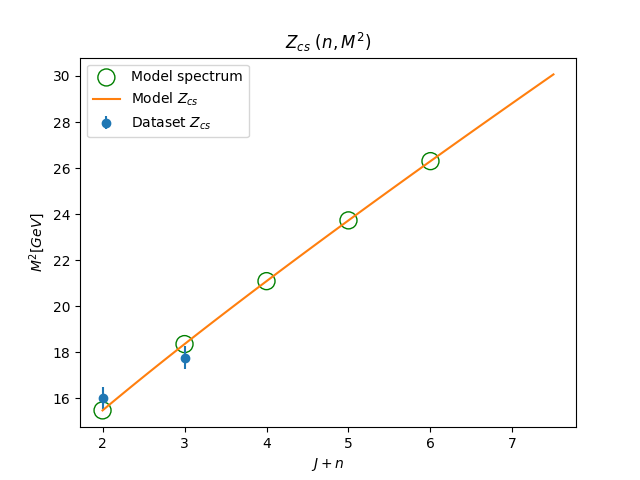}
\twographsintable{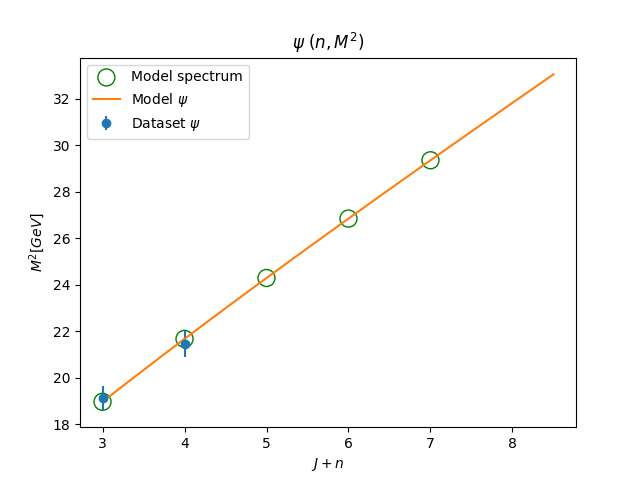}{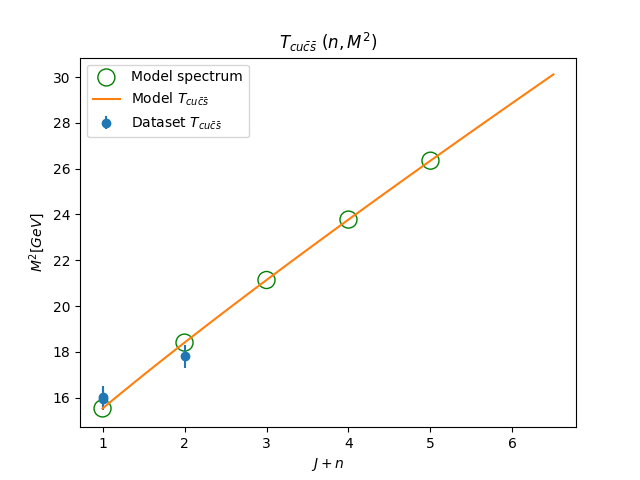}
\twographsintable{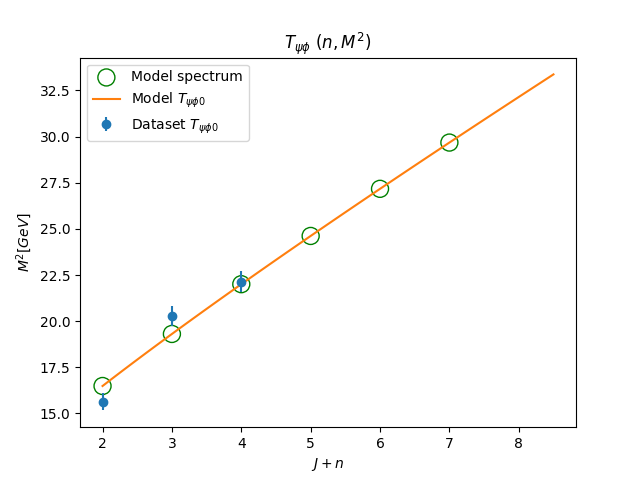}{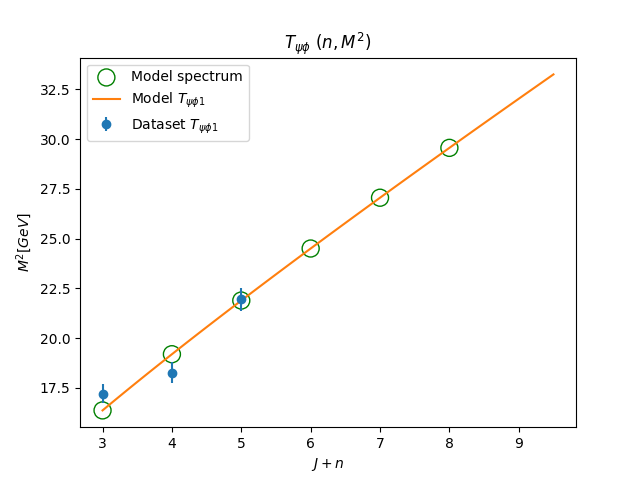}
\twographsintable{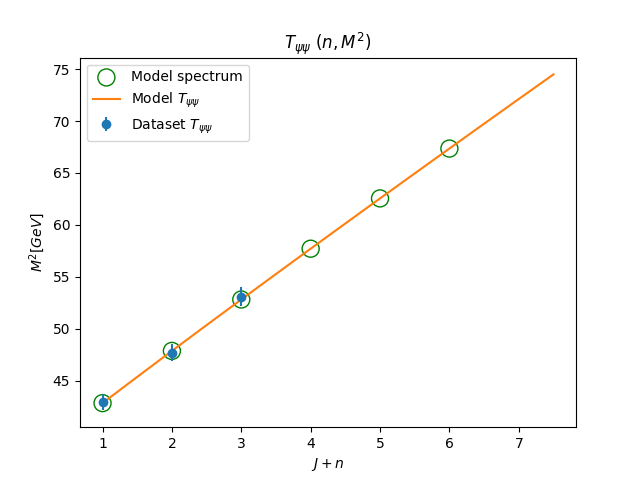}{tetraquarks/global/fit_34}
\twographsintable{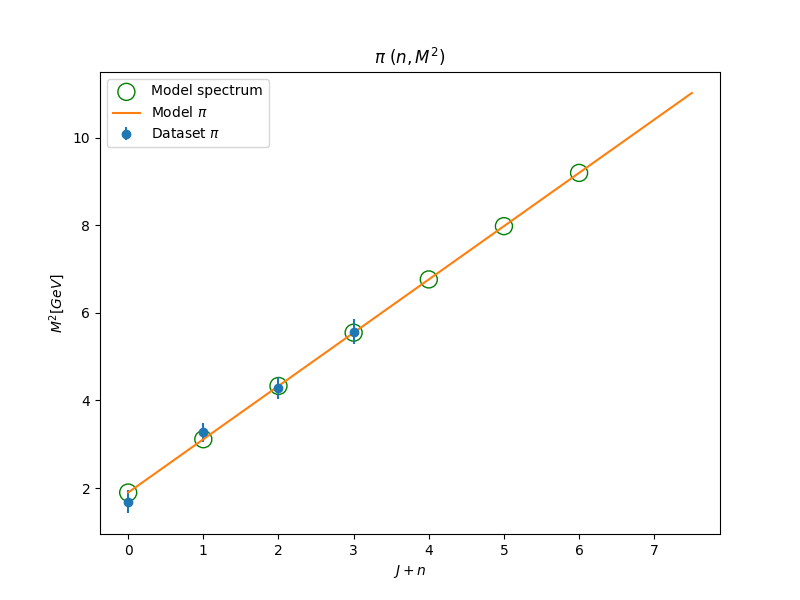}{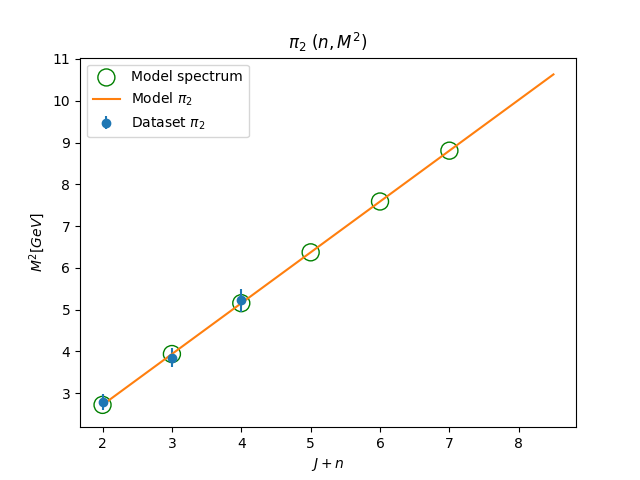}
\twographsintable{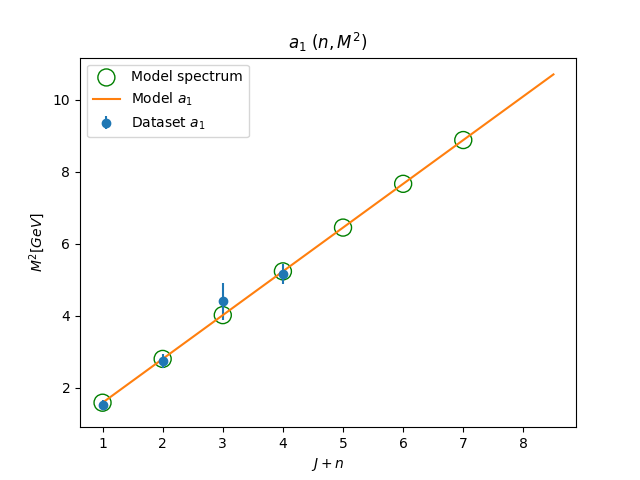}{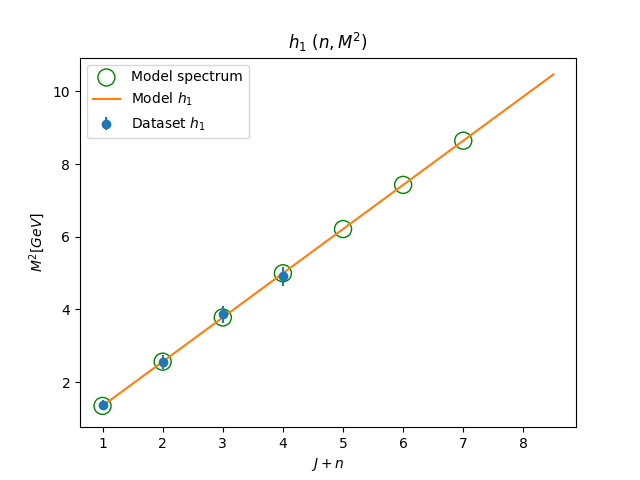}
\twographsintable{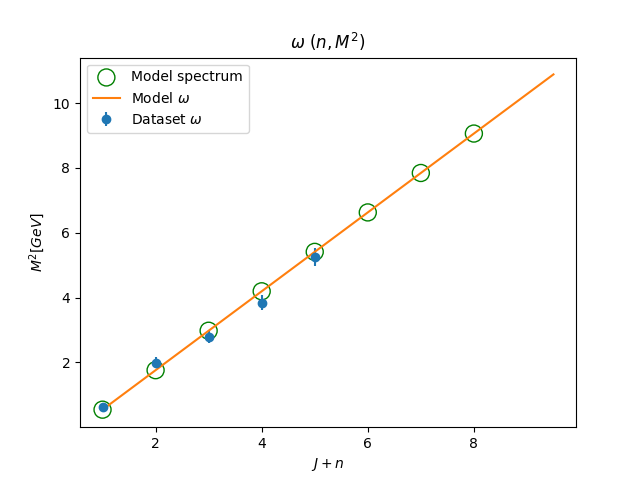}{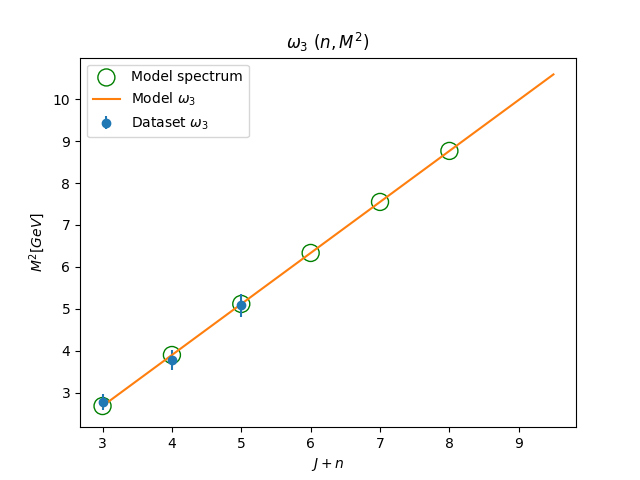}
\twographsintable{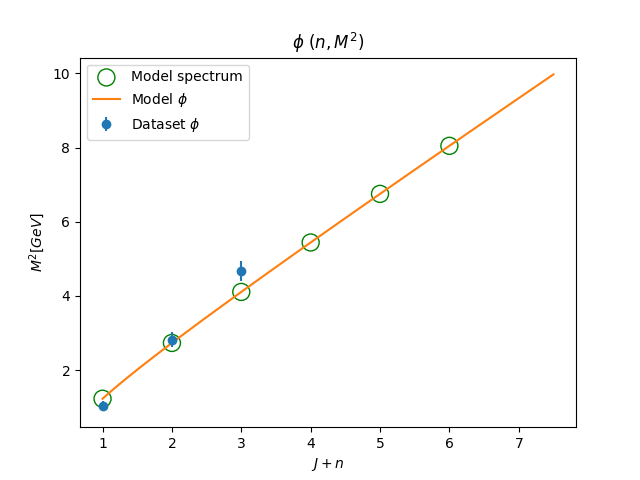}{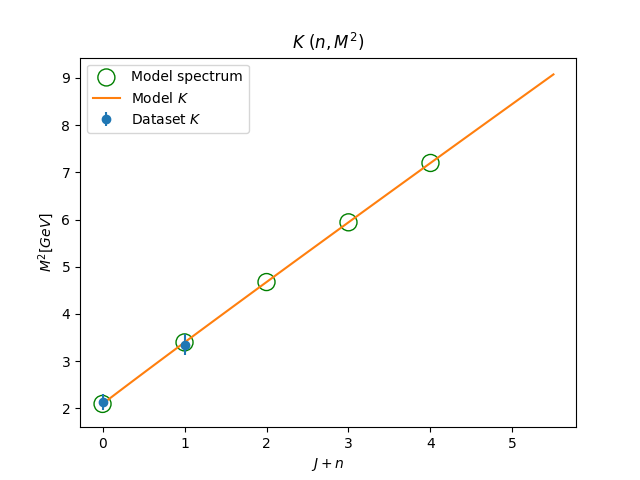}
\twographsintable{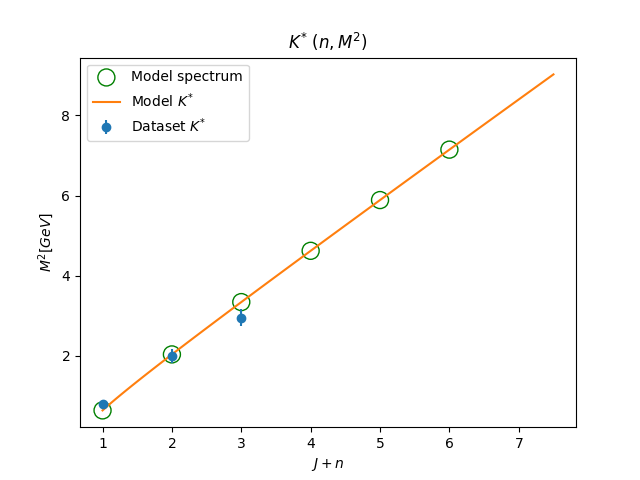}{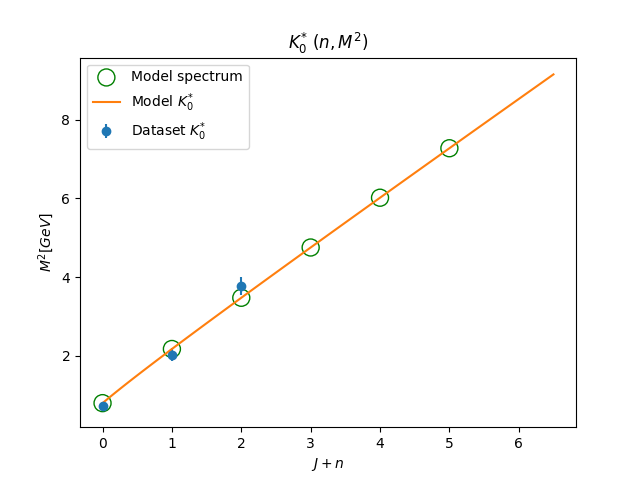}
\twographsintable{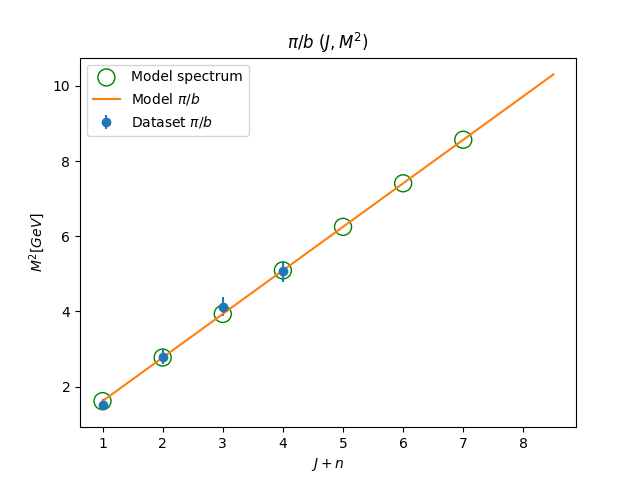}{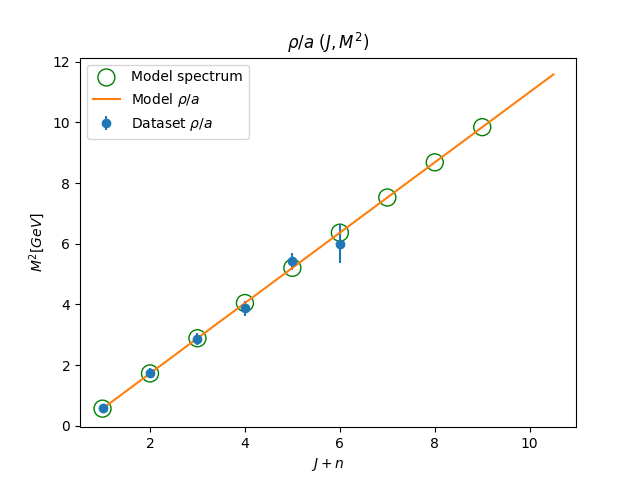}
\twographsintable{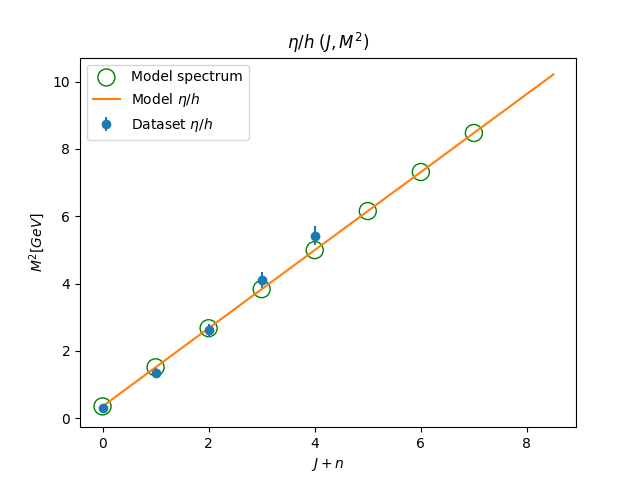}{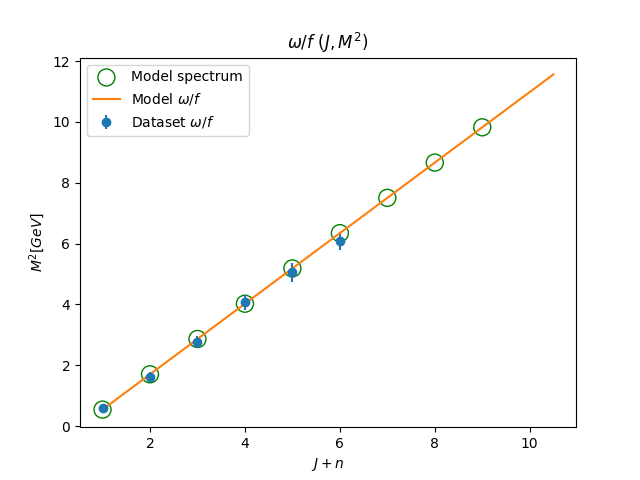}
\twographsintable{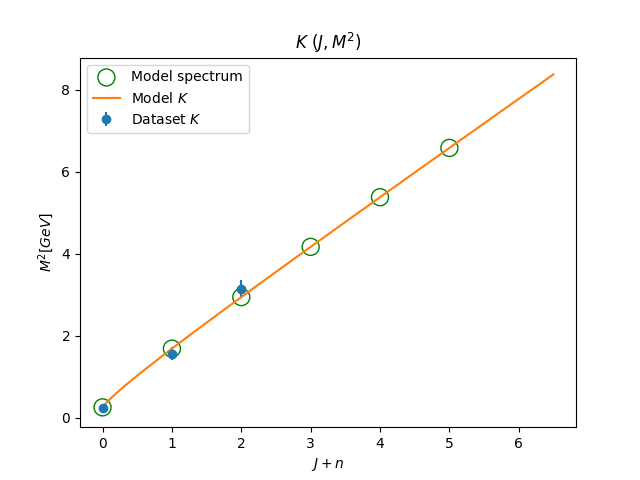}{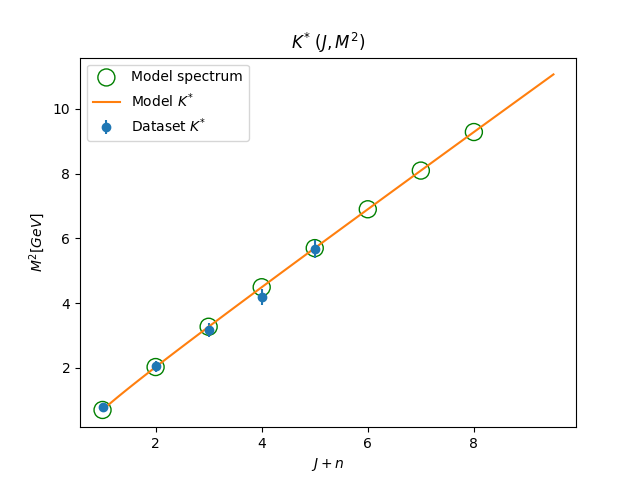}
\twographsintable{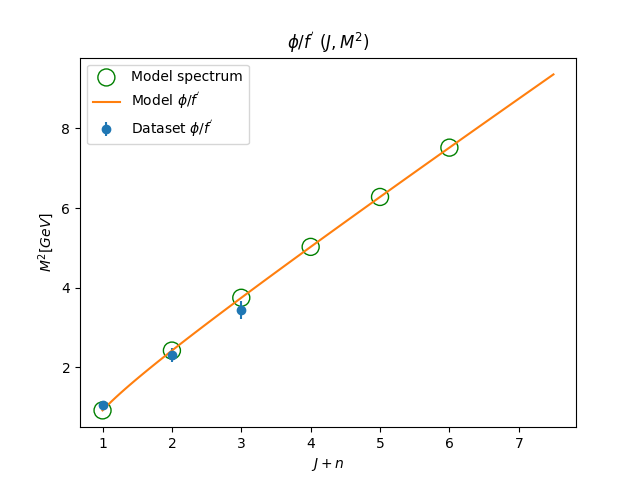}{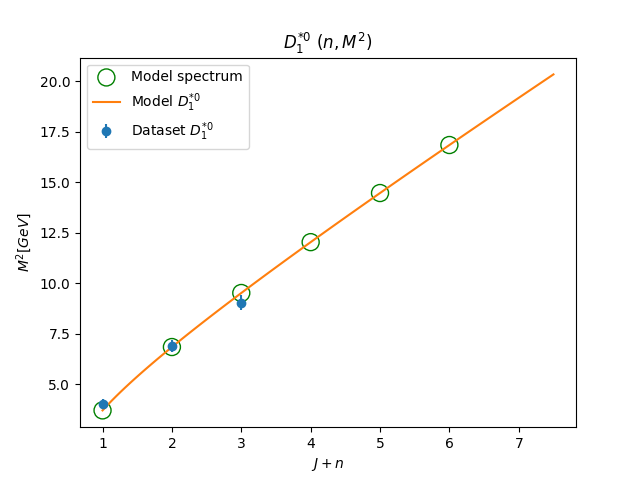}
\twographsintable{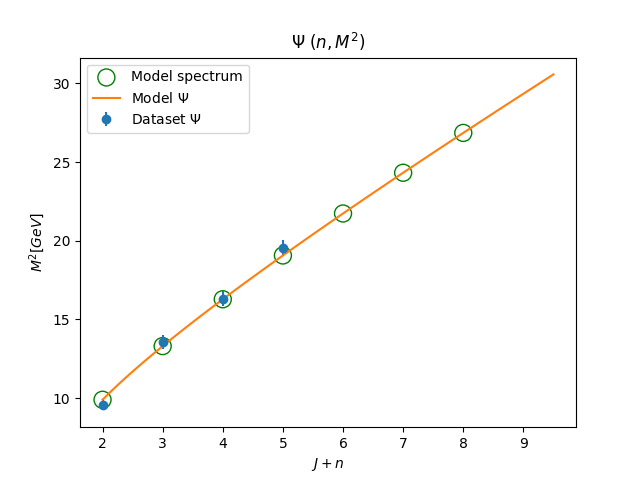}{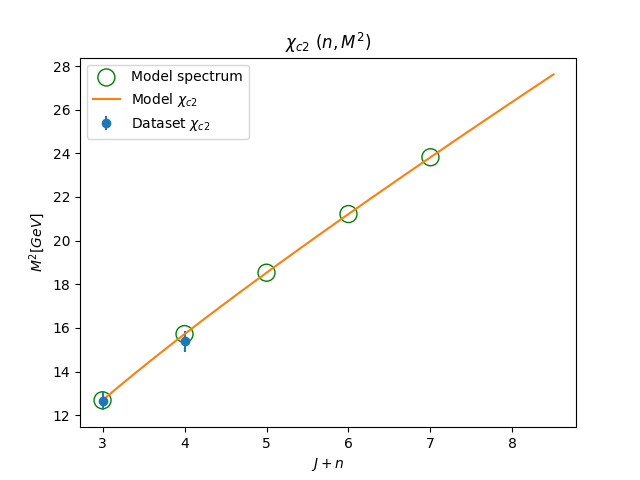}
\twographsintable{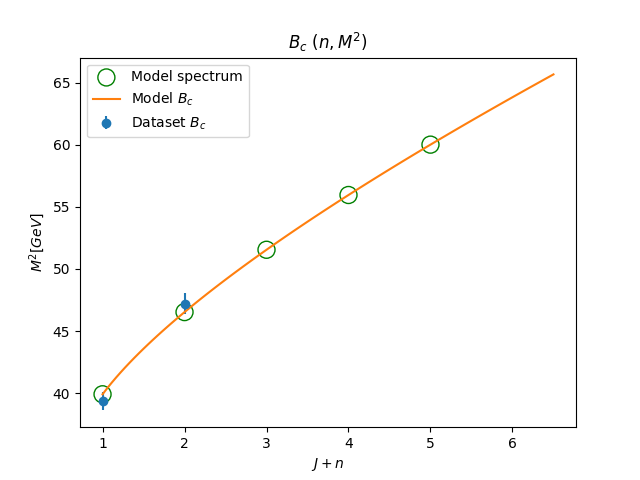}{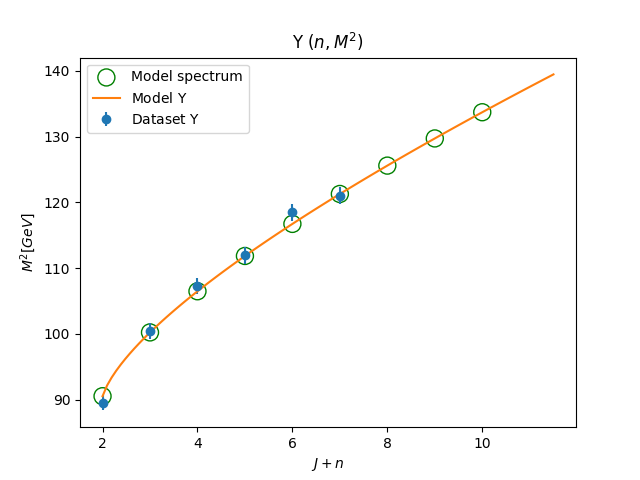}
\twographsintable{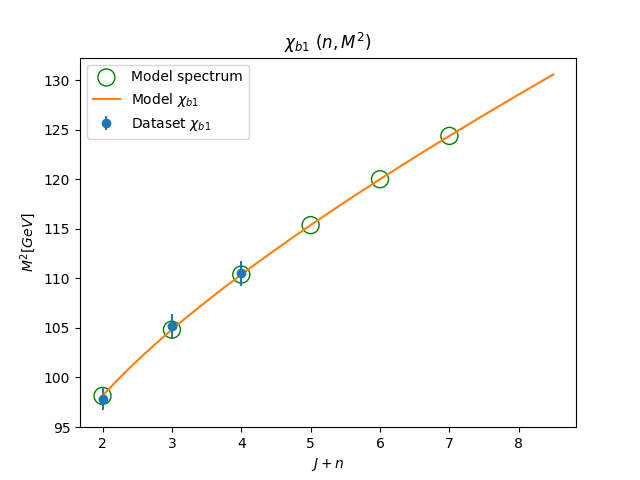}{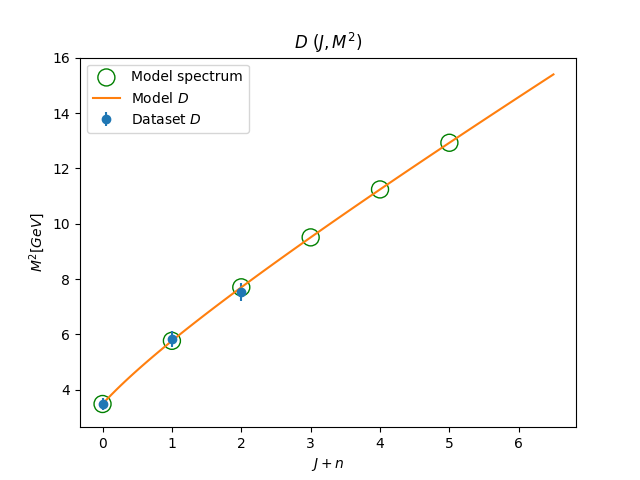}
\twographsintable{tetraquarks/global/fit_22}{tetraquarks/global/fit_23}
\twographsintable{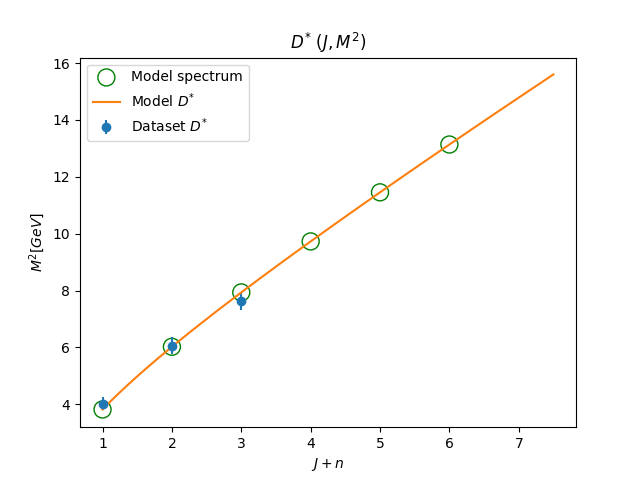}{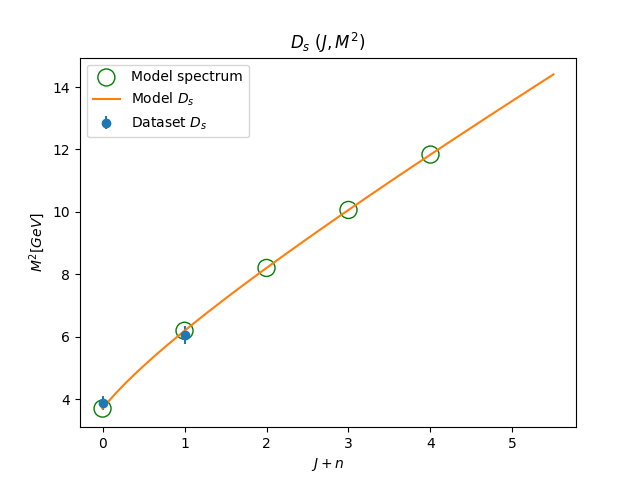}
\twographsintable{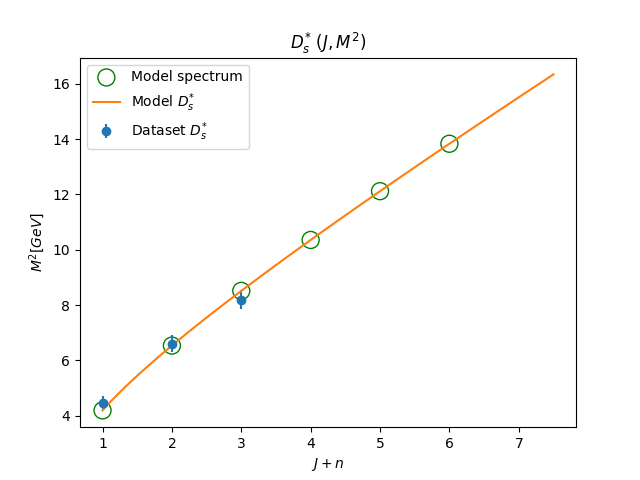}{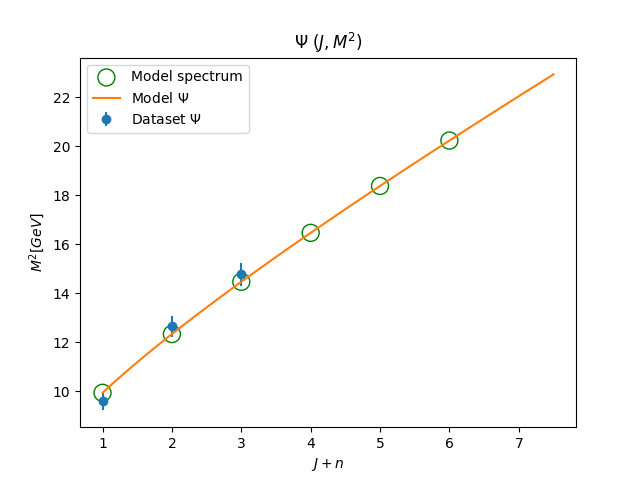}
\twographsintable{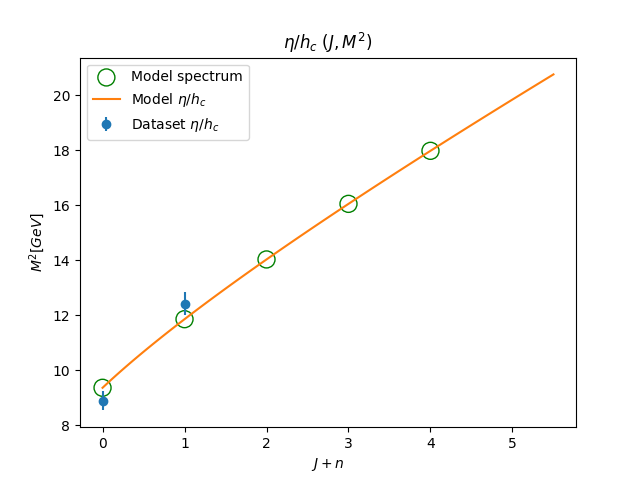}{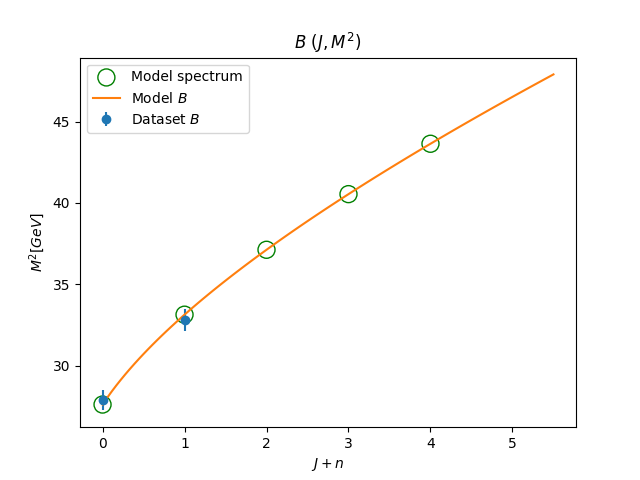}
\twographsintable{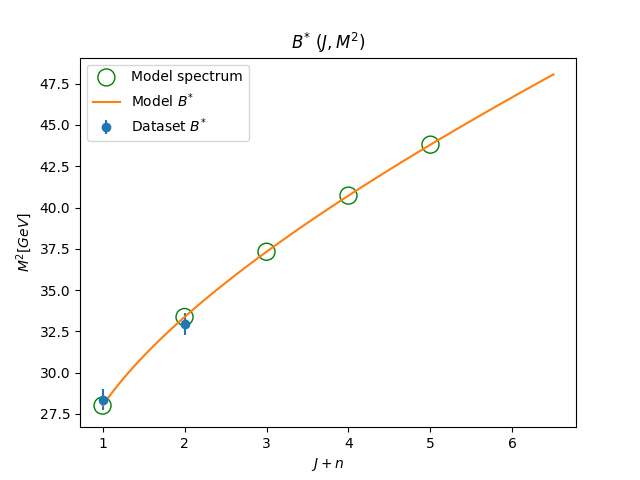}{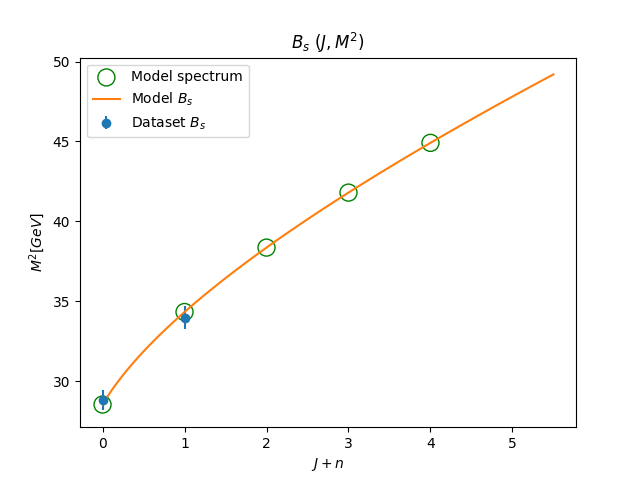}
\twographsintable{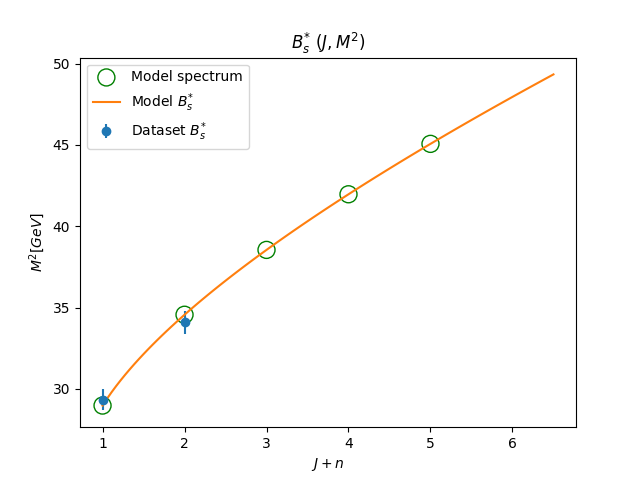}{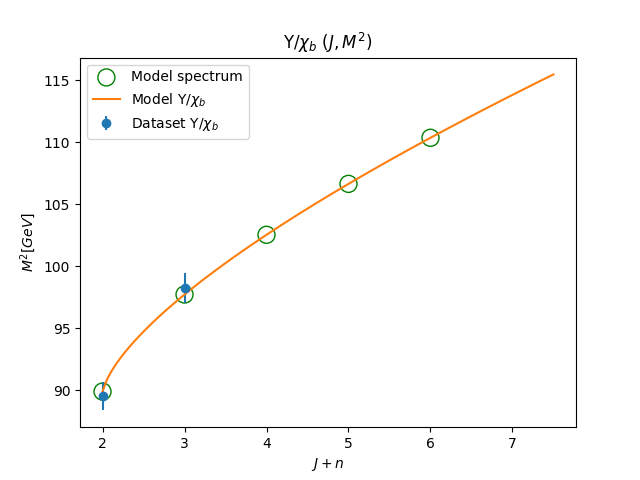}
\centering
\includegraphics[width=0.48\textwidth]{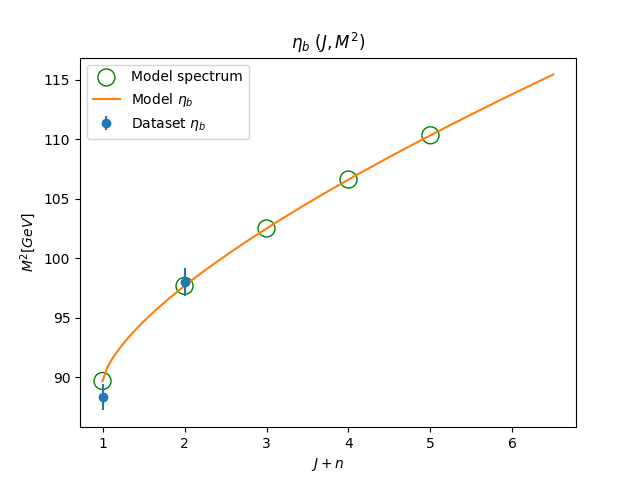}
\label{fig:mesons_tetra_global_fit}
\end{longtable}

\begin{longtable}[c]{|c|c|c|c|c|c|c|c|c|c|c|}
    \hline
    \makecell{$\chi^{2}_{r}$} & \makecell{$\alpha^{'}_{J,L}$\\$[GeV^{-2}]$} & \makecell{$\alpha^{'}_{J,H}$\\$[GeV^{-2}]$} & \makecell{$\alpha^{'}_{n,L}$\\$[GeV^{-2}]$} & \makecell{$\alpha^{'}_{n,H}$\\$[GeV^{-2}]$} & \makecell{$m_{u/d}$\\$[GeV]$} & \makecell{$m_{s}$\\$[GeV]$} & \\
    \hline
    \hline
    $1.95$&$0.86$&$0.65$&$0.83$&$0.48$&$0.011$&$0.36$ \\
    \hline
    \makecell{$m_{c}$\\$[GeV]$} & \makecell{$m_{b}$\\$[GeV]$} & \makecell{$\alpha_{T_{\psi\psi}}$} & \makecell{$m_{BV_{u/d,u/d}}$} & \makecell{$m_{BV_{s,u/d}}$} & \makecell{$m_{BV_{c,u/d}}$} & \makecell{$m_{BV_{c,s}}$} & \makecell{$m_{BV_{c,c}}$} \\
    \hline
    \hline
    $1.27$&$4.73$&$0.23$&$0.93$&$0.35$&$1.32$&$1.31$&$1.53$ \\
    \hline
    \caption{The parameters results from the tetraquarks and mesons global fit. We constraint the heavy slopes to equal in mesons and tetraquarks, aside from the $T_{c\bar{c}c\bar{c}}$, which clearly have a lower $\alpha^{'}$.}
    \label{table:mesons_tetra_global_fit}
\end{longtable}

\begin{longtable}[c]{|c|c|c|c|c|c|}
\hline
\makecell{Traj.} & \makecell{$a$\\$n/J$} & \makecell{Traj.} & \makecell{$a$\\$n/J$ \\ only} & \makecell{Traj.} & \makecell{$a$\\$n/J$} \\
\hline
\hline
$\pi/b$&$a_{J}=-0.38$&$\eta_{c}/h_{c}$&$a_{J}=-0.71$&$\omega_{3}$&$a_{n}=0.80$\\
\hline
$\rho/a$&$a_{J}=0.52$&$B$&$a_{J}=-0.36$&$\phi$&$a_{n}=0.67$\\
\hline
$\eta/h$&$a_{J}=-0.30$&$B^{*}$&$a_{J}=0.58$&$K$&$a_{n}=-1.29$\\
\hline
$\omega/f$&$a_{J}=0.53$&$B_{s}$&$a_{J}=-0.19$&$K^{*}$&$a_{n}=0.75$\\
\hline
$K$&$a_{J}=-0.03$&$B_{s}^{*}$&$a_{J}=0.75$&$K^{*}_{0}$&$a_{n}=-0.35$\\
\hline
$K^{*}$&$a_{J}=0.70$&$\Upsilon/\chi_{b}$&$a_{J}=1.97$&$D^{*0}_{1}$&$a_{n}=0.65$\\
\hline
$\phi/f^{'}$&$a_{J}=0.84$&$\eta_{b}$&$a_{J}=1.00$&$\Psi$&$a_{n}=1.38$\\
\hline
$D$&$a_{J}=-0.42$&$\pi$&$a_{n}=-1.56$&$B_{c}$&$a_{n}=0.72$\\
\hline
$D^{*}$&$a_{J}=0.45$&$\pi_{2}$&$a_{n}=-0.23$&$\chi_{c2}$&$a_{n}=1.58$\\
\hline
$D_{s}$&$a_{J}=-0.21$&$a_{1}$&$a_{n}=-0.29$&$\Upsilon$&$a_{n}=1.98$\\
\hline
$D^{*}_{s}$&$a_{J}=0.62$&$h_{1}$&$a_{n}=-0.10$&$\chi_{b1}$&$a_{n}=1.26$\\
\hline
$\Psi/\chi_{c}$&$a_{J}=0.08$&$\omega$&$a_{n}=0.55$&$f_{2}$&$a_{n}=1.72$\\
\hline
$X(3350)$&$a_{n}=1.35$&$Z_{c}$&$a_{n}=0.06$&$Z_{cs}$&$a_{n}=-0.18$\\
\hline
$\Psi(4630)$&$a_{n}=-0.41$&$T_{cu\bar{c}\bar{s}}$&$a_{n}=0.06$&$T_{\psi\phi0}$&$a_{n}=-0.54$\\
\hline
$T_{\psi\phi1}$&$a_{n}=-4.47$&$T_{\psi\psi}$&$a_{n}=1.53$&&\\
\hline
\caption{Mesons and tetraquarks global fits intercepts.
    $a_{J}$ marks that the HMRT was done in the $(J,M^{2})$ plane. $a_{n}$ is
    the intercept of a $(n, M^{2})$ HMRT.}\label{table:mesons_tetra_global_intercepts}
\end{longtable}

\subsection{Tetraquarks HMRTs predictions}\label{TetraquarksHMRTs}
The following table summarizes all the thresholds and predicted trajectories for all possible charm tetraquarks configurations. The thresholds were calculated for each possible product according to the HISH model through the annihilation and string tear mechanisms. The third mechanism that is mentioned in \ref{Decay mechanisms of tetraquarks} is not calculated in the table.

The table predictions were calculated based on the mesons fit parameters.
We assumed that the diquarks (and anti-diquarks) masses are the sum of masses of the two quarks.
The slopes for each trajectory ($(J, M^{2})$ and $(n, M^{2})$), was taken to be the heavy-slope from \ref{table:mesons_global_fit}\label{paragraph:tetra_pred_params}.

The 'Candidates' column contains the resonances that were already detected.
We elaborate on all of these at the previous sub sections,
including the decay channels through which they were observed.
The 'Thry Width' column contains two values, one for the states at the
$(J,M^{2})$ MHRT, and the second row is for the $(n,M^{2})$ MHRT.
The 'Mesons Pair' and 'Meson Threshold' contains the mesons symbols and the
threshold for their creation respectively (for example, in the first row of
the table, $D^{0}\pi^{0}_{d}$ has a threshold of 2000, while $D^{+}\pi^{-}$
has threshold of 2010). The same applies for the 'Baryon-Antibaryon' and
'Baryonic Threshold' columns.

We calculated the 'J Spec' and 'n Spec' in the following steps:

\begin{itemize}
    \item We first calculated the baryonic threshold and let it vary between $40-150MeV$. This was done by looking at the available data that we collected, where the created tetraquarks candidates are above the baryonic threshold. We saw that this is the range of the distance from the threshold.
    \item The value for $J+n$ is calculated according to the number of states that are potentially on the trajectories between the mesonic and the baryonic thresholds. We calculated how many states are there, assuming the other parameters are similar to the mesons as explained above \ref{paragraph:tetra_pred_params}.
    \item We calculated the range of the baryonic spectrum accordingly.
    \item We derived the range of the spectrum for two states above, and two states below the expected $J+n$ of the baryonic threshold for each of the two spectra.
    For example, in the first row the in the 'J Spec' column, the $3264-3374$ is the range of the spectrum of the first state that we expect to decay to baryon-antibaryon pair. The $3526-3626$ is the first state above that should still decay to a baryonic pair and should have $J+n=4$, whereas the $2972-3095$ should decay through annihilation only, and should have $J+n=2$.
    \item When we weren't able to calculate the baryonic threshold (since the baryons weren't measured yet), we started from the mesonic threshold and calculated the trajectory from there up to 12 states.
\end{itemize}

The 'Genuine' column indicates whether the quark content is a {\it genuine} tetraquark or a {\it not obviously genuine}. This was decided as follows:

\begin{itemize}
    \item If the content includes a light quark ($u$, $d$, $s$) and its counterpart, we assume its source could be a meson that decay through string tear. Thus this is  {\it not obviously genuine} case. This is marked at the table as an empty cell.
    \item If the content include a heavy quark ($c$, $b$) and its counterpart, which is less likely to be a product of meson string break, it is likely to be genuine. This was marked with one $*$.
    \item $**$ means it is a pure tetraquark. These do not contain quark-antiquark pair of the same flavour.
\end{itemize}

Empty baryonic threshold means these baryons weren't found yet.

\setlength\LTleft{-1.4cm}
\setlength\LTpre{0.5cm}
{\scriptsize\tabcolsep=1pt{
\begin{longtable}[!c]{|c|c|c|c|c|c|c|c|c|c|c|c|c|c|}
  \hline
  \thead{$\#$} & \thead{Candi-\\dates} & \thead{$J^{PC}$} & \thead{Quarks} & \thead{Thry \\ Width \\ $[MeV]$} & \thead{Mesons \\ Pair} & \thead{Meson \\ Threshold \\ $[MeV]$} & \thead{Baryon \\ Anti-\\baryon} & \thead{Baryonic \\ Threshold \\ $[MeV]$} & \thead{$J$ \\ Spec} & \thead{$n$ \\ Spec} & \thead{$a_{T_{j}}$ \\ $a_{T_{n}}$} & \thead{Thry \\ $J + n$} & \thead{Gen-\\uine} \\
  \hline
  \hline
1 & \makecell{} & \makecell{} & $c\bar{u}u\bar{u}$ & \makecell{$79-118$\\$30-61$} & \makecell{$D^{0} \pi^{0}_{u}$} & \makecell{$2000$} & \makecell{$\Lambda^{+}_{c} \bar{p}$ \\ $\Xi^{+}_{c} \bar{\Sigma}^{-}$ \\ $\Sigma^{++}_{c} \bar{\Delta}^{--}$} & \makecell{$3224$ \\ $3657$ \\ $3686$} & \makecell{$2634-2779$\\$2972-3095$\\$3264-3374$\\$3526-3626$\\$3766-3858$} & \makecell{$2270-2448$\\$2829-2961$\\$3264-3374$\\$3635-3731$\\$3964-4051$} & \makecell{\makecell{$-1.6$-$-1.2$} \\ \makecell{$-1.2$-$-0.92$}} & \makecell{$3$ \\ $2$} &  \\
\hline
2 & \makecell{$X(3500)$\\$X(3800)$} & \makecell{(${0}$ or ${1}$)$^{??}$\\$?^{??}$} & $c\bar{d}d\bar{u}$ & \makecell{$79-118$\\$30-61$} & \makecell{$D^{0} \pi^{0}_{d}$ \\ $D^{+} \pi^{-}$} & \makecell{$2000$ \\ $2010$} & \makecell{$\Lambda^{+}_{c} \bar{p}$ \\ $\Sigma^{0}_{c} \bar{n}$ \\ $\Xi^{0}_{c} \bar{\Lambda}^{0}$} & \makecell{$3224$ \\ $3394$ \\ $3586$} & \makecell{$2634-2779$\\$2972-3095$\\$3264-3374$\\$3526-3626$\\$3766-3858$} & \makecell{$2270-2448$\\$2829-2961$\\$3264-3374$\\$3635-3731$\\$3964-4051$} & \makecell{\makecell{$-1.6$-$-1.2$} \\ \makecell{$-1.2$-$-0.92$}} & \makecell{$3$ \\ $2$} &  \\
\end{longtable}}\normalsize}

\subsubsection{Detailed Predictions for Genuine and Semi-Genuine Tetraquarks}

In the following sections we present the graphs and extended predicted spectra of each of the genuine and semi-genuine tetraquarks configurations that haven't been analysed yet in section \ref{section:tetra_cands}.

\subsubsection{\texorpdfstring{$T_{c\bar{u}d\bar{u}}$}{}}\label{T_cduu}

\twographs[(a) is $(J, M^{2})$ $T_{c\bar{u}d\bar{u}}$ HMRT, (b) $(n, M^{2})$ $T_{c\bar{u}d\bar{u}}$ HMRT.][T_cduu]{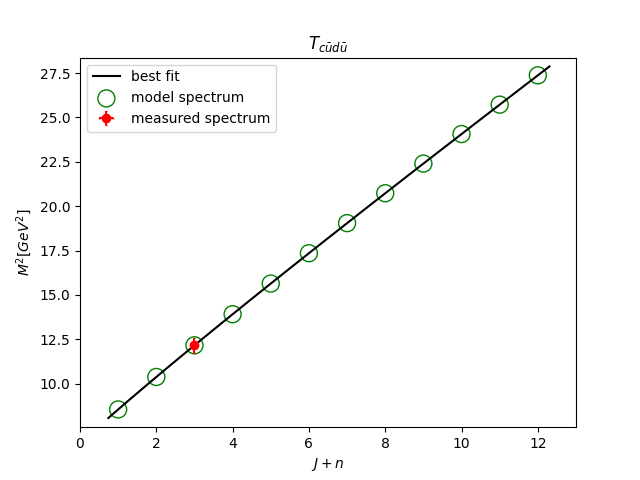}{tetraquarks/predictions/cd-uu/by_n/T_cduu_one_by_n}

    \begin{center}
    \begin{minipage}{.49\linewidth}
        \begin{center}
\begin{longtable}[c]{|c|c|}
  \hline
  \thead{$J$ \\ Spec} & \thead{$a_{T_{j}}$} \\
  \hline
  \hline
\makecell{$2854-2992$\\$3160-3282$\\$3432-3542$\\$3679-3780$\\$3908-4002$\\$4121-4210$\\$4322-4406$\\$4513-4593$\\$4695-4771$\\$4869-4942$\\$5036-5107$\\$5198-5266$} & \makecell{$-2.3$-$-1.8$} \\
\hline
  \caption{$T_{c\bar{u}d\bar{u}}$ $(J, M^2)$ predictions.} \label{table:Tcduu_j_predictions}
 \end{longtable}
        \end{center}
    \end{minipage}
    \begin{minipage}{.49\linewidth}
        \begin{center}
            \begin{longtable}[c]{|c|c|}
  \hline
  \thead{$n$ \\ Spec} & \thead{$a_{T_{n}}$} \\
  \hline
  \hline
\makecell{$2537-2698$\\$3029-3157$\\$3432-3542$\\$3783-3881$\\$4098-4187$\\$4387-4469$\\$4655-4732$\\$4907-4979$\\$5144-5214$\\$5371-5437$\\$5587-5650$\\$5794-5855$} & \makecell{$-1.7$-$-1.4$} \\
\hline
  \caption{$T_{c\bar{u}d\bar{u}}$ $(n, M^2)$ predictions.} \label{table:Tcduu_n_predictions}
 \end{longtable}
        \end{center}
    \end{minipage}
    \end{center}

\subsubsection{\texorpdfstring{$T_{c\bar{d}u\bar{d}}$}{}}\label{T_cudd}

\twographs[(a) is $(J, M^{2})$ $T_{c\bar{d}u\bar{d}}$ HMRT, (b) $(n, M^{2})$ $T_{c\bar{d}u\bar{d}}$ HMRT.][T_cudd]{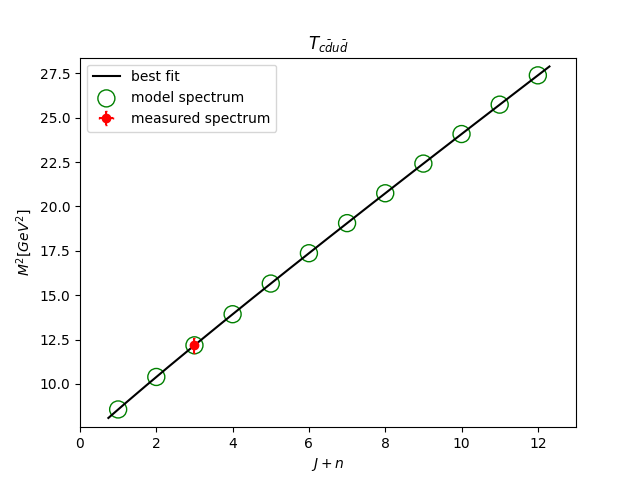}{tetraquarks/predictions/cu-dd/by_n/T_cudd_one_by_n}

    \begin{center}
    \begin{minipage}{.49\linewidth}
        \begin{center}
            \begin{longtable}[c]{|c|c|}
  \hline
  \thead{$J$ \\ Spec} & \thead{$a_{T_{j}}$} \\
  \hline
  \hline
\makecell{$2856-2994$\\$3162-3284$\\$3434-3544$\\$3681-3782$\\$3909-4004$\\$4122-4211$\\$4323-4408$\\$4514-4594$\\$4696-4773$\\$4870-4944$\\$5038-5108$\\$5199-5267$} & \makecell{$-2.3$-$-1.9$} \\
\hline
  \caption{$T_{c\bar{d}u\bar{d}}$ $(J, M^2)$ predictions.} \label{table:Tcudd_j_predictions}
 \end{longtable}
        \end{center}
    \end{minipage}
    \begin{minipage}{.49\linewidth}
        \begin{center}
            \begin{longtable}[c]{|c|c|}
  \hline
  \thead{$n$ \\ Spec} & \thead{$a_{T_{n}}$} \\
  \hline
  \hline
\makecell{$2540-2701$\\$3031-3160$\\$3434-3544$\\$3784-3883$\\$4099-4189$\\$4388-4471$\\$4656-4734$\\$4908-4981$\\$5146-5215$\\$5372-5438$\\$5588-5651$\\$5795-5856$} & \makecell{$-1.7$-$-1.4$} \\
\hline
  \caption{$T_{c\bar{d}u\bar{d}}$ $(n, M^2)$ predictions.} \label{table:Tcudd_n_predictions}
 \end{longtable}
        \end{center}
    \end{minipage}
    \end{center}

\subsubsection{\texorpdfstring{$T_{c\bar{d}u\bar{s}}$}{}}\label{T_cuds}

\twographs[(a) is $(J, M^{2})$ $T_{c\bar{d}u\bar{s}}$ HMRT, (b) $(n, M^{2})$ $T_{c\bar{d}u\bar{s}}$ HMRT.][T_cuds]{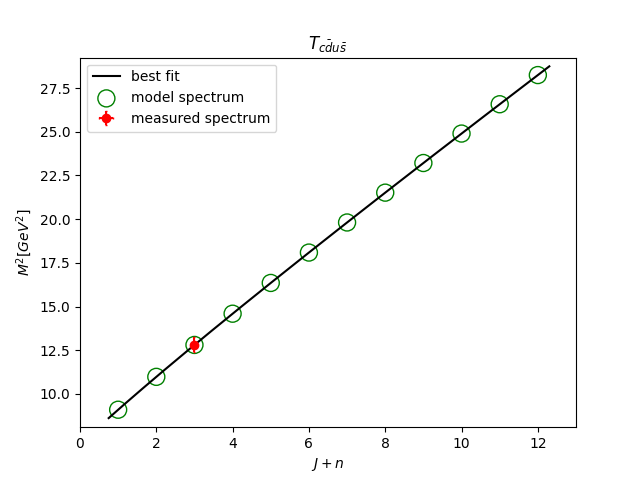}{tetraquarks/predictions/cu-ds/by_n/T_cuds_one_by_n}

    \begin{center}
    \begin{minipage}{.49\linewidth}
        \begin{center}
            \begin{longtable}[c]{|c|c|}
  \hline
  \thead{$J$ \\ Spec} & \thead{$a_{T_{j}}$} \\
  \hline
  \hline
\makecell{$2946-3085$\\$3252-3374$\\$3523-3633$\\$3769-3870$\\$3996-4091$\\$4209-4298$\\$4409-4493$\\$4599-4679$\\$4780-4856$\\$4953-5027$\\$5120-5191$\\$5281-5349$} & \makecell{$-2.1$-$-1.6$} \\
\hline
  \caption{$T_{c\bar{d}u\bar{s}}$ $(J, M^2)$ predictions.} \label{table:Tcuds_j_predictions}
 \end{longtable}
        \end{center}
    \end{minipage}
    \begin{minipage}{.49\linewidth}
        \begin{center}
            \begin{longtable}[c]{|c|c|}
  \hline
  \thead{$n$ \\ Spec} & \thead{$a_{T_{n}}$} \\
  \hline
  \hline
\makecell{$2628-2791$\\$3121-3250$\\$3523-3633$\\$3872-3970$\\$4186-4275$\\$4473-4556$\\$4740-4818$\\$4991-5064$\\$5228-5297$\\$5453-5519$\\$5668-5731$\\$5874-5935$} & \makecell{$-1.5$-$-1.2$} \\
\hline
  \caption{$T_{c\bar{d}u\bar{s}}$ $(n, M^2)$ predictions.} \label{table:Tcuds_n_predictions}
 \end{longtable}
        \end{center}
    \end{minipage}
    \end{center}

\subsubsection{\texorpdfstring{$T_{c\bar{s}d\bar{u}}$}{}}\label{T_cdsu}

\twographs[(a) is $(J, M^{2})$ $T_{c\bar{s}d\bar{u}}$ HMRT, (b) $(n, M^{2})$ $T_{c\bar{s}d\bar{u}}$ HMRT.][T_cdsu]{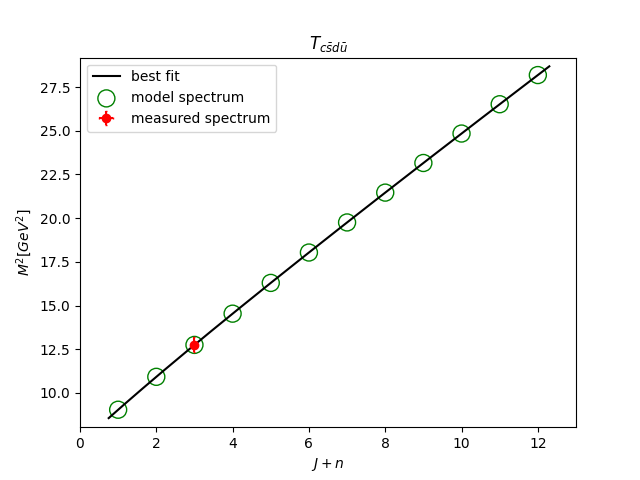}{tetraquarks/predictions/cd-su/by_n/T_cdsu_one_by_n}

    \begin{center}
    \begin{minipage}{.49\linewidth}
        \begin{center}
            \begin{longtable}[c]{|c|c|}
  \hline
  \thead{$J$ \\ Spec} & \thead{$a_{T_{j}}$} \\
  \hline
  \hline
\makecell{$2936-3075$\\$3243-3365$\\$3515-3625$\\$3762-3863$\\$3990-4084$\\$4202-4291$\\$4403-4487$\\$4593-4673$\\$4774-4851$\\$4948-5021$\\$5115-5185$\\$5276-5344$} & \makecell{$-2.0$-$-1.6$} \\
\hline
  \caption{$T_{c\bar{s}d\bar{u}}$ $(J, M^2)$ predictions.} \label{table:Tcdsu_j_predictions}
 \end{longtable}
        \end{center}
    \end{minipage}
    \begin{minipage}{.49\linewidth}
        \begin{center}
            \begin{longtable}[c]{|c|c|}
  \hline
  \thead{$n$ \\ Spec} & \thead{$a_{T_{n}}$} \\
  \hline
  \hline
\makecell{$2616-2780$\\$3112-3240$\\$3515-3625$\\$3865-3963$\\$4179-4269$\\$4467-4550$\\$4735-4812$\\$4986-5058$\\$5223-5292$\\$5448-5514$\\$5663-5726$\\$5870-5931$} & \makecell{$-1.5$-$-1.2$} \\
\hline
  \caption{$T_{c\bar{s}d\bar{u}}$ $(n, M^2)$ predictions.} \label{table:Tcdsu_n_predictions}
 \end{longtable}
        \end{center}
    \end{minipage}
    \end{center}

\subsubsection{\texorpdfstring{$T_{c\bar{u}s\bar{u}}$}{}}\label{T_csuu}

\twographs[(a) is $(J, M^{2})$ $T_{c\bar{u}s\bar{u}}$ HMRT, (b) $(n, M^{2})$ $T_{c\bar{u}s\bar{u}}$ HMRT.][T_csuu]{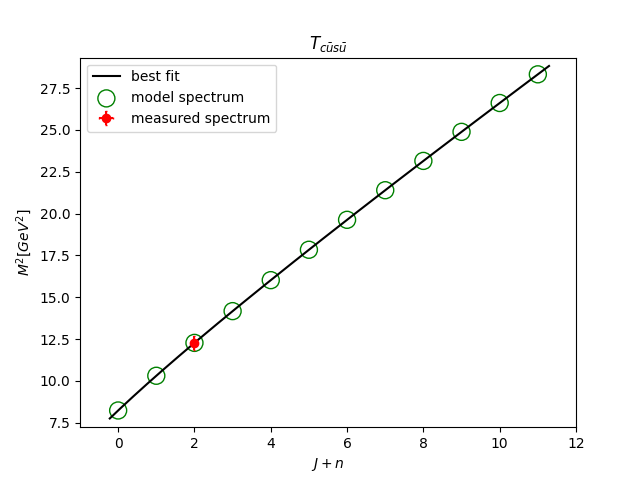}{tetraquarks/predictions/cs-uu/by_n/T_csuu_one_by_n}

    \begin{center}
    \begin{minipage}{.49\linewidth}
        \begin{center}
            \begin{longtable}[c]{|c|c|}
  \hline
  \thead{$J$ \\ Spec} & \thead{$a_{T_{j}}$} \\
  \hline
  \hline
\makecell{$2793-2943$\\$3147-3273$\\$3448-3558$\\$3715-3814$\\$3957-4048$\\$4181-4266$\\$4390-4471$\\$4588-4664$\\$4775-4848$\\$4954-5023$\\$5125-5192$\\$5290-5354$} & \makecell{$-2.2$-$-1.8$} \\
\hline
  \caption{$T_{c\bar{u}s\bar{u}}$ $(J, M^2)$ predictions.} \label{table:Tcsuu_j_predictions}
 \end{longtable}
        \end{center}
    \end{minipage}
    \begin{minipage}{.49\linewidth}
        \begin{center}
            \begin{longtable}[c]{|c|c|}
  \hline
  \thead{$n$ \\ Spec} & \thead{$a_{T_{n}}$} \\
  \hline
  \hline
\makecell{$3448-3558$\\$3825-3921$\\$4157-4243$\\$4457-4536$\\$4734-4808$\\$4993-5062$\\$5236-5301$\\$5466-5528$\\$5686-5745$\\$5896-5953$\\$6098-6152$\\$6292-6345$} & \makecell{$-1.9$-$-1.6$} \\
\hline
  \caption{$T_{c\bar{u}s\bar{u}}$ $(n, M^2)$ predictions.} \label{table:Tcsuu_n_predictions}
 \end{longtable}
        \end{center}
    \end{minipage}
    \end{center}

\subsubsection{\texorpdfstring{$T_{c\bar{d}s\bar{u}}$}{}}\label{T_csdu}

\twographs[(a) is $(J, M^{2})$ $T_{c\bar{d}s\bar{u}}$ HMRT, (b) $(n, M^{2})$ $T_{c\bar{d}s\bar{u}}$ HMRT.][T_csdu]{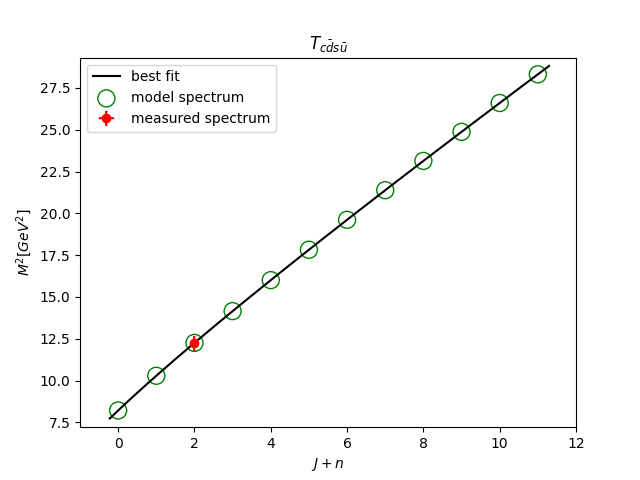}{tetraquarks/predictions/cs-du/by_n/T_csdu_one_by_n}

    \begin{center}
    \begin{minipage}{.49\linewidth}
        \begin{center}
            \begin{longtable}[c]{|c|c|}
  \hline
  \thead{$J$ \\ Spec} & \thead{$a_{T_{j}}$} \\
  \hline
  \hline
\makecell{$2790-2941$\\$3145-3270$\\$3446-3556$\\$3713-3812$\\$3955-4047$\\$4179-4265$\\$4389-4469$\\$4587-4663$\\$4774-4846$\\$4953-5022$\\$5124-5191$\\$5289-5353$} & \makecell{$-2.2$-$-1.8$} \\
\hline
  \caption{$T_{c\bar{d}s\bar{u}}$ $(J, M^2)$ predictions.} \label{table:Tcsdu_j_predictions}
 \end{longtable}
        \end{center}
    \end{minipage}
    \begin{minipage}{.49\linewidth}
        \begin{center}
            \begin{longtable}[c]{|c|c|}
  \hline
  \thead{$n$ \\ Spec} & \thead{$a_{T_{n}}$} \\
  \hline
  \hline
\makecell{$3446-3556$\\$3823-3919$\\$4155-4241$\\$4456-4535$\\$4733-4806$\\$4992-5060$\\$5235-5300$\\$5465-5527$\\$5685-5744$\\$5895-5952$\\$6097-6151$\\$6291-6344$} & \makecell{$-1.9$-$-1.6$} \\
\hline
  \caption{$T_{c\bar{d}s\bar{u}}$ $(n, M^2)$ predictions.} \label{table:Tcsdu_n_predictions}
 \end{longtable}
        \end{center}
    \end{minipage}
    \end{center}

\subsubsection{\texorpdfstring{$T_{c\bar{d}s\bar{d}}$}{}}\label{T_csdd}

\twographs[(a) is $(J, M^{2})$ $T_{c\bar{d}s\bar{d}}$ HMRT, (b) $(n, M^{2})$ $T_{c\bar{d}s\bar{d}}$ HMRT.][T_csdd]{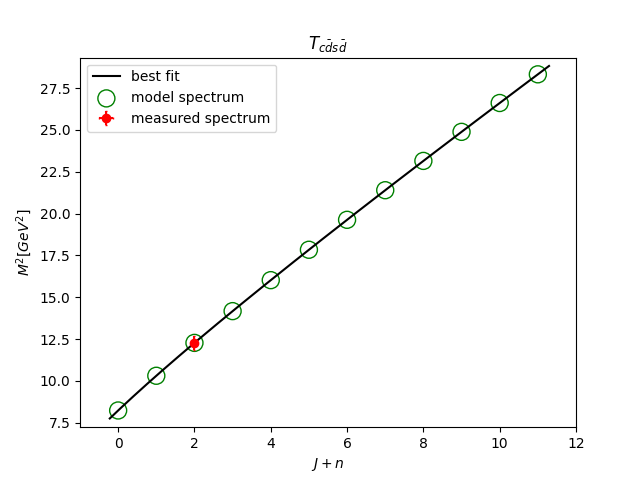}{tetraquarks/predictions/cs-dd/by_n/T_csdd_one_by_n}

    \begin{center}
    \begin{minipage}{.49\linewidth}
        \begin{center}
            \begin{longtable}[c]{|c|c|}
  \hline
  \thead{$J$ \\ Spec} & \thead{$a_{T_{j}}$} \\
  \hline
  \hline
\makecell{$2793-2943$\\$3147-3273$\\$3448-3558$\\$3715-3814$\\$3957-4048$\\$4181-4266$\\$4390-4471$\\$4588-4664$\\$4775-4848$\\$4954-5023$\\$5125-5192$\\$5290-5354$} & \makecell{$-2.2$-$-1.8$} \\
\hline
  \caption{$T_{c\bar{d}s\bar{d}}$ $(J, M^2)$ predictions.} \label{table:Tcsdd_j_predictions}
 \end{longtable}
        \end{center}
    \end{minipage}
    \begin{minipage}{.49\linewidth}
        \begin{center}
            \begin{longtable}[c]{|c|c|}
  \hline
  \thead{$n$ \\ Spec} & \thead{$a_{T_{n}}$} \\
  \hline
  \hline
\makecell{$3448-3558$\\$3825-3921$\\$4157-4243$\\$4457-4536$\\$4734-4808$\\$4993-5062$\\$5236-5301$\\$5466-5528$\\$5686-5745$\\$5896-5953$\\$6098-6152$\\$6292-6345$} & \makecell{$-1.9$-$-1.6$} \\
\hline
  \caption{$T_{c\bar{d}s\bar{d}}$ $(n, M^2)$ predictions.} \label{table:Tcsdd_n_predictions}
 \end{longtable}
        \end{center}
    \end{minipage}
    \end{center}

\subsubsection{\texorpdfstring{$T_{c\bar{s}u\bar{s}}$}{}}\label{T_cuss}

\twographs[(a) is $(J, M^{2})$ $T_{c\bar{s}u\bar{s}}$ HMRT, (b) $(n, M^{2})$ $T_{c\bar{s}u\bar{s}}$ HMRT.][T_cuss]{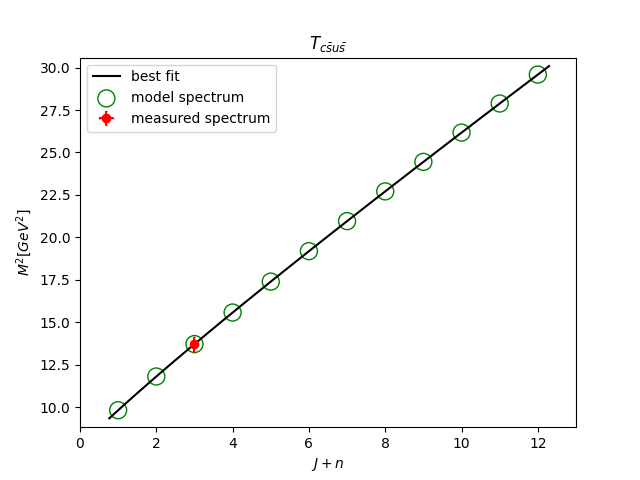}{tetraquarks/predictions/cu-ss/by_n/T_cuss_one_by_n}

    \begin{center}
    \begin{minipage}{.49\linewidth}
        \begin{center}
            \begin{longtable}[c]{|c|c|}
  \hline
  \thead{$J$ \\ Spec} & \thead{$a_{T_{j}}$} \\
  \hline
  \hline
\makecell{$3062-3203$\\$3374-3497$\\$3648-3758$\\$3896-3996$\\$4124-4217$\\$4336-4424$\\$4536-4619$\\$4726-4805$\\$4906-4982$\\$5079-5152$\\$5246-5316$\\$5406-5473$} & \makecell{$-1.7$-$-1.2$} \\
\hline
  \caption{$T_{c\bar{s}u\bar{s}}$ $(J, M^2)$ predictions.} \label{table:Tcuss_j_predictions}
 \end{longtable}
        \end{center}
    \end{minipage}
    \begin{minipage}{.49\linewidth}
        \begin{center}
            \begin{longtable}[c]{|c|c|}
  \hline
  \thead{$n$ \\ Spec} & \thead{$a_{T_{n}}$} \\
  \hline
  \hline
\makecell{$2732-2901$\\$3241-3371$\\$3648-3758$\\$3999-4097$\\$4313-4402$\\$4600-4682$\\$4867-4944$\\$5117-5189$\\$5353-5421$\\$5577-5643$\\$5792-5854$\\$5997-6057$} & \makecell{$-1.2$-$-0.94$} \\
\hline
  \caption{$T_{c\bar{s}u\bar{s}}$ $(n, M^2)$ predictions.} \label{table:Tcuss_n_predictions}
 \end{longtable}
        \end{center}
    \end{minipage}
    \end{center}

\subsubsection{\texorpdfstring{$T_{c\bar{s}d\bar{s}}$}{}}\label{T_cdss}

\twographs[(a) is $(J, M^{2})$ $T_{c\bar{s}d\bar{s}}$ HMRT, (b) $(n, M^{2})$ $T_{c\bar{s}d\bar{s}}$ HMRT.][T_cdss]{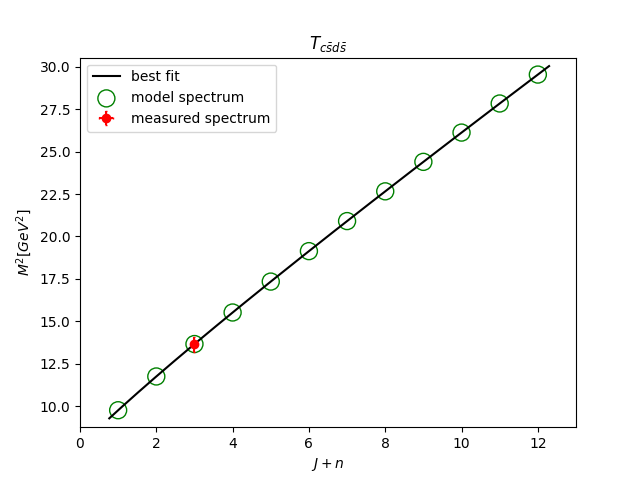}{tetraquarks/predictions/cd-ss/by_n/T_cdss_one_by_n}

    \begin{center}
    \begin{minipage}{.49\linewidth}
        \begin{center}
            \begin{longtable}[c]{|c|c|}
  \hline
  \thead{$J$ \\ Spec} & \thead{$a_{T_{j}}$} \\
  \hline
  \hline
\makecell{$3053-3194$\\$3366-3489$\\$3641-3751$\\$3889-3990$\\$4118-4211$\\$4331-4419$\\$4531-4614$\\$4721-4800$\\$4902-4977$\\$5075-5147$\\$5241-5311$\\$5401-5469$} & \makecell{$-1.7$-$-1.2$} \\
\hline
  \caption{$T_{c\bar{s}d\bar{s}}$ $(J, M^2)$ predictions.} \label{table:Tcdss_j_predictions}
 \end{longtable}
        \end{center}
    \end{minipage}
    \begin{minipage}{.49\linewidth}
        \begin{center}
            \begin{longtable}[c]{|c|c|}
  \hline
  \thead{$n$ \\ Spec} & \thead{$a_{T_{n}}$} \\
  \hline
  \hline
\makecell{$2721-2891$\\$3233-3363$\\$3641-3751$\\$3993-4090$\\$4308-4396$\\$4595-4677$\\$4862-4939$\\$5112-5184$\\$5349-5417$\\$5573-5638$\\$5788-5850$\\$5994-6054$} & \makecell{$-1.2$-$-0.92$} \\
\hline
  \caption{$T_{c\bar{s}d\bar{s}}$ $(n, M^2)$ predictions.} \label{table:Tcdss_n_predictions}
 \end{longtable}
        \end{center}
    \end{minipage}
    \end{center}

\subsubsection{\texorpdfstring{$T_{c\bar{c}d\bar{u}}$}{}}\label{T_cdcu}

\twographs[(a) is $(J, M^{2})$ $T_{c\bar{c}d\bar{u}}$ HMRT, (b) $(n, M^{2})$ $T_{c\bar{c}d\bar{u}}$ HMRT.][T_cdcu]{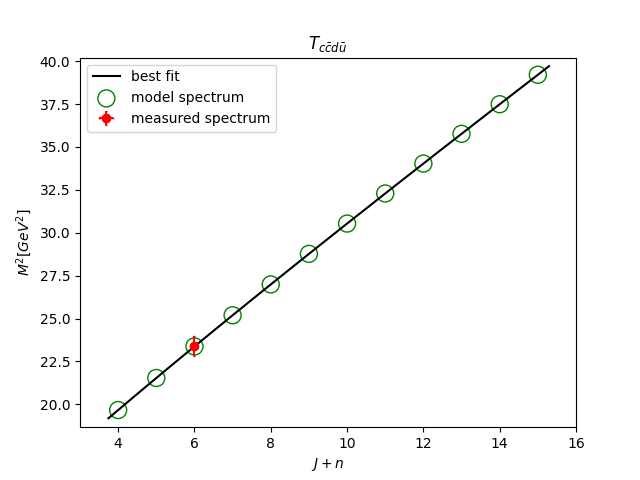}{tetraquarks/predictions/cd-cu/by_n/T_cdcu_one_by_n}

    \begin{center}
    \begin{minipage}{.49\linewidth}
        \begin{center}
            \begin{longtable}[c]{|c|c|}
  \hline
  \thead{$J$ \\ Spec} & \thead{$a_{T_{j}}$} \\
  \hline
  \hline
\makecell{$4373-4497$\\$4583-4699$\\$4780-4890$\\$4967-5072$\\$5145-5245$\\$5316-5412$\\$5480-5572$\\$5638-5727$\\$5790-5877$\\$5938-6023$\\$6082-6164$\\$6222-6301$} & \makecell{$-2.6$-$-2.0$} \\
\hline
  \caption{$T_{c\bar{c}d\bar{u}}$ $(J, M^2)$ predictions.} \label{table:Tcdcu_j_predictions}
 \end{longtable}
        \end{center}
    \end{minipage}
    \begin{minipage}{.49\linewidth}
        \begin{center}
            \begin{longtable}[c]{|c|c|}
  \hline
  \thead{$n$ \\ Spec} & \thead{$a_{T_{n}}$} \\
  \hline
  \hline
\makecell{$4174-4306$\\$4491-4611$\\$4780-4890$\\$5047-5150$\\$5297-5394$\\$5533-5625$\\$5757-5844$\\$5971-6054$\\$6176-6256$\\$6373-6450$\\$6563-6638$\\$6747-6819$} & \makecell{$-2.0$-$-1.6$} \\
\hline
  \caption{$T_{c\bar{c}d\bar{u}}$ $(n, M^2)$ predictions.} \label{table:Tcdcu_n_predictions}
 \end{longtable}
        \end{center}
    \end{minipage}
    \end{center}

\subsubsection{\texorpdfstring{$T_{c\bar{c}s\bar{u}}$}{}}\label{T_cscu}

\twographs[(a) is $(J, M^{2})$ $T_{c\bar{c}s\bar{u}}$ HMRT, (b) $(n, M^{2})$ $T_{c\bar{c}s\bar{u}}$ HMRT.][T_cscu]{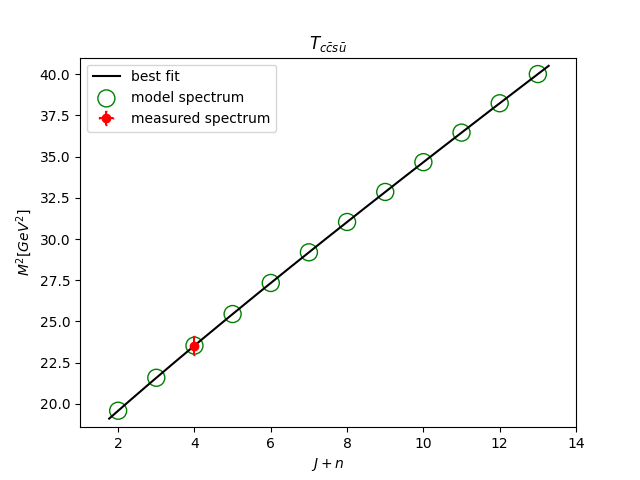}{tetraquarks/predictions/cs-cu/by_n/T_cscu_one_by_n}

    \begin{center}
    \begin{minipage}{.49\linewidth}
        \begin{center}
            \begin{longtable}[c]{|c|c|}
  \hline
  \thead{$J$ \\ Spec} & \thead{$a_{T_{j}}$} \\
  \hline
  \hline
\makecell{$4362-4488$\\$4587-4704$\\$4796-4906$\\$4993-5097$\\$5179-5278$\\$5356-5450$\\$5525-5616$\\$5688-5776$\\$5845-5930$\\$5997-6079$\\$6144-6223$\\$6287-6364$} & \makecell{$-3.3$-$-2.7$} \\
\hline
  \caption{$T_{c\bar{c}s\bar{u}}$ $(J, M^2)$ predictions.} \label{table:Tcscu_j_predictions}
 \end{longtable}
        \end{center}
    \end{minipage}
    \begin{minipage}{.49\linewidth}
        \begin{center}
            \begin{longtable}[c]{|c|c|}
  \hline
  \thead{$n$ \\ Spec} & \thead{$a_{T_{n}}$} \\
  \hline
  \hline
\makecell{$4145-4282$\\$4489-4610$\\$4796-4906$\\$5076-5178$\\$5337-5432$\\$5580-5670$\\$5811-5896$\\$6030-6111$\\$6240-6317$\\$6441-6515$\\$6634-6706$\\$6821-6891$} & \makecell{$-2.0$-$-1.6$} \\
\hline
  \caption{$T_{c\bar{c}s\bar{u}}$ $(n, M^2)$ predictions.} \label{table:Tcscu_n_predictions}
 \end{longtable}
        \end{center}
    \end{minipage}
    \end{center}

\subsubsection{\texorpdfstring{$T_{c\bar{c}s\bar{d}}$}{}}\label{T_cscd}

\twographs[(a) is $(J, M^{2})$ $T_{c\bar{c}s\bar{d}}$ HMRT, (b) $(n, M^{2})$ $T_{c\bar{c}s\bar{d}}$ HMRT.][T_cscd]{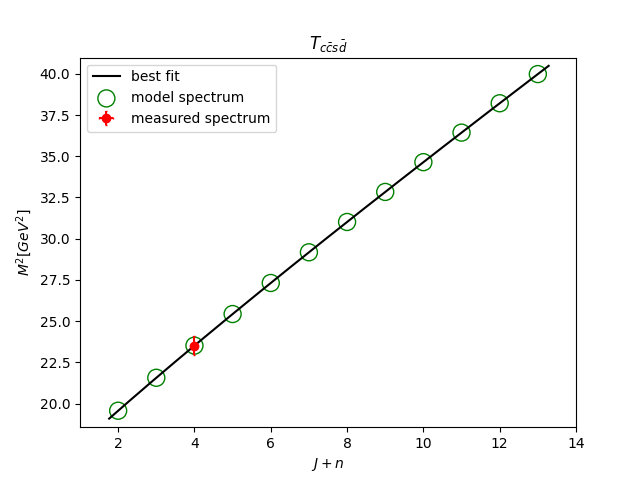}{tetraquarks/predictions/cs-cd/by_n/T_cscd_one_by_n}

    \begin{center}
    \begin{minipage}{.49\linewidth}
        \begin{center}
            \begin{longtable}[c]{|c|c|}
  \hline
  \thead{$J$ \\ Spec} & \thead{$a_{T_{j}}$} \\
  \hline
  \hline
\makecell{$4359-4486$\\$4585-4702$\\$4794-4904$\\$4991-5095$\\$5177-5276$\\$5354-5449$\\$5524-5615$\\$5687-5774$\\$5844-5928$\\$5996-6077$\\$6143-6222$\\$6285-6362$} & \makecell{$-3.2$-$-2.7$} \\
\hline
  \caption{$T_{c\bar{c}s\bar{d}}$ $(J, M^2)$ predictions.} \label{table:Tcscd_j_predictions}
 \end{longtable}
        \end{center}
    \end{minipage}
    \begin{minipage}{.49\linewidth}
        \begin{center}
            \begin{longtable}[c]{|c|c|}
  \hline
  \thead{$n$ \\ Spec} & \thead{$a_{T_{n}}$} \\
  \hline
  \hline
\makecell{$4142-4279$\\$4487-4608$\\$4794-4904$\\$5075-5176$\\$5335-5430$\\$5579-5668$\\$5809-5894$\\$6029-6110$\\$6238-6316$\\$6439-6514$\\$6633-6705$\\$6820-6889$} & \makecell{$-2.0$-$-1.6$} \\
\hline
  \caption{$T_{c\bar{c}s\bar{d}}$ $(n, M^2)$ predictions.} \label{table:Tcscd_n_predictions}
 \end{longtable}
        \end{center}
    \end{minipage}
    \end{center}

\subsubsection{\texorpdfstring{$T_{c\bar{c}d\bar{s}}$}{}}\label{T_cdcs}

\twographs[(a) is $(J, M^{2})$ $T_{c\bar{c}d\bar{s}}$ HMRT, (b) $(n, M^{2})$ $T_{c\bar{c}d\bar{s}}$ HMRT.][T_cdcs]{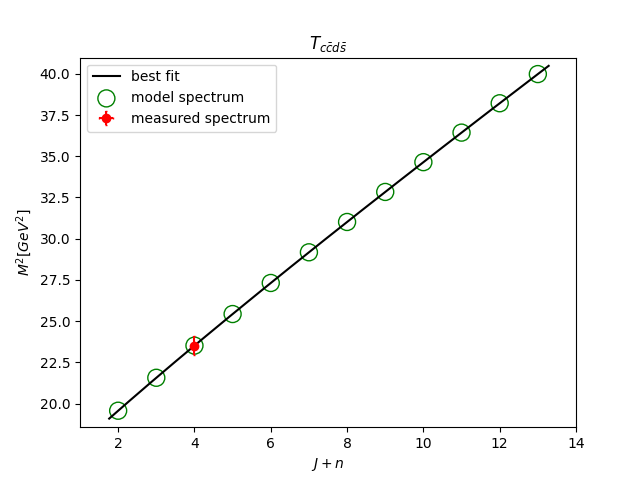}{tetraquarks/predictions/cd-cs/by_n/T_cdcs_one_by_n}

    \begin{center}
    \begin{minipage}{.49\linewidth}
        \begin{center}
            \begin{longtable}[c]{|c|c|}
  \hline
  \thead{$J$ \\ Spec} & \thead{$a_{T_{j}}$} \\
  \hline
  \hline
\makecell{$4359-4486$\\$4585-4702$\\$4794-4904$\\$4991-5095$\\$5177-5276$\\$5354-5449$\\$5524-5615$\\$5687-5774$\\$5844-5928$\\$5996-6077$\\$6143-6222$\\$6285-6362$} & \makecell{$-3.2$-$-2.7$} \\
\hline
  \caption{$T_{c\bar{c}d\bar{s}}$ $(J, M^2)$ predictions.} \label{table:Tcdcs_j_predictions}
 \end{longtable}
        \end{center}
    \end{minipage}
    \begin{minipage}{.49\linewidth}
        \begin{center}
            \begin{longtable}[c]{|c|c|}
  \hline
  \thead{$n$ \\ Spec} & \thead{$a_{T_{n}}$} \\
  \hline
  \hline
\makecell{$4142-4279$\\$4487-4608$\\$4794-4904$\\$5075-5176$\\$5335-5430$\\$5579-5668$\\$5809-5894$\\$6029-6110$\\$6238-6316$\\$6439-6514$\\$6633-6705$\\$6820-6889$} & \makecell{$-2.0$-$-1.6$} \\
\hline
  \caption{$T_{c\bar{c}d\bar{s}}$ $(n, M^2)$ predictions.} \label{table:Tcdcs_n_predictions}
 \end{longtable}
        \end{center}
    \end{minipage}
    \end{center}

\subsubsection{\texorpdfstring{$T_{c\bar{u}c\bar{u}}$}{}}\label{T_ccuu}

\twographs[(a) is $(J, M^{2})$ $T_{c\bar{u}c\bar{u}}$ HMRT, (b) $(n, M^{2})$ $T_{c\bar{u}c\bar{u}}$ HMRT.][T_ccuu]{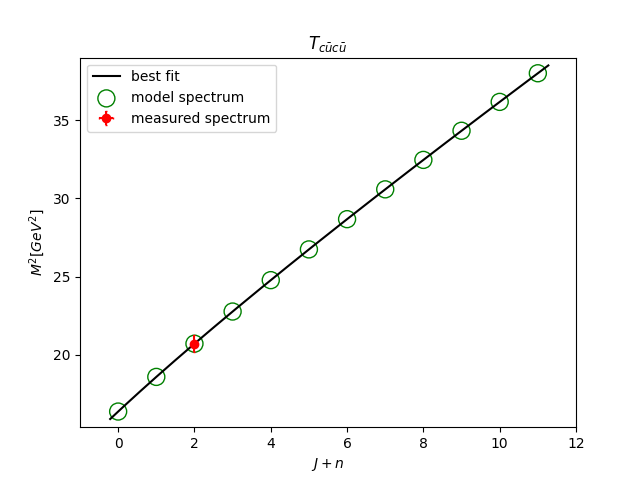}{tetraquarks/predictions/cc-uu/by_n/T_ccuu_one_by_n}

    \begin{center}
    \begin{minipage}{.49\linewidth}
        \begin{center}
            \begin{longtable}[c]{|c|c|}
  \hline
  \thead{$J$ \\ Spec} & \thead{$a_{T_{j}}$} \\
  \hline
  \hline
\makecell{$3981-4115$\\$4253-4373$\\$4497-4607$\\$4721-4823$\\$4930-5026$\\$5126-5217$\\$5312-5398$\\$5489-5571$\\$5658-5737$\\$5820-5896$\\$5977-6050$\\$6128-6199$} & \makecell{$-3.7$-$-3.3$} \\
\hline
  \caption{$T_{c\bar{u}c\bar{u}}$ $(J, M^2)$ predictions.} \label{table:Tccuu_j_predictions}
 \end{longtable}
        \end{center}
    \end{minipage}
    \begin{minipage}{.49\linewidth}
        \begin{center}
            \begin{longtable}[c]{|c|c|}
  \hline
  \thead{$n$ \\ Spec} & \thead{$a_{T_{n}}$} \\
  \hline
  \hline
\makecell{$4497-4607$\\$4816-4915$\\$5105-5196$\\$5372-5456$\\$5621-5700$\\$5856-5931$\\$6078-6150$\\$6291-6359$\\$6494-6560$\\$6690-6753$\\$6879-6940$\\$7061-7121$} & \makecell{$-3.0$-$-2.6$} \\
\hline
  \caption{$T_{c\bar{u}c\bar{u}}$ $(n, M^2)$ predictions.} \label{table:Tccuu_n_predictions}
 \end{longtable}
        \end{center}
    \end{minipage}
    \end{center}

\subsubsection{\texorpdfstring{$T_{c\bar{d}c\bar{d}}$}{}}\label{T_ccdd}

\twographs[(a) is $(J, M^{2})$ $T_{c\bar{d}c\bar{d}}$ HMRT, (b) $(n, M^{2})$ $T_{c\bar{d}c\bar{d}}$ HMRT.][T_ccdd]{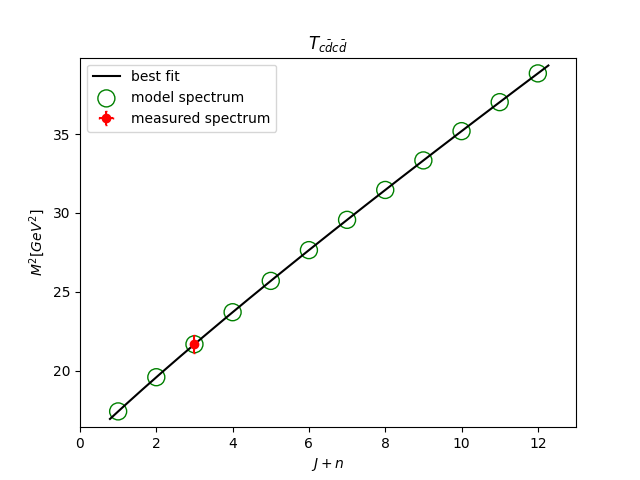}{tetraquarks/predictions/cc-dd/by_n/T_ccdd_one_by_n}

    \begin{center}
    \begin{minipage}{.49\linewidth}
        \begin{center}
            \begin{longtable}[c]{|c|c|}
  \hline
  \thead{$J$ \\ Spec} & \thead{$a_{T_{j}}$} \\
  \hline
  \hline
\makecell{$4109-4241$\\$4367-4487$\\$4602-4712$\\$4819-4921$\\$5021-5118$\\$5212-5304$\\$5394-5481$\\$5567-5651$\\$5733-5814$\\$5893-5970$\\$6047-6122$\\$6196-6269$} & \makecell{$-3.2$-$-2.7$} \\
\hline
  \caption{$T_{c\bar{d}c\bar{d}}$ $(J, M^2)$ predictions.} \label{table:Tccdd_j_predictions}
 \end{longtable}
        \end{center}
    \end{minipage}
    \begin{minipage}{.49\linewidth}
        \begin{center}
            \begin{longtable}[c]{|c|c|}
  \hline
  \thead{$n$ \\ Spec} & \thead{$a_{T_{n}}$} \\
  \hline
  \hline
\makecell{$3852-4000$\\$4256-4381$\\$4602-4712$\\$4910-5010$\\$5192-5284$\\$5453-5539$\\$5697-5778$\\$5928-6004$\\$6147-6220$\\$6356-6427$\\$6557-6625$\\$6751-6816$} & \makecell{$-2.3$-$-2.0$} \\
\hline
  \caption{$T_{c\bar{d}c\bar{d}}$ $(n, M^2)$ predictions.} \label{table:Tccdd_n_predictions}
 \end{longtable}
        \end{center}
    \end{minipage}
    \end{center}

\subsubsection{\texorpdfstring{$T_{c\bar{s}c\bar{u}}$}{}}\label{T_ccsu}

\twographs[(a) is $(J, M^{2})$ $T_{c\bar{s}c\bar{u}}$ HMRT, (b) $(n, M^{2})$ $T_{c\bar{s}c\bar{u}}$ HMRT.][T_ccsu]{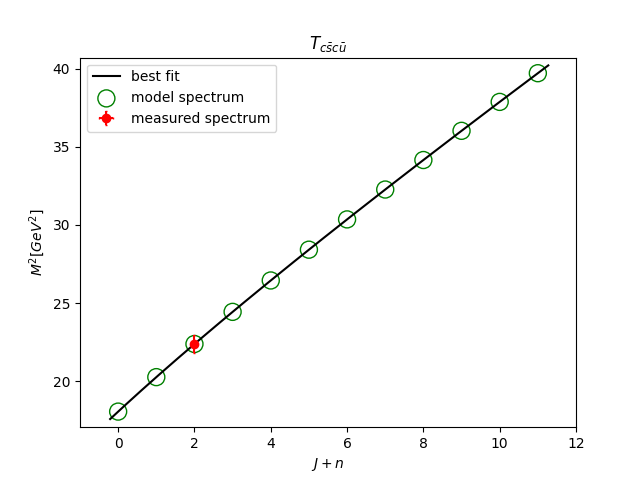}{tetraquarks/predictions/cc-su/by_n/T_ccsu_one_by_n}

    \begin{center}
    \begin{minipage}{.49\linewidth}
        \begin{center}
            \begin{longtable}[c]{|c|c|}
  \hline
  \thead{$J$ \\ Spec} & \thead{$a_{T_{j}}$} \\
  \hline
  \hline
\makecell{$4181-4314$\\$4440-4560$\\$4675-4785$\\$4892-4994$\\$5094-5191$\\$5285-5377$\\$5466-5553$\\$5639-5723$\\$5805-5885$\\$5964-6041$\\$6118-6193$\\$6267-6339$} & \makecell{$-3.9$-$-3.4$} \\
\hline
  \caption{$T_{c\bar{s}c\bar{u}}$ $(J, M^2)$ predictions.} \label{table:Tccsu_j_predictions}
 \end{longtable}
        \end{center}
    \end{minipage}
    \begin{minipage}{.49\linewidth}
        \begin{center}
            \begin{longtable}[c]{|c|c|}
  \hline
  \thead{$n$ \\ Spec} & \thead{$a_{T_{n}}$} \\
  \hline
  \hline
\makecell{$3922-4071$\\$4328-4453$\\$4675-4785$\\$4983-5083$\\$5264-5356$\\$5525-5611$\\$5769-5849$\\$5999-6076$\\$6218-6291$\\$6427-6497$\\$6627-6694$\\$6820-6885$} & \makecell{$-2.1$-$-1.8$} \\
\hline
  \caption{$T_{c\bar{s}c\bar{u}}$ $(n, M^2)$ predictions.} \label{table:Tccsu_n_predictions}
 \end{longtable}
        \end{center}
    \end{minipage}
    \end{center}

\subsubsection{\texorpdfstring{$T_{c\bar{d}c\bar{s}}$}{}}\label{T_ccds}

\twographs[(a) is $(J, M^{2})$ $T_{c\bar{d}c\bar{s}}$ HMRT, (b) $(n, M^{2})$ $T_{c\bar{d}c\bar{s}}$ HMRT.][T_ccds]{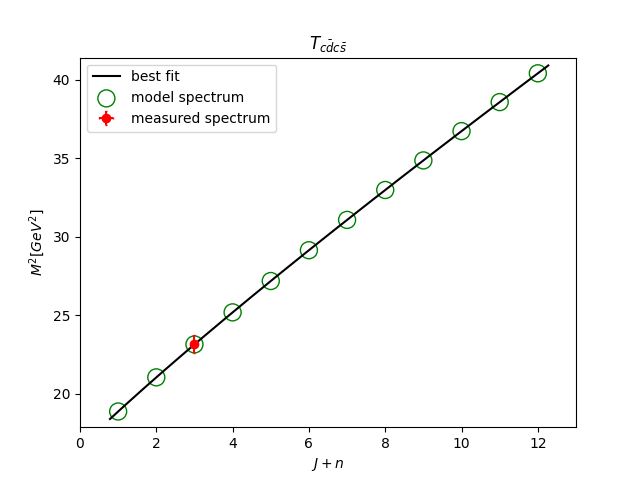}{tetraquarks/predictions/cc-ds/by_n/T_ccds_one_by_n}

    \begin{center}
    \begin{minipage}{.49\linewidth}
        \begin{center}
            \begin{longtable}[c]{|c|c|}
  \hline
  \thead{$J$ \\ Spec} & \thead{$a_{T_{j}}$} \\
  \hline
  \hline
\makecell{$4279-4410$\\$4528-4647$\\$4756-4866$\\$4967-5070$\\$5165-5262$\\$5352-5445$\\$5530-5618$\\$5700-5785$\\$5864-5945$\\$6021-6099$\\$6173-6249$\\$6320-6393$} & \makecell{$-3.3$-$-2.8$} \\
\hline
  \caption{$T_{c\bar{d}c\bar{s}}$ $(J, M^2)$ predictions.} \label{table:Tccds_j_predictions}
 \end{longtable}
        \end{center}
    \end{minipage}
    \begin{minipage}{.49\linewidth}
        \begin{center}
            \begin{longtable}[c]{|c|c|}
  \hline
  \thead{$n$ \\ Spec} & \thead{$a_{T_{n}}$} \\
  \hline
  \hline
\makecell{$4032-4178$\\$4421-4544$\\$4756-4866$\\$5057-5157$\\$5332-5425$\\$5588-5675$\\$5828-5910$\\$6055-6133$\\$6271-6346$\\$6478-6549$\\$6676-6745$\\$6868-6934$} & \makecell{$-2.4$-$-2.0$} \\
\hline
  \caption{$T_{c\bar{d}c\bar{s}}$ $(n, M^2)$ predictions.} \label{table:Tccds_n_predictions}
 \end{longtable}
        \end{center}
    \end{minipage}
    \end{center}

\subsubsection{\texorpdfstring{$T_{c\bar{s}c\bar{s}}$}{}}\label{T_ccss}

\twographs[(a) is $(J, M^{2})$ $T_{c\bar{s}c\bar{s}}$ HMRT, (b) $(n, M^{2})$ $T_{c\bar{s}c\bar{s}}$ HMRT.][T_ccss]{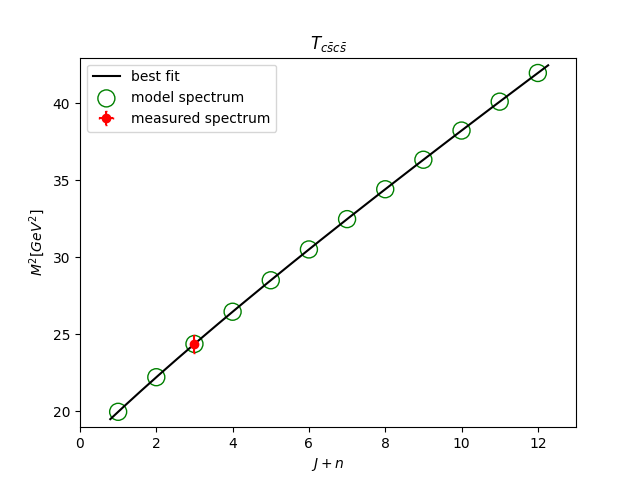}{tetraquarks/predictions/cc-ss/by_n/T_ccss_one_by_n}

    \begin{center}
    \begin{minipage}{.49\linewidth}
        \begin{center}
            \begin{longtable}[c]{|c|c|}
  \hline
  \thead{$J$ \\ Spec} & \thead{$a_{T_{j}}$} \\
  \hline
  \hline
\makecell{$4401-4533$\\$4652-4772$\\$4881-4991$\\$5093-5195$\\$5291-5387$\\$5477-5569$\\$5655-5743$\\$5825-5909$\\$5988-6068$\\$6144-6222$\\$6296-6371$\\$6442-6515$} & \makecell{$-2.9$-$-2.4$} \\
\hline
  \caption{$T_{c\bar{s}c\bar{s}}$ $(J, M^2)$ predictions.} \label{table:Tccss_j_predictions}
 \end{longtable}
        \end{center}
    \end{minipage}
    \begin{minipage}{.49\linewidth}
        \begin{center}
            \begin{longtable}[c]{|c|c|}
  \hline
  \thead{$n$ \\ Spec} & \thead{$a_{T_{n}}$} \\
  \hline
  \hline
\makecell{$4151-4298$\\$4544-4668$\\$4881-4991$\\$5182-5282$\\$5457-5550$\\$5712-5799$\\$5952-6033$\\$6178-6256$\\$6394-6468$\\$6600-6671$\\$6797-6866$\\$6988-7054$} & \makecell{$-2.1$-$-1.7$} \\
\hline
  \caption{$T_{c\bar{s}c\bar{s}}$ $(n, M^2)$ predictions.} \label{table:Tccss_n_predictions}
 \end{longtable}
        \end{center}
    \end{minipage}
    \end{center}

\subsubsection{\texorpdfstring{$T_{c\bar{c}c\bar{u}}$}{}}\label{T_cccu}

\twographs[(a) is $(J, M^{2})$ $T_{c\bar{c}c\bar{u}}$ HMRT, (b) $(n, M^{2})$ $T_{c\bar{c}c\bar{u}}$ HMRT.][T_cccu]{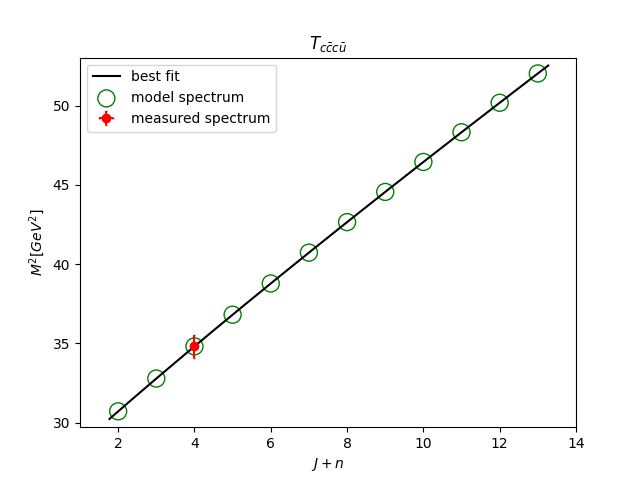}{tetraquarks/predictions/cc-cu/by_n/T_cccu_one_by_n}

    \begin{center}
    \begin{minipage}{.49\linewidth}
        \begin{center}
            \begin{longtable}[c]{|c|c|}
  \hline
  \thead{$J$ \\ Spec} & \thead{$a_{T_{j}}$} \\
  \hline
  \hline
\makecell{$5481-5602$\\$5668-5783$\\$5845-5955$\\$6014-6120$\\$6176-6278$\\$6333-6430$\\$6483-6578$\\$6629-6721$\\$6770-6859$\\$6908-6994$\\$7041-7126$\\$7172-7254$} & \makecell{$-5.5$-$-4.9$} \\
\hline
  \caption{$T_{c\bar{c}c\bar{u}}$ $(J, M^2)$ predictions.} \label{table:Tcccu_j_predictions}
 \end{longtable}
        \end{center}
    \end{minipage}
    \begin{minipage}{.49\linewidth}
        \begin{center}
            \begin{longtable}[c]{|c|c|}
  \hline
  \thead{$n$ \\ Spec} & \thead{$a_{T_{n}}$} \\
  \hline
  \hline
\makecell{$5305-5434$\\$5586-5704$\\$5845-5955$\\$6087-6191$\\$6316-6414$\\$6532-6626$\\$6739-6829$\\$6938-7024$\\$7129-7212$\\$7313-7393$\\$7491-7569$\\$7664-7740$} & \makecell{$-3.6$-$-3.2$} \\
\hline
  \caption{$T_{c\bar{c}c\bar{u}}$ $(n, M^2)$ predictions.} \label{table:Tcccu_n_predictions}
 \end{longtable}
        \end{center}
    \end{minipage}
    \end{center}

\subsubsection{\texorpdfstring{$T_{c\bar{c}c\bar{d}}$}{}}\label{T_cccd}

\twographs[(a) is $(J, M^{2})$ $T_{c\bar{c}c\bar{d}}$ HMRT, (b) $(n, M^{2})$ $T_{c\bar{c}c\bar{d}}$ HMRT.][T_cccd]{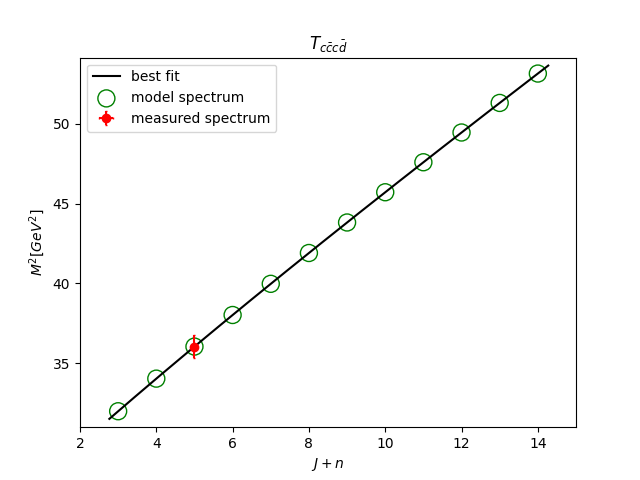}{tetraquarks/predictions/cc-cd/by_n/T_cccd_one_by_n}

    \begin{center}
    \begin{minipage}{.49\linewidth}
        \begin{center}
            \begin{longtable}[c]{|c|c|}
  \hline
  \thead{$J$ \\ Spec} & \thead{$a_{T_{j}}$} \\
  \hline
  \hline
\makecell{$5595-5716$\\$5776-5891$\\$5948-6058$\\$6113-6219$\\$6271-6373$\\$6424-6522$\\$6572-6667$\\$6715-6807$\\$6854-6944$\\$6989-7076$\\$7120-7206$\\$7249-7332$} & \makecell{$-5.2$-$-4.5$} \\
\hline
  \caption{$T_{c\bar{c}c\bar{d}}$ $(J, M^2)$ predictions.} \label{table:Tcccd_j_predictions}
 \end{longtable}
        \end{center}
    \end{minipage}
    \begin{minipage}{.49\linewidth}
        \begin{center}
            \begin{longtable}[c]{|c|c|}
  \hline
  \thead{$n$ \\ Spec} & \thead{$a_{T_{n}}$} \\
  \hline
  \hline
\makecell{$5425-5552$\\$5696-5814$\\$5948-6058$\\$6184-6288$\\$6407-6506$\\$6620-6714$\\$6823-6914$\\$7018-7105$\\$7206-7290$\\$7388-7469$\\$7564-7643$\\$7735-7811$} & \makecell{$-4.0$-$-3.6$} \\
\hline
  \caption{$T_{c\bar{c}c\bar{d}}$ $(n, M^2)$ predictions.} \label{table:Tcccd_n_predictions}
 \end{longtable}
        \end{center}
    \end{minipage}
    \end{center}

\subsubsection{\texorpdfstring{$T_{c\bar{c}c\bar{s}}$}{}}\label{T_cccs}

\twographs[(a) is $(J, M^{2})$ $T_{c\bar{c}c\bar{s}}$ HMRT, (b) $(n, M^{2})$ $T_{c\bar{c}c\bar{s}}$ HMRT.][T_cccs]{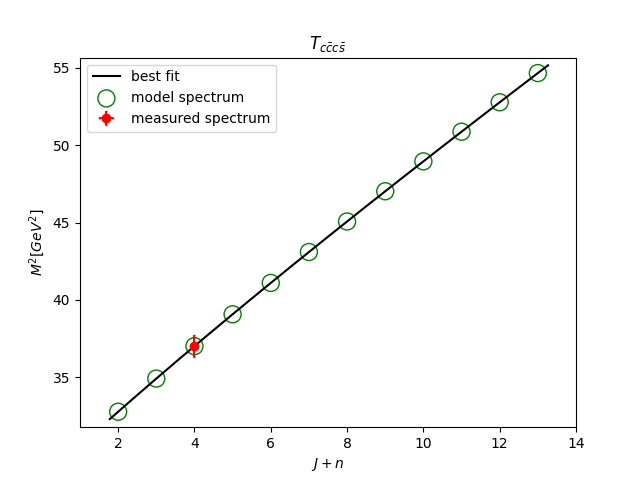}{tetraquarks/predictions/cc-cs/by_n/T_cccs_one_by_n}

    \begin{center}
    \begin{minipage}{.49\linewidth}
        \begin{center}
            \begin{longtable}[c]{|c|c|}
  \hline
  \thead{$J$ \\ Spec} & \thead{$a_{T_{j}}$} \\
  \hline
  \hline
\makecell{$5664-5786$\\$5851-5967$\\$6029-6139$\\$6198-6304$\\$6360-6462$\\$6516-6614$\\$6667-6761$\\$6812-6903$\\$6953-7042$\\$7090-7176$\\$7223-7307$\\$7353-7435$} & \makecell{$-5.1$-$-4.4$} \\
\hline
  \caption{$T_{c\bar{c}c\bar{s}}$ $(J, M^2)$ predictions.} \label{table:Tcccs_j_predictions}
 \end{longtable}
        \end{center}
    \end{minipage}
    \begin{minipage}{.49\linewidth}
        \begin{center}
            \begin{longtable}[c]{|c|c|}
  \hline
  \thead{$n$ \\ Spec} & \thead{$a_{T_{n}}$} \\
  \hline
  \hline
\makecell{$5487-5616$\\$5769-5888$\\$6029-6139$\\$6271-6375$\\$6499-6597$\\$6716-6809$\\$6922-7011$\\$7120-7206$\\$7310-7393$\\$7493-7573$\\$7671-7748$\\$7843-7918$} & \makecell{$-3.3$-$-2.8$} \\
\hline
  \caption{$T_{c\bar{c}c\bar{s}}$ $(n, M^2)$ predictions.} \label{table:Tcccs_n_predictions}
 \end{longtable}
        \end{center}
    \end{minipage}
    \end{center}

\subsubsection{\texorpdfstring{$T_{c\bar{b}d\bar{u}}$}{}}\label{T_cdbu}

\twographs[(a) is $(J, M^{2})$ $T_{c\bar{b}d\bar{u}}$ HMRT, (b) $(n, M^{2})$ $T_{c\bar{b}d\bar{u}}$ HMRT.][T_cdbu]{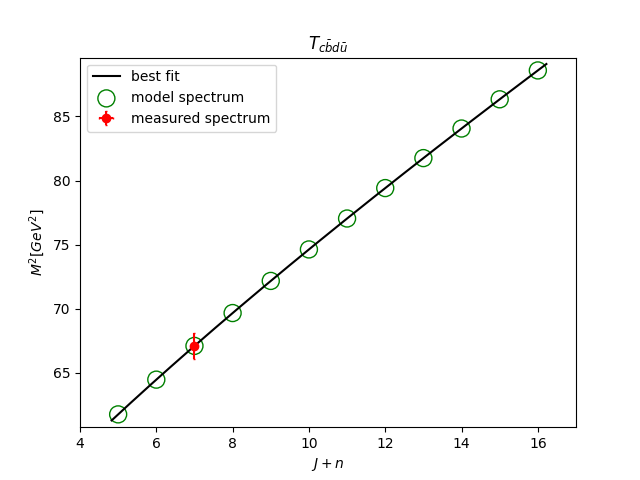}{tetraquarks/predictions/cd-bu/by_n/T_cdbu_one_by_n}

    \begin{center}
    \begin{minipage}{.49\linewidth}
        \begin{center}
            \begin{longtable}[c]{|c|c|}
  \hline
  \thead{$J$ \\ Spec} & \thead{$a_{T_{j}}$} \\
  \hline
  \hline
\makecell{$7798-7920$\\$7972-8088$\\$8137-8247$\\$8294-8400$\\$8445-8546$\\$8590-8688$\\$8730-8825$\\$8866-8958$\\$8997-9086$\\$9125-9212$\\$9250-9334$\\$9371-9454$} & \makecell{$-2.5$-$-1.8$} \\
\hline
  \caption{$T_{c\bar{b}d\bar{u}}$ $(J, M^2)$ predictions.} \label{table:Tcdbu_j_predictions}
 \end{longtable}
        \end{center}
    \end{minipage}
    \begin{minipage}{.49\linewidth}
        \begin{center}
            \begin{longtable}[c]{|c|c|}
  \hline
  \thead{$n$ \\ Spec} & \thead{$a_{T_{n}}$} \\
  \hline
  \hline
\makecell{$7635-7764$\\$7896-8014$\\$8137-8247$\\$8362-8466$\\$8574-8672$\\$8776-8869$\\$8968-9058$\\$9153-9239$\\$9331-9414$\\$9503-9583$\\$9669-9747$\\$9830-9906$} & \makecell{$-1.5$-$-1.1$} \\
\hline
  \caption{$T_{c\bar{b}d\bar{u}}$ $(n, M^2)$ predictions.} \label{table:Tcdbu_n_predictions}
 \end{longtable}
        \end{center}
    \end{minipage}
    \end{center}

\subsubsection{\texorpdfstring{$T_{c\bar{b}u\bar{d}}$}{}}\label{T_cubd}

\twographs[(a) is $(J, M^{2})$ $T_{c\bar{b}u\bar{d}}$ HMRT, (b) $(n, M^{2})$ $T_{c\bar{b}u\bar{d}}$ HMRT.][T_cubd]{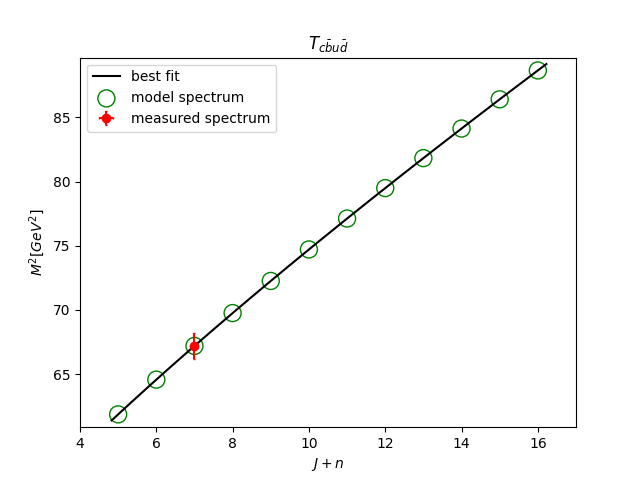}{tetraquarks/predictions/cu-bd/by_n/T_cubd_one_by_n}

    \begin{center}
    \begin{minipage}{.49\linewidth}
        \begin{center}
            \begin{longtable}[c]{|c|c|}
  \hline
  \thead{$J$ \\ Spec} & \thead{$a_{T_{j}}$} \\
  \hline
  \hline
\makecell{$7804-7926$\\$7977-8093$\\$8142-8252$\\$8299-8404$\\$8450-8551$\\$8595-8692$\\$8734-8829$\\$8870-8962$\\$9001-9091$\\$9129-9216$\\$9253-9338$\\$9375-9457$} & \makecell{$-2.5$-$-1.8$} \\
\hline
  \caption{$T_{c\bar{b}u\bar{d}}$ $(J, M^2)$ predictions.} \label{table:Tcubd_j_predictions}
 \end{longtable}
        \end{center}
    \end{minipage}
    \begin{minipage}{.49\linewidth}
        \begin{center}
            \begin{longtable}[c]{|c|c|}
  \hline
  \thead{$n$ \\ Spec} & \thead{$a_{T_{n}}$} \\
  \hline
  \hline
\makecell{$7641-7769$\\$7902-8020$\\$8142-8252$\\$8367-8470$\\$8579-8677$\\$8780-8874$\\$8972-9062$\\$9157-9243$\\$9335-9418$\\$9506-9587$\\$9672-9751$\\$9834-9910$} & \makecell{$-1.6$-$-1.1$} \\
\hline
  \caption{$T_{c\bar{b}u\bar{d}}$ $(n, M^2)$ predictions.} \label{table:Tcubd_n_predictions}
 \end{longtable}
        \end{center}
    \end{minipage}
    \end{center}

\subsubsection{\texorpdfstring{$T_{c\bar{b}s\bar{u}}$}{}}\label{T_csbu}

\twographs[(a) is $(J, M^{2})$ $T_{c\bar{b}s\bar{u}}$ HMRT, (b) $(n, M^{2})$ $T_{c\bar{b}s\bar{u}}$ HMRT.][T_csbu]{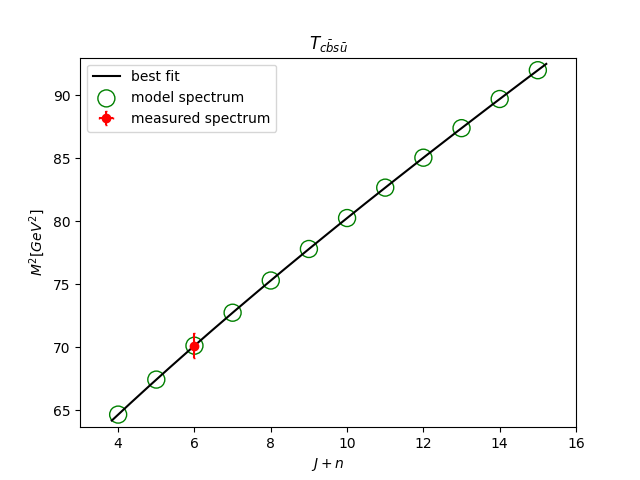}{tetraquarks/predictions/cs-bu/by_n/T_csbu_one_by_n}

    \begin{center}
    \begin{minipage}{.49\linewidth}
        \begin{center}
            \begin{longtable}[c]{|c|c|}
  \hline
  \thead{$J$ \\ Spec} & \thead{$a_{T_{j}}$} \\
  \hline
  \hline
\makecell{$7980-8102$\\$8154-8270$\\$8319-8429$\\$8476-8581$\\$8627-8728$\\$8771-8869$\\$8911-9005$\\$9046-9137$\\$9177-9266$\\$9304-9391$\\$9428-9512$\\$9549-9631$} & \makecell{$-3.0$-$-2.3$} \\
\hline
  \caption{$T_{c\bar{b}s\bar{u}}$ $(J, M^2)$ predictions.} \label{table:Tcsbu_j_predictions}
 \end{longtable}
        \end{center}
    \end{minipage}
    \begin{minipage}{.49\linewidth}
        \begin{center}
            \begin{longtable}[c]{|c|c|}
  \hline
  \thead{$n$ \\ Spec} & \thead{$a_{T_{n}}$} \\
  \hline
  \hline
\makecell{$7816-7945$\\$8078-8196$\\$8319-8429$\\$8544-8647$\\$8755-8853$\\$8956-9050$\\$9148-9238$\\$9332-9418$\\$9509-9592$\\$9680-9760$\\$9845-9923$\\$10006-10081$} & \makecell{$-2.2$-$-1.7$} \\
\hline
  \caption{$T_{c\bar{b}s\bar{u}}$ $(n, M^2)$ predictions.} \label{table:Tcsbu_n_predictions}
 \end{longtable}
        \end{center}
    \end{minipage}
    \end{center}

\subsubsection{\texorpdfstring{$T_{c\bar{b}u\bar{s}}$}{}}\label{T_cubs}

\twographs[(a) is $(J, M^{2})$ $T_{c\bar{b}u\bar{s}}$ HMRT, (b) $(n, M^{2})$ $T_{c\bar{b}u\bar{s}}$ HMRT.][T_cubs]{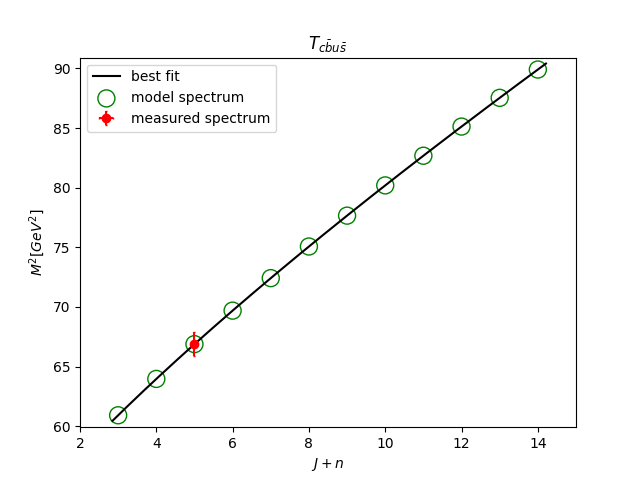}{tetraquarks/predictions/cu-bs/by_n/T_cubs_one_by_n}

    \begin{center}
    \begin{minipage}{.49\linewidth}
        \begin{center}
            \begin{longtable}[c]{|c|c|}
  \hline
  \thead{$J$ \\ Spec} & \thead{$a_{T_{j}}$} \\
  \hline
  \hline
\makecell{$7742-7868$\\$7939-8056$\\$8123-8233$\\$8296-8400$\\$8460-8559$\\$8617-8712$\\$8767-8858$\\$8911-8999$\\$9050-9136$\\$9185-9268$\\$9316-9396$\\$9443-9521$} & \makecell{$-2.3$-$-1.7$} \\
\hline
  \caption{$T_{c\bar{b}u\bar{s}}$ $(J, M^2)$ predictions.} \label{table:Tcubs_j_predictions}
 \end{longtable}
        \end{center}
    \end{minipage}
    \begin{minipage}{.49\linewidth}
        \begin{center}
            \begin{longtable}[c]{|c|c|}
  \hline
  \thead{$n$ \\ Spec} & \thead{$a_{T_{n}}$} \\
  \hline
  \hline
\makecell{$7552-7688$\\$7853-7974$\\$8123-8233$\\$8370-8472$\\$8600-8695$\\$8815-8906$\\$9020-9106$\\$9214-9297$\\$9401-9480$\\$9580-9656$\\$9753-9827$\\$9920-9992$} & \makecell{$-2.1$-$-1.6$} \\
\hline
  \caption{$T_{c\bar{b}u\bar{s}}$ $(n, M^2)$ predictions.} \label{table:Tcubs_n_predictions}
 \end{longtable}
        \end{center}
    \end{minipage}
    \end{center}

\subsubsection{\texorpdfstring{$T_{c\bar{b}d\bar{s}}$}{}}\label{T_cdbs}

\twographs[(a) is $(J, M^{2})$ $T_{c\bar{b}d\bar{s}}$ HMRT, (b) $(n, M^{2})$ $T_{c\bar{b}d\bar{s}}$ HMRT.][T_cdbs]{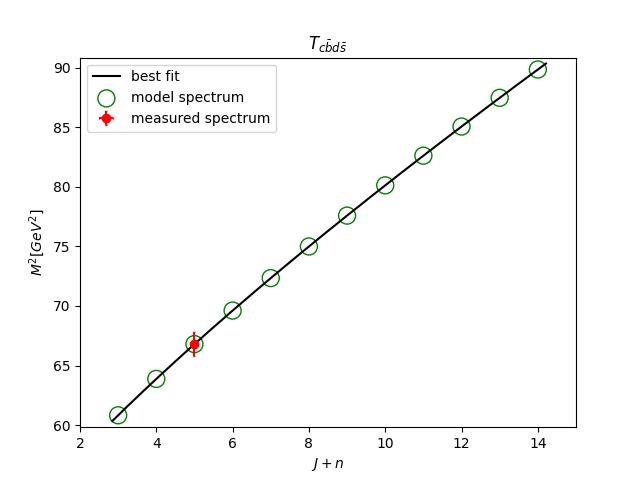}{tetraquarks/predictions/cd-bs/by_n/T_cdbs_one_by_n}

    \begin{center}
    \begin{minipage}{.49\linewidth}
        \begin{center}
            \begin{longtable}[c]{|c|c|}
  \hline
  \thead{$J$ \\ Spec} & \thead{$a_{T_{j}}$} \\
  \hline
  \hline
\makecell{$7736-7862$\\$7934-8051$\\$8118-8228$\\$8291-8395$\\$8456-8555$\\$8612-8707$\\$8763-8854$\\$8907-8995$\\$9046-9132$\\$9181-9264$\\$9312-9392$\\$9439-9518$} & \makecell{$-2.3$-$-1.7$} \\
\hline
  \caption{$T_{c\bar{b}d\bar{s}}$ $(J, M^2)$ predictions.} \label{table:Tcdbs_j_predictions}
 \end{longtable}
        \end{center}
    \end{minipage}
    \begin{minipage}{.49\linewidth}
        \begin{center}
            \begin{longtable}[c]{|c|c|}
  \hline
  \thead{$n$ \\ Spec} & \thead{$a_{T_{n}}$} \\
  \hline
  \hline
\makecell{$7545-7682$\\$7848-7969$\\$8118-8228$\\$8365-8467$\\$8595-8691$\\$8811-8902$\\$9016-9102$\\$9211-9293$\\$9397-9476$\\$9577-9653$\\$9750-9823$\\$9917-9988$} & \makecell{$-2.1$-$-1.6$} \\
\hline
  \caption{$T_{c\bar{b}d\bar{s}}$ $(n, M^2)$ predictions.} \label{table:Tcdbs_n_predictions}
 \end{longtable}
        \end{center}
    \end{minipage}
    \end{center}

\subsubsection{\texorpdfstring{$T_{c\bar{b}s\bar{d}}$}{}}\label{T_csbd}

\twographs[(a) is $(J, M^{2})$ $T_{c\bar{b}s\bar{d}}$ HMRT, (b) $(n, M^{2})$ $T_{c\bar{b}s\bar{d}}$ HMRT.][T_csbd]{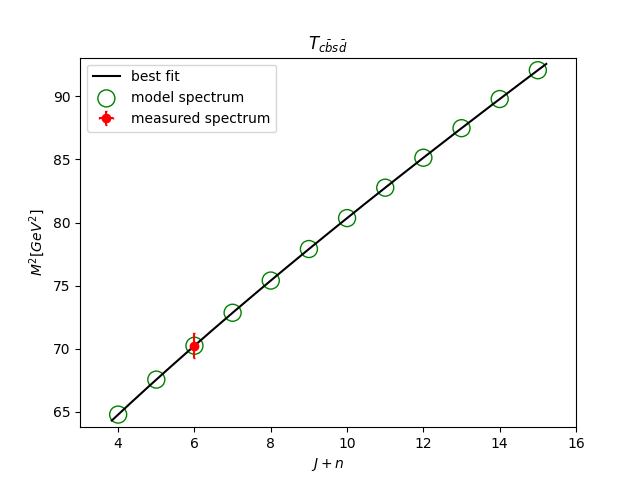}{tetraquarks/predictions/cs-bd/by_n/T_csbd_one_by_n}

    \begin{center}
    \begin{minipage}{.49\linewidth}
        \begin{center}
            \begin{longtable}[c]{|c|c|}
  \hline
  \thead{$J$ \\ Spec} & \thead{$a_{T_{j}}$} \\
  \hline
  \hline
\makecell{$7988-8110$\\$8161-8277$\\$8326-8436$\\$8483-8588$\\$8633-8734$\\$8777-8875$\\$8917-9011$\\$9052-9143$\\$9183-9272$\\$9310-9396$\\$9433-9518$\\$9554-9637$} & \makecell{$-3.0$-$-2.3$} \\
\hline
  \caption{$T_{c\bar{b}s\bar{d}}$ $(J, M^2)$ predictions.} \label{table:Tcsbd_j_predictions}
 \end{longtable}
        \end{center}
    \end{minipage}
    \begin{minipage}{.49\linewidth}
        \begin{center}
            \begin{longtable}[c]{|c|c|}
  \hline
  \thead{$n$ \\ Spec} & \thead{$a_{T_{n}}$} \\
  \hline
  \hline
\makecell{$7824-7953$\\$8086-8204$\\$8326-8436$\\$8550-8654$\\$8762-8860$\\$8962-9056$\\$9154-9243$\\$9338-9424$\\$9514-9597$\\$9685-9765$\\$9850-9928$\\$10011-10086$} & \makecell{$-2.2$-$-1.8$} \\
\hline
  \caption{$T_{c\bar{b}s\bar{d}}$ $(n, M^2)$ predictions.} \label{table:Tcsbd_n_predictions}
 \end{longtable}
        \end{center}
    \end{minipage}
    \end{center}

\subsubsection{\texorpdfstring{$T_{b\bar{u}c\bar{u}}$}{}}\label{T_bcuu}

\twographs[(a) is $(J, M^{2})$ $T_{b\bar{u}c\bar{u}}$ HMRT, (b) $(n, M^{2})$ $T_{b\bar{u}c\bar{u}}$ HMRT.][T_bcuu]{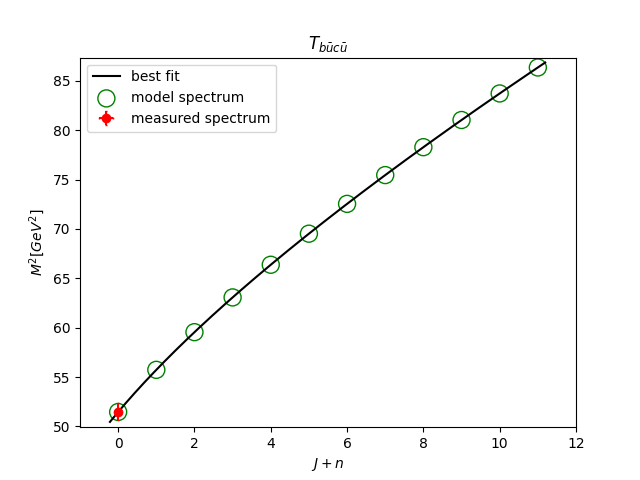}{tetraquarks/predictions/bc-uu/by_n/T_bcuu_one_by_n}

    \begin{center}
    \begin{minipage}{.49\linewidth}
        \begin{center}
            \begin{longtable}[c]{|c|c|}
  \hline
  \thead{$J$ \\ Spec} & \thead{$a_{T_{j}}$} \\
  \hline
  \hline
\makecell{$7154-7194$\\$7449-7482$\\$7702-7732$\\$7928-7955$\\$8135-8159$\\$8326-8349$\\$8506-8528$\\$8676-8697$\\$8838-8857$\\$8992-9011$\\$9141-9159$\\$9284-9301$} & \makecell{$-2.3$-$-2.1$} \\
\hline
  \caption{$T_{b\bar{u}c\bar{u}}$ $(J, M^2)$ predictions.} \label{table:Tbcuu_j_predictions}
 \end{longtable}
        \end{center}
    \end{minipage}
    \begin{minipage}{.49\linewidth}
        \begin{center}
            \begin{longtable}[c]{|c|c|}
  \hline
  \thead{$n$ \\ Spec} & \thead{$a_{T_{n}}$} \\
  \hline
  \hline
\makecell{$7154-7194$\\$7565-7597$\\$7904-7931$\\$8200-8224$\\$8467-8489$\\$8713-8733$\\$8942-8961$\\$9157-9175$\\$9361-9378$\\$9556-9572$\\$9742-9758$\\$9921-9937$} & \makecell{$-1.6$-$-1.5$} \\
\hline
  \caption{$T_{b\bar{u}c\bar{u}}$ $(n, M^2)$ predictions.} \label{table:Tbcuu_n_predictions}
 \end{longtable}
        \end{center}
    \end{minipage}
    \end{center}

\subsubsection{\texorpdfstring{$T_{b\bar{d}c\bar{u}}$}{}}\label{T_bcdu}

\twographs[(a) is $(J, M^{2})$ $T_{b\bar{d}c\bar{u}}$ HMRT, (b) $(n, M^{2})$ $T_{b\bar{d}c\bar{u}}$ HMRT.][T_bcdu]{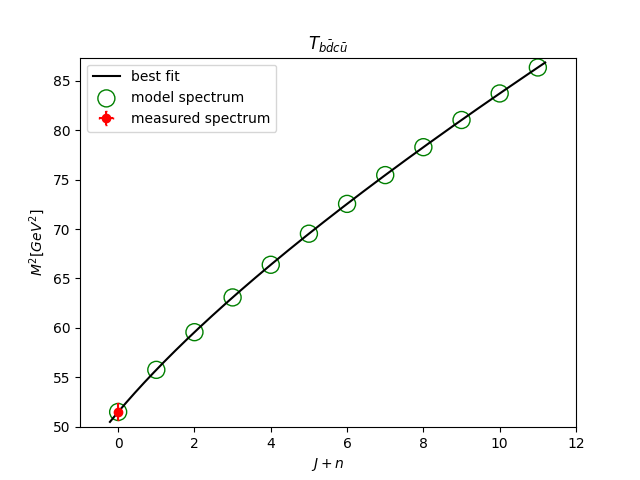}{tetraquarks/predictions/bc-du/by_n/T_bcdu_one_by_n}

    \begin{center}
    \begin{minipage}{.49\linewidth}
        \begin{center}
            \begin{longtable}[c]{|c|c|}
  \hline
  \thead{$J$ \\ Spec} & \thead{$a_{T_{j}}$} \\
  \hline
  \hline
\makecell{$7155-7195$\\$7450-7483$\\$7703-7732$\\$7929-7956$\\$8135-8160$\\$8327-8350$\\$8507-8528$\\$8677-8697$\\$8838-8858$\\$8993-9012$\\$9141-9159$\\$9284-9302$} & \makecell{$-2.3$-$-2.1$} \\
\hline
  \caption{$T_{b\bar{d}c\bar{u}}$ $(J, M^2)$ predictions.} \label{table:Tbcdu_j_predictions}
 \end{longtable}
        \end{center}
    \end{minipage}
    \begin{minipage}{.49\linewidth}
        \begin{center}
            \begin{longtable}[c]{|c|c|}
  \hline
  \thead{$n$ \\ Spec} & \thead{$a_{T_{n}}$} \\
  \hline
  \hline
\makecell{$7155-7195$\\$7566-7598$\\$7905-7932$\\$8201-8225$\\$8468-8490$\\$8713-8734$\\$8942-8961$\\$9157-9175$\\$9361-9379$\\$9556-9573$\\$9743-9758$\\$9922-9937$} & \makecell{$-1.6$-$-1.5$} \\
\hline
  \caption{$T_{b\bar{d}c\bar{u}}$ $(n, M^2)$ predictions.} \label{table:Tbcdu_n_predictions}
 \end{longtable}
        \end{center}
    \end{minipage}
    \end{center}

\subsubsection{\texorpdfstring{$T_{b\bar{d}c\bar{d}}$}{}}\label{T_bcdd}

\twographs[(a) is $(J, M^{2})$ $T_{b\bar{d}c\bar{d}}$ HMRT, (b) $(n, M^{2})$ $T_{b\bar{d}c\bar{d}}$ HMRT.][T_bcdd]{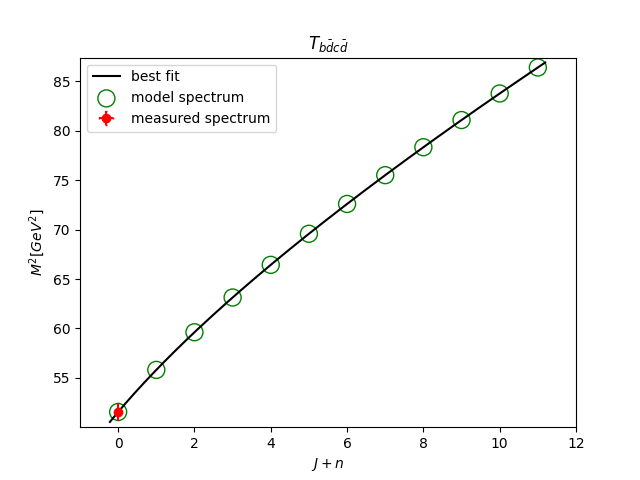}{tetraquarks/predictions/bc-dd/by_n/T_bcdd_one_by_n}

    \begin{center}
    \begin{minipage}{.49\linewidth}
        \begin{center}
            \begin{longtable}[c]{|c|c|}
  \hline
  \thead{$J$ \\ Spec} & \thead{$a_{T_{j}}$} \\
  \hline
  \hline
\makecell{$7160-7200$\\$7454-7487$\\$7706-7736$\\$7932-7959$\\$8138-8163$\\$8330-8353$\\$8509-8531$\\$8679-8700$\\$8841-8860$\\$8995-9014$\\$9143-9162$\\$9286-9304$} & \makecell{$-2.3$-$-2.1$} \\
\hline
  \caption{$T_{b\bar{d}c\bar{d}}$ $(J, M^2)$ predictions.} \label{table:Tbcdd_j_predictions}
 \end{longtable}
        \end{center}
    \end{minipage}
    \begin{minipage}{.49\linewidth}
        \begin{center}
            \begin{longtable}[c]{|c|c|}
  \hline
  \thead{$n$ \\ Spec} & \thead{$a_{T_{n}}$} \\
  \hline
  \hline
\makecell{$7160-7200$\\$7570-7602$\\$7908-7935$\\$8204-8228$\\$8470-8493$\\$8716-8736$\\$8944-8964$\\$9160-9178$\\$9364-9381$\\$9558-9575$\\$9744-9760$\\$9924-9939$} & \makecell{$-1.6$-$-1.5$} \\
\hline
  \caption{$T_{b\bar{d}c\bar{d}}$ $(n, M^2)$ predictions.} \label{table:Tbcdd_n_predictions}
 \end{longtable}
        \end{center}
    \end{minipage}
    \end{center}

\subsubsection{\texorpdfstring{$T_{b\bar{s}c\bar{u}}$}{}}\label{T_bcsu}

\twographs[(a) is $(J, M^{2})$ $T_{b\bar{s}c\bar{u}}$ HMRT, (b) $(n, M^{2})$ $T_{b\bar{s}c\bar{u}}$ HMRT.][T_bcsu]{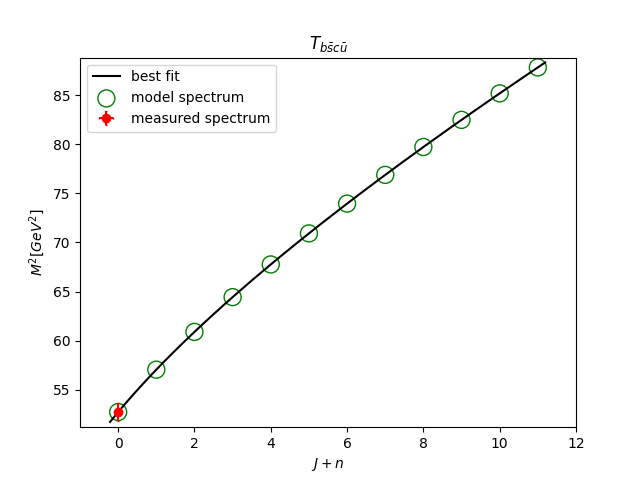}{tetraquarks/predictions/bc-su/by_n/T_bcsu_one_by_n}

    \begin{center}
    \begin{minipage}{.49\linewidth}
        \begin{center}
            \begin{longtable}[c]{|c|c|}
  \hline
  \thead{$J$ \\ Spec} & \thead{$a_{T_{j}}$} \\
  \hline
  \hline
\makecell{$7242-7282$\\$7537-7570$\\$7789-7818$\\$8014-8040$\\$8219-8244$\\$8410-8433$\\$8589-8610$\\$8758-8778$\\$8918-8938$\\$9072-9091$\\$9220-9238$\\$9362-9380$} & \makecell{$-2.0$-$-1.9$} \\
\hline
  \caption{$T_{b\bar{s}c\bar{u}}$ $(J, M^2)$ predictions.} \label{table:Tbcsu_j_predictions}
 \end{longtable}
        \end{center}
    \end{minipage}
    \begin{minipage}{.49\linewidth}
        \begin{center}
            \begin{longtable}[c]{|c|c|}
  \hline
  \thead{$n$ \\ Spec} & \thead{$a_{T_{n}}$} \\
  \hline
  \hline
\makecell{$7242-7282$\\$7653-7684$\\$7990-8017$\\$8284-8308$\\$8550-8572$\\$8794-8814$\\$9022-9041$\\$9236-9254$\\$9439-9456$\\$9633-9649$\\$9818-9834$\\$9997-10012$} & \makecell{$-1.4$-$-1.3$} \\
\hline
  \caption{$T_{b\bar{s}c\bar{u}}$ $(n, M^2)$ predictions.} \label{table:Tbcsu_n_predictions}
 \end{longtable}
        \end{center}
    \end{minipage}
    \end{center}

\subsubsection{\texorpdfstring{$T_{b\bar{d}c\bar{s}}$}{}}\label{T_bcds}

\twographs[(a) is $(J, M^{2})$ $T_{b\bar{d}c\bar{s}}$ HMRT, (b) $(n, M^{2})$ $T_{b\bar{d}c\bar{s}}$ HMRT.][T_bcds]{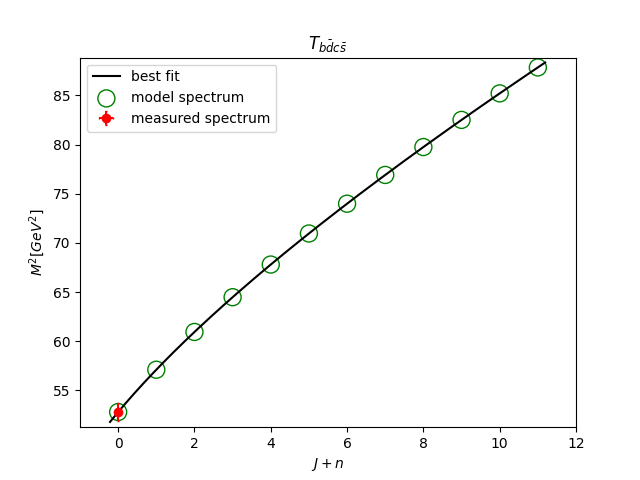}{tetraquarks/predictions/bc-ds/by_n/T_bcds_one_by_n}

    \begin{center}
    \begin{minipage}{.49\linewidth}
        \begin{center}
            \begin{longtable}[c]{|c|c|}
  \hline
  \thead{$J$ \\ Spec} & \thead{$a_{T_{j}}$} \\
  \hline
  \hline
\makecell{$7247-7287$\\$7541-7574$\\$7792-7822$\\$8017-8044$\\$8222-8247$\\$8413-8436$\\$8591-8613$\\$8760-8781$\\$8921-8940$\\$9075-9093$\\$9222-9240$\\$9364-9382$} & \makecell{$-2.0$-$-1.9$} \\
\hline
  \caption{$T_{b\bar{d}c\bar{s}}$ $(J, M^2)$ predictions.} \label{table:Tbcds_j_predictions}
 \end{longtable}
        \end{center}
    \end{minipage}
    \begin{minipage}{.49\linewidth}
        \begin{center}
            \begin{longtable}[c]{|c|c|}
  \hline
  \thead{$n$ \\ Spec} & \thead{$a_{T_{n}}$} \\
  \hline
  \hline
\makecell{$7247-7287$\\$7657-7688$\\$7993-8020$\\$8287-8311$\\$8553-8574$\\$8797-8817$\\$9024-9043$\\$9238-9256$\\$9441-9458$\\$9635-9651$\\$9820-9836$\\$9999-10014$} & \makecell{$-1.4$-$-1.3$} \\
\hline
  \caption{$T_{b\bar{d}c\bar{s}}$ $(n, M^2)$ predictions.} \label{table:Tbcds_n_predictions}
 \end{longtable}
        \end{center}
    \end{minipage}
    \end{center}

\subsubsection{\texorpdfstring{$T_{b\bar{s}c\bar{s}}$}{}}\label{T_bcss}

\twographs[(a) is $(J, M^{2})$ $T_{b\bar{s}c\bar{s}}$ HMRT, (b) $(n, M^{2})$ $T_{b\bar{s}c\bar{s}}$ HMRT.][T_bcss]{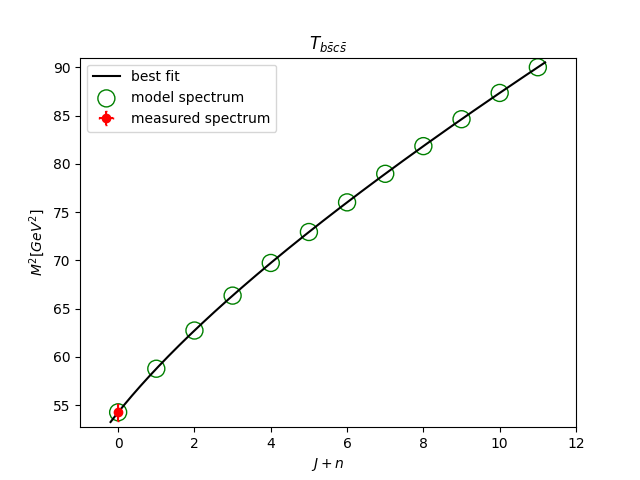}{tetraquarks/predictions/bc-ss/by_n/T_bcss_one_by_n}

    \begin{center}
    \begin{minipage}{.49\linewidth}
        \begin{center}
            \begin{longtable}[c]{|c|c|}
  \hline
  \thead{$J$ \\ Spec} & \thead{$a_{T_{j}}$} \\
  \hline
  \hline
\makecell{$7345-7385$\\$7649-7682$\\$7906-7934$\\$8132-8158$\\$8339-8362$\\$8529-8551$\\$8708-8729$\\$8877-8896$\\$9037-9056$\\$9191-9209$\\$9338-9355$\\$9480-9496$} & \makecell{$-1.6$-$-1.5$} \\
\hline
  \caption{$T_{b\bar{s}c\bar{s}}$ $(J, M^2)$ predictions.} \label{table:Tbcss_j_predictions}
 \end{longtable}
        \end{center}
    \end{minipage}
    \begin{minipage}{.49\linewidth}
        \begin{center}
            \begin{longtable}[c]{|c|c|}
  \hline
  \thead{$n$ \\ Spec} & \thead{$a_{T_{n}}$} \\
  \hline
  \hline
\makecell{$7345-7385$\\$7768-7798$\\$8108-8134$\\$8404-8427$\\$8669-8690$\\$8913-8933$\\$9140-9158$\\$9354-9371$\\$9556-9573$\\$9749-9765$\\$9934-9949$\\$10112-10126$} & \makecell{$-1.1$-$-1.0$} \\
\hline
  \caption{$T_{b\bar{s}c\bar{s}}$ $(n, M^2)$ predictions.} \label{table:Tbcss_n_predictions}
 \end{longtable}
        \end{center}
    \end{minipage}
    \end{center}

\subsubsection{\texorpdfstring{$T_{b\bar{c}c\bar{u}}$}{}}\label{T_bccu}

\twographs[(a) is $(J, M^{2})$ $T_{b\bar{c}c\bar{u}}$ HMRT, (b) $(n, M^{2})$ $T_{b\bar{c}c\bar{u}}$ HMRT.][T_bccu]{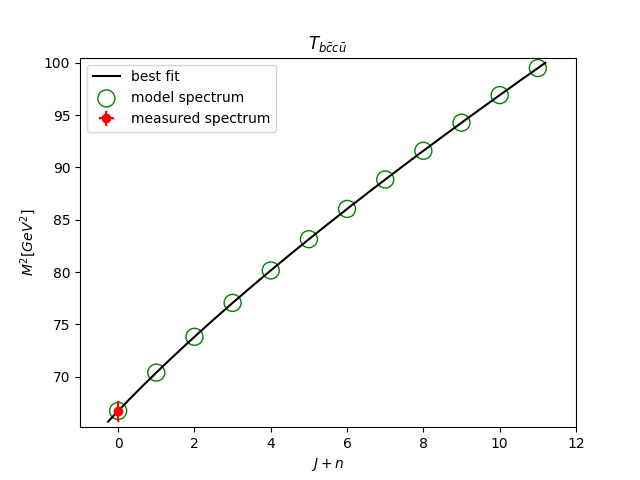}{tetraquarks/predictions/bc-cu/by_n/T_bccu_one_by_n}

    \begin{center}
    \begin{minipage}{.49\linewidth}
        \begin{center}
            \begin{longtable}[c]{|c|c|}
  \hline
  \thead{$J$ \\ Spec} & \thead{$a_{T_{j}}$} \\
  \hline
  \hline
\makecell{$8149-8189$\\$8372-8408$\\$8575-8608$\\$8763-8794$\\$8939-8968$\\$9105-9132$\\$9262-9289$\\$9413-9439$\\$9558-9583$\\$9698-9722$\\$9833-9856$\\$9964-9986$} & \makecell{$-3.4$-$-3.2$} \\
\hline
  \caption{$T_{b\bar{c}c\bar{u}}$ $(J, M^2)$ predictions.} \label{table:Tbccu_j_predictions}
 \end{longtable}
        \end{center}
    \end{minipage}
    \begin{minipage}{.49\linewidth}
        \begin{center}
            \begin{longtable}[c]{|c|c|}
  \hline
  \thead{$n$ \\ Spec} & \thead{$a_{T_{n}}$} \\
  \hline
  \hline
\makecell{$8149-8189$\\$8465-8499$\\$8743-8774$\\$8995-9023$\\$9228-9255$\\$9446-9471$\\$9652-9676$\\$9848-9870$\\$10035-10057$\\$10215-10235$\\$10387-10408$\\$10555-10574$} & \makecell{$-2.3$-$-2.2$} \\
\hline
  \caption{$T_{b\bar{c}c\bar{u}}$ $(n, M^2)$ predictions.} \label{table:Tbccu_n_predictions}
 \end{longtable}
        \end{center}
    \end{minipage}
    \end{center}

\subsubsection{\texorpdfstring{$T_{c\bar{b}u\bar{c}}$}{}}\label{T_cubc}

\twographs[(a) is $(J, M^{2})$ $T_{c\bar{b}u\bar{c}}$ HMRT, (b) $(n, M^{2})$ $T_{c\bar{b}u\bar{c}}$ HMRT.][T_cubc]{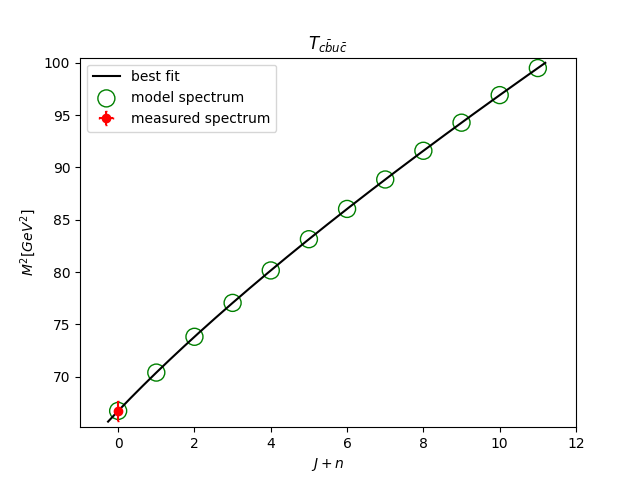}{tetraquarks/predictions/cu-bc/by_n/T_cubc_one_by_n}

    \begin{center}
    \begin{minipage}{.49\linewidth}
        \begin{center}
            \begin{longtable}[c]{|c|c|}
  \hline
  \thead{$J$ \\ Spec} & \thead{$a_{T_{j}}$} \\
  \hline
  \hline
\makecell{$8149-8189$\\$8372-8408$\\$8575-8608$\\$8763-8794$\\$8939-8968$\\$9105-9132$\\$9262-9289$\\$9413-9439$\\$9558-9583$\\$9698-9722$\\$9833-9856$\\$9964-9986$} & \makecell{$-3.4$-$-3.2$} \\
\hline
  \caption{$T_{c\bar{b}u\bar{c}}$ $(J, M^2)$ predictions.} \label{table:Tcubc_j_predictions}
 \end{longtable}
        \end{center}
    \end{minipage}
    \begin{minipage}{.49\linewidth}
        \begin{center}
            \begin{longtable}[c]{|c|c|}
  \hline
  \thead{$n$ \\ Spec} & \thead{$a_{T_{n}}$} \\
  \hline
  \hline
\makecell{$8149-8189$\\$8465-8499$\\$8743-8774$\\$8995-9023$\\$9228-9255$\\$9446-9471$\\$9652-9676$\\$9848-9870$\\$10035-10057$\\$10215-10235$\\$10387-10408$\\$10555-10574$} & \makecell{$-2.3$-$-2.2$} \\
\hline
  \caption{$T_{c\bar{b}u\bar{c}}$ $(n, M^2)$ predictions.} \label{table:Tcubc_n_predictions}
 \end{longtable}
        \end{center}
    \end{minipage}
    \end{center}

\subsubsection{\texorpdfstring{$T_{c\bar{b}c\bar{u}}$}{}}\label{T_ccbu}

\twographs[(a) is $(J, M^{2})$ $T_{c\bar{b}c\bar{u}}$ HMRT, (b) $(n, M^{2})$ $T_{c\bar{b}c\bar{u}}$ HMRT.][T_ccbu]{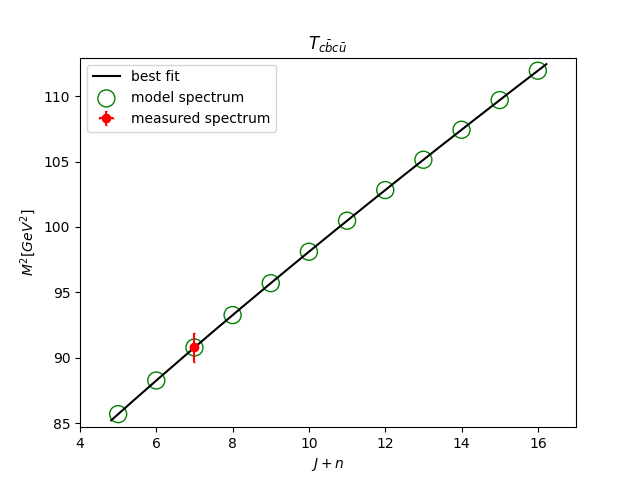}{tetraquarks/predictions/cc-bu/by_n/T_ccbu_one_by_n}

    \begin{center}
    \begin{minipage}{.49\linewidth}
        \begin{center}
            \begin{longtable}[c]{|c|c|}
  \hline
  \thead{$J$ \\ Spec} & \thead{$a_{T_{j}}$} \\
  \hline
  \hline
\makecell{$9198-9316$\\$9338-9451$\\$9473-9583$\\$9604-9711$\\$9731-9835$\\$9854-9955$\\$9975-10073$\\$10092-10188$\\$10207-10301$\\$10319-10411$\\$10429-10519$\\$10536-10625$} & \makecell{$-5.9$-$-5.1$} \\
\hline
  \caption{$T_{c\bar{b}c\bar{u}}$ $(J, M^2)$ predictions.} \label{table:Tccbu_j_predictions}
 \end{longtable}
        \end{center}
    \end{minipage}
    \begin{minipage}{.49\linewidth}
        \begin{center}
            \begin{longtable}[c]{|c|c|}
  \hline
  \thead{$n$ \\ Spec} & \thead{$a_{T_{n}}$} \\
  \hline
  \hline
\makecell{$9069-9191$\\$9276-9392$\\$9473-9583$\\$9661-9766$\\$9841-9942$\\$10014-10112$\\$10181-10276$\\$10343-10435$\\$10501-10590$\\$10654-10740$\\$10803-10887$\\$10948-11031$} & \makecell{$-3.9$-$-3.4$} \\
\hline
  \caption{$T_{c\bar{b}c\bar{u}}$ $(n, M^2)$ predictions.} \label{table:Tccbu_n_predictions}
 \end{longtable}
        \end{center}
    \end{minipage}
    \end{center}

\subsubsection{\texorpdfstring{$T_{c\bar{b}c\bar{d}}$}{}}\label{T_ccbd}

\twographs[(a) is $(J, M^{2})$ $T_{c\bar{b}c\bar{d}}$ HMRT, (b) $(n, M^{2})$ $T_{c\bar{b}c\bar{d}}$ HMRT.][T_ccbd]{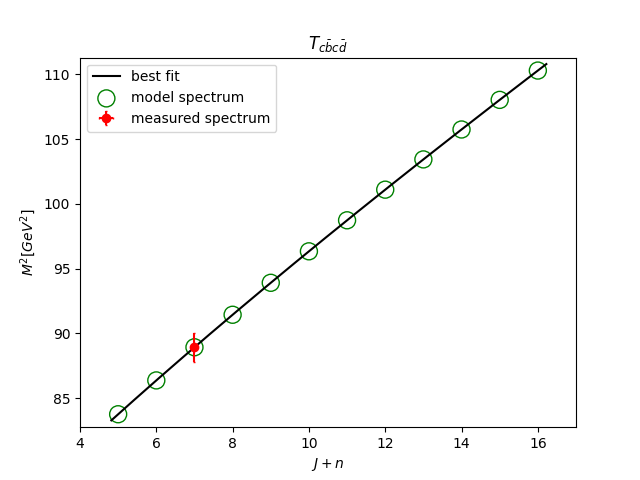}{tetraquarks/predictions/cc-bd/by_n/T_ccbd_one_by_n}

    \begin{center}
    \begin{minipage}{.49\linewidth}
        \begin{center}
            \begin{longtable}[c]{|c|c|}
  \hline
  \thead{$J$ \\ Spec} & \thead{$a_{T_{j}}$} \\
  \hline
  \hline
\makecell{$9093-9211$\\$9236-9350$\\$9375-9485$\\$9509-9615$\\$9639-9742$\\$9765-9865$\\$9887-9985$\\$10007-10102$\\$10123-10217$\\$10237-10329$\\$10349-10438$\\$10458-10546$} & \makecell{$-5.2$-$-4.3$} \\
\hline
  \caption{$T_{c\bar{b}c\bar{d}}$ $(J, M^2)$ predictions.} \label{table:Tccbd_j_predictions}
 \end{longtable}
        \end{center}
    \end{minipage}
    \begin{minipage}{.49\linewidth}
        \begin{center}
            \begin{longtable}[c]{|c|c|}
  \hline
  \thead{$n$ \\ Spec} & \thead{$a_{T_{n}}$} \\
  \hline
  \hline
\makecell{$8960-9082$\\$9173-9289$\\$9375-9485$\\$9567-9672$\\$9751-9852$\\$9927-10025$\\$10098-10192$\\$10262-10353$\\$10422-10510$\\$10577-10663$\\$10728-10812$\\$10875-10957$} & \makecell{$-4.4$-$-3.8$} \\
\hline
  \caption{$T_{c\bar{b}c\bar{d}}$ $(n, M^2)$ predictions.} \label{table:Tccbd_n_predictions}
 \end{longtable}
        \end{center}
    \end{minipage}
    \end{center}

\subsubsection{\texorpdfstring{$T_{c\bar{b}d\bar{c}}$}{}}\label{T_cdbc}

\twographs[(a) is $(J, M^{2})$ $T_{c\bar{b}d\bar{c}}$ HMRT, (b) $(n, M^{2})$ $T_{c\bar{b}d\bar{c}}$ HMRT.][T_cdbc]{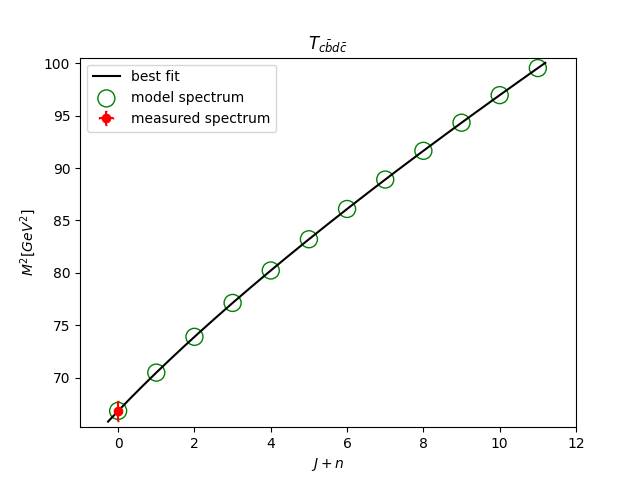}{tetraquarks/predictions/cd-bc/by_n/T_cdbc_one_by_n}

    \begin{center}
    \begin{minipage}{.49\linewidth}
        \begin{center}
            \begin{longtable}[c]{|c|c|}
  \hline
  \thead{$J$ \\ Spec} & \thead{$a_{T_{j}}$} \\
  \hline
  \hline
\makecell{$8154-8194$\\$8377-8413$\\$8579-8612$\\$8767-8798$\\$8942-8971$\\$9108-9136$\\$9266-9292$\\$9417-9442$\\$9561-9586$\\$9701-9724$\\$9836-9859$\\$9967-9989$} & \makecell{$-3.4$-$-3.2$} \\
\hline
  \caption{$T_{c\bar{b}d\bar{c}}$ $(J, M^2)$ predictions.} \label{table:Tcdbc_j_predictions}
 \end{longtable}
        \end{center}
    \end{minipage}
    \begin{minipage}{.49\linewidth}
        \begin{center}
            \begin{longtable}[c]{|c|c|}
  \hline
  \thead{$n$ \\ Spec} & \thead{$a_{T_{n}}$} \\
  \hline
  \hline
\makecell{$8154-8194$\\$8469-8504$\\$8746-8778$\\$8998-9027$\\$9231-9258$\\$9449-9474$\\$9655-9679$\\$9851-9873$\\$10038-10059$\\$10217-10238$\\$10390-10410$\\$10557-10576$} & \makecell{$-2.4$-$-2.2$} \\
\hline
  \caption{$T_{c\bar{b}d\bar{c}}$ $(n, M^2)$ predictions.} \label{table:Tcdbc_n_predictions}
 \end{longtable}
        \end{center}
    \end{minipage}
    \end{center}

\subsubsection{\texorpdfstring{$T_{b\bar{c}c\bar{d}}$}{}}\label{T_bccd}

\twographs[(a) is $(J, M^{2})$ $T_{b\bar{c}c\bar{d}}$ HMRT, (b) $(n, M^{2})$ $T_{b\bar{c}c\bar{d}}$ HMRT.][T_bccd]{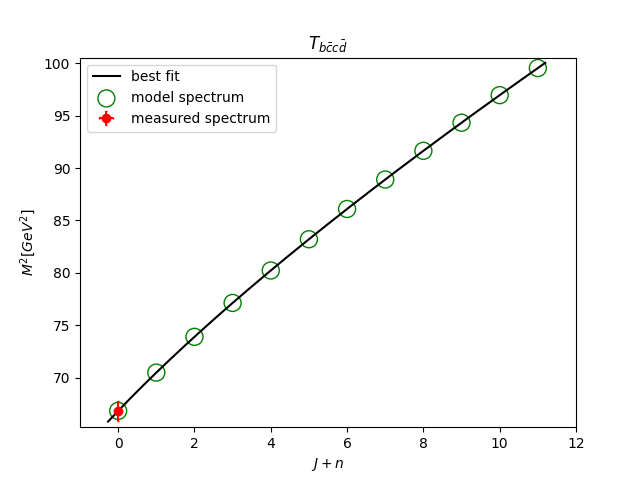}{tetraquarks/predictions/bc-cd/by_n/T_bccd_one_by_n}

    \begin{center}
    \begin{minipage}{.49\linewidth}
        \begin{center}
            \begin{longtable}[c]{|c|c|}
  \hline
  \thead{$J$ \\ Spec} & \thead{$a_{T_{j}}$} \\
  \hline
  \hline
\makecell{$8154-8194$\\$8377-8413$\\$8579-8612$\\$8767-8798$\\$8942-8971$\\$9108-9136$\\$9266-9292$\\$9417-9442$\\$9561-9586$\\$9701-9724$\\$9836-9859$\\$9967-9989$} & \makecell{$-3.4$-$-3.2$} \\
\hline
  \caption{$T_{b\bar{c}c\bar{d}}$ $(J, M^2)$ predictions.} \label{table:Tbccd_j_predictions}
 \end{longtable}
        \end{center}
    \end{minipage}
    \begin{minipage}{.49\linewidth}
        \begin{center}
            \begin{longtable}[c]{|c|c|}
  \hline
  \thead{$n$ \\ Spec} & \thead{$a_{T_{n}}$} \\
  \hline
  \hline
\makecell{$8154-8194$\\$8469-8504$\\$8746-8778$\\$8998-9027$\\$9231-9258$\\$9449-9474$\\$9655-9679$\\$9851-9873$\\$10038-10059$\\$10217-10238$\\$10390-10410$\\$10557-10576$} & \makecell{$-2.4$-$-2.2$} \\
\hline
  \caption{$T_{b\bar{c}c\bar{d}}$ $(n, M^2)$ predictions.} \label{table:Tbccd_n_predictions}
 \end{longtable}
        \end{center}
    \end{minipage}
    \end{center}

\subsubsection{\texorpdfstring{$T_{b\bar{c}c\bar{s}}$}{}}\label{T_bccs}

\twographs[(a) is $(J, M^{2})$ $T_{b\bar{c}c\bar{s}}$ HMRT, (b) $(n, M^{2})$ $T_{b\bar{c}c\bar{s}}$ HMRT.][T_bccs]{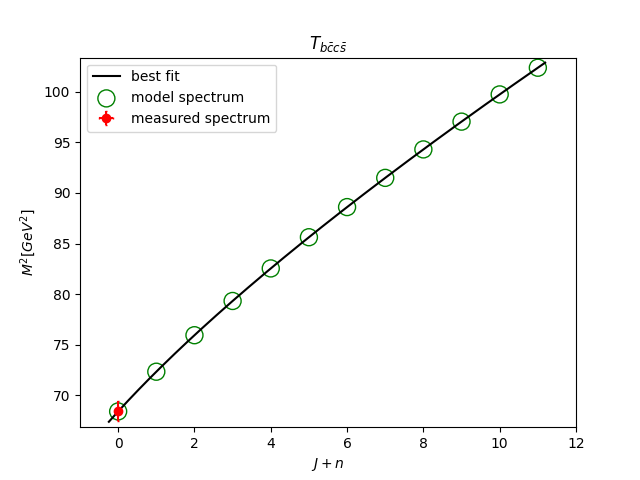}{tetraquarks/predictions/bc-cs/by_n/T_bccs_one_by_n}

    \begin{center}
    \begin{minipage}{.49\linewidth}
        \begin{center}
            \begin{longtable}[c]{|c|c|}
  \hline
  \thead{$J$ \\ Spec} & \thead{$a_{T_{j}}$} \\
  \hline
  \hline
\makecell{$8252-8292$\\$8488-8523$\\$8699-8731$\\$8892-8922$\\$9072-9099$\\$9240-9266$\\$9400-9425$\\$9553-9577$\\$9699-9722$\\$9840-9862$\\$9976-9997$\\$10107-10128$} & \makecell{$-2.6$-$-2.4$} \\
\hline
  \caption{$T_{b\bar{c}c\bar{s}}$ $(J, M^2)$ predictions.} \label{table:Tbccs_j_predictions}
 \end{longtable}
        \end{center}
    \end{minipage}
    \begin{minipage}{.49\linewidth}
        \begin{center}
            \begin{longtable}[c]{|c|c|}
  \hline
  \thead{$n$ \\ Spec} & \thead{$a_{T_{n}}$} \\
  \hline
  \hline
\makecell{$8252-8292$\\$8584-8618$\\$8871-8901$\\$9129-9156$\\$9365-9391$\\$9586-9609$\\$9793-9816$\\$9990-10012$\\$10178-10199$\\$10358-10378$\\$10532-10550$\\$10699-10717$} & \makecell{$-1.8$-$-1.7$} \\
\hline
  \caption{$T_{b\bar{c}c\bar{s}}$ $(n, M^2)$ predictions.} \label{table:Tbccs_n_predictions}
 \end{longtable}
        \end{center}
    \end{minipage}
    \end{center}

\subsubsection{\texorpdfstring{$T_{c\bar{b}c\bar{s}}$}{}}\label{T_ccbs}

\twographs[(a) is $(J, M^{2})$ $T_{c\bar{b}c\bar{s}}$ HMRT, (b) $(n, M^{2})$ $T_{c\bar{b}c\bar{s}}$ HMRT.][T_ccbs]{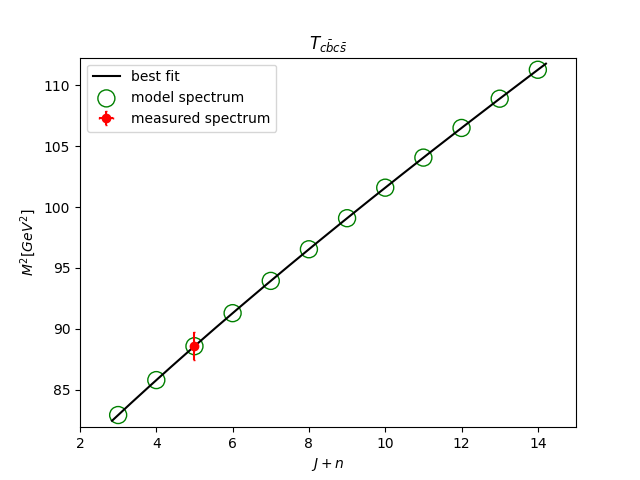}{tetraquarks/predictions/cc-bs/by_n/T_ccbs_one_by_n}

    \begin{center}
    \begin{minipage}{.49\linewidth}
        \begin{center}
            \begin{longtable}[c]{|c|c|}
  \hline
  \thead{$J$ \\ Spec} & \thead{$a_{T_{j}}$} \\
  \hline
  \hline
\makecell{$9045-9166$\\$9204-9319$\\$9356-9466$\\$9501-9607$\\$9641-9743$\\$9776-9875$\\$9907-10002$\\$10033-10126$\\$10156-10247$\\$10276-10364$\\$10393-10479$\\$10507-10592$} & \makecell{$-4.6$-$-3.9$} \\
\hline
  \caption{$T_{c\bar{b}c\bar{s}}$ $(J, M^2)$ predictions.} \label{table:Tccbs_j_predictions}
 \end{longtable}
        \end{center}
    \end{minipage}
    \begin{minipage}{.49\linewidth}
        \begin{center}
            \begin{longtable}[c]{|c|c|}
  \hline
  \thead{$n$ \\ Spec} & \thead{$a_{T_{n}}$} \\
  \hline
  \hline
\makecell{$8896-9023$\\$9134-9252$\\$9356-9466$\\$9564-9668$\\$9761-9860$\\$9949-10044$\\$10129-10220$\\$10302-10390$\\$10470-10555$\\$10631-10714$\\$10788-10868$\\$10941-11019$} & \makecell{$-2.7$-$-2.1$} \\
\hline
  \caption{$T_{c\bar{b}c\bar{s}}$ $(n, M^2)$ predictions.} \label{table:Tccbs_n_predictions}
 \end{longtable}
        \end{center}
    \end{minipage}
    \end{center}

\subsubsection{\texorpdfstring{$T_{c\bar{b}s\bar{c}}$}{}}\label{T_csbc}

\twographs[(a) is $(J, M^{2})$ $T_{c\bar{b}s\bar{c}}$ HMRT, (b) $(n, M^{2})$ $T_{c\bar{b}s\bar{c}}$ HMRT.][T_csbc]{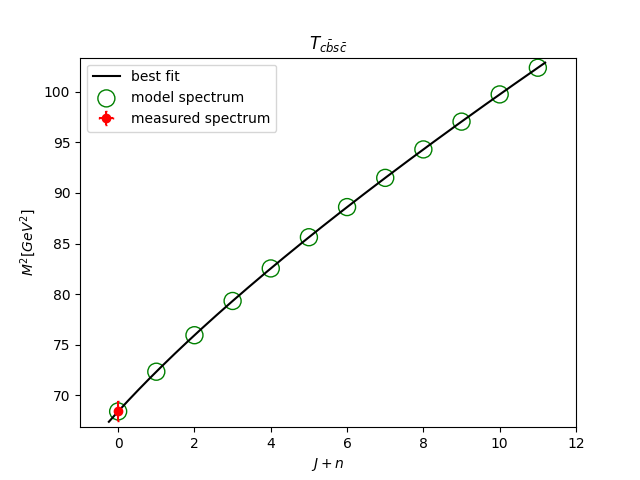}{tetraquarks/predictions/cs-bc/by_n/T_csbc_one_by_n}

    \begin{center}
    \begin{minipage}{.49\linewidth}
        \begin{center}
            \begin{longtable}[c]{|c|c|}
  \hline
  \thead{$J$ \\ Spec} & \thead{$a_{T_{j}}$} \\
  \hline
  \hline
\makecell{$8252-8292$\\$8488-8523$\\$8699-8731$\\$8892-8922$\\$9072-9099$\\$9240-9266$\\$9400-9425$\\$9553-9577$\\$9699-9722$\\$9840-9862$\\$9976-9997$\\$10107-10128$} & \makecell{$-2.6$-$-2.4$} \\
\hline
  \caption{$T_{c\bar{b}s\bar{c}}$ $(J, M^2)$ predictions.} \label{table:Tcsbc_j_predictions}
 \end{longtable}
        \end{center}
    \end{minipage}
    \begin{minipage}{.49\linewidth}
        \begin{center}
            \begin{longtable}[c]{|c|c|}
  \hline
  \thead{$n$ \\ Spec} & \thead{$a_{T_{n}}$} \\
  \hline
  \hline
\makecell{$8252-8292$\\$8584-8618$\\$8871-8901$\\$9129-9156$\\$9365-9391$\\$9586-9609$\\$9793-9816$\\$9990-10012$\\$10178-10199$\\$10358-10378$\\$10532-10550$\\$10699-10717$} & \makecell{$-1.8$-$-1.7$} \\
\hline
  \caption{$T_{c\bar{b}s\bar{c}}$ $(n, M^2)$ predictions.} \label{table:Tcsbc_n_predictions}
 \end{longtable}
        \end{center}
    \end{minipage}
    \end{center}

\subsubsection{\texorpdfstring{$T_{c\bar{b}c\bar{c}}$}{}}\label{T_ccbc}

\twographs[(a) is $(J, M^{2})$ $T_{c\bar{b}c\bar{c}}$ HMRT, (b) $(n, M^{2})$ $T_{c\bar{b}c\bar{c}}$ HMRT.][T_ccbc]{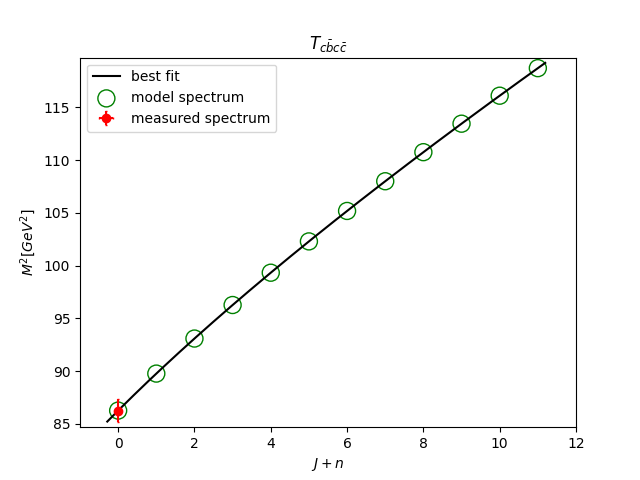}{tetraquarks/predictions/cc-bc/by_n/T_ccbc_one_by_n}

    \begin{center}
    \begin{minipage}{.49\linewidth}
        \begin{center}
            \begin{longtable}[c]{|c|c|}
  \hline
  \thead{$J$ \\ Spec} & \thead{$a_{T_{j}}$} \\
  \hline
  \hline
\makecell{$9268-9308$\\$9456-9493$\\$9631-9666$\\$9795-9828$\\$9951-9982$\\$10099-10129$\\$10241-10270$\\$10378-10406$\\$10510-10537$\\$10638-10664$\\$10762-10787$\\$10883-10907$} & \makecell{$-4.2$-$-4.0$} \\
\hline
  \caption{$T_{c\bar{b}c\bar{c}}$ $(J, M^2)$ predictions.} \label{table:Tccbc_j_predictions}
 \end{longtable}
        \end{center}
    \end{minipage}
    \begin{minipage}{.49\linewidth}
        \begin{center}
            \begin{longtable}[c]{|c|c|}
  \hline
  \thead{$n$ \\ Spec} & \thead{$a_{T_{n}}$} \\
  \hline
  \hline
\makecell{$9268-9308$\\$9535-9571$\\$9777-9810$\\$10001-10032$\\$10210-10239$\\$10408-10435$\\$10596-10622$\\$10776-10801$\\$10949-10973$\\$11115-11138$\\$11276-11299$\\$11432-11454$} & \makecell{$-2.9$-$-2.8$} \\
\hline
  \caption{$T_{c\bar{b}c\bar{c}}$ $(n, M^2)$ predictions.} \label{table:Tccbc_n_predictions}
 \end{longtable}
        \end{center}
    \end{minipage}
    \end{center}

\subsubsection{\texorpdfstring{$T_{b\bar{b}c\bar{u}}$}{}}\label{T_bcbu}

\twographs[(a) is $(J, M^{2})$ $T_{b\bar{b}c\bar{u}}$ HMRT, (b) $(n, M^{2})$ $T_{b\bar{b}c\bar{u}}$ HMRT.][T_bcbu]{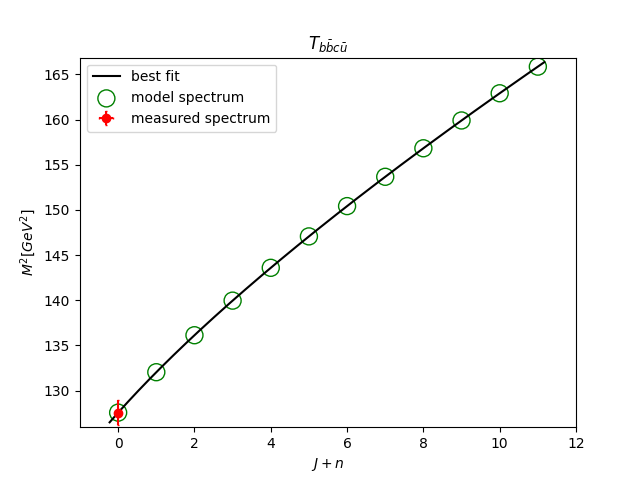}{tetraquarks/predictions/bc-bu/by_n/T_bcbu_one_by_n}

    \begin{center}
    \begin{minipage}{.49\linewidth}
        \begin{center}
            \begin{longtable}[c]{|c|c|}
  \hline
  \thead{$J$ \\ Spec} & \thead{$a_{T_{j}}$} \\
  \hline
  \hline
\makecell{$11274-11314$\\$11473-11508$\\$11651-11683$\\$11815-11845$\\$11969-11997$\\$12114-12141$\\$12252-12277$\\$12384-12409$\\$12511-12535$\\$12634-12657$\\$12753-12775$\\$12868-12890$} & \makecell{$-2.5$-$-2.3$} \\
\hline
  \caption{$T_{b\bar{b}c\bar{u}}$ $(J, M^2)$ predictions.} \label{table:Tbcbu_j_predictions}
 \end{longtable}
        \end{center}
    \end{minipage}
    \begin{minipage}{.49\linewidth}
        \begin{center}
            \begin{longtable}[c]{|c|c|}
  \hline
  \thead{$n$ \\ Spec} & \thead{$a_{T_{n}}$} \\
  \hline
  \hline
\makecell{$11274-11314$\\$11554-11588$\\$11798-11828$\\$12018-12046$\\$12222-12248$\\$12413-12437$\\$12593-12617$\\$12766-12788$\\$12931-12952$\\$13089-13110$\\$13243-13263$\\$13391-13411$} & \makecell{$-1.7$-$-1.6$} \\
\hline
  \caption{$T_{b\bar{b}c\bar{u}}$ $(n, M^2)$ predictions.} \label{table:Tbcbu_n_predictions}
 \end{longtable}
        \end{center}
    \end{minipage}
    \end{center}

\subsubsection{\texorpdfstring{$T_{b\bar{b}c\bar{d}}$}{}}\label{T_bcbd}

\twographs[(a) is $(J, M^{2})$ $T_{b\bar{b}c\bar{d}}$ HMRT, (b) $(n, M^{2})$ $T_{b\bar{b}c\bar{d}}$ HMRT.][T_bcbd]{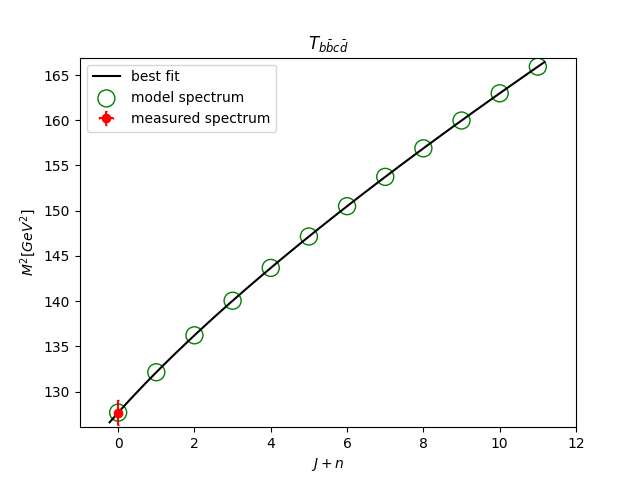}{tetraquarks/predictions/bc-bd/by_n/T_bcbd_one_by_n}

    \begin{center}
    \begin{minipage}{.49\linewidth}
        \begin{center}
            \begin{longtable}[c]{|c|c|}
  \hline
  \thead{$J$ \\ Spec} & \thead{$a_{T_{j}}$} \\
  \hline
  \hline
\makecell{$11279-11319$\\$11477-11513$\\$11655-11687$\\$11819-11849$\\$11972-12001$\\$12117-12144$\\$12255-12281$\\$12387-12412$\\$12514-12538$\\$12637-12660$\\$12755-12778$\\$12871-12892$} & \makecell{$-2.5$-$-2.3$} \\
\hline
  \caption{$T_{b\bar{b}c\bar{d}}$ $(J, M^2)$ predictions.} \label{table:Tbcbd_j_predictions}
 \end{longtable}
        \end{center}
    \end{minipage}
    \begin{minipage}{.49\linewidth}
        \begin{center}
            \begin{longtable}[c]{|c|c|}
  \hline
  \thead{$n$ \\ Spec} & \thead{$a_{T_{n}}$} \\
  \hline
  \hline
\makecell{$11279-11319$\\$11558-11592$\\$11801-11832$\\$12021-12049$\\$12225-12251$\\$12416-12440$\\$12596-12619$\\$12768-12791$\\$12933-12955$\\$13092-13113$\\$13245-13265$\\$13394-13413$} & \makecell{$-1.7$-$-1.6$} \\
\hline
  \caption{$T_{b\bar{b}c\bar{d}}$ $(n, M^2)$ predictions.} \label{table:Tbcbd_n_predictions}
 \end{longtable}
        \end{center}
    \end{minipage}
    \end{center}

\subsubsection{\texorpdfstring{$T_{b\bar{b}c\bar{s}}$}{}}\label{T_bcbs}

\twographs[(a) is $(J, M^{2})$ $T_{b\bar{b}c\bar{s}}$ HMRT, (b) $(n, M^{2})$ $T_{b\bar{b}c\bar{s}}$ HMRT.][T_bcbs]{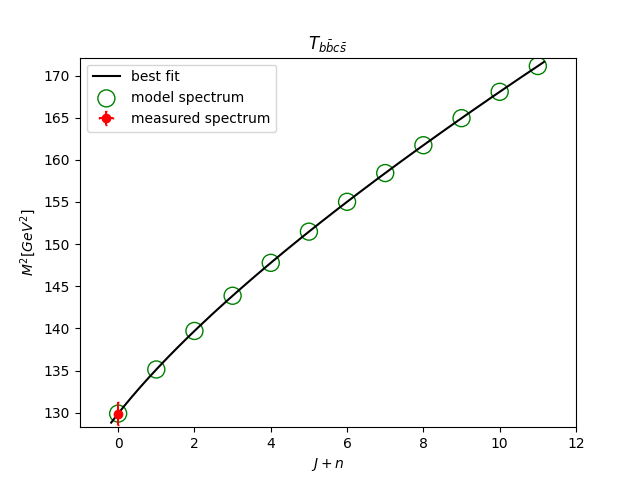}{tetraquarks/predictions/bc-bs/by_n/T_bcbs_one_by_n}

    \begin{center}
    \begin{minipage}{.49\linewidth}
        \begin{center}
            \begin{longtable}[c]{|c|c|}
  \hline
  \thead{$J$ \\ Spec} & \thead{$a_{T_{j}}$} \\
  \hline
  \hline
\makecell{$11377-11417$\\$11608-11641$\\$11805-11834$\\$11981-12008$\\$12144-12169$\\$12296-12319$\\$12440-12462$\\$12576-12598$\\$12707-12728$\\$12833-12853$\\$12955-12974$\\$13073-13091$} & \makecell{$-1.4$-$-1.3$} \\
\hline
  \caption{$T_{b\bar{b}c\bar{s}}$ $(J, M^2)$ predictions.} \label{table:Tbcbs_j_predictions}
 \end{longtable}
        \end{center}
    \end{minipage}
    \begin{minipage}{.49\linewidth}
        \begin{center}
            \begin{longtable}[c]{|c|c|}
  \hline
  \thead{$n$ \\ Spec} & \thead{$a_{T_{n}}$} \\
  \hline
  \hline
\makecell{$11377-11417$\\$11699-11730$\\$11963-11990$\\$12196-12220$\\$12408-12431$\\$12606-12627$\\$12792-12812$\\$12968-12987$\\$13136-13155$\\$13298-13315$\\$13454-13470$\\$13604-13620$} & \makecell{$-0.98$-$-0.87$} \\
\hline
  \caption{$T_{b\bar{b}c\bar{s}}$ $(n, M^2)$ predictions.} \label{table:Tbcbs_n_predictions}
 \end{longtable}
        \end{center}
    \end{minipage}
    \end{center}

\subsubsection{\texorpdfstring{$T_{c\bar{b}u\bar{b}}$}{}}\label{T_cubb}

\twographs[(a) is $(J, M^{2})$ $T_{c\bar{b}u\bar{b}}$ HMRT, (b) $(n, M^{2})$ $T_{c\bar{b}u\bar{b}}$ HMRT.][T_cubb]{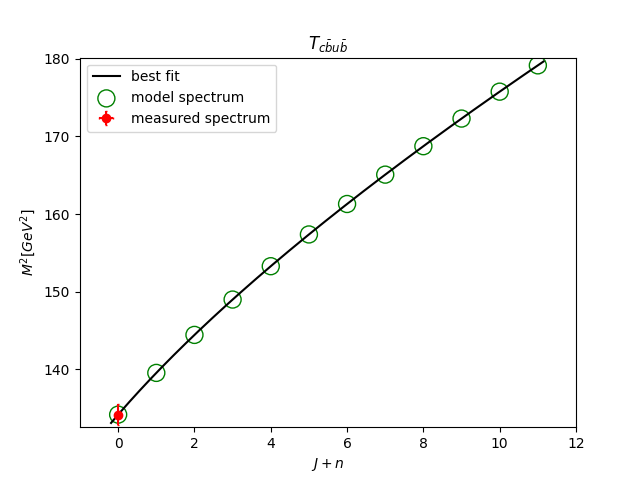}{tetraquarks/predictions/cu-bb/by_n/T_cubb_one_by_n}

    \begin{center}
    \begin{minipage}{.49\linewidth}
        \begin{center}
            \begin{longtable}[c]{|c|c|}
  \hline
  \thead{$J$ \\ Spec} & \thead{$a_{T_{j}}$} \\
  \hline
  \hline
\makecell{$11563-11603$\\$11795-11830$\\$12002-12034$\\$12191-12221$\\$12367-12395$\\$12532-12558$\\$12688-12713$\\$12837-12860$\\$12979-13001$\\$13116-13137$\\$13247-13268$\\$13375-13395$} & \makecell{$-2.7$-$-2.6$} \\
\hline
  \caption{$T_{c\bar{b}u\bar{b}}$ $(J, M^2)$ predictions.} \label{table:Tcubb_j_predictions}
 \end{longtable}
        \end{center}
    \end{minipage}
    \begin{minipage}{.49\linewidth}
        \begin{center}
            \begin{longtable}[c]{|c|c|}
  \hline
  \thead{$n$ \\ Spec} & \thead{$a_{T_{n}}$} \\
  \hline
  \hline
\makecell{$11563-11603$\\$11890-11923$\\$12171-12201$\\$12423-12450$\\$12654-12679$\\$12869-12892$\\$13071-13093$\\$13262-13283$\\$13444-13464$\\$13618-13637$\\$13786-13804$\\$13948-13965$} & \makecell{$-1.9$-$-1.8$} \\
\hline
  \caption{$T_{c\bar{b}u\bar{b}}$ $(n, M^2)$ predictions.} \label{table:Tcubb_n_predictions}
 \end{longtable}
        \end{center}
    \end{minipage}
    \end{center}

\subsubsection{\texorpdfstring{$T_{c\bar{b}d\bar{b}}$}{}}\label{T_cdbb}

\twographs[(a) is $(J, M^{2})$ $T_{c\bar{b}d\bar{b}}$ HMRT, (b) $(n, M^{2})$ $T_{c\bar{b}d\bar{b}}$ HMRT.][T_cdbb]{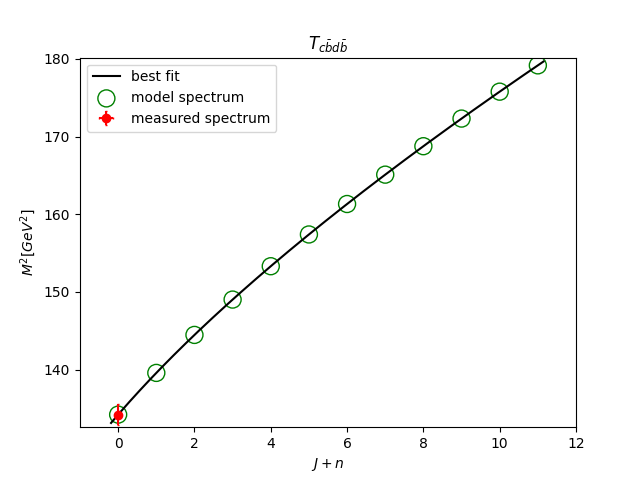}{tetraquarks/predictions/cd-bb/by_n/T_cdbb_one_by_n}

    \begin{center}
    \begin{minipage}{.49\linewidth}
        \begin{center}
            \begin{longtable}[c]{|c|c|}
  \hline
  \thead{$J$ \\ Spec} & \thead{$a_{T_{j}}$} \\
  \hline
  \hline
\makecell{$11564-11604$\\$11796-11831$\\$12003-12035$\\$12192-12222$\\$12368-12395$\\$12533-12559$\\$12689-12713$\\$12837-12861$\\$12979-13002$\\$13116-13138$\\$13248-13269$\\$13375-13396$} & \makecell{$-2.7$-$-2.6$} \\
\hline
  \caption{$T_{c\bar{b}d\bar{b}}$ $(J, M^2)$ predictions.} \label{table:Tcdbb_j_predictions}
 \end{longtable}
        \end{center}
    \end{minipage}
    \begin{minipage}{.49\linewidth}
        \begin{center}
            \begin{longtable}[c]{|c|c|}
  \hline
  \thead{$n$ \\ Spec} & \thead{$a_{T_{n}}$} \\
  \hline
  \hline
\makecell{$11564-11604$\\$11891-11924$\\$12172-12202$\\$12424-12451$\\$12655-12680$\\$12869-12893$\\$13071-13093$\\$13262-13283$\\$13444-13464$\\$13619-13638$\\$13786-13805$\\$13948-13966$} & \makecell{$-1.9$-$-1.8$} \\
\hline
  \caption{$T_{c\bar{b}d\bar{b}}$ $(n, M^2)$ predictions.} \label{table:Tcdbb_n_predictions}
 \end{longtable}
        \end{center}
    \end{minipage}
    \end{center}

\subsubsection{\texorpdfstring{$T_{c\bar{b}s\bar{b}}$}{}}\label{T_csbb}

\twographs[(a) is $(J, M^{2})$ $T_{c\bar{b}s\bar{b}}$ HMRT, (b) $(n, M^{2})$ $T_{c\bar{b}s\bar{b}}$ HMRT.][T_csbb]{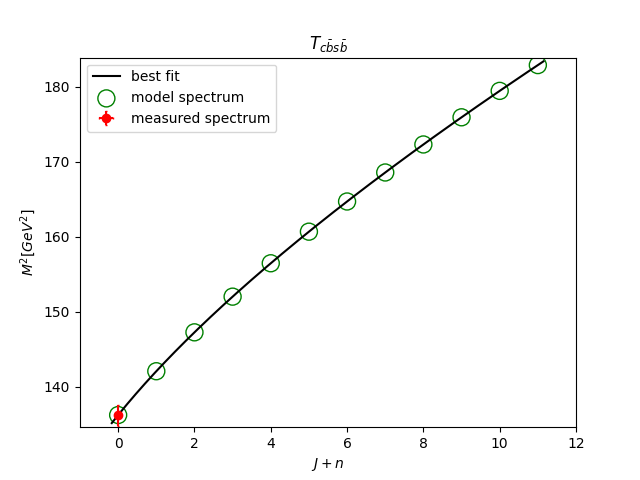}{tetraquarks/predictions/cs-bb/by_n/T_csbb_one_by_n}

    \begin{center}
    \begin{minipage}{.49\linewidth}
        \begin{center}
            \begin{longtable}[c]{|c|c|}
  \hline
  \thead{$J$ \\ Spec} & \thead{$a_{T_{j}}$} \\
  \hline
  \hline
\makecell{$11651-11691$\\$11902-11936$\\$12120-12150$\\$12316-12343$\\$12496-12521$\\$12664-12688$\\$12822-12845$\\$12973-12995$\\$13117-13137$\\$13254-13274$\\$13387-13406$\\$13515-13534$} & \makecell{$-1.9$-$-1.8$} \\
\hline
  \caption{$T_{c\bar{b}s\bar{b}}$ $(J, M^2)$ predictions.} \label{table:Tcsbb_j_predictions}
 \end{longtable}
        \end{center}
    \end{minipage}
    \begin{minipage}{.49\linewidth}
        \begin{center}
            \begin{longtable}[c]{|c|c|}
  \hline
  \thead{$n$ \\ Spec} & \thead{$a_{T_{n}}$} \\
  \hline
  \hline
\makecell{$11651-11691$\\$12002-12034$\\$12295-12322$\\$12553-12578$\\$12788-12811$\\$13005-13027$\\$13209-13229$\\$13401-13421$\\$13584-13603$\\$13759-13777$\\$13927-13944$\\$14089-14106$} & \makecell{$-1.3$-$-1.2$} \\
\hline
  \caption{$T_{c\bar{b}s\bar{b}}$ $(n, M^2)$ predictions.} \label{table:Tcsbb_n_predictions}
 \end{longtable}
        \end{center}
    \end{minipage}
    \end{center}

\subsubsection{\texorpdfstring{$T_{b\bar{b}c\bar{c}}$}{}}\label{T_bcbc}

\twographs[(a) is $(J, M^{2})$ $T_{b\bar{b}c\bar{c}}$ HMRT, (b) $(n, M^{2})$ $T_{b\bar{b}c\bar{c}}$ HMRT.][T_bcbc]{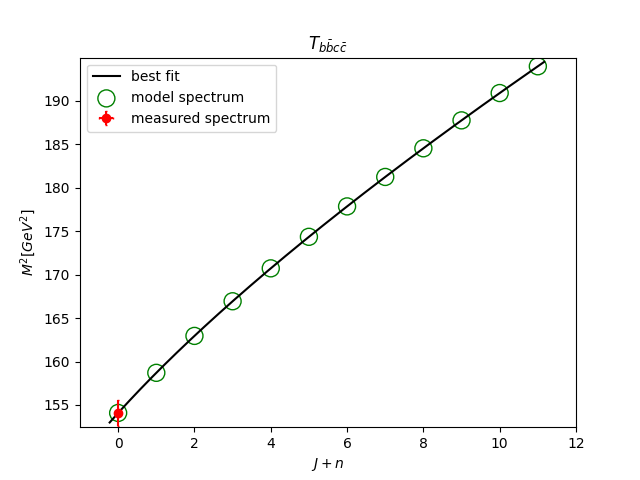}{tetraquarks/predictions/bc-bc/by_n/T_bcbc_one_by_n}

    \begin{center}
    \begin{minipage}{.49\linewidth}
        \begin{center}
            \begin{longtable}[c]{|c|c|}
  \hline
  \thead{$J$ \\ Spec} & \thead{$a_{T_{j}}$} \\
  \hline
  \hline
\makecell{$12393-12433$\\$12580-12616$\\$12749-12782$\\$12905-12936$\\$13052-13081$\\$13191-13218$\\$13323-13349$\\$13450-13475$\\$13572-13597$\\$13691-13714$\\$13805-13828$\\$13916-13939$} & \makecell{$-2.7$-$-2.5$} \\
\hline
  \caption{$T_{b\bar{b}c\bar{c}}$ $(J, M^2)$ predictions.} \label{table:Tbcbc_j_predictions}
 \end{longtable}
        \end{center}
    \end{minipage}
    \begin{minipage}{.49\linewidth}
        \begin{center}
            \begin{longtable}[c]{|c|c|}
  \hline
  \thead{$n$ \\ Spec} & \thead{$a_{T_{n}}$} \\
  \hline
  \hline
\makecell{$12393-12433$\\$12657-12691$\\$12888-12919$\\$13099-13127$\\$13294-13321$\\$13478-13503$\\$13652-13676$\\$13818-13841$\\$13977-13999$\\$14131-14152$\\$14279-14300$\\$14423-14443$} & \makecell{$-1.9$-$-1.7$} \\
\hline
  \caption{$T_{b\bar{b}c\bar{c}}$ $(n, M^2)$ predictions.} \label{table:Tbcbc_n_predictions}
 \end{longtable}
        \end{center}
    \end{minipage}
    \end{center}

\subsubsection{\texorpdfstring{$T_{c\bar{b}c\bar{b}}$}{}}\label{T_ccbb}

\twographs[(a) is $(J, M^{2})$ $T_{c\bar{b}c\bar{b}}$ HMRT, (b) $(n, M^{2})$ $T_{c\bar{b}c\bar{b}}$ HMRT.][T_ccbb]{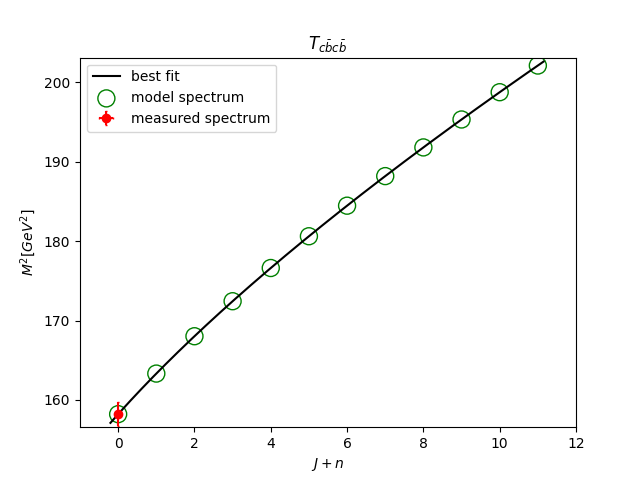}{tetraquarks/predictions/cc-bb/by_n/T_ccbb_one_by_n}

    \begin{center}
    \begin{minipage}{.49\linewidth}
        \begin{center}
            \begin{longtable}[c]{|c|c|}
  \hline
  \thead{$J$ \\ Spec} & \thead{$a_{T_{j}}$} \\
  \hline
  \hline
\makecell{$12558-12598$\\$12762-12798$\\$12946-12979$\\$13116-13147$\\$13276-13304$\\$13426-13453$\\$13569-13595$\\$13706-13731$\\$13838-13862$\\$13965-13988$\\$14087-14110$\\$14207-14228$} & \makecell{$-2.9$-$-2.7$} \\
\hline
  \caption{$T_{c\bar{b}c\bar{b}}$ $(J, M^2)$ predictions.} \label{table:Tccbb_j_predictions}
 \end{longtable}%
        \end{center}
    \end{minipage}
    \end{center}

\subsubsection{\texorpdfstring{$T_{b\bar{b}c\bar{b}}$}{}}\label{T_bcbb}

\twographs[(a) is $(J, M^{2})$ $T_{b\bar{b}c\bar{b}}$ HMRT, (b) $(n, M^{2})$ $T_{b\bar{b}c\bar{b}}$ HMRT.][T_bcbb]{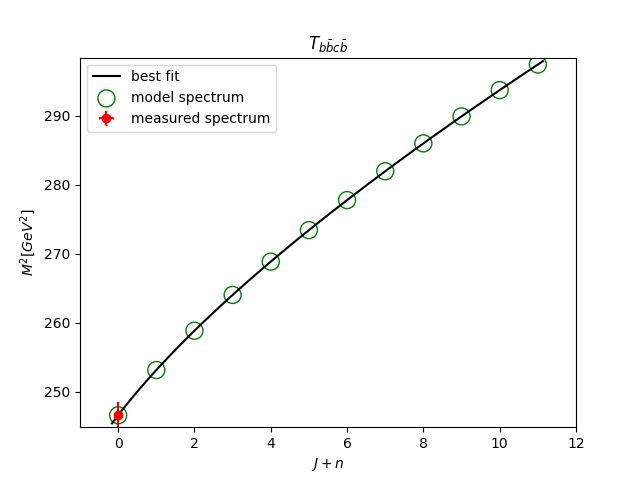}{tetraquarks/predictions/bc-bb/by_n/T_bcbb_one_by_n}

    \begin{center}
    \begin{minipage}{.49\linewidth}
        \begin{center}
            %
        \end{center}
    \end{minipage}
    \end{center}

\subsection{Pentaquarks Candidates}\label{section:pentaquarks_results}

\begin{center}
\small
\centering
%
\end{center}

\subsection{\texorpdfstring{$P_{\psi}^{N^{+}}$ (quark content $c\bar{c}uud$)}{}}

The pentaquark candidates were observed in \cite{P_ccuud_2015} and in \cite{Pc_2019}. The state $P_{c}(4380)$ requires further confirmation and is not included in the analysis of \cite{Pc_2019}. This state does not fit the trajectory. The decay channel in which these states were observed is $J/\psi p$ which is compatible with the annihilation mechanism.

\subsubsection{\texorpdfstring{$P_{\psi}^{N^{+}}$ baryonic configuration fit}{}}{

For the baryonic configuration, the fit that we got is displayed in figure \ref{fig:Pccuud_baryon_all}, alongside with the best fit. The spectrum for the best fit is in table \ref{table:Pccuud_baryon_spec}.

\begin{longtable}[c]{ m{8cm} m{8cm} }
\twographsintable{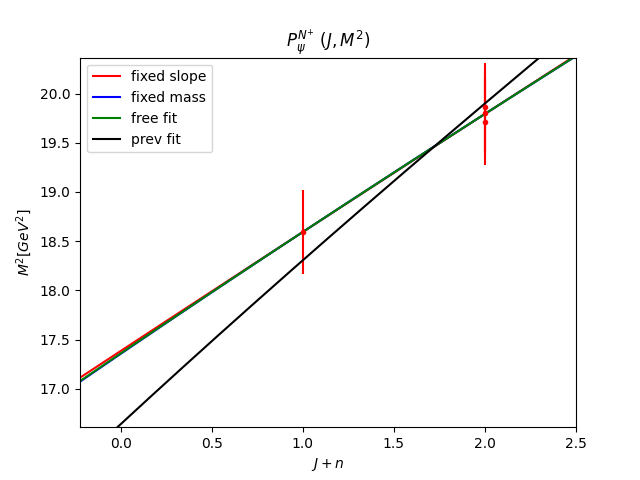}{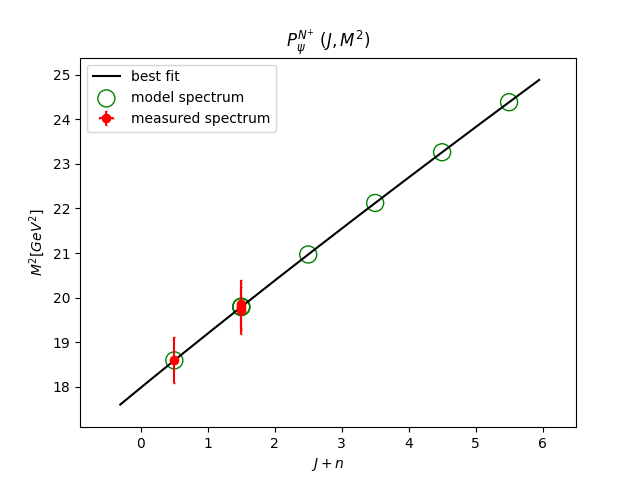}
\caption{On the left - $P_{\psi}^{N^{+}}$ all fits, on the right is the best fit.}\label{fig:Pccuud_baryon_all}
\end{longtable}

\begin{longtable}[c]{|c|c|c|c|c|c|}
  \hline
  \thead{$State$} & \thead{$M$ \\ $[MeV]$} & \thead{Thry \\ $M$ \\ $[MeV]$} & \thead{$J$} & \thead{Thry \\ $n$} & \thead{Thry \\ $J + n$} \\
  \hline
  \hline
$P(4312)^{N^{+}}_{\psi 0}$ & $4312$ & $4312$ & $0$ & $0$ & $0$ \\
\hline
$P(4440)^{N^{+}}_{\psi 0}$ & $4440$ & $4449$ & $0$ & $1$ & $2$ \\
\hline
$P(4450)^{N^{+}}_{\psi 0}$ & $4450$ & $4449$ & $0$ & $1$ & $2$ \\
\hline
$P(4457)^{N^{+}}_{\psi 0}$ & $4457$ & $4449$ & $0$ & $1$ & $2$ \\
\hline
 &  & $4579$ &  &  & $2$ \\
\hline
 &  & $4703$ &  &  & $4$ \\
\hline
 &  & $4823$ &  &  & $4$ \\
\hline
 &  & $4938$ &  &  & $6$ \\
\hline
  \caption{$P_{\psi}^{N^{+}}$ $(J, M^{2})$ HMRT.} \label{table:Pccuud_baryon_spec}
 \end{longtable}
}

\subsubsection{\texorpdfstring{$P_{\psi}^{N^{+}}$ tetraquark-like configuration fit}{}}{

For the tetraquark-like configuration, the fit that we got is displayed in figure \ref{fig:Pccuud_tetra_all}, alongside with the best fit. The spectrum for the best fit is in table \ref{table:Pccuud_tetra_spec}.

\begin{longtable}[c]{ m{8cm} m{8cm} }
\twographsintable{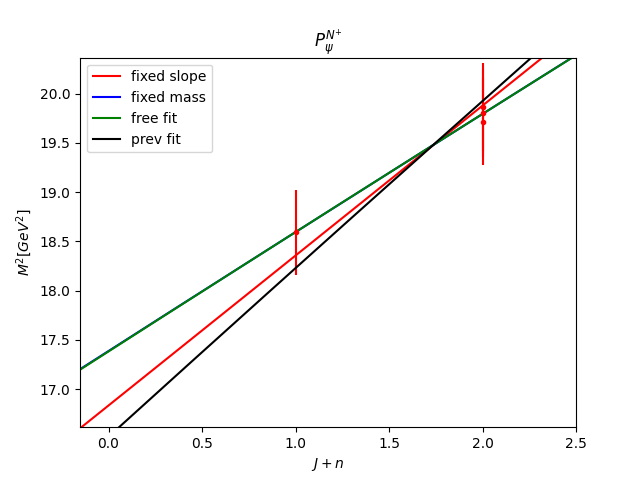}{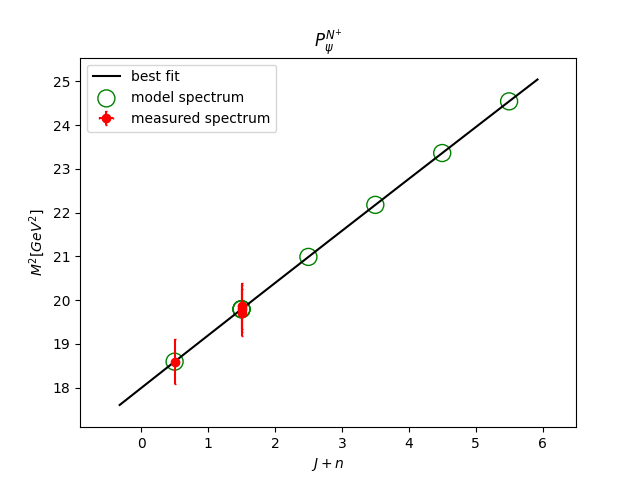}
\caption{On the left - $P_{\psi}^{N^{+}}$ all fits, on the right is the best fit.}\label{fig:Pccuud_tetra_all}
\end{longtable}

\begin{longtable}[c]{|c|c|c|c|c|c|}
  \hline
  \thead{$State$} & \thead{$M$ \\ $[MeV]$} & \thead{Thry \\ $M$ \\ $[MeV]$} & \thead{$J$} & \thead{Thry \\ $n$} & \thead{Thry \\ $J + n$} \\
  \hline
  \hline
$P(4312)^{N^{+}}_{\psi 0}$ & $4312$ & $4312$ & $0$ & $0$ & $0$ \\
\hline
$P(4440)^{N^{+}}_{\psi 0}$ & $4440$ & $4449$ & $0$ & $1$ & $2$ \\
\hline
$P(4450)^{N^{+}}_{\psi 0}$ & $4450$ & $4449$ & $0$ & $1$ & $2$ \\
\hline
$P(4457)^{N^{+}}_{\psi 0}$ & $4457$ & $4449$ & $0$ & $1$ & $2$ \\
\hline
 &  & $4581$ &  &  & $2$ \\
\hline
 &  & $4709$ &  &  & $4$ \\
\hline
 &  & $4833$ &  &  & $4$ \\
\hline
 &  & $4954$ &  &  & $6$ \\
\hline
  \caption{$P_{\psi}^{N^{+}}$ HMRT.} \label{table:Pccuud_tetra_spec}
 \end{longtable}
}

\subsection{\texorpdfstring{$P_{c\Bar{c}sud}$}{}}

This state was observed in \cite{PhysRevLett.131.031901}. In this seciton we built the trajectories for both baryonic and tetraquark configurations.

\subsubsection{\texorpdfstring{$P_{c\Bar{c}sud}$}{} baryon configuration HMRTs}

    \begin{center}
    \begin{minipage}{.49\linewidth}
        \begin{center}
            \begin{longtable}[c]{|c|c|c|c|c|c|}
  \hline
  \thead{$State$} & \thead{$M$ \\ $[MeV]$} & \thead{Thry \\ $M$ \\ $[MeV]$} & \thead{$J$} & \thead{Thry \\ $n$} & \thead{Thry \\ $J + n$} \\
  \hline
  \hline
$P(4338)^{0}_{\psi s0}$ & $4338$ & $4338$ &  &  & $0$ \\
\hline
 &  & $4482$ &  &  & $2$ \\
\hline
 &  & $4619$ &  &  & $2$ \\
\hline
 &  & $4748$ &  &  & $4$ \\
\hline
 &  & $4873$ &  &  & $4$ \\
\hline
  \caption{$P^{\Lambda}_{\psi s} (J, M^{2})$ HMRT.} \label{table:Pccuds_baryon_J_HMRT}
 \end{longtable}
        \end{center}
    \end{minipage}
    \begin{minipage}{.49\linewidth}
        \begin{center}
            \begin{longtable}[c]{|c|c|c|c|c|c|}
  \hline
  \thead{$State$} & \thead{$M$ \\ $[MeV]$} & \thead{Thry \\ $M$ \\ $[MeV]$} & \thead{$J$} & \thead{Thry \\ $n$} & \thead{Thry \\ $J + n$} \\
  \hline
  \hline
$P(4338)^{0}_{\psi s0}$ & $4338$ & $4338$ &  &  & $0$ \\
\hline
 &  & $4496$ &  &  & $2$ \\
\hline
 &  & $4645$ &  &  & $2$ \\
\hline
 &  & $4787$ &  &  & $4$ \\
\hline
 &  & $4922$ &  &  & $4$ \\
\hline
  \caption{$P^{\Lambda}_{\psi s} (n, M^{2})$ HMRT.} \label{table:Pccuds_baryon_n_HMRT}
 \end{longtable}
        \end{center}
    \end{minipage}
    \end{center}

\begin{longtable}[c]{ m{8cm} m{8cm} }
\twographsintable{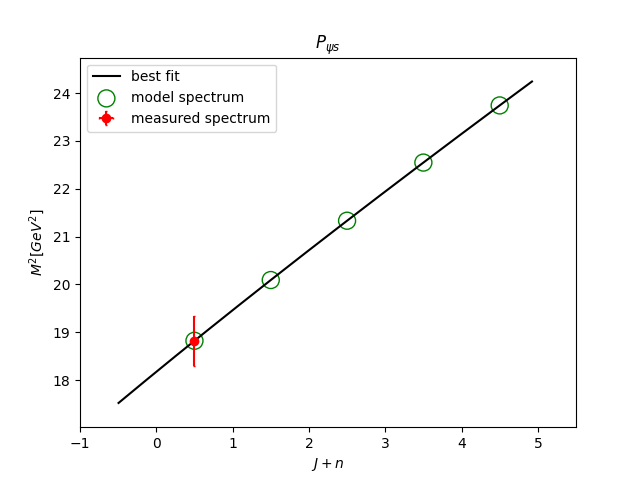}{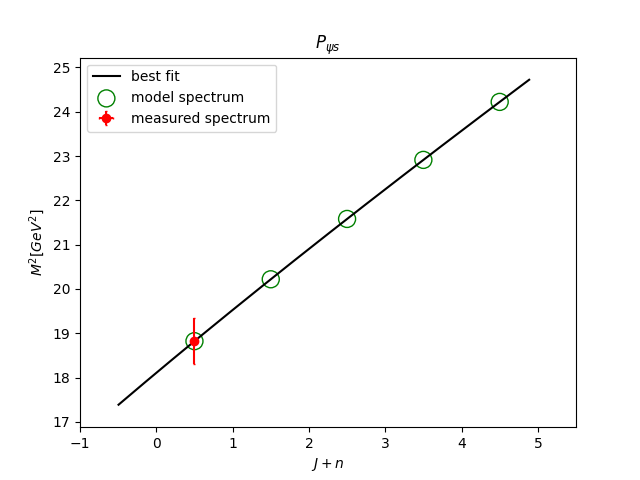}
\end{longtable}

\subsubsection{\texorpdfstring{$P_{c\Bar{c}sud}$}{} tetra configuration fit}

    \begin{center}
    \begin{minipage}{.49\linewidth}
        \begin{center}
            \begin{longtable}[c]{|c|c|c|c|c|c|}
  \hline
  \thead{$State$} & \thead{$M$ \\ $[MeV]$} & \thead{Thry \\ $M$ \\ $[MeV]$} & \thead{$J$} & \thead{Thry \\ $n$} & \thead{Thry \\ $J + n$} \\
  \hline
  \hline
$P(4338)^{0}_{\psi s0}$ & $4338$ & $4338$ &  &  & $0$ \\
\hline
 &  & $4532$ &  &  & $2$ \\
\hline
 &  & $4716$ &  &  & $2$ \\
\hline
 &  & $4892$ &  &  & $4$ \\
\hline
 &  & $5061$ &  &  & $4$ \\
\hline
  \caption{$P^{\Lambda}_{\psi s} (J, M^{2})$ HMRT.} \label{table:Pccuds_tetra_J_HMRT}
 \end{longtable}
        \end{center}
    \end{minipage}
    \begin{minipage}{.49\linewidth}
        \begin{center}
            \begin{longtable}[c]{|c|c|c|c|c|c|}
  \hline
  \thead{$State$} & \thead{$M$ \\ $[MeV]$} & \thead{Thry \\ $M$ \\ $[MeV]$} & \thead{$J$} & \thead{Thry \\ $n$} & \thead{Thry \\ $J + n$} \\
  \hline
  \hline
$P(4338)^{0}_{\psi s0}$ & $4338$ & $4338$ &  &  & $0$ \\
\hline
 &  & $4614$ &  &  & $2$ \\
\hline
 &  & $4872$ &  &  & $2$ \\
\hline
 &  & $5116$ &  &  & $4$ \\
\hline
 &  & $5346$ &  &  & $4$ \\
\hline
  \caption{$P^{\Lambda}_{\psi s} (n, M^{2})$ HMRT.} \label{table:Pccuds_tetra_n_HMRT}
 \end{longtable}
        \end{center}
    \end{minipage}
    \end{center}

\begin{longtable}[c]{ m{8cm} m{8cm} }
\twographsintable{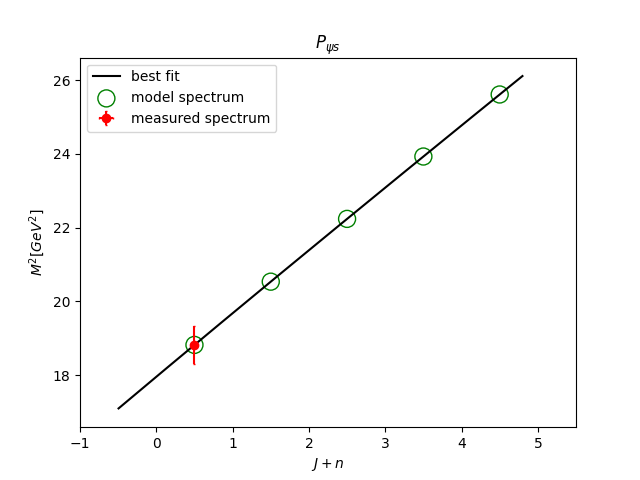}{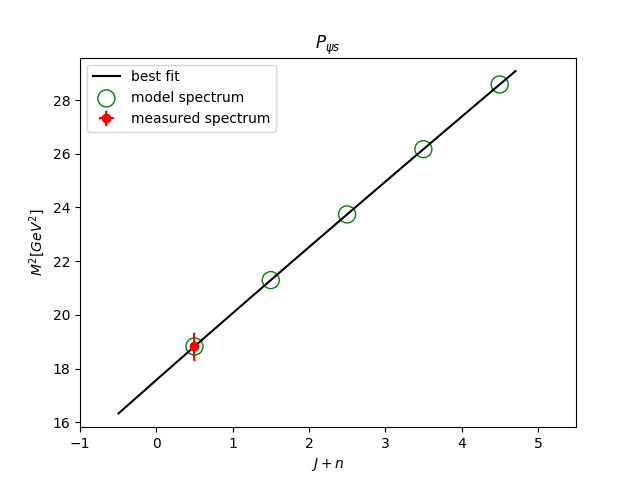}
\end{longtable}

\subsection{Pentaquarks HMRTs Predictions}\label{section:pentaquarks_predictions}
The following table summarizes all the thresholds and predicted trajectories for all possible charm pentaquarks configurations. The thresholds were calculated for each possible product according to the HISH model through the annihilation mechanisms. This is because in the case of pentaquarks, there is not yet any obsereved state that matches the string-tear mechanism.

The 'Candidates' column contains the resonances that were already detected. We elaborate on all of these at the following sub sections, including the decay channels through which they were observed.

The table spectrum predictions were calculated based on the mesons and baryons fit parameters. We assumed that the endpoint masses are the sum of masses of the quarks (or antiquarks). We calculated four spectrum ranges - the first two ('J Spec Baryon' and 'n Spec Baryon') are based on the pentaquarks baryon configuration as described in \ref{fig:pentaquark_simple}. For these spectra we used the baryons charm slopes. For the second two ('J Spec Tetra' and 'n Spec Tetra') are based on the tetraquark-like configuration as described in \ref{fig:pentaquark_simple2}. The slopes for the latter are the mesonic slopes for heavy mesons for each plane as summarized at \ref{table:mesons_global_fit}.

We did assume that the first \ref{fig:pentaquark_simple} configuration is allowed only when there is quark that is of the same flavor as the antiquark. Therefore the 'baryonic' spectrum is not calculated for all pentaquarks.

We calculated the 'J Spec (Baryon)' and 'n Spec (Baryon)' in the following steps:

\begin{itemize}
    \item We first calculated the annihilation threshold and let it vary between $175-275MeV$ above the threshold. This was done by looking at the available data that we collected, where the created pentaquarks candidates are above the annihilation threshold.
    \item We assumed that this is the first resonance on the trajectory, so we set $J + n = 0$.
    \item Since we do not know the arrangement of the quarks within the endpoints, we went over all the possible arrangement.
    \item For each arrangement, we calculated the endpoint masses and deduced the spectrum using the baryons slope of the plane ($J$ or $n$).
    \item The final spectrum range for each plane was taken to be between the minimum and maximum of all the possible arrangements.
    \item We derived the range of the spectrum for four more states that were above the lowest mass.
\end{itemize}

We calculated the 'J Spec (Tetra)' and 'n Spec (Tetra)' in the following steps:

\begin{itemize}
    \item We first calculated the annihilation threshold and let it vary between $175-275MeV$ above the threshold. This was done by looking at the available data that we collected, where the created pentaquarks candidates are above the annihilation threshold.
    \item We assumed that this is the first resonance on the trajectory, so we set $J + n = 0$.
    \item Since we do not know the arrangement of the quarks within the endpoints, we went over all the possible arrangement.
    \item We assumed that the antiquark is always around the center of mass, so it does not contribute to the $J+n$ (its $\beta$ is close to $0$). This means that it can be treated as part of the energy of the string itself, and not as part of the endpoint masses.
    \item For each arrangement, we calculated the endpoint masses and deduced the spectrum using the heavy mesons slope of the plane ($J$ or $n$).
    \item The final spectrum range for each plane was taken to be between the minimum and maximum of all the possible arrangements.
    \item We derived the range of the spectrum for four more states that were above the lowest mass.
\end{itemize}

The 'Genuine' column indicates whether the quark content is a pure pentaquark or not. This was decided as follows:

\begin{itemize}
    \item If the content includes a light quark ($u$, $d$, $s$) and its counterpart, we assume its source could be a meson that decay through string tear. This is marked at the table as an empty cell. 
    \item If the content include a heavy quark ($c$, $b$) and its counterpart, which is less likely to be a product of meson string break, it is likely to be genuine. This was marked with one $*$.
    \item $**$ means it is a pure pentaquark. These does not contain quark-antiquark pair of the same flavour.
\end{itemize} 

Empty spectrum/threshold, means these baryons weren't found yet, therefore we couldn't predict the HMRTs.

\setlength\LTleft{-0.7cm}
\setlength\LTpre{0.5cm}
{\scriptsize{
}\normalsize}

\section{Summary and Conclusions}
In this paper we use the HISH model in order to tam the zoo of exotic hadrons in particular tetraquarks and pentaquarks. We enlist 71 candidates of tetraquark states and 210  candidates of pentaquarks. For  a great part of them we determine the corresponding HISH modeified Regge trajectory. Out of the 71 tetraquark states we compute a trajectory with 5 states, thus predicting 240 new tetraquarks. We added another additional 5 states for genuine and semi-genuine tetraquarks, with another ~350 states. For 160 candidate states of the pentaquarks we determine for each 5 states on the J trajectory and 5 in the n trajectory thus predicting altogether 1600 new pentaquark states. 

The basic structure of the stringy mesons occurs also  for the stringy tetraquarks. However, whereas for the mesons the predominantly decay mechanism is the breakup of the string into two strings, namely a transition from a meson to two mesons, for the tetraquarks there is an additional mechanism of the annihilation of a BV and an anti-BV. In the latter case the result of the decay of the tetraquark are two mesons, in the former the output is a baryon and an anti-baryon. Thus, for the tetraquarks, unlike for the mesons, the states on a trajecory are divided into two classes the lower mass one where the decay is  via annihilation into two mesons and the higher mass one where in addition there is the mechanism of string breaking.  In section (\ref{Decay mechanisms of tetraquarks}) we have mentioned another mechanism that involves an annihilation of a quark from the di-quark and an anti-quark from the anti-diquark. This is the analog of a similar mechanism that takes place in the Zweig suppressed decay of heavy quarkonia \cite{Sonnenschein_decay_width_2018}. 

The stringy structure of baryons is based on a single string connecting a quark to a BV which is attached to a di-qaurk. On the other hand for the pentaquarks there are two possible structures: one  similar to the baryon with a single BV and the other with two strings connecting an anti-BV in the center to two BVs on both sides. These two structures provide a large number of options of the spectra and of the decay processes.

Our procedure of determining the HMRTs is based on the assumption that the first tetraquark state that can decay into a baryon and an anti-baryon is of a mass that is higher than that of the baryonic threshold by 40-150 MEV. This is based on the few cases with such a decay  that have been already discovered. Using this assumption we can determine the approximated value of the intercept.  It is clear that once there will be more such states found, we will be able to improve the determination of the corresponding intercept and slope and hence improvement of  the whole trajectory. 

According to our survey  there are of the order of thousand candidates of tetraquarks and pentaquarks  This obviously calls for an intensive effort to experimentally identify these resonance states. This should be done in the running experiments and may also call  for a new experiment designed specially for the exotic hadrons. This can obviously be of much lower energy than the one used in the LHC. 

This was a data driven project, were we tried to unify all the available exotic candidates under the HISH framework in order to systematically predict tetraquark and pentaquark modified Regge trajectories. The overall result allows us to provide a "map" that locates exotic states energy-wise, guiding future searches for exotic candidates. As more data will be accumulated, we will be able to improve our model predictions.
The assumptions that tetraquarks $\alpha^{'}$ is similar to that of mesons, was in general good as can be seen from the global tetraquarks-mesons fit \ref{section:tetra_mesons_global}. The slope of $T_{c\bar{c}c\bar{c}}$ was the only one that didn't fit this assumption, and we fit it separately. This requires further research.
Another interesting outcome, is that the predictions of the HMRTs of tetraquarks that decay to baryons that weren't found yet could potentially allow us to predict where to search for these baryons. This is also a result that should be analysed in future work.
One important results from past papers \cite{Sonnenschein_2014_mesons}, \cite{Sonnenschein_2014_baryons} is that the excitation in the two planes $J$ and $n$ are different. This implies multiple HMRTs for each state and could explain resonances that are close in energy, but doesn't rely exactly on the same trajectory.
The results for the pentaquarks candidates were not far from the baryonic trajectories as we expected, but they remain inconclusive since there isn't enough data to determine if they rely on a trajectory. Another observation to note is that the pentaquarks configurations, as baryons, are ambiguous in their internal division (in which quarks occupy each of the endpoints). The multiple resonances observed with content $c\bar{c}uud$ may be explained as a mix of some of these configurations.

\section{Open Questions}\label{section:open_questions}
Applying the HISH model to the study of exotic hadrons and in particular tetraquarks and pentaquarks is still in a preliminary phase and there are  many open questions related to it.  Here we enumerate several such questions.
\begin{itemize}
\item  Probably the most important question is whether the stringy description of exotic hadrons   is the ``right" way  to view them and not  just as a bound states  of elementary particles. Of course the string picture includes particles at the endpoints of the strings but it is only with the strings that one can get a faithful description of the spectra and decay widths.
\item 
The HISH model, both for ordinary hadrons, as well as for exotic ones, lacks a theoretical account of the intercept, namely, the Casimir energy. As was explained in section 2 hadron phenomenology implies a negative $\tilde a =a -S$, which provides a repulsion that  balances the string tension for non-rotating strings. Ordinary bosonic string admits a positive intercept.
The theory of  QCD strings,namly, flux tubes has been only partially revealed.  One may hope that acquiring more data about the experimental values of the intercepts will be useful in determining the full quantum  theory of the stringy hadrons including  controlling the repulsive Casimir forces.
\item 
Whereas for the breaking of the string decay process we can determine the decay width using a string calculation, for the annihilation process we do not  have yet an exact procedure which  provides such a calculation and so far  we have been using a ``crude approximation" of it.
\item 
In fact the zoo of exotic hadrons is divided into two classes: genuine exotic hadrons and molecules. In the stringy description the former are described by connected stringy diagrams whereas the latter  by disconnected diagrams. The simplest molecules, in the HISH picture take the form of two disconnected strings.
In this paper we have discussed only genuine exotic states. In our future research we intend to address the description of molecules in the HISH framework in particular the analog of the Van der Waals forces that bind separate  "atom strings" into molecules. 
\item 
One of the most interesting questions is: could there exist in nature a stable (against strong decays) exotic hadron. This means,  for instance for a tetraquark, with mass that  is lower than that of the two possible lightest daughter mesons. In \cite{Karliner:2017qjm} it was argued that  the  tetraquark $(b, b, \bar u, \bar d)$ should be stable. This statement got a support from lattice \cite{Eichten:2017ffp} and large $N_{c}$ \cite{Czarnecki:2017vco} calculations. In this paper we have investigated charmed tetraquark and hence did not address this particular state. 
In the future, with a better insight on the intercepts, one should in principle be able to determine whether, and in what conditions, can a stable  exotic charmed hadron exist.
\item The HISH model that we have utilized is a simplified HISH model where we have not taken into account explicitly  the  electric charges\cite{Sonnenschein:2019bca} and spins \cite{Berezin:1976eg} and  \cite{Pisarski:1987jb} of the endpoint particles. These properties, in particular the spin, play an important role in the determinations of the trajectories of the exotic hadrons. In the  present work we used the data about the intercepts, which depend on these properties, from the experimental results about non-exotic hadrons.  The structure of the di-quarks attached to the BVs  where not play any significant role apart from its total mass. The simplified model that we used  should be improved by  taking into account the latter properties.
\item 
For the pentaquarks unlike the tetraquarks there are two possible structures. One based on a single BV and the other with an anti-BV and two BVs. It will be interesting, based on theoretical calculations or on observational data to get a better understanding of who out or the two options is preferable. Distinguishing between the two options will be relevant for the constructions of  stringy states with higher baryon number. This question was discussed recently  in \cite{Andreev:2023hmh}.
\item 
At present exotic hadrons have been discovered in the LHC and in low energy charm and bottom factories.  It is very probable that many more exotic hadrons will be discovered if special purpose experiments  are constructed for that mission. For instance the breaking process of tetraquarks producing a baryon anti-baryon pairs, may hint  that a proton anti-proton machine of low energies or  may be adequate for that purpose. 
\item 
We have mentioned in section (\ref{Exotic hadrons}) that there may exist hadrons which are  even more exotic than tetraquarks and pentaquarks. These include exotic glueballs, hadrons of higher number of quarks like heptaquarks etc and string states of hadrons with baryon number $B>1$ like the sexaquarks \cite{Farrar:2017eqq}.
\item 
The phenomena of  hadronization of quark gluon fluid, for instance in heavy ion collisions or in the evolution of the early universe incorporate productions of hadrons in the form of glueballs and hadrons. If the description of hadrons in terms of strings, in particular in the HISH model. there should be also exotic hadrons like tetraquarks and pentaquarks coming out of the quark gluon fluid. It will be interesting to search for them in these experiments and to determine their abundance in comparison with ordinary hadrons
\item 
A very important aspect of elementary particle physics  that the HISH model has  not been  capable yet  to account for is the electro-weak interaction. This  is a drawback of the model in the context of ordinary hadrons as well as for exotic ones. They challenge is use the holographic setup  of multiflavor QCD-like theories and deduce from it the way the HISH model has to be modified to accommodate  also the weak interactions. Related to his problem is the  incorporation of leptons in the stringy picture of hadrons.
\end{itemize}

\section{Acknowledgment}
We would like to thank O. Andreev, S. Dubovsky, G. Farrar, M. Karliner,  M. Peskin and especially  D. Weissman, for useful discussions. We also thank O. Andreev, S. Nussinov, A. Sofer, D. Weissman and specially M. Karliner  for  their comments on the manuscript.
This work  was supported by a
center of excellence of the Israel Science Foundation (grant number 2289/18).
J.S would like to thank the c.c.p.p of NYU and the Simons Center at Stony Brook,  where part of this work was done, for support. 
    

\bibliographystyle{unsrt} 
\bibliography{References} 


\appendix
\section{Mesons Fits}\label{appendix:mesons_global_fit}

In this section we detail the global fits that were done to mesons in order to use the results for predicting tetraquarks spectrum and widths as detailed in \ref{fitting_models}. The states that were used for the fits are detailed in table \ref{table:mesons_states_J}, which are the states for $(J, M^{2})$ plane. Table \ref{table:mesons_states_n} contain the states used for the fits in the $(n, M^{2})$ plane\footnote{\label{excluded_states}Some of the states in the trajectories above were omitted, see \cite{Sonnenschein_2014_mesons} A.1 for full explanation.}.

As explained in \ref{mesons_confronting_data}, for the tetraquarks predictions, we did a global for each of the planes - $(J, M^{2})$, $(n, M^{2})$ and once for both. In the first two, we constrained the endpoint masses according to the quarks content (e.g. $m_{s}$ was forced as $m_{1}$ and $m_{2}$ of $\phi/f^{'}$ as well as $m_{1}$ of $K^{*}$). We also constrained $\alpha^{'}$ of all light mesons to be the same, as well as the heavy mesons. In the common global fit, we constrained only the endpoint masses, but allowed the $\alpha^{'}$ parameters to differ between the two planes, yet preserving the same light/heavy separation for each.

The degree-of-conformity of each global fit, as well as the parameters the fits emitted, are in table \ref{table:DOC_params_mesons_global}. It can be seen that the global fit used didn't change most of the parameters significantly, nor worsen $\chi^{2}_{r}$ significantly with respect to $(n, M^{2})$ fit. The only parameter that changed significantly is $m_{u/d}$, but its mass wasn't strict in the first place as discussed in \cite{Sonnenschein_2014_mesons}.

For the global fit for both planes,
we performed the fit once with the light/heavy slopes
constrain and once without it. The parameters emitted
almost didn't change. The results are presented in
\ref{table:DOC_params_mesons_global}.

\caption{}\label{table:mesons_J_universal_fit}
\end{center} 
\end{table}

\begin{table}[!ht]
\begin{center}
\begin{tabular}{|c|c|c|c|c|c|c|c|c|}
\hline
\makecell{$a_{\pi/b}$} & \makecell{$a_{\rho/a}$} & \makecell{$a_{\eta/h}$} & \makecell{$a_{\omega/f}$} & \makecell{$a_{K^{*}}$} & \makecell{$a_{\phi/f^{'}}$} & \makecell{$a_{D}$} & \makecell{$a_{D^{*}_{s}}$} & \makecell{$a_{\Psi}$}\\
\hline
\hline
$-0.33$&$0.52$&$-0.22$&$0.53$&$-0.33$&$0.52$&$-0.22$&$0.53$&$-0.06$\\
\hline
\end{tabular}\caption{}\label{table:mesons_J_universal_intercepts}
\end{center}
\end{table}

\begin{center}
%
\end{center}

\section{Baryons Fits}\label{appendix:baryons_global_fit}

In this section we summarize the results from previous papers \cite{Sonnenschein_2014_baryons} and \cite{Sonnenschein_2019_predictions} for the HISH model parameters. We used these for the predictions and fits when \hyperlink{section:pentaquarks_results}{confronting the pentaquarks data}.

\begin{center}
%
\end{center}

\subsection{\texorpdfstring{Baryons $(n, M^{2})$ Global Fit.}{}}\label{appendix:baryons_n_global_fit}

The fit procedure for the results of this section are at \ref{section:baryons_n_global_fit_procedure}. Table \ref{table:baryons_global_fit_n} summarizes the output parameters for the results in this section. The $\chi^{2}_{r}$ value is $2.32$ which is a good result for the accuracy that we expect from this model. The fits are displayed in \ref{fig:baryon_n_global_fit}, and the states used are detailed in \ref{table:baryons_n_states}.

%

\end{document}